\documentstyle[preprint,floats,prd,aps]{revtex}

\def\PRref#1&#2&#3(#4){\unskip\ #1~\bf #2\rm, #3 (#4)}

\def\NIMA{Nucl. Inst. and Meth. A}
\def\NPB{Nucl. Phys. B}
\def\PLB{Phys. Lett. B}
\def\PRL{Phys. Rev. Lett.}
\def\PRD{Phys. Rev. D}
\def\ZPC{Z. Phys. C}
\def\etal{{\it et al.}}
\def\psfig#1{} %disabled by xxx
\begin{document}
\tighten

\preprint{\vbox{\hbox{UH 511-816-95 \hfill}
                \hbox{OHSTPY-HEP-E-95-010 \hfill}
                \hbox{\today        \hfill}}}

\title{B Mesons}

\author{ Thomas E. Browder}
\address{University of Hawaii at Manoa, Honolulu, Hawaii 96822}
\author{Klaus Honscheid}
\address{Ohio State University, Columbus, Ohio 43210}

\bigskip
\bigskip
\bigskip

\maketitle

\vfill

\begin{abstract}
$B$ mesons are bound states of a $b$ quark and a light anti-quark. While the
binding is provided by the strong interaction $B$ mesons can only decay
by the weak interaction.
Since the top quark mass is large, $B$ mesons are  the
only mesons containing quarks of the third generation and thus
their decays provide a unique opportunity to measure the
Cabibbo-Kobayashi-Maskawa (CKM) matrix elements $V_{cb}$, $V_{ub}$, $V_{ts}$,
and $V_{td}$ which describe the couplings of the third generation of quarks to
the lighter quarks.

We review experimental results on masses, lifetimes
 and decays of
$B$ mesons. These include
measurements of the inclusive production of charmed and non-charmed mesons and
baryons, observations of semileptonic $B$ decays, $B-\bar{B}$ mixing, $B$ meson
lifetimes,
measurements of exclusive hadronic final
states with charmed mesons, the search for exclusive hadronic
final states without charmed mesons, the first observation of the
decay $B\to K^*\gamma$ which is described by an electromagnetic penguin
diagram and measurement of the inclusive $b\to s \gamma$ rate.
The theoretical implications of these results will be considered.
\end{abstract}

\vspace{0.5cm}

\newpage

\tableofcontents

\newpage
%%%%%%\section{INTRODUCTION}
\section{INTRODUCTION}

\label{intro}
One intriguing puzzle in physics is the regular pattern
of the three
fermion and quark families. The existence of families gives rise to many of the
free parameters of the Standard Model,
in particular the fermion masses and the
elements of the  Cabibbo-Kobayashi-Maskawa matrix (CKM) \cite{KM}
that describe the
mixing between the quark generations. The determination of all of
these parameters is required to fully define the Standard Model and
may also reveal an underlying structure that will point to new physics.
In the Standard Model of three
generations the CKM matrix is defined by three real  parameters and one complex
phase. It relates the eigenstates of the strong and
weak interactions and can be written
 \begin{equation}
  V =\pmatrix{V_{ud}&V_{us}&V_{ub}\cr
                V_{cd}&V_{cs}&V_{cb}\cr
                V_{td}&V_{ts}&V_{tb}\cr}
  \end{equation}
The matrix $V$ can be expressed approximately as
   \begin{equation}
   \simeq \pmatrix{1-\lambda^2/2&\lambda&A\lambda^3(\rho-i\eta)\cr
                      -\lambda&1-\lambda^2/2&A\lambda^2\cr
                      A\lambda^3(1-\rho-i\eta)&-A\lambda^2&1\cr}+
                      O(\lambda^4)
   \end{equation}
This empirical parameterization, suggested by Wolfenstein\cite{Wolfie},
is correct to terms of order $\lambda^4$ with
$\lambda = \sin{\theta_{Cabibbo}}\approx 0.22$.
In the case of the two generations, the matrix $V$ is a simple
rotation matrix where $\theta_{Cabibbo}$ is the angle of rotation.
For three generations, $V$ may contain complex elements and allows
for CP violation if the parameter $\eta$ is non-zero.
Although readily accommodated in the Standard
Model, CP violation remains one of the least well
understood phenomena in physics.
So far it has only been observed in the decays of kaons.
While the results from the kaon sector are consistent with
the Standard Model, the complications introduced by
strong interaction effects make it nearly impossible to
ascertain whether the complex CKM phase is the sole source
for the observed asymmetries.

The only other observational constraint
 on CP violation comes from cosmology. As was first
noted by Sakharov, there is an important connection between
the observed baryon asymmetry in the universe
and CP violation in fundamental
processes \cite{Sakharov}.
He postulated that CP violation in fundamental processes
in the early universe, C and baryon number violation,
and the absence of thermal equilibrium gave rise to the observed
baryon asymmetry.
However, recent work suggests that the Standard Model and
the complex phase in the CKM
matrix cannot provide sufficient CP violation to account for the magnitude
of the baryon asymmetry so that other sources of CP violation
must be present\cite{Dine}.
An experimental effort to determine the magnitude of the CKM matrix elements
and to measure the CP violating phase is therefore of
fundamental importance.

$B$ meson decay provides an ideal opportunity to pursue such a program.
Since the dominant $B$ meson decay mechanisms involve generation changing
transitions which are suppressed by the small CKM matrix element $V_{cb}$,
rare processes such as $b \to s$, $b \to u$, and $b \to d$
transitions are expected to be observable.
Several of these $B$ decay mechanisms
 are shown in Figure \ref{Fdiag}. Measurements of
$B$ meson decay rates
are used to determine the couplings between quarks of the third
generation and lighter quarks,
the CKM  elements $|V_{cb}|$, $|V_{ub}|$, $|V_{ts}|$, and
$|V_{td}|$. In addition,
the Standard Model predicts large CP asymmetries for the $B$ system.
Experiments with $B$ mesons may lead to the first precise determination of
the complex CKM phase.

In the framework of the Standard Model
the CKM matrix must be unitary, {\it ie.}
$VV^{\dagger}\; = \; 1$.
This gives rise to the following relationships between the
matrix elements.
$$ V_{ud}^{*} V_{us} + V_{cd}^* V_{cs} + V_{td}^{*} V_{ts} = 0$$
$$ V_{us}^{*} V_{ub} + V_{cs}^* V_{cb} + V_{ts}^{*} V_{tb} = 0$$
$$ V_{ub}^{*} V_{ud} + V_{cb}^* V_{cd} + V_{tb}^{*} V_{td} = 0$$
Chau, Keung \cite{ChauK} and Bjorken have noted that the last
equation can be visualized as a triangle in the complex plane
with vertices at $(0,0)$, $(0,1)$ and $(\rho, \eta)$.
% Measurements of the CKM matrix elements which
%involve b quarks constrain the allowed range of CP violation
%in the $\rho - \eta$ plane.
Measurements of the magnitudes of the CKM elements determine
the lengths of the sides of the triangle, while measurements
of CP asymmetries determine the interior angles of the triangle.

In recent years,
there have been major advances in our understanding of $B$
meson decay. However, data samples at least one order of magnitude larger
than those available at present are required to observe
CP violating asymmetries in the $B$ meson system and to provide
fundamental consistency checks of the Standard Model.
This is the justification for the
construction of high luminosity $e^+ e^-$
storage rings in the US at SLAC(PEP~II/BABAR)\cite{BABAR}, at Cornell
(CESR PHASE~III, CLEO~III)\cite{PHASEIII}, and in
Japan(TRISTAN-II/BELLE)\cite{BELLE},
as well as a dedicated fixed target experiment
at the HERA ring at DESY\cite{DESY}, and
proposals for hadron collider experiments
at Fermilab\cite{CDFB} and at CERN\cite{LHC}.

In this paper we will review the current status of experimental
$B$ physics and then briefly discuss CP violation.
Most of our present knowledge on $B$ mesons comes from experiments
performed on the $\Upsilon (4S)$ resonance at a center of mass energy of
10.58 GeV and in recent times from the large data sample of about
3 fb$^{-1}$ that has been collected by the CLEO II
collaboration at the Cornell Electron Storage Ring (CESR).
Older results from the ARGUS experiment, which
operated at the DORIS storage ring
and from the CLEO~1.5 experiment, which preceded the CLEO~II detector are
also included.
We note that the LEP experiments and the CDF experiment at the Tevatron
Collider have recently provided precise measurements of B meson lifetimes.
The LEP experiments have also
directly observed the time dependence of
$B_{d}-\bar{B_d}$ mixing and set limits on the $B_s-\bar{B_s}$ mixing.
They have also
observed exclusive hadronic decays of both $B$ and $B_s$ mesons.

\begin{figure}[htb]
\begin{center}
\unitlength 1.0in
\begin{picture}(5.,5.)(0,0)
\put(1.01,0.1){\psfig{width=3.5in,height=5.0in,%
file=bdecays.eps}}
\end{picture}
\bigskip
\vskip 3 mm
\caption{$B$ meson decay mechanisms: (a)
external spectator diagram and (b) color suppressed spectator diagram
(c) $b\to u$ spectator diagram (d) $b\to s\gamma$ electromagnetic
penguin (e) W-exchange diagram (f) W-annihilation diagram
and (g),(h) box diagrams for $B-\bar{B}$ mixing.}
\label{Fdiag}
\end{center}
\end{figure}

\subsection{Hadronic and Semileptonic $B$ Decays}

$B$ meson decays occur primarily
through the CKM favored $b\to c$ transition.
In such decays
the dominant weak decay diagram is the spectator diagram, shown in
Fig.~\ref{Fdiag}(a), where the virtual $W^-$ materializes into either a
lepton and anti-neutrino or a
$\bar u d$ or $\bar c s$ quark
pair.
In hadronic decays, the quark pair becomes one of the final state
hadrons while the $c$ quark pairs with the spectator anti-quark to form the
other hadron, while in
semileptonic decays, the c quark and spectator antiquark
hadronize independently of the leptonic current.

The extraction of Standard Model parameters from experimental results
is complicated by the fact that only $B$ mesons can be studied and not
free $b$ quarks. The light quarks and the gluons surrounding the $b$
quark in the $B$ meson lead to significant corrections that have
to be taken into account.
Since leptons do not interact strongly, semileptonic $B$ meson decays are
less affected by these QCD corrections and the theoretical calculations
are believed to be more reliable.
We discuss several measurements of both
inclusive and exclusive semileptonic decay rates that
have been used to determine
the strength of the $b \to c$ coupling, {\it ie.} $|V_{cb}|$.

In hadronic decays,
the spectator diagram is modified by hard gluon exchanges
between the initial and final quark lines.
This leads to the ``color suppressed'' diagram shown in Fig.~\ref{Fdiag}(b),
which has a different set of quark pairings.
Observation of $B \to \psi X_s$ decays, where $X_s$ is a strange meson, gives
experimental evidence for the presence of this diagram.
Further information on the size of the color suppressed contribution can be
obtained from $\bar{B^0} \to D^0$ (or $D^{*0} ) X^0$ transitions,
where $X^0$ is a neutral meson. In $B^-$ decays, both types of
diagrams are present and can interfere.
By comparing the rates for $B^-$ and $\bar{B^0}$ decays,
the size and the relative sign
of the color suppressed amplitude can be determined.

It was suggested that in
analogy to semileptonic decays, two body
hadronic decays of $B$ mesons can be expressed
as the product of two independent hadronic currents, one
describing the formation of a charm meson and the other the hadronization of
the $\bar{u} d$ (or $\bar c s$) system from the virtual $W^-$\cite{early}.
Qualitatively, for a $B$ decay with a large energy release the
$\bar{u} d$ pair, which is produced as a color singlet, travels fast enough to
leave the interaction region without influencing the
second hadron formed from the $c$ quark and the spectator anti-quark.
The  assumption that the amplitude can be expressed
as the product of two hadronic currents is called ``factorization'' in this
paper. It is expected that the simple approximation of the strong interaction
effects by the factorization hypothesis will be more reliable in $B$ meson
decays than in the equivalent $D$ meson decays due to the larger
characteristic energy transfers and the consequent suppression
of final state interactions.
We will discuss several tests of the factorization hypothesis based on
the comparison of semileptonic and hadronic $B$ meson decays\cite{Bjorken}.

\subsection{Rare $B$ Decays}

All $B$ meson decays that do not occur through the usual $b\to c$ transition
are known as rare $B$ decays. The simplest diagram for a rare $B$ decay is
obtained by replacing the $b\to c$ transition by
a CKM suppressed $b\to u$ transition. These decays probe the small
CKM matrix element $V_{ub}$, the magnitude of which sets bounds on
the combination $\rho^2 + \eta^2$ in the Wolfenstein
parameterization of the CKM matrix. So far the only
measurement of the magnitude of $V_{ub}$ has been obtained from
measurements of inclusive semileptonic
$B$ decays\cite{btoulnu}. We will discuss the status of the search for rare
hadronic $B$ decays, and in particular the possibility of measuring the decay
$B^0\to\pi^+\pi^-$ which is important for the study of CP violation in
$B$ decays.

Since $b \to u$ transitions are suppressed by the small value of $|V_{ub}|$
it is expected that additional diagrams will make observable
contributions to some hadronic decay modes.
The most significant of these diagrams
is the one-loop flavor-changing neutral current diagram known as the
``penguin'' diagram (Fig. \ref{Fdiag} d). The CKM favoured part of
this diagram, corresponding to a $b\to s$ transition, is expected to dominate
the amplitude of rare decays to final states with one or three $s$-quarks.
There is also a CKM suppressed $b\to d$ amplitude which may not be
negligible in decays to final states with no $c$ or $s$ quarks.
It should be noted that the
loop diagram is much more significant in $B$ decays
than in $D$ decays because the $b\to s$ loop is sensitive to
the large t quark
couplings $V_{tb}$ and $V_{ts}$, whereas contributions to
the equivalent $c\to u$ loop are suppressed either by the small couplings
$V_{cb}$ and $V_{ub}$, or by the small $s$ and $d$ quark masses.

The observation of the decay $B\to K^*(892) \gamma$,
reported in 1993 by the CLEO~II experiment,
is the first direct evidence for the penguin diagram \cite{PRLbsg}.
This decay is described by the electromagnetic transition
$b\to s\gamma$, which is a $b\to s$ penguin loop
accompanied by the radiation of a photon from either the loop, or the initial
or final state quarks.
This important new result will be discussed in some detail.
We will also comment on the recent discussion about the sensitivity of
the $b\to s\gamma$ process to non-standard model contributions within the
loop\cite{Hewett}.
In many extensions of the standard model an additional
contribution to $b\to s\gamma$ is expected to come from a charged Higgs.
We will discuss the extent to which recent data from the CLEO~II experiment
allow bounds to be set on such non-standard model contributions.

To date, no evidence has been found that
either annihilation or W-exchange processes are
present in $B$ meson decay.
The annihilation diagram (shown in Fig. \ref{Fdiag} (f)) would
result in purely leptonic decays
such as $B^+ \to \tau^+ \bar{\nu}$. These modes have
been searched for and  provide constraints
on the $B$ meson decay constant, $f_B$.

In 1955, Gell-Mann and Pais predicted oscillations between neutral meson
and their
antiparticles which were later  observed in the neutral kaon
system \cite{gellmann}.
In the Standard Model such particle-antiparticle oscillations are described
by the box diagrams shown in Fig. \ref{Fdiag} (g),(h).
Historically, it was a great surprise in 1987
when ARGUS observed $B^0\bar{B}^0$
oscillations at a rate nearly two orders of magnitude larger
than the theoretical expectation at that time \cite{argusdis}.
This result was the first indication that the
top quark was more massive than the
30-50 GeV range anticipated by the theorists
and indicated by the UA1 experiment at the time.
It also demonstrates how the
study of $B$ mesons has provided insight into
the physics of higher mass scales.

To summarize, the detailed study of B mesons is driven by
the need to determine the elements of the
CKM matrix which are
 fundamental parameters of the Standard Model. A complete set
of measurements will overconstrain the Standard Model and will check
its internal consistency.
In  light of recent developments in cosmology, it is conceivable
that this will lead us
to new physics outside of the framework of the Standard Model.
Efforts are now underway to measure additional rare $B$
decays and to observe CP Violation in the $B$ sector. As noted above,
every major high energy physics laboratory has embarked on this program.
%%%%%%\section{THE EXPERIMENTAL STUDY OF B DECAY}
\section{THE EXPERIMENTAL STUDY OF B DECAY}
Experimental $b$ physics began in
1977 when the CFS collaboration at Fermilab observed a narrow resonance
at an energy of
about 9.5 GeV in the reaction p~+~nucleus $\to$
$\mu^+ \mu^-$~+~~X\cite{discovery}.
This resonance was named $\Upsilon (9460)$ and was subsequently
identified as the $1\,^3S_1$ state of the $b\bar{b}$ or bottomonium system.
A second resonance at
a mass near 10.0 GeV was later isolated in the Fermilab
data and later identified as a radial excitation
of the $b \bar{b}$ state. For almost two decades which followed,
this was the last significant contribution to $B$ physics by a
hadron machine as $e^+e^-$ storage rings took over.
Within a year of its
discovery, the $\Upsilon$ resonance was confirmed by experiments at
DORIS \cite{doris1,doris2,doris3} and  at
CESR \cite{cesr1,cesr2}. Most of our current knowledge of $B$ mesons
is based on analyses of data collected at these two machines.
In recent years, advances in detector technology, in particular the
introduction
of high resolution silicon vertex detectors have allowed experiments
at high energy colliders (i.e. LEP, SLC and the TEVATRON) to efficiently tag
$b$ quarks. This has led to precise lifetime measurements and to the
discovery of new $b$-flavored hadrons.

\subsection{$\Upsilon (4S)$ Experiments}
\label{y4sexp}

The total $e^+e^-$ annihilation cross section
as a function of center of mass energy in the region of the
$\Upsilon$ resonances is shown in Fig. \ref{upsilon}.
The width of the $\Upsilon (4S)$ state is $23.8 \pm 2.2$ MeV \cite{PDG}
which is significantly larger than the width of the three lighter resonances.
OZI (Okubo-Zweig-Iizuka) suppression of
hadronic decays, which is responsible for
the narrow width of the $\Upsilon(1S) \;
\Upsilon(2S)$ and $\Upsilon(3S)$ states
is no longer operative. This is the first indication
that the $\Upsilon (4S)$ resonance
lies above the threshold for $B \bar{B}$ production.
Further evidence for $B$ meson production came from
the observation of a dramatic increase
in the lepton yield at the $\Upsilon (4S)$
resonance. The momentum spectrum of the leptons was found to be consistent
with production of a heavy quark.
\begin{figure}[htb]
\begin{center}
\unitlength 1.0in
\begin{picture}(4.,2.5)(0,0)
\put(-0.31,-0.8){\psfig{width=4.5in,height=3.0in,%
bbllx=0pt,bblly=0pt,bburx=567pt,bbury=567pt,file=upsilon.ps}}
\end{picture}
\vskip 3 mm
\caption{$e^+e^-$ cross-section measured by CLEO and CUSB showing the
masses of the $\Upsilon$ resonances.(Broken horizontal scale)}
\label{upsilon}
\end{center}
\end{figure}

The first fully reconstructed $B$ mesons were reported in 1983 by the
CLEO~I collaboration \cite{FirstB}.
Since then the CLEO~1.5 experiment \cite{anotherB,SecondB}
has collected a sample with an integrated
luminosity of $212 ~\rm{pb}^{-1}$ \cite{only4s},
the ARGUS experiment \cite{ThirdB,FourthB,FifthB,ARGUSDDs}
has collected $246 ~\rm{pb}^{-1}$,
and the CLEO~II experiment
has collected about $3 ~\rm{fb}^{-1}$, of which between
$0.9$ and $2.0 ~\rm{fb}^{-1}$ have been used
to obtain the results
described in this review\cite{SixthB,PRLkpi,cleodds,PRLbsg}.
All of these
experiments at $e^+ e^-$ colliders
record data on the $\Upsilon (4S)$ resonance, which is
only 20 MeV above $B\bar{B}$ threshold.
The observed
events originate from the decay of either the $B$ or the $\bar{B}$ meson
as there is not sufficient energy
to produce additional particles. The
$B$ mesons are also produced nearly at rest. The average momentum is about
$330$~MeV so the average decay length is approximately 30 $\mu m$.
The $\Upsilon (4S)$ resonance decays
only to $B^0\bar{B}^0$ or to $B^+B^-$
pairs, while
heavier states such as $B_s$ or $B_c$ mesons and $b$-flavored baryons are
not accessible.

For quantitative studies of $B$ decays the initial composition of the data
sample must be known. The ratio of
the production of neutral and charged decays of the
$\Upsilon (4S)$ is therefore an important
parameter for these experiments. The ratio is denoted
$$
\frac{f_+}{f_0}\; = \; \frac{\Upsilon(4S) \to B^+B^-}{\Upsilon(4S) \to
B^0\bar{B}^0}.
$$
CLEO has measured this ratio and found\cite{dstlnu}
$$
\frac{f_+}{f_0}\; = \; 1.04 \pm 0.13 \pm 0.12 \pm 0.10
$$
The last error is due to the uncertainties in the ratio of $B^0$ and
$B^+$ lifetimes.
This is consistent with equal production of
$B^+ B^-$ and $B^0 \bar{B^0}$ pairs and unless explicitly stated otherwise
we will assume that $f_+/f_0\, = \, 1$.
Older results which assumed other values of $f_+$ and $f_0$ have been
rescaled.

$B$ meson pairs from $\Upsilon(4S)$ decays are produced in a state with
quantum numbers $J^{PC}\, = \, 1^{--}$.
As a result the direction of flight of the $B$
meson will follow a $\sin^2 \theta_B$ distribution, whereas most of
the background has a flat distribution in this variable.
An important consequence
of production with these quantum numbers is that
 a $B^0\bar{B}^0$ meson pair will evolve coherently until one
of the mesons decays.

\subsubsection {Continuum Background Suppression}
The $\Upsilon (4 S)$ resonance sits on a continuum background
consisting of $e^+ e^- \to q\bar{q}$, where $q$ can be any of $u,d,s,c$.
The ratio of the resonance to continuum cross section is approximately 1:3.
The continuum background is studied by taking a significant amount of data at
an energy just below the $\Upsilon (4 S)$ resonance, e.g. CLEO~II records a
third of its data at an energy 55~MeV below the resonance.
Using this data sample, and Monte Carlo simulations of $q\bar{q}$ jets,
cuts have been devised to suppress the continuum background.
\begin{figure}[htb]
\begin{center}
\unitlength 1.0in
\begin{picture}(3.,2.8)(0,0)
\put(-0.5,0.3){\psfig{width=3.5in,height=2.5in,%
file=sphericity.eps}}
\end{picture}
\bigskip
\caption{Absolute value of the
cosine of the angle between the the direction of the $B$ candidate
and the sphericity axis of the remaining event for (a) continuum data and (b)
$B\bar{B}$ Monte Carlo.}
\label{sphericity}
\end{center}
\end{figure}
In $\Upsilon (4S)$ production of $B\bar{B}$ pairs, the $B$ mesons are produced
almost at rest, and their decay axes are uncorrelated. These events are rather
spherical in shape, and can be distinguished from jetlike continuum events
using a variety of event shape variables. For the study of inclusive production
in $B$ decays a particularly useful variable is
 $R_2\, = \, H_2/H_0$,
where $H_{2}$ and $H_{0}$ are the
second and zeroth Fox-Wolfram moments\cite{fw}.
This variable is 0 for a perfectly spherical
event, and 1 for an event completely collimated around the jet axis.
For the study of exclusive $B$ decay modes it is often
more useful to compare the
axis of the reconstructed $B$ candidate with the axis of the
rest of the event.
Examples of variables used are the direction of the sphericity axis (see
Fig. \ref{sphericity}) or the
thrust axis of the rest of the event with respect to the $B$ candidate,
$\theta_S$ or $\theta_T$, and the sum of the momenta transverse
to the axis of the $B$ candidate, known as $s_{\perp}$.

In some cases, we will discuss the
use of these cuts, and their effectiveness for particular analyses, but refer
the reader to other references for
a more detailed discussion of the
shape variables\cite{Artuso},\cite{Roe}.

\subsubsection{Selection of $B$ Candidates}
\label{B-recon}

As an example of the techniques of $B$ reconstruction we will briefly describe
the procedure used by the CLEO~II experiment to reconstruct the decay modes
$B\to D^{(*)}(n\pi)^-$.
The CLEO~II detector is described in detail elsewhere \cite{TRA}.
It has a momentum resolution for charged tracks given by
$(\delta p/p)^2 = (0.0015p)^2 + (0.005)^2$,
and an energy resolution for isolated photons
from the CsI barrel calorimeter of
$\delta E/E [\%] = 0.35/E^{0.75} + 1.9 - 0.1E$, where $p$ and $E$ are in GeV.
Charged tracks are identified as pions or kaons if they have
ionization loss information $(dE/dx)$, and/or time-of-flight information (ToF),
consistent with the correct particle hypothesis.
Photon candidates are selected from showers in the calorimeter barrel
with a minimum energy of 30~MeV, which are not matched to charged tracks, and
which have a lateral energy distribution consistent with that expected for a
photon. Neutral pions are selected from pairs of photons with an invariant
mass within $2.5\sigma$ of the known $\pi^0$ mass.

Candidate $D^0$ mesons are identified in the decay modes $D^0\to K^-\pi^+$,
$D^0\to K^-\pi^+\pi^0$, and $D^0\to K^-\pi^+\pi^+\pi^-$, while $D^+$ mesons
are identified in the decay mode $D^+\to K^-\pi^+\pi^+$.
Charged $D^*$ candidates are found using the decay $D^{*+}\to D^0\pi^+$,
while neutral $D^*$ candidates are found using the decay $D^{*0}\to D^0\pi^0$.
Other $D$ and $D^*$ decay modes are not used because of poorer signal to
background ratios, or because of lower yields\cite{xdfeg}.
The reconstructed $D$ masses and $D^*-D^0$ mass differences are required to
be within $2.5\sigma$ of the known values.

The $D$ meson candidates are combined with one or more additional pions
to form a $B$ candidate.
Cuts on the topology of the rest of the event are made in order
to distinguish
$B \bar{B}$ events from continuum background, as discussed in the previous
section. The following requirements are imposed:
$R_2<0.5$, and $|\cos(\theta_S)|<0.9(0.8,0.7)$ depending on
whether there are one(two,three) pions added to the $D$ meson.
As is shown in Fig. \ref{sphericity}, the cosine of the sphericity
angle, $\theta_S$, is uniformly distributed for signal, but
peaks near $\pm 1$ for continuum background.
Requiring that  $|\cos(\theta_S)|$ be less than 0.7 typically
removes  80\% of the continuum background, while retaining
70\% of the $B$ decays.

\begin{figure}[htb]
\begin{center}
\unitlength 1.0in
\vskip 12 mm
\begin{picture}(3.,3.)(0,0)
\put(-.35,-.3){\psfig{width=2.8in,height=2.8in,%
file=bexcl_dnorm.ps}}
\end{picture}
\bigskip
\bigskip
\bigskip
\bigskip
\vskip 15 mm
\caption[]{The beam constrained mass distributions
from CLEO~II {\protect{\cite{SixthB}}} for:
 (a) $ B^- \to D^0 \pi^-$ decays.
(b) $ B^- \to D^0\rho^-$ decays for $|\cos\Theta_{\rho}|>0.4$.
 (c) $ \bar{B^0} \to D^+\pi^-$ decays .
(d) $ \bar{B}^0 \to D^+ \rho^-$
decays for
$|\cos\Theta_{\rho}|>0.4$.}
\label{dpi}
\end{center}
\end{figure}

The measured sum of charged and neutral energies, $E_{meas}$,
of correctly reconstructed $B$ mesons produced at the
$\Upsilon (4S)$, must equal the
beam energy, $E_{beam}$, to within the experimental resolution.
Depending on the $B$ decay mode,
$\sigma_{\Delta E}$, the resolution on the energy difference
$\Delta E\; = \; E_{beam} - E_{meas}$
varies between 14 and 46~MeV.
Note that this resolution is usually sufficient to distinguish
the correct $B$ decay mode from a mode that differs by one pion.
For final states with a fast $\rho^-$ the energy resolution
depends on the momenta of the final state pions from the $\rho$ meson.
This dependence is conveniently parameterized as a function
of the angle between the $\pi^-$ direction in the $\rho^-$ rest frame
and the $\rho^-$ direction in the lab frame, which we
denote as the $\rho$ helicity angle, $\Theta_{\rho}$.
When $\cos\Theta_{\rho} = +1$, the error in the energy measurement
is dominated by the momentum resolution on the fast $\pi^-$,
whereas at $\cos\Theta_{\rho}= -1$
the largest contribution to the error in the energy measurement comes from
the calorimeter energy resolution on the fast $\pi^0$.

To determine the signal yield and
display the data the beam constrained mass is formed
\begin{equation} M_B^2=E_{beam}^2 - \left(\sum_i{\vec{p_i}}\right)^2,
\label{EBmass}
\end{equation}
where $\vec{p_i}$ is the momentum of the $i$-th daughter of the $B$ candidate.
The resolution in this variable is determined by the beam energy spread,
and is about 2.7~MeV for CLEO~II, and about 4.0~MeV for ARGUS.
\cite{mbrange}
These resolutions are  a factor of ten better
than the resolution in invariant mass obtained without the beam energy
constraint. The invariant mass
spectra from the CLEO II analysis of  $B \to D (n\pi)^-$
decays are shown in Fig. \ref{dpi} \cite{SixthB}.

For a specific $B$ decay chain,
such as $B^- \to D^0 \pi^-, D^0 \to K^- \pi^+ \pi^0$ there may be
multiple combinations in a given decay chain.
In the CLEO~II analysis,
if there are multiple candidates only the entry with the smallest absolute
value of $\Delta E$ is selected for events with $M_{B} > 5.2 $ GeV.
An alternative method is to select the candidate with the highest total
probability as calculated from the sum of all $\chi^2$ contributions
from particle identification, kinematical fits and the beam energy
constraint\cite{FifthB}.

\subsubsection{Background Studies}
\label{bkg-studies}

In order to extract the number of signal events
it is crucial to understand
the shape of the background in the $M_B$ distribution.
There are two contributions to this background, continuum
and other $B \bar{B}$ decays. The fraction of continuum
background varies between $58 \%$ and $91\%$ depending on the
B decay mode\cite{cfrac}. The shape of the continuum background is
well understood since it depends primarily
on the transverse momentum distributions of
the final state particles relative to the jet axis. This has been studied using
the off-resonance data sample, and using Monte Carlo techniques.

The shape of the $B \bar{B}$ background is more
difficult to understand since it is mode dependent.
It also has a tendency to peak in the signal region, since the combinatorial
background comes mostly from combinations in which the true final state is
altered by one low energy particle. A particularly troublesome background
occurs when the decay $D^{*0}\to D^0\gamma$ is replaced by the decay
$D^{*0}\to D^0\pi^0$. To determine the correct background shape
for each $B$ decay mode, CLEO~II has studied the
$M_B$ distributions for $\Delta E$
sidebands, and for combinations in which the charged particles have the wrong
charges for the expected spectator decay diagram, e.g. $D^+\pi^+$ and
$\bar{D^0}\pi^+$.

It is found that all of the background distributions can be
fitted with a linear background below M$_B$=5.282~GeV,
and a smooth kinematical cutoff at the endpoint, which is chosen to be
parabolic. For each $B$ decay mode CLEO~II
uses this background function
and a Gaussian signal with a  fixed width of $2.64$ MeV to determine the
yield of signal events. In the ARGUS and CLEO~1.5 experiments slightly
different background parameterizations were used \cite{ARGUSback}.

\subsection {High Energy Collider Experiments}

While the $\Upsilon(4S)$ machines are well suited to study many aspects
of $B$ physics some questions can only be investigated by experiments
at higher center of mass energies.
These include lifetime
measurements and the search for heavier $b$-flavored mesons and baryons.

The four LEP experiments and SLD operate on the $Z^0$ resonance. At this
energy, the cross section for $b\bar{b}$ production  is about
6.6 nb and the signal-to-noise ratio for hadronic events is 1:4, comparable
to the $\Upsilon(4S)$ resonance. The kinematic constraints available
on the $\Upsilon(4S)$ cannot be used on the $Z^0$ but due to
the large boost the
$b$ quarks travel $\approx$ 2.5 mm before they decay and the decay products of
the two $b$-hadrons are clearly separated in the detector.
The large boost makes precise lifetime measurements possible
as well as observations of time dependent phenomena such as
$B_{d}-\bar{B_{d}}$ mixing.

Compared to $e^+e^-$ annihilation, the $b\bar{b}$ production cross section
at hadron colliders is enormous, about $\, 50 \mu b$ at 1.8 TeV.
However, a signal-to-background ratio of about 1:50 makes it difficult to
extract $b$ quark signals and to fully reconstruct $B$ mesons.

In the past, evidence for the production of $b$ quarks
in high energy experiments has been
deduced from the presence of high $p_{\perp}$ leptons.
At hadron colliders, selecting final states containing leptons
also provides a powerful tool to suppress QCD backgrounds.
Recently, significant progress in the isolation of events
containing $b$ quarks
has been made possible by the installation of silicon vertex detectors
near the interaction point at several collider experiments.

\begin{figure}[htb]
\unitlength 1.0in
\vskip 10 mm
\begin{center}
\begin{picture}(2.5,2.2)(0.0,0.0)
\put(-.01,0.0){\psfig{width=1.8in,height=1.8in,file=cdf_bd.ps}}
\end{picture}
\vskip 17 mm
\caption[]{(a) The $\psi K^+$ mass distribution
from the CDF experiment
 (b)  The $\psi K^{*0}$ mass
distribution from the CDF experiment. The solid
line indicates the fitted region
\protect{\cite{cdf_cc}}.}
\label{cdfbd}
\end{center}
\end{figure}

These $b$ quarks hadronize as $B_d$, $B_u$, $B_s$, and $B_c$ mesons or
as baryons containing $b$ quarks.
With the improvement in background suppression provided by these
solid state detectors, signals for exclusive hadronic
$B_d$, $B_u$ and $B_s$
 meson decays have been isolated in the invariant mass spectra for low
multiplicity final states. Examples include the decay modes
$B^{+(0)} \to \psi K^{+(0*)}$ shown in Fig.
\ref{cdfbd} and $B_s \to \psi \phi$ shown in Fig. \ref{cdfbs}.
However, the resolution in invariant mass
for high energy experiments, which is of order 10-20 MeV,
is poorer than the resolution in beam constrained mass in threshold
experiments. At CDF, the mass resolution, about 10 MeV, is
sufficient to separate modes such as $B^-\to \psi \pi^-$ from
$B^-\to \psi K^-$. However,
for experiments at LEP the mass resolution
 is frequently not sufficient
to clearly separate B meson decay
 modes with an additional photon or modes where one kaon
is replaced with a pion.
Evidence for the production of $b$-flavored baryons
has also been reported recently but the relative
production fractions are not well known \cite{Sharma}.

Although collider experiments cannot determine absolute branching
fractions without making further assumptions or using information from
experiments at the $\Upsilon (4S)$, they can measure ratios of
branching fractions such as
the ratio ${\cal B}(B\to \psi K^*)/{\cal B}(B\to \psi K)$.
Some high energy
experiments have also obtained inclusive signals
for $D^0, D^{*+}, \psi$ mesons in $B$ decay.
%However, it is often difficult
%to distinguish the contribution of $B_d$, $B_u$ and $B_s$ mesons.

\subsection {Determination of Average $B$ Meson Branching Fractions}
\label{thatsit}
To extract $B$ meson branching fractions, the detection efficiencies are
determined from a Monte Carlo simulation and the yields are corrected
for the charmed meson branching fractions.
In order to determine new average branching ratios for $B$ meson
decays the results from individual experiments must be normalized
with respect to a common set of charm meson and baryon absolute branching
fractions.
The branching fractions for
the $D^0$ and $D^+$ modes used to calculate the $B$ branching fractions
are given in Tables~\ref{Tbd0br},~\ref{Tbdpbr}.
We have chosen the average of values
for the $D^0 \to K^- \pi^+$ branching fraction recently reported by
CLEO~II and ALEPH to normalize the results \cite{DKpi}.
The branching fractions of other $D^0$
decay modes relative to $D^0 \to K^- \pi^+$ are taken from the PDG
compilation\cite{PDG}. The $D^+$ branching fractions are also taken
from the PDG compilation \cite{PDG}.

\begin{table}[htb]
\caption{$D^0$ branching fractions [\%] used in previous publications and
this review.}
\label{Tbd0br}
\begin{tabular}{lllll}
Mode & ARGUS, CLEO 1.5 \cite{SecondB},\cite{FourthB}
 & ARGUS ($DD_s$) \cite{ARGUSDDs} & CLEO II \cite{SixthB} & This review\\
\hline
$K^-\pi^+$
 & $4.2 \pm 0.6 $ & $ 3.7 \pm 0.3 $ & $ 3.9 \pm 0.2 $ & $3.9 \pm 0.2 $\\
$K^-\pi^+ \pi^- \pi^+$
& $ 9.1 \pm 1.1 $ & $ 7.5 \pm 0.5 $ & $ 8.0 \pm 0.5 $ & $ 7.9\pm 0.6$\\
$K^-\pi^+ \pi^0$
& $ 13.3 \pm 1.8 $ & $ 11.3 \pm 1.1 $ & $ 12.1 \pm 1.1 $& $13.4\pm 1.15$\\
$K^0\pi^+ \pi^-$
 & $ 6.4 \pm 1.1 $ & $ 5.4 \pm 0.5 $ &         & $5.1\pm 0.6$
\end{tabular}
\end{table}

\begin{table}[htb]
\caption{$D^+$ branching fractions [\%] used in previous publications and
this review.}
\label{Tbdpbr}
\begin{tabular}{lllll}
Mode & ARGUS, CLEO 1.5 \cite{SecondB},\cite{FourthB}
 & ARGUS ($DD_s$) \cite{ARGUSDDs} & CLEO II \cite{SixthB} & This review\\
\hline
$K^-\pi^+ \pi^+$ & $9.1 \pm 1.4 $ & $ 7.7 \pm 1.0 $ & $ 9.1 \pm 1.4 $ &
$9.1 \pm 0.6$\\
$K^0\pi^+ $& $ 3.2 \pm 0.5 $ & $ 2.6 \pm 0.4 $ & & $2.7\pm 0.3$\\
$K^0\pi^+ \pi^+ \pi^-$ &  & $ 6.9 \pm 1.1 $ &  & $7.0\pm 1.0$ \\
\end{tabular}
\end{table}

Branching ratios for all $D_s$ decay modes are normalized relative to
${\cal{B}}(D_s^+\rightarrow \phi \pi^+)$. There are no model-independent
measurements of the absolute branching fraction for
$D_s^+\rightarrow \phi \pi^+$.
The currently favored method uses measurements of
$\Gamma(D_s^+\rightarrow \phi \ell^+\nu)$/$\Gamma(D_s^+
\rightarrow \phi \pi^+)$.
The rate $\Gamma(D_s^+\rightarrow \phi^+ \ell\nu)$
is determined from measurements of $\tau_{D_s^+}/\tau_{D^+}$,
$\Gamma(D^+\rightarrow K^* \ell\nu)$, and using
$\Gamma(D^+\rightarrow K^* \ell\nu)$/
$\Gamma(D_s^+\rightarrow \phi \ell^+\nu)$ obtained from theory.
We use the value of ${\cal{B}}(D_s^+\rightarrow \phi \pi^+)$
derived in reference \cite{CLNS9314}.
Other methods include using ${\cal B}(D_s^+\to X\ell^+\nu)$ obtained by
combining measurements of ${\cal B}(D_s^+\to \phi\ell^+\nu)/
{\cal B}(D_s^+\to \phi\pi^+)$, ${\cal B}(D_s^+\to \eta\ell^+\nu)/
{\cal B}(D_s^+\to \phi\pi^+)$, ${\cal B}(D_s^+\to \eta^{'}\ell^+\nu)/
{\cal B}(D_s^+\to \phi\pi^+)$ and comparing to ${\cal B}(D_s\to
X\ell \nu)$ obtained from ${\cal B}(D^0\to X\ell\nu)$ and the
ratio of $\tau_{D^0}/\tau_{D_s}$. This method also gives
 values in the same range.
We believe, however, that the PDG group
has underestimated the error on this value.
\begin{table}[htb]
\caption{$D_s$ branching fraction [\%] used in previous publications and
this review.}
\label{Tbdsbr}
\begin{tabular}{lllll}
Mode & CLEO 1.5 \cite{SecondB}
 & ARGUS ($DD_s$) \cite{ARGUSDDs} & CLEO II \cite{cleodds}
& This review \\ \hline
$\phi \pi^+ $& $2.7 \pm 0.7 $ & $ 3.0 \pm 1.1 $&  $ 3.5 \pm 0.4 $ &
$ 3.7 \pm 0.9 $
\end{tabular}
\end{table}

Since the publication of the original
ARGUS and CLEO~1.5 papers on hadronic
decays, the branching fractions for the $D^{*} \to D \pi (\gamma)$
modes have been significantly improved by more
precise measurements from CLEO~II \cite{CLEODSTAR}.
For modes which contain $D^{*}$ mesons
we have recalculated the branching ratios using the CLEO~II measurements.

\begin{table}[htb]
\caption{$D^{*}$ branching fractions [\%] used in previous publications and
this review.}
\label{Tbdstarbr}
\begin{tabular}{llll}
Mode & ARGUS, CLEO 1.5 \cite{SecondB},\cite{FourthB}
 & CLEO II \cite{SixthB} & This review\\ \hline
$D^{*0}\rightarrow D^0\pi^0 $& $55.0 \pm 6 $ & $ 63.6 \pm 4.0 $ & $63.6 \pm
4.0$\\
$D^{*0}\rightarrow D^0\gamma $& $45.0 \pm 6 $ & $ 36.4 \pm 4.0 $ & $36.4 \pm
4.0$\\
$D^{*+}\rightarrow D^0\pi^+ $& $57.0 \pm 6 $ & $ 68.1 \pm 1.6 $ & $68.1 \pm
1.6$
\end{tabular}
\end{table}

\begin{table}[htb]
\caption{Charmonium branching fractions [\%] used in previous publications and
this review.}
\label{Tbccbr}
\begin{tabular}{llll}
Mode & ARGUS, CLEO 1.5 \cite{SecondB},\cite{FourthB}
 & CLEO II \cite{SixthB} & This review \cite{BRpsi}\\ \hline
$\psi \rightarrow e^+e^- $&
 $6.9 \pm 0.9$ & $ 5.91\pm 0.25 $ & $ 5.91 \pm 0.25 $\\
$\psi \rightarrow \mu^+\mu^- $&
$6.9 \pm 0.9$ & $ 5.91\pm 0.25 $ & $ 5.91 \pm 0.25 $\\
$\psi' \rightarrow e^+e^-$ and $\mu^+\mu^- $& $1.7 \pm 0.3$ & $ 1.7 \pm 0.3 $ &
$ 1.7 \pm 0.3 $\\
$\psi' \rightarrow \psi \pi^+\pi^-$& $32.4 \pm 2.6$ & $ 32.4 \pm 2.6 $ &
$ 32.4 \pm 2.6 $\\
$\chi_{c1} \rightarrow \psi \gamma $& $ 27.3 \pm 1.6$ & $ 27.3 \pm 1.6 $ &
$ 27.3 \pm 1.6 $ \\
$\chi_{c2} \rightarrow \psi \gamma $&  & $ 13.5 \pm 1.1 $ &
$ 13.5 \pm 1.1 $
\end{tabular}
\end{table}

We also give the old and new values assumed for the branching
fractions of the decays
$\psi\to e^+e^-$ and $\psi\to \mu^+\mu^-$.
We have chosen to use the precise measurement of these decays
performed by the MARK III collaboration \cite{BRpsi}.
The modes $\psi^{'}\to \ell ^+ \ell ^-$ and
$\psi^{'}\to \psi \pi^+ \pi^-$ are used to form $B$ meson candidates in modes
involving
$\psi'$ mesons.
Decays of $B$ meson into final states
containing $\chi_{c}$ mesons are reconstructed
using the channel $\chi_{c1,~(c2)}\to \psi \gamma$.
Product branching ratios for all modes containing $\psi$ mesons have been
rescaled
to account for the improved  $\psi$ branching fractions.

In the cases where only one $D^0$ decay mode was used to reconstruct the $B$
meson
the published branching ratio is simply rescaled.
The procedure for re-calculating the branching ratios becomes more involved
when more than one $D$ decay channel is used.
CLEO 1.5 and ARGUS used the following procedure to obtain their results
\[\displaystyle
{\cal{B}}(B)\; = \; \frac{N_{observed}}{\epsilon \times N_B \times
({\cal{B}}(D^*)) \times
\sum{{\cal{B}}_i(D^0)}}\]
where $N_B$ is the number of $B$ mesons.
The efficiency $\epsilon$ is defined as
\[\epsilon \; = \; \frac{\sum{{\cal{B}}_i(D^0)\epsilon
_i}}{\sum{{\cal{B}}_i(D^0)}}\]
The index $i$ refers to the $D$ meson decay channel.
Therefore the rescaled branching ratio is given by
\[{\cal{B}}\; = \; \frac{N_{observed}}{N_B \times ({\cal{B}}(D^*))\times
\sum{{\cal{B}}_i(D^0)\epsilon _i}}\]

The CLEO collaboration published enough information, including the
yields and the efficiencies for the individual $D^0$ decay channels, so that
rescaling their $B$ branching ratios is straightforward.

The CLEO II branching ratios for $B \to D^{(*)}$ decays are calculated as
weighted average of the results determined in each $D$ sub-mode. Since
yield and efficiencies were provided in the original publication, the results
could easily be rescaled to accommodate the improved $D^0$ and $D^{+}$
branching ratios.

Although the $D^0$ reconstruction efficiencies depend slightly on the $B$ meson
decay channel under study, the only information available from the ARGUS
collaboration are average $D^0$ reconstruction efficiencies $<\epsilon >_i$.
Therefore we had to  make the assumption that the correct way to
renormalize the ARGUS results is to multiply their branching ratios by
the scale factor $F$
where
\[
F\;=\;\frac{\sum{<\epsilon >_i \times {\cal{B}}_i(D^0)_{old}}}
{\sum{<\epsilon > _i \times {\cal{B}}_i(D^0)_{new}}}
\]
The validity of this assumption has been checked using CLEO~1.5 data.
A similar procedure had to be employed for the CLEO II results on $B \to DD_s$.

Statistical errors are recalculated in the same way as the branching ratios.
For the results from individual experiments on $B$ decays to final states with
$D$
mesons two systematic errors are quoted.
The second systematic error contains the contribution
due to the uncertainties in the $D^0\to K^-\pi^+$, $D^+\to K^-\pi^+\pi^+$
or $D_s \to \phi \pi$
branching fractions.
This will allow easier rescaling when these
branching ratios  are measured more precisely.
The first systematic error includes the experimental uncertainties and when
relevant the uncertainties in the
ratios of charm branching ratios, e.g. $\Gamma(D^0 \rightarrow
K^- \pi^+ \pi^+ \pi^-)/\Gamma(D^0 \rightarrow K^- \pi^+)$ and the error in the
$D^*$ branching fractions. For modes involving $D_s$ mesons, the first
systematic error also includes the uncertainties
due to the the $D^0$ and $D^+$ branching ratios.
For all other modes only one systematic error is given.
For the world averages, the
 statistical and the first systematic error are combined
in quadrature while the errors due to the
$D^0$, $D^+$ and $D_s$ branching ratio scales
are still listed separately.
With the improvement in the precision
of the $D^0$ and $D^*$ branching fractions these are no longer the
dominant source of systematic error in the study of hadronic $B$
meson decay. The error on the $D_s^+$ branching ratio scale remains large.

%%%%\section{B MESON MASSES AND LIFETIMES}
\section{B MESON MASSES AND LIFETIMES}
\label{mass-diff}

Particles are characterized by their mass, lifetimes
and internal quantum numbers.
Only four $b$-flavored mesons have been established to date.
The bound states with a $b$ quark and a
$\bar{u}$ or $\bar{d}$ anti-quark
are referred to as the $B_{d}$ ($\bar{B^0}$) and the $B_u$ ($B^-$)
mesons, respectively.
The first radial excitation is called the $B^*$ meson. The $B_s$ meson,
a bound state containing a $b$ quark and
$\bar{s}$ anti-quark, has been discovered in the past two years.
The spectrum of known and predicted $B$ states is shown in Fig.
\ref{bspectrum}. These predictions are taken from a recent potential model
calculation by Eichten, Hill and Quigg (EHQ)\cite{ehq}.
\begin{figure}[htb]
\unitlength 1.0in
\begin{center}
\begin{picture}(3.0,3.0)(0.0,0.0)
\put(-0.001,0.01){\psfig{width=2.8in,height=2.8in,file=bspectrum.eps}}
\end{picture}
\bigskip
\bigskip
\caption{Measured and predicted masses of the low-lying $B$ meson states.
The predictions are taken from the EHQ model.}
\label{bspectrum}
\end{center}
\end{figure}
The spectrum reflects the remarkable feature of heavy quark spin symmetry.
In the limit of large heavy quark masses, the spin of the heavy quark decouples
from the dynamics and the pseudoscalar and vector states will be degenerate.
In the $B$ system,
an indication that this symmetry is respected is the experimental ratio
\begin{equation}
\frac{m_{B^*}\, - \, m_B}{m_B}\; \;\sim 0.9\, \%.
\end{equation}
%
% **** Golowich suggests slightly different ratios and values
% from charm and rho.

Since the ground state mesons containing
heavy quarks decay weakly, their lifetimes are typical of the weak
interaction scale, in the range of 0.1-2 ps.
Ten years ago, before the MAC \cite{macblife} and MARK II \cite{mkiiblife}
collaborations presented the first measurement of the $b$ lifetime, the only
phenomenological guide to the strength of the coupling between the quark
generations was the Cabibbo angle. If the coupling between the third and
second generations had the same strength as the coupling between the second and
first, the $b$ lifetime would be about 0.1 ps. The measurements of
lifetimes from the PEP experiments that indicated a value
longer than 1 ps were not anticipated. Since to first order
the $b$ lifetime can be expressed in analogy to muon decay
$$
\tau \; = \; \frac{1}{\Gamma_{tot}} \; = \; \frac{{\cal{B}}_{sl}}{\Gamma_{sl}}
\; = \; \frac{\displaystyle
{\cal{B}}_{sl}}{\displaystyle\frac{G_F^2}{192\pi^3}m_b^5 |V_{cb}|^2\times
phase\; space}
$$
it was concluded that the CKM matrix element $|V_{cb}|$ was very small.

In the naive spectator model,
all mesons and baryons containing $b$ quarks have the same
lifetime.
Differences in the hadronic decay channels and interference between
contributing amplitudes, {\it {i.e.}} if the same final state can be
reached through an external and internal spectator decay such as
$B^- \to D^0\pi^-$, will modify this simple picture
and give rise to a hierarchy of lifetimes.
For the $b$ system we expect \cite{bigimarch}
\begin{equation}
\tau(B^-)\; \geq \; \tau(\bar{B}^0)\; \approx \;
 \tau(B_s)\; >\; \tau(\Lambda_b^0)
\end{equation}
A similar lifetime hierarchy has been observed in charm decay.
However, since the lifetime differences are expected to
scale as $1/m_Q^2$, where $m_{Q}$ is the mass
of the heavy quark, the variation in the $b$ system
should be less than $10 \%$.
Measurements
of lifetimes for the various b-flavored hadrons
thus provide a way to determine
the importance of non-spectator mechanisms.

\subsection{$\bar{B^0}$ and $B^-$ Masses.}
\label{bmss}

We now discuss measurements of the $\bar{B^0}$ and $B^-$ masses
and the mass difference between them.
For these analyses only fully reconstructed $B$ decays in modes with good
signal to background are used. As an example,
CLEO~II \cite{SixthB} uses the modes $B^-\to\psi K^-$,
$\bar{B^0}\to\psi K^{*0}$,
$B^-\to D^0\pi^-$, $B^-\to D^0\rho^-$,
$B^-\to D^{*0}\pi^-$, $B^-\to D^{*0}\rho^-$,
$\bar{B^0}\to D^+\pi^-$, $\bar{B^0}\to D^+\rho^-$,
$\bar{B^0}\to D^{*+}\pi^-$, and $\bar{B^0}\to D^{*+}\rho^-$.
\begin{figure}[htb]
\unitlength 1.0in
\begin{center}
\begin{picture}(3.0,3.0)(0.0,0.0)
\put(-.55,-1.13){\psfig{bbllx=0pt,bblly=0pt,bburx=567pt,bbury=567pt,width=3.8in,height=3.8in,file=bexcl_mball.ps}}
\end{picture}
\bigskip
\bigskip
\caption{Beam constrained mass distributions from CLEO~II
for (a) $B^-$ events and (b) $\bar{B^0}$ events.}
\label{FBM}
\end{center}
\end{figure}
With tight cuts,
362 $B^-$ and 340 $B^0$ candidates have been reconstructed.
The beam constrained mass distributions for the sum of these modes are
shown in Fig.~\ref{FBM}.

\begin{table}[htb]
\caption{Measurements of the $\bar{B^0}$ and $B^-$  Masses [MeV].}
\label{Tmabs}
\begin{tabular}{lccc}
Experiment& $M_{\bar{B^0}}$ & $M_{B^-}$ & $M_{\bar{B^0}}-M_{B^-}$ \\ \hline
ARGUS   & $5279.6\pm 0.7 \pm 2.0 $  & $ 5280.5\pm 1.0 \pm 2.0$ &
$-0.9\pm 1.2 \pm 0.5$  \\
CLEO 87 & $5278.0 \pm 0.4 \pm 2.0$  & $5278.3 \pm 0.4 \pm 2.0$ &
$-0.4\pm 0.6 \pm 0.5$  \\
CLEO 93 & $5279.2 \pm 0.2 \pm 2.0$  & $5278.8 \pm 0.2 \pm 2.0$ &
$0.41\pm 0.25 \pm 0.19$  \\ \hline
Average &
$5278.9 \pm 0.2 \pm 2.0   $ & $5278.7 \pm 0.2 \pm 2.0$ & $  0.2\pm 0.3$
\end{tabular}
\end{table}

The absolute values of the $B^-$ and $\bar{B^0}$ masses are limited in
accuracy by the knowledge of the beam energy.
A correction of (-1.1$\pm$0.5) MeV is made for initial state radiation
as described in Ref.~\cite{Cdsr}.
The systematic error from the uncertainty in the absolute value of the
CESR/DORIS energy scale is determined by calibrating to the known
$\Upsilon (1S)$ mass.
The mass difference is determined more accurately than the masses themselves,
because the beam energy uncertainty cancels, as do many systematic errors
associated with the measurement errors on the charged tracks and neutral pions.
There are several models which predict the isospin mass
difference\cite{MDone}, which give values between 1.2 and 2.3 MeV
which are larger than the experimental results given in Table \ref{Tmabs}.
However, Goity and Hou as well as Lebed \cite{quatsch} have found
models that can lead to small values of the mass difference.
That the $\bar{B^0} - B^-$ mass difference is much smaller than the
corresponding mass differences in the $K$ and $D$ mesons is
surprising.

We conclude this section on the $B$ meson masses with  a result that will
be discussed in greater detail in Section \ref{mixing}.
In the neutral $B$ meson system, the eigenstates of the strong and
electromagnetic interaction (production) do not coincide with the
weak eigenstates (decay). According to the CPT theorem, the flavor
eigenstates $B^0\; = \; (d\bar{b})$ and $\bar{B}^0 \; = \; (\bar{d}b)$
must have equal mass and lifetime but this is not required for the
weak eigenstates $B_1$ and $B_2$. The situation is similar to the neutral kaon
system. The $B_1-B_2$ mass difference is too small to be measured directly but
it can be determined from the rate of $B^0\bar{B}^0$ oscillations (see Section
\ref{bdbdbmix})
$$
|\Delta m_{(B_1-B_2)}| \; = \; {
\displaystyle\frac{x_d}{\tau_{B^0}}} \; = \; (3.07 \pm 0.17)
\times 10^{-4}\; {\rm eV}
$$
$x_d$ is the $B^0\bar{B}^0$ mixing parameter.
This is about 100 times larger than the corresponding mass difference in the
$K^0\bar{K}^0$ system.

\subsection{Measurement of the $B^*$ Mass}

The vector partner of the pseudoscalar $B$ meson is called $B^*$. It
has been observed by the CUSB and CLEO collaborations through the
electromagnetic transition $B^* \to \gamma B$ \cite{cleobs},\cite{cusbbs}
which gives a quasi-monochromatic photon.
CLEO has determined the $B^*-B$ mass difference to be $(46.4 \pm 0.3 \pm 0.8)$
MeV which combined with the average $B$ mass gives
\begin{center}
\begin{tabular}{ll}
$m_{B^*} \; = \; 5325.2 \pm 0.5 \pm 2.8$ & MeV
\end{tabular}
\end{center}
$B^*$ production has also been been observed by the LEP experiment
\cite{l3bs},\cite{alephbs},\cite{delphibs}.

\begin{figure}[htb]
\unitlength 1.0in
\begin{center}
\begin{picture}(3.0,3.0)(0.0,0.0)
\put(0.0,-0.0){\psfig{bbllx=0pt,bblly=0pt,bburx=567pt,bbury=567pt,%
width=3.8in,height=2.8in,file=delphi_bstarstar_fig1.ps}}
\end{picture}
\bigskip
\bigskip
\caption{DELPHI results on $B^{**}$ production.
(a) Distribution of the Q-value for $B^{(*)}\pi$ pairs (data points) along with
the Monte Carlo expectation without $B^{**}$ production (shaded area). Q is
defined as $Q \, = \, m_{B^{(*)}\pi} - m_{B^{(*)}} - m_{\pi}$.
(b) Background subtracted $B^{(*)}\pi$ pair Q-value distribution with a
Gaussian fit overlaid.}
\label{bstarstar}
\end{center}
\end{figure}
\subsection{Observation of $B^{**}$ Production}
Evidence for production of orbitally
excited B meson states has recently been reported
by the OPAL and DELPHI collaborations
\cite{opalbstarstar},\cite{delphibstarstar}.
As noted by Gronau, Nippe and Rosner \cite{bstarstar_theory},
$B^{**}$ mesons, if produced in sufficient quantity,
allow self-tagging flavor identification at production time.
They could provide a powerful tool
to study flavor oscillations and CP violation in the neutral
$B$ system.
$B^{**}$ mesons containing a $u$ or $d$ quark are expected to decay to
$B \pi$ or $B^*\pi$ while the dominant decay for $B_s^{**}$ states is
$B_s^{**} \to B^{(*)}K$.

Inclusive correlations between charged pions and b quark jets
as well the displacement of
the decay products of the b quark jet from
the interaction point are used by the LEP experiments.
This method does not distinguish between
$B$ and $B^*$ mesons.
Evidence for production of orbitally excited $B_s^{**}$ mesons
is found using inclusive correlations between the fragmentation
kaon and the products of the $b$ quark jet. The results of the DELPHI
analysis are shown in Figure \ref{bstarstar}. They observe $2157 \pm 120
\pm 323$ candidates and extract a fragmentation ratio for $b$ quark jets
to a $B_{u,d}^{**}$ meson of
\cite{delphibstarstar}
$$
\sigma_{B_{u,d}^{**}/\sigma_{b-jet}} \; = \; 0.27 \pm 0.02 \pm 0.06
$$
The width of the measured signal is consistent  with the predictions for
orbital excitations. Assuming a $B^*\pi$ to $B\pi$ ratio of 2:1, the
$B_{u,d}^{**}$ mass averaged over the four expected states is determined to
$$
m_{B_{u,d}^{**}} \; = \; 5732 \pm 5 \pm 20 \; \rm MeV
$$
OPAL has studied $B\pi$ and $BK$ correlations and found the probability that
the pion charge correctly tags the b quark flavor to be $0.706 \pm 0.013$
\cite{opalbstarstar}.

\subsection{Measurement of the $B_s$ Mass}
\label{Bs-mass}

Evidence for exclusive $B_s$ decays has been reported
by the CDF\cite{cdfbs}, OPAL\cite{opalbs}, ALEPH\cite{alephbs},
and DELPHI collaborations\cite{delphibs}.
CDF observes a signal of $32.9\pm 6.8 $ events
in the $B_s\to \psi \phi$ mode (see Fig.~\ref{cdfbs})
and determines
the $B_s$ mass to be $5367.7 \pm 2.4 \pm 4.8$~MeV.
\begin{figure}[htb]
\unitlength 1.0in
\vskip 10 mm
\begin{center}
\begin{picture}(2.5,2.2)(0.0,0.0)
\put(-0.2,0.0){\psfig{width=3.0in,height=3.0in,file=cdf_bs_new.ps}}
\end{picture}
\vskip 10 mm
\caption[]{The $\psi K^+ K^-$ mass distribution
from the CDF experiment for events with
 $K^+ K^-$ mass within 10 MeV$/c^2$ of the nominal $\phi$ mass.}
\label{cdfbs}
\end{center}
\end{figure}
OPAL finds one $B_s$ candidate in the $\psi\phi$ mode which is used
to extract a mass of $5359 \pm 19\pm 7$ MeV for the $B_s$ meson.
ALEPH finds one unambiguous $B_s$ event
in the $B_s \to \psi' \phi$
mode and obtains a mass of $5368.6\pm 5.6 \pm 1.5$~MeV.
 Finally, DELPHI has three
candidates of which  one is in the $B_s\to \psi \phi$ mode and obtains
a mass of $5374\pm 16 \pm 2$ MeV \cite{delphibs}.
By reconstructing
exclusive $B^-$ and $\bar{B}^0$ decays (see  Fig.~\ref{cdfbd}),
the high energy experiments calibrate their $B_s$ measurements
relative to the known $B^-$ and $\bar{B}^0$ masses.
The four $B_s$ mass measurements are consistent
with each other. The average value
\begin{center}
\begin{tabular}{ll}
$m_{B_s}\; = \; 5368.1 \pm 3.8$ & MeV
\end{tabular}
\end{center}
is consistent with quark model predictions\cite{ehq,Kwros}.

\subsection{$b$-Baryon Masses}
Evidence for the production of $\Lambda_b$ and $\Xi_b$ baryons has been
found in $\Lambda - \ell^+$ and $\Xi - \ell^+$ correlations, respectively
\cite{forty}.
OPAL has reported preliminary evidence for
$\Lambda_b^0\to\Lambda_c^+\pi^-$ \cite{opal_lambdab}.
However, the value originally quoted for the $\Lambda_b$ mass,
 m$_{\Lambda_b^0}\, =\, 5620 \pm 30$ MeV, is not confirmed by later
data from OPAL which show no evidence for this decay mode.
The current upper limit at 90\%  C.L. for the
product branching ratio is \cite{hassan}
$$
{\cal {B}}(b\to \Lambda_b)\times {\cal{B}}(\Lambda_b \to \Lambda_c^+ \pi^-)
\; < \: 1.9 \times 10^{-3}
$$
Earlier claims by the UA1 and ISR experiments
for the decay $\Lambda_b \to \Lambda \psi$ in $p \bar{p}$ collisions
have been ruled out by the CDF and LEP experiments\cite{stochi} who find
\begin{center}
\begin{tabular}{lll}
${\cal{B}}(b\to \Lambda_b)\times {\cal{B}}(\Lambda_b \to \Lambda \psi)$
 & $<\, 1.8 \times 10^{-4}$, &  CDF + LEP\\
&$=(1.8 \pm 1.0)\times 10^{-3}$, & UA1
\end{tabular}
\end{center}

\subsection{Techniques of $b$ Lifetime Measurements}

The pioneering measurements by the experiments at PEP established a
lifetime  greater than $1$ ps for a mixture of $b$ flavored hadrons.
This is far
too short to be directly measured in threshold experiments.
At higher $b$ momenta when
combined with relativistic time dilation, however, the hadron containing
the $b$ quark travels a measurable distance before decaying.
The lifetime of a particle is related to its decay length by
\begin{equation}
\tau_b \; = \; \frac{L_b}{\gamma \beta c}
\label{tau_eq}
\end{equation}
At LEP energies, for example, the average $b$ momentum is about 30 GeV
which yields an average decay length of $2.5$ mm for $<\tau_b> \, = \,
1.5$ ps. Similarly, at CDF the mean vertex displacement in the transverse
direction is about $0.9$ mm.

A variety of methods has been developed to measure the decay length and to
determine the $b$ lifetime. They all follow the same principal steps.
A purified sample is selected and the decay length is either measured
directly or determined indirectly
 by using the impact parameter. The resulting decay length is
then corrected for the Lorentz boost. An additional correction for
background contamination is applied as well.

\subsubsection{Selection of an enriched $b$ sample}

Ideally, one would like to have a sample of fully reconstructed decays to
determine the lifetime of a specific $b$ hadron. The $b$ vertex could then be
reconstructed allowing a measurement of the decay length. The momentum of
the $b$ hadron gives the $\gamma\beta$ factor in equation (\ref{tau_eq})
without any further assumptions. The resulting proper time distribution would
be an exponential convoluted with a Gaussian
resolution function representing the
experimental measurement errors.
Although currently limited by statistics
this procedure will ultimately yield the most precise measurements of
individual $b$ hadron lifetimes.

The best statistical precision
in the determination of lifetimes of hadrons
containing $b$ quarks is currently obtained
from measurements using partial reconstruction of semileptonic
decays. These decays
represent about 21\% of the total $b$ decay rate and have
the experimental advantage
that both electrons and muons can be efficiently identified
with low background.
The purity of the sample can be enhanced by kinematical cuts which
take advantage of the large mass of the $b$ quark e.g.
selecting leptons with large transverse momentum
with respect to the $b$ direction.
Event samples with purities above 90\% have been obtained
at LEP. However, in such
semileptonic decays the neutrino is not detected so the
$b$ hadron is not completely reconstructed.
One then has to rely on Monte Carlo simulations
to estimate the $b$ momentum
and to extract the proper time distribution from the
decay length measurements.

At the TEVATRON the inclusive
$b$ samples with the best signal to background ratios
are obtained by selecting events with two energetic
leptons from $\psi$ decay and a detached vertex.

\begin{figure}[htb]
\unitlength 1.0in
\begin{center}
\begin{picture}(2.0,2.0)(0.0,0.0)
\put(-0.6,0.01){\psfig{width=3.8in,height=1.8in,file=impact.eps}}
\end{picture}
\bigskip
\bigskip
\caption{Lifetime measurements using the impact parameter method (a) and
reconstruction of the $b$ decay vertex (b).}
\label{impact}
\end{center}
\end{figure}

For inclusive lifetime measurements,
the presence of a high $p_{\perp}$ lepton  or a $\psi$
meson is usually sufficient to demonstrate the presence of
a b quark,
while for exclusive measurements of individual $b$ hadron
lifetimes an additional particle
in the decay has to be reconstructed in order to establish a signature
characteristic for the decaying $b$ hadron (Fig \ref{impact}(b)).
The $\Lambda_b$ lifetime, for example, is measured
using a sample of events
containing $\Lambda_c^+\ell^-$ or $\Lambda \ell^-$ combinations.

\subsubsection {Impact parameter method}

In early experiments
the vertexing precision was not sufficient to measure the
decay length, $l\, = \, \gamma \beta c \tau$, directly.
The impact parameter method
shown schematically in Fig. \ref{impact}(a)
was developed as alternative.
Due to the finite lifetime of the $b$ hadron, a
lepton from the semileptonic decay of the heavy quark
will miss the primary
vertex where the $b$ hadron was produced.
The miss distance or impact parameter,
$\delta$, is given by
\begin{equation}
\delta \; = \; \gamma \beta c \tau_b \sin{\alpha} \sin{\theta}.
\end{equation}
where $\alpha$ is the angle
between the lepton and $b$ directions and $\theta$ is the polar angle.
The $b$ direction is usually approximated by the axis of the
hadronic jet. A negative sign is assigned
to the impact parameter
if the lepton track crosses the jet axis behind the the beam spot
indicating a mismeasured
lepton or a background event. The main advantage of the
impact parameter method is that it is
rather insensitive to the unknown boost of the parent;
as $\gamma\beta$ increases
with the $b$ momentum, $\sin{\alpha}$ decreases approximately as
$1/\gamma\beta$ for $\beta \approx 1$ \cite{Roe}.

Improvements in lifetime measurements have come about from
larger data samples, smaller beam spots, the use of
neutral energy in the jet finding algorithms as well as
three dimensional vertex reconstruction.
 In the best LEP measurements the average impact parameter
uncertainty has been reduced to about 80 $\mu m$ \cite{Sharma}.
The disadvantage of this
method is, that a single track is a
relative poor estimator for the $b$ decay length and not all
the available information in the event is used.
Today, impact parameter measurements are
used only for inclusive lifetime measurements and for cases in which
the $b$ decay length cannot be easily reconstructed
\cite{Sharma}.

\subsubsection{Decay length measurements}

New interaction region designs with smaller radius beampipes and beam
spots\footnote{Typical beam spot sizes. LEP: $\sigma_x\approx 150\,
\mu m\, , \, \sigma_y \approx 10\, \mu m$,
CDF: $\sigma_x\approx 40\, \mu m\, , \, \sigma_y \approx 40\, \mu m$,\\
SLD: $\sigma_x\approx 2\,
\mu m\, , \, \sigma_y \approx 1\, \mu m$ \cite{Sharma}}
combined with high resolution silicon vertex detector allow for decay
length measurements with a
precision better than 300 $\mu m$. This is a factor of 10 smaller
than the average $b$ flight distance
at LEP. As indicated in Fig \ref{impact}(b)
the $b$ vertex is reconstructed using the
lepton track and the direction of the reconstructed
charm meson. The momentum of the $b$ hadron
is estimated using the observed decay
products, missing momentum and a correction factor
determined from a Monte Carlo simulation. The proper
time distribution is then given by
an exponential convoluted with a Gaussian resolution
function and the momentum
correction factor. A maximum likelihood fit is used to extract
the lifetime \cite{aleph_bs_plb}.

\subsubsection{Averaging lifetime measurements}
\label{aver_life}

In order to obtain
the most precise
value for inclusive and exclusive $b$ lifetimes the results of
lifetime measurements from different experiments have been combined.
Using the
conventional approach of weighting
the measurements according to their error does not
take into account the
underlying exponential (lifetime) distribution. If a measurement
fluctuates low then its weight in the average
will increase, leading to a bias towards
low values. This is particularly
relevant for low statistics measurements such as
the $B_s$ lifetime. According to a study by Forty\cite{forty},
this bias can be
avoided
if the weight is calculated using the
relative error $\sigma_i/\tau_i$.
\footnote{This procedure assumes good vertex resolution,
{\it {i.e.}} $\sigma<\tau/10$.}
We find a 1-3\% difference in the average lifetimes computed,
with the second method giving the larger
value. A slight bias of the latter method  towards
higher lifetime values could be avoided
by taking into account asymmetric errors.
This effect has been found empirically to be rather small and
we omit this additional complication in the calculation
of our lifetime averages.

\subsection{Inclusive $b$ Lifetime}

%
% *** add some comment on the difference between a quark and hadron
% lifetime
Inclusive measurements of the $b$ lifetime were important historically
to establish the long $b$ lifetime and provided the first evidence that the
coupling between the second and third quark generation is quite small.
They are still needed for some electroweak studies such as the determination
of the forward-backward asymmetry in $Z \to b\bar{b}$
where the different hadrons containing $b$ quarks are not distinguished.
For $B$ physics, {\it {i.e.}} the study
of $B$ meson decays, the exclusive measurements of individual $b$ hadron
lifetimes are preferable. For example,
to extract the  value of the CKM matrix element $|V_{cb}|$
from measurements of semileptonic $B$ decays
the average of the $B^+$ and $\bar{B}^0$ lifetimes should be used rather than
the inclusive $b$ lifetime which contains additional
contributions from $B_s$ mesons and $b$ baryons.
Inclusive $b$ lifetime measurements are performed using one of
the following three methods:

\begin{itemize}
\item{Impact parameter method $\; b \to c\ell \nu$
e.g. $$<\tau_b> \, = \, 1.487 \pm 0.023 \pm 0.0384 \rm{~ps~(ALEPH}
\cite{aleph_tau_inc})$$ }
\item{Decay length reconstruction in $b \to \psi X$
e.g. $$<\tau_b> \, = \, 1.46 \pm 0.06 \pm 0.06 \rm{~ps~(CDF }
\cite{cdf_tau_inc})$$ }
\item{Vertex topology in hadronic $b$ decays
e.g. $$<\tau_b> \, = \, 1.599 \pm 0.014 \pm 0.035 \rm{~ps~(DELPHI}
\cite{delphi_tau_top})$$ }
\end{itemize}
The third method uses displaced multi-prong vertices to reconstruct
the decay point of the $b$ hadron. Extensive Monte Carlo simulations
are used to estimate the $b$ momentum and to determine the proper time
spectrum.

The current world average for the inclusive $b$ lifetime
which includes many measurements including those given above
 is \cite{forty},
\begin{center}
\begin{tabular}{ll}
$<\tau_b> \; = \; 1.524 \pm 0.027$ &ps.
\end{tabular}
\end{center}
The  world average for this quantity in 1992 was $(1.29\pm 0.05)$ ps.
The substantial
change in the value has been attributed to several improvements:
the use of neutral energy when calculating the b jet direction,
and better knowledge of the resolution function as a result of the
use of silicon vertex detectors\cite{Sharma},\cite{forty}.

\begin{figure}[htb]
\unitlength 1.0in
\begin{center}
\begin{picture}(3.0,3.0)(0.0,0.0)
\put(-0.4,0.01){\psfig{width=3.8in,height=3.8in,file=cdf_dstarlnu.ps}}
\end{picture}
\vskip -10 mm
\caption{$D^{*+}$-lepton correlations from $B$ decays observed by the CDF
collaboration. A clear signal is present in  right sign combinations while
no signal is present when same sign charged leptons are combined with
$D^{*+}$ candidates.}
\label{cdf_dstarlnu}
\end{center}
\end{figure}

\subsection{Exclusive Lifetime Measurements}

Precise measurements of exclusive lifetimes for b-flavored hadrons
have been carried out by CDF and by some of the LEP experiments
\cite{cdfglas},\cite{alephglas1},\cite{alephglas2},\cite{opalglas},
\cite{delphiglas}.
The most recent results and the techniques used are given in Table
\ref{Tblife}.

\begin{table}[htb]
\caption{Measurements of exclusive lifetimes for b flavored hadrons.}
\medskip
\label{Tblife}
\begin{tabular}{llllll}
Particle & Technique &CDF  & ALEPH & OPAL & DELPHI  \\  \hline \medskip
$ \bar{B}^0$& $D^{*}-l$ & $1.62\pm0.16\pm0.15$& $1.71^{+0.12}_{-0.11}\pm
0.06$& $1.62\pm 0.10\pm0.10$& $1.17^{+0.29}_{-0.23}\pm 0.15\pm 0.05$    \\
$ \bar{B}^0$& excl &$1.57\pm0.18\pm0.08$& $1.17^{+0.24}_{-0.19}\pm 0.06$
      &        &   \\
$ \bar{B}^0$& topol. & & & & $1.68\pm 0.15^{+0.13}_{-0.17}$ \\ \hline\medskip
$ B^-$& $D-l$ & $1.63\pm 0.20\pm 0.16$ & $1.71\pm0.15\pm0.08$
 & $1.53\pm0.14\pm0.11$ & $1.30^{+0.33}_{-0.29}\pm 0.15\pm 0.05$ \\
$ B^-$&  excl & $1.61\pm 0.16\pm 0.05$ &
$1.30^{+0.25}_{-0.20}\pm 0.06$ &   & \\
$ B^-$&  topol. & &  &   & $1.72\pm0.08\pm0.06$ \\ \hline \medskip
$ B_s^0$&$D_s -l$&$1.42^{+0.27}_{-0.23}\pm 0.11$&
$1.90^{+0.46}_{-0.36}\pm0.05$&
$1.33^{+0.26}_{-0.21}\pm 0.06$& $1.34^{+0.37}_{-0.29}\pm 0.14$ \\
$ B_s^0$&    $D_s-h$ &      & $1.75^{+0.30+0.18}_{-0.28-0.23}$&
  & $1.56\pm 0.35\pm 0.23$  \\
$ B_s^0$&$\psi\phi$&$1.74^{+1.08}_{-0.69}\pm 0.07$ &&   & \\ \hline\medskip
$ \Lambda_b$  & $\Lambda - l$&      & $1.05^{+0.12}_{-0.11}\pm 0.09$& $1.26^
{+0.16}_{-0.15}\pm 0.07$ &  $1.13^{+0.30+0.05}_{-0.23-0.08}$   \\
$ \Lambda_b$  & $\Lambda_c -l$&      & $1.06^{+0.40}_{-0.27}\pm 0.07$&
                     &  $1.33^{+0.71+0.08}_{-0.42-0.09}$   \\
$ \Lambda_b$  & $p-l$ &      &                               &
                     &  $1.28^{+0.35+0.11}_{-0.29-0.12}$   \\ \hline\medskip
$ \Xi_b^0$  & $\Xi-l$ &      &                               &
                     &  $1.5^{+0.7}_{-0.4}\pm0.3$   \\
\end{tabular}
\end{table}

\subsubsection{$B^-$ and $\bar{B}^0$ lifetimes}
The best statistical precision in the determination of exclusive lifetimes
is obtained from measurements using
lepton-particle correlations. For example, a sample of
$B^0$ candidates can be
obtained from events with lepton-$D^{*+}$ correlations of the correct
sign which originate from the decay $\bar{B^0}\to D^{*+} \ell^-\nu$,
$D^{*+}\to D^0 \pi^+$ and $D^0 \to K^- \pi^+$ (see Fig. \ref{impact} (b) for
the method and Fig. \ref{cdf_dstarlnu} for the CDF results).
The pion from the strong decay and the lepton
form a detached vertex.
This information combined with the direction of the reconstructed
$D^0$ meson determines the location of the $B$ decay vertex from which
the decay length can be measured. For a LEP experiment the
$D^{(*)}\ell$ sample typically contains ${\cal{O}}(100)$ events.
To obtain the lifetime from the decay length, requires knowledge
of the $\gamma\beta$ factor which is estimated from the
momenta of the observed decay products. Since the neutrino is not
observed, a correction is made in the boost factor. The uncertainty
in the size of this correction is included in the systematic error
and is typically of order $3\%$.
Another systematic problem is the contamination from decays
$B^-\to D^{**}~l^- \nu$, followed by $D^{**}\to D^{*+}\pi^-$
where the $\pi^-$
from the strong decay of the $D^{**}$ (p-wave) meson is not detected.
These backgrounds will lead to a $B^-$ meson contamination
in the $\bar{B}^0$ lifetime sample (and vice-versa).
Since the branching fractions for such decays are poorly measured, this
is another important systematic limitation and gives a contribution
of order 5\% to the systematic error. Significant contributions
to the systematic error
also result from the uncertainty in the level of background and its
lifetime spectrum. A detailed discussion of exclusive lifetime measurements
can be found in a recent review  by Sharma and Weber \cite{Sharma}.

The systematic problems associated with the boost correction and
the contamination from poorly measured backgrounds can be avoided
by using fully reconstructed decays such as $\bar{B^0}\to D^+ \pi^-$
or $B^- \to \psi K^-$.
However, since exclusive $B$ branching ratios are
small, this method has much poorer statistical precision.
In hadron collider experiments, this approach has been successfully
used to determine the $\bar{B^0}$, $B^-$, and $B_s$ lifetimes from
exclusive modes with $\psi$ mesons e.g. $\bar{B^0}\to \psi K^{*0}$,
$B^-\to \psi K^-$\cite{cdflife} and $B_s\to \psi \phi$ \cite{cdf_bs_life}.

A topological vertexing method has been used by
the DELPHI experiment. Candidate $\bar{B^0}$ and
$B^+$ mesons
are distinguished on the basis of the net charge of the tracks at
the decay vertex. This method has small statistical errors however
care must be taken to assure that systematic uncertainties from mistracking
and incorrect assignments of decay
vertices are controlled. The neutral B lifetime that is extracted
is an average over the lifetimes over all neutral b flavored hadrons
including $B_d^0$, $B_s^0$, and $\Lambda_b^0$. With good knowledge
of the production fractions, the exclusive $B^0$ lifetime can be extracted.

A topological vertexing technique has also been used by the
Fermilab fixed target experiment E653 \cite{E653},
which has observed 11 charged
$B$ candidates and 17 neutral $B$ candidates in their emulsion data.
They find $\tau_{B^+}=3.25^{+1.50+0.27}_{-0.90-0.10}$ ps  and $\tau_{B^0}
=0.91^{+0.27+0.10}_{-0.20-0.04}$ ps. These results have been omitted
from the determination of world average lifetimes.

Using the procedure
for averaging measurements described in Section \ref{aver_life}, we combine
the individual $B^-$ and $\bar{B}^0$ lifetime measurements and obtain
\begin{center}
\begin{tabular}{ll}
$\tau_{B^-} \; = \; 1.646 \, \pm \, 0.063 $ &ps
\end{tabular}
\end{center}
and
\begin{center}
\begin{tabular}{ll}
$ \tau_{\bar{B}^0} \; = \; 1.621 \, \pm \, 0.067 $ & ps
\end{tabular}
\end{center}
When  averaging the results obtained by studying $D-\ell$ correlations
a common systematic error of 3\% has been assumed.

\subsubsection{$B_s$ lifetime measurements}

The $B_s$ lifetime was measured by CDF \cite{cdf_bs_life}
and the LEP experiments using
partial reconstruction of the
semileptonic decay $\bar{B}_s^0 \to D_s^- \ell^+ \nu$.
Candidate $D_s^-$ mesons were reconstructed in the  $\phi \pi^-$ or
$K^{*0}K^-$ final states.
\begin{figure}[htb]
\unitlength 1.0in
\begin{center}
\begin{picture}(3.0,3.0)(0.0,0.0)
\put(-0.5,0.01){\psfig{width=4.8in,height=8.8in,file=aleph_bs_life.ps}}
\end{picture}
\bigskip
\bigskip
\caption{$B_s$ lifetime measurement by ALEPH.
a) $K^-K^+\pi^+$ invariant mass distribution for right-sign $D_s^+\ell^-$
combinations.
b) $K^-K^+\pi^+$ invariant mass distribution for wrong-sign $D_s^+\ell^+$
combinations.
c) Proper time distribution of the right-sign $D_s^+\ell^-$ sample.
d) Proper time distribution of the combinatorial background.}
\label{aleph_bs_life}
\end{center}
\end{figure}
Fig. \ref{aleph_bs_life}(a) shows the $K^-K^+\pi^+$ invariant mass spectrum
obtained by ALEPH \cite{aleph_bs_life}
for right-sign and wrong-sign $D_s\ell$ combinations.
This spectrum contains
47 $D_s$ candidates which were used for the lifetime measurement.
The $B_s$ decay length was measured and converted to the $B_s$ proper
time using a $B_s$ momentum estimator based on the reconstructed lepton
and the $D_s$ momentum as well as an estimated neutrino energy obtained by
using a missing mass technique. $D_s$ backgrounds from $D_s^*$
decays were treated using a Monte Carlo simulation. The $B_s$ lifetime
was extracted from the proper time distribution using a maximum
likelihood fit. The result of such a procedure is shown in Fig.
\ref{aleph_bs_life}(b).

A recent CDF result \cite{cdf_bs_life} also uses
$D_s$-lepton correlations as well as exclusive reconstruction
of $B_s \to \psi \phi$ decays to extract the $B_s$ lifetime\cite{cdf_bs_life}.

The uncertainty in the $B_s$ lifetime is still dominated by the statistical
error. Assuming a common systematic error of 2\% \cite{Sharma} for the
uncertainty in the vertex resolution and the neutrino energy estimate we
obtain
\begin{center}
\begin{tabular}{ll}
$\tau_{B_s} \; = \; 1.55 \, \pm \, 0.13 $ & ps
\end{tabular}
\end{center}

\subsubsection{$b$ baryon lifetime measurements}

Studies of $\Lambda_c^+ \ell^-$ and $\Lambda \ell^-$ correlations at LEP
provided the first evidence for the production of the $\Lambda_b$
baryon. Using the decay chain
\begin{eqnarray*}
\Lambda_b & \to & \Lambda_c^+ \ell^- \bar{\nu} \\
  & & \; ^|\hspace{-2.0mm}\rightarrow \Lambda X\\
  & & \; \; \; \; \; \; \; ^|\hspace{-2.0mm}\rightarrow p\pi^-
\end{eqnarray*}
OPAL \cite{opal_lambdab_life}
found the invariant $p\pi^-$ mass distribution shown in Fig.
\ref{opal_lambdab_life}(a).
\begin{figure}[htb]
\unitlength 1.0in
\begin{center}
\begin{picture}(3.0,2.8)(0.0,0.0)
\put(-1.01,-0.35){\psfig{width=5.0in,height=7.8in,file=opal_lambdab_life.ps}}
\end{picture}
\caption{$\Lambda_b$ lifetime measurement by OPAL.
a) $p\pi^-$ invariant mass distribution for right-sign and wrong-sign
$\Lambda\ell$ combinations.
b) Decay length distribution of the right-sign $\Lambda\ell^-$ sample.
The inset shows the corresponding distribution for the wrong-sign
$\Lambda\ell^+$ candidates.}
\label{opal_lambdab_life}
\end{center}
\end{figure}
Although the composition of the $b$ baryon sample is not known, it is
expected that the $\Lambda_b$ baryon is the most copiously produced.
The production of $\Xi_b$ and $\Omega_b$ baryons
is suppressed due to the additional strange quarks required for their
formation. Both impact parameter and decay length measurements are
used to determine $\tau_{\Lambda_b}$. Since the $\Lambda_c^+$ lifetime
is short, the $\Lambda_b$ decay length can be estimated using the
$\Lambda \ell^-$ vertex. The resulting
time distribution from the OPAL analysis is
shown in Fig. \ref{opal_lambdab_life}(b).

A better estimate of the $\Lambda_b$ decay point is obtained from fully
reconstructing the $\Lambda_c^+$ baryon and finding the $\Lambda_c^+ \ell^-$
vertex. However, the sample sizes become very small. Using this method,
DELPHI finds $\tau_{\Lambda_b} \, = \, 1.33 ^{+0.71}_{-0.42} \pm 0.13$ ps.
Combining the results listed in Table \ref{Tblife} we determine the
world average $\Lambda_b$ lifetime to be
\begin{center}
\begin{tabular}{ll}
$\tau_{\Lambda_b} \; = \; 1.17 \, \pm \, 0.09 $ & ps.
\end{tabular}
\end{center}
DELPHI \cite{Roudeau} has searched for $\Xi^- \ell^-$ correlations
and found 10 $\Xi_b$ candidates. These are expected to come from
$\Xi_b^- \to \Xi_c^0 \ell^- \bar{\nu}X$ and
$\Xi_b^0 \to \Xi_c^+ \ell^- \bar{\nu}X$ followed by
$\Xi_x \to \Xi^- X'$. A simple average of the proper time of the 10
candidates gives the $\Xi_b$ lifetime estimate of $1.5^{+0.7}_{-0.4}\pm 0.3$
ps.

\subsubsection{Lifetime Ratios}
\label{liferat}

The ratio of the $B^-$ and $\bar{B^0}$ lifetimes has been measured
by a number of experiments. These measurements are performed
either by using
correlations between $D$ mesons and leptons
or  by using exclusive final states such as
$B^-\to \psi K^-$ and $\bar{B^0}\to \psi K^{*0}$. The CLEO~II experiment has
measured ${\cal B}(B^0\to X~l^-\nu)$ and ${\cal B}(B^-\to X~l^-\nu)$
using the yield of leptons found opposite fully
and partially reconstructed B decays \cite{cleoiitptz}.
{}From isospin invariance, the ratio of the two branching fractions
is the ratio of the lifetimes.
\begin{table}[htb]
\caption{Measurements of lifetime ratios for b flavored hadrons.}
\medskip
\label{Tbratio}
\begin{tabular}{clllll}
 Method &CDF  & ALEPH & OPAL & DELPHI& CLEO~II \\ \hline \medskip
 $D-l$& $1.01\pm0.19\pm0.17$ & $1.00^{+0.14}_{-0.13}\pm
0.08$ & $0.94\pm0.12 \pm 0.07$ & $1.11^{+0.55}_{-0.39}\pm 0.11$&   \\
 $excl$& $1.02\pm 0.16\pm 0.05$& $1.11^{+0.31}_{-0.25}
\pm 0.03$& & &    \\
  topol. & & & &  $1.02^{+0.13+0.13}_{-0.10-0.10}$& \\
  $\rm{B~tags}$& & & & &   $0.93\pm 0.18\pm 0.12$\\
\end{tabular}
\end{table}
Averaging the results listed in Table \ref{Tbratio} we obtain
$$
\frac{\tau_{B^-}}{\tau_{\bar{B}^0}} \; = \; 0.995 \, \pm \, 0.068
$$
Note that this value is not exactly equal to the ratio
of the world averages for the
$B^-$ and $\bar{B}^0$ lifetimes since the average value of
$\tau_{B^-}/\tau_{\bar{B}^0}$ is calculated directly
from the ratios reported by the experiments.

\subsection{Lifetime Summary}

A summary of the
measurements of all the $b$ hadron lifetimes
can be found in Fig.~\ref{blifetime}.
The pattern of measured lifetimes follows the
theoretical expectations outlined in the introduction to this chapter.
However,
the $\Lambda_b$ meson lifetime is unexpectedly short.
Scaling from the observed
$\Lambda_c - D^0$ lifetime difference, $\tau_{\Lambda_b}$ should not deviate
from the average $b$ lifetime by more than 10\%. A more precise
determination of this lifetime would be of great interest.

\bigskip
\begin{figure}[htb]
\unitlength 1.0in
\begin{center}
\begin{picture}(2.5,3.2)(0.0,0.0)
\put(-0.2,0.01){\psfig{bbllx=0pt,bblly=0pt,bburx=567pt,bbury=567pt,%
            width=2.8in,height=2.4in,file=blifetime.ps}}
\end{picture}
\caption{Summary of exclusive $b$ lifetime measurements.}
\label{blifetime}
\end{center}
\end{figure}

Assuming a $Z \to B^-,\,\bar{B}^0,\,B_s,\,\Lambda_b$ production ratio of
$0.39\,:\,0.39\,:\,0.12\,:\,0.10$ we can average the exclusive lifetime
measurements and find $<\tau_{excl.}>\; = \; 1.58\pm0.08$ ps;
consistent with the
inclusive $b$ lifetime, $\tau_b \; = \; 1.524\pm 0.027$ ps.
Further improvements in the determination of $b$ lifetimes can be expected
as the data samples available at the TEVATRON increase and from
reduced systematic errors in the $D^{(*)}\ell$ correlation measurements
at LEP.

%%%%\section{SEMILEPTONIC B MESON DECAYS}
%\input semileptonic.tex
\section{SEMILEPTONIC B MESON DECAYS}

Semileptonic transitions are the simplest $B$
decays: the heavy $b$ quark decays to either a $c$ or an $u$ quark
and the virtual $W$ boson becomes a lepton pair.
These decays are described by the  external spectator diagram shown
in Fig.~\ref{Fdiag}(a), (c).
Measurements of semileptonic B decays are used to determine
the weak couplings
$|V_{cb}|$ and $|V_{ub}|$.
In addition, detailed measurements
of these decays
test models of the dynamics of heavy quark decay.
The leptonic current can be calculated exactly while
corrections due to the strong interaction are restricted to the
$b\to c$ and $b \to u$ vertices, respectively.

Experimentally, semileptonic decays have the advantage of large
branching ratios and the characteristic
 signature of the energetic charged lepton.
The neutrino, however, escapes undetected so a full reconstruction
of the decaying $B$ meson is impossible. Various techniques
which take advantage of production at threshold or
the hermiticity of the detector have been developed by
the ARGUS, CLEO and LEP
experiments to overcome this difficulty.

Semileptonic decays are also useful for the  study of other phenomena
in $B$ physics.
The charge of the  lepton in a semileptonic decay is directly correlated
with the
flavor of the $B$ meson. A negative lepton comes only from the decay
of a $b$ quark while the decay of a $\bar{b}$ anti-quark yields positive
leptons. Tagging the $b$ flavor has been essential to the discovery of
$B^0\bar{B}^0$ oscillations and will be equally important in searches for
CP violation in the $B$ system.

We begin with
inclusive measurements and then discuss the results on exclusive $b\to c$
transitions.
The dynamics of semileptonic decays is considered in the following
section in which results
on polarization and form factor measurements are given.
Both inclusive and exclusive measurements of semileptonic decays of
$B$ mesons have been used to determine $|V_{cb}|$. We summarize these
results in Section \ref{brsl_vcb}. In the final section,
 inclusive and exclusive
$b \to u$ transitions and the extraction  of the CKM element $|V_{ub}|$
are discussed.

\subsection{Inclusive Semileptonic b$\to$ c Transitions}
\label{inclbc}

There are three types of measurements of inclusive semileptonic
B decays. These are
measurements of the inclusive single lepton
momentum spectrum, measurements of dilepton events using charge
and angular correlations, and  measurements of the separate $B^-$ and
$\bar{B}^0$ branching ratios by using events which contain
a lepton and a reconstructed B meson.

Measurements of the semileptonic $B$ branching ratio, ${\cal{B}}_{sl}$,
have been
performed on the $\Upsilon(4S)$ resonance and at higher energies by the
PEP, PETRA and LEP experiments. In all cases, the
branching fraction of semileptonic
decays is determined from the inclusive lepton yield. The primary
difficulty in these analyses is distinguishing between leptons from $B$ decay
and leptons from other sources.
Once the fraction of direct or primary $b \to c\ell \nu$ leptons
 is obtained, the semileptonic branching ratio is extracted
from the integral over the momentum distribution.

\subsubsection{Measurements of ${\cal{B}}_{sl}$ on the $\Upsilon(4S)$}

The momentum spectrum of electrons and muons from $B$
decays as measured by the CLEO II collaboration \cite{Bart1l}
is shown in Fig.~\ref{accmm}. It cuts off at the
kinematical limit around 2.4 GeV. Leptons from continuum
$e^+e^-$ annihilation and other background sources such as
$\psi \to \ell^+\ell^-$ decays have  been subtracted.
Corrections for final state radiation
have been applied to the electron spectrum following
the prescription by Atwood and Marciano \cite{radcorr}.
The electron spectrum is measured down to
momenta of 0.5 GeV while the muon detection system is fully
efficient at about 1.4 GeV. Using similar techniques,
measurements of the inclusive semileptonic branching
fraction have also been published by the
ARGUS\cite{ARGUS1l}, CRYSTAL BALL\cite{crystal1l},
CUSB\cite{CUSB1l}, and CLEO 1.5\cite{Hend1l} experiments.

The spectrum in Fig.~\ref{accmm} contains two components:
primary leptons from direct semileptonic decays ({\it {i.e.}}
$b \to  c\ell \nu$) and
leptons from cascade decays, $b\to cX,\; c \to s\ell \nu$.
The experimental challenge is to separate the two so that the
inclusive semileptonic branching fraction can be determined from the
direct component.
Several methods have been devised to accomplish this.
These include measurements of the inclusive single lepton
momentum spectrum together with a model of semileptonic
decays and measurements of dilepton events using charge
and angular correlations. Separate $B^-$ and
$\bar{B}^0$ branching ratios have been determined using events which contain
a lepton and a reconstructed B meson.

In the single lepton analyses, the primary difficulty
is separating the contributions
from direct semileptonic decay (i.e. $b\to c~l~\nu$) and cascade
semileptonic decays, $b \to c~X, ~c\to s~l~\nu$. If only leptons
above $1.4$ GeV are considered, then there is a negligible
contribution from
cascade decays, which have a soft spectrum, but a large extrapolation
to lower momenta is then required to obtain the
branching ratio.
If the full lepton momentum range is used, then the spectrum
after background subtraction must be
fitted to the sum of the two components.
The shape of the cascade component
is obtained by convoluting the measured $B\to D X$ momentum distribution
with the experimental $D\to l~Y$ spectrum\cite{Delco}.
% reference W. Bacino \etal (DELCO Collaboration),
% Phys. Rev. Lett 43, 1073 (1979). Summing exclusive
% measurements will give a better result.
The shape of the primary
spectrum is taken from a model of semileptonic B decay. Since the
two components are not orthogonal, the separation of background from
cascade decays introduces a significant model dependence in the
determination of the semileptonic branching ratio.

\begin{figure}[htb]
\begin{center}
\unitlength 1.0in
\begin{picture}(3.,2.5)(0,0)
\put(-1.4,-0.0)
{\psfig{bbllx=0pt,bblly=0pt,width=6.0in,height=6.0in,file=bexcl_lepyield.ps}}
\end{picture}
\vskip 25 mm
\caption{CLEO II measurement of the inclusive lepton yield from $B$ decays.
a) Fit to the lepton momentum spectrum for muons and electrons
using the ACCMM model.
b) Fit to the lepton momentum spectrum for muons and electrons
using the modified ISGW model
with the $D^{**}$ fraction allowed to float.}
\label{isgws}
\label{accmm}
\bigskip
\end{center}
\end{figure}

Two classes of models are used to parameterize the shape of the primary
$b\to c  \ell \nu$ contribution.
Parton models, such as the ACCMM model\cite{ACCM}, assume a Gaussian smearing
in momentum space for the parton in the meson, and leave the Fermi
momentum of the b quark and the charm quark  mass as free parameters
to be determined by the data.
The model of ISGW \cite{ISGW}
is an example of the second class. These models are
called exclusive models since here the $b\to c  \ell \nu$
transition is described  as the sum of $B\to D \ell \nu$,
$B\to D^* \ell \nu$ and
$B\to D^{**}  \ell \nu$ channels.
The form factors and rate for each channel are
calculated using a simple quark model\cite{ISGW}.

A fit to the lepton spectrum for muons and electrons from the CLEO II
experiment using the ACCMM model is shown in Fig~\ref{accmm}(a).
% CANT FIND THESE VALUES: TEB
%The two free parameter of the model, the Fermi momentum
%and the ratio of the $c$ quark to $b$ quark mass are
%determined to $230\pm **$ MeV and $0.34\pm **$, respectively.
%
The ACCMM model gives a good fit to the data while the  ISGW model
gives a  fit with a somewhat poor $\chi^2$/dof.
The CLEO~II experiment chooses to remedy this
defect of the exclusive model (as did CLEO~1.5) by fixing the ratio of
the vector to pseudoscalar
contributions and allowing the normalization of the $B\to D^{**} \ell \nu$
component to float in the fit. The result is a fit with an improved
$\chi^2/$dof and with the
fraction $B\to D^{**} \ell \nu$/$b\to c \ell\nu$ determined to be
$21.2\pm 1.6\pm 8.0\%$. This result
will be compared in section~\ref{exclbdstl}
to other methods of measuring the $B\to D^{**}\ell\nu$ branching fraction.

A new measurement technique using events with two leptons
was introduced by the ARGUS experiment\cite{Argus2l}
which significantly reduces the  model dependence associated with the
subtraction of the cascade component.
A high momentum lepton
is selected ($p_l>1.4$ GeV) which tags a primary decay. This primary
lepton is then combined with an additional lepton candidate which has
a momentum above $0.5$ GeV.
In the absence of mixing, if the second lepton has the
a sign opposite to the tagging lepton it is a primary lepton,
while if the second lepton has the same sign
as the tag it is a secondary lepton.
Since the threshold for muon detection is about $1.4$ GeV,
dielectrons are used in this technique.

A small background
from dileptons which originate from a single B can be removed by
using the angular correlation between the leptons.
By momentum conservation, dileptons from
the same B at the $\Upsilon(4S)$ will be approximately
back to back while dileptons from different B's
will be uncorrelated (Fig.~\ref{primsec} a).
A more refined angular correlation as a function
of momentum is used in the CLEO~II analysis\cite{Cleo2l}.

Including the effect of mixing gives the following relation between
the unlike and like sign spectra and the primary and secondary branching
fractions ($\displaystyle{{\cal B}(b)\over {dp}}$,
$\displaystyle{{\cal B}(c)\over {dp}}$),
$$ {d N(\ell^{\pm} e^{\mp}) \over dp} = N_\ell \epsilon_1 (p) \epsilon_2 (p)
[ {d {\cal B}(b) \over dp} (1- \chi) + {d {\cal B}(c) \over dp} \chi ]$$

$$ {d N(\ell^{\pm} e^{\pm}) \over dp} = N_\ell \epsilon_1 (p)
[ {d {\cal B}(b) \over dp} \chi + {d {\cal B}(c) \over dp} (1-\chi) ]$$
Here $\epsilon_1(p)$ is the efficiency of lepton identification,
$\epsilon_2 (p)$ is the efficiency of the angular correlation cut,
$\chi$ accounts for  $B-\bar{B}$ mixing.

Note that if the second lepton has the opposite charge of the
first one, it must be a primary lepton while
if second lepton has the same charge it is
a cascade lepton. By applying this method it is therefore possible to
determine the yield of primary and cascade leptons for each momentum bin.
\begin{figure}[htb]
\begin{center}
\unitlength 1.0in
\begin{picture}(3.,2.5)(0,0)
\put(-3.5,-1.3)
{\psfig{width=10.0in,height=12.5in,file=primsec_fig_new.ps}}
\end{picture}
\vskip 8 mm
\caption{Model independent analysis of
the dielectron momentum spectrum
from the CLEO~II experiment:
(a) Distribution of the angle between the two leptons in a Monte
Carlo simulation.
b) The electron momentum spectrum in data.
The contributions from primary (filled circles)
and secondary electrons (open circles) are shown separately.}
\label{primsec}
\end{center}
\end{figure}
The results of the CLEO~II analysis \cite{Cleo2l},
based on a data sample of 2.07 fb$^{-1}$ taken on the $\Upsilon(4S)$
and 0.99 fb$^{-1}$ taken just below the resonance,
are shown in Fig.~\ref{primsec}(b). The measured electron
momentum spectrum extends down to 600 MeV and there is
only a small extrapolation to  zero momentum. The unmeasured
part of the spectrum amounts to
only $5.8\pm 0.5$\% of the total semileptonic rate and hence the model
dependence is small. A
correction has to be applied for the small
contamination of cascade
leptons with momenta above 1.4 GeV ($\approx 2.8$\%).

Once the leptons from $B$ decays have been isolated using any
of the methods discussed above,
the semileptonic branching ratio is determined by integrating over the
$b \to c\ell \nu$ momentum spectrum. The results for the single lepton
measurements and for the dilepton analyses are given in
Tables~\ref{Tbbsl} and \ref{Tbbsll}.

CLEO~II finds
$$
{\cal{B}}_{sl} \; = \; (10.36 \, \pm \, 0.17 \, \pm \, 0.40)\, \%
$$
where the systematic error includes the uncertainties in the electron
identification efficiency, tracking efficiency,
and the $B^0\bar{B}^0$ mixing rate.
The small model dependence introduced when extrapolating
from $0.5$ GeV to zero momentum is determined by
comparing the results obtained using the ACCMM and
ISGW models and is  included in the quoted systematic error.

\begin{table}[htb]
\caption{Inclusive semileptonic branching ratios in [\%]
determined from an analysis of the yield of single leptons ($1 \ell$).}
\label{Tbbsl}
\begin{tabular}{llll}
Experiment & ACCMM  & ISGW & ISGW$^{**}$  \\ \hline
ARGUS (1 $\ell$)& $10.2\pm 0.5 \pm 0.2$ & $ 9.8\pm 0.5$ & \\
CRYSTAL BALL (1 $\ell$)& $12.0\pm 0.5\pm 0.7$ & $ 11.9\pm 0.4\pm 0.7$ & \\
CUSB~II (1 $\ell$)& $10.0\pm 0.4 \pm 0.3$ & $ 10.0\pm 0.4\pm 0.3$ & \\
CLEO 1.5 (1 $\ell$)& $10.5\pm 0.2 \pm 0.4$ & $ 9.9\pm 0.1\pm 0.4$ &
                                                  $11.2\pm 0.3\pm 0.4$ \\
CLEO~II (1 $\ell$)& $10.65\pm 0.05 \pm 0.33$ & $ 10.41\pm 0.07\pm 0.33$ &
                                            $10.87\pm 0.10 \pm 0.33$\\ \hline
Average (1 $\ell$)& $10.51\pm0.21$& $10.21\pm 0.20$& $10.98\pm 0.28$ \\
\end{tabular}
\end{table}
\begin{table}[htb]
\caption{Inclusive semileptonic branching ratios in [\%]
determined using dilepton events ($2 \ell$)
which has less statistical power but much reduced model dependence.}
\label{Tbbsll}
\begin{tabular}{ll}
Experiment & ${\cal{B}}_{sl}$\\ \hline
ARGUS (2 $\ell$)& $9.1\pm 0.5\pm 0.4$ \\
CLEO~II (2 $\ell$)& $10.36\pm 0.17\pm 0.40$ \\ \hline
Average (2 $\ell$)& $9.96\pm 0.36 $\\
\end{tabular}
\end{table}

The results obtained from the dilepton method
are consistent with the results obtained using the single
lepton technique and show that there is no large
systematic problem associated with the subtraction of
the cascade component.

Unlike the single lepton measurement, the measurement using the
dilepton technique does not require the assumption that the
$\Upsilon (4S)$ resonance always decays to pairs of B mesons.
The agreement between the CLEO~II
results for the dilepton analysis and the
single lepton result can also be used to constrain possible
non-$B\bar{B}$ decays of the $\Upsilon(4S)$.  The 95\% confidence
level upper limit on the fraction of these uncoventional
decays is 0.05~\cite{Cleo2l}.

The dilepton method also gives a measurement
of the cascade electron spectrum.
This can be compared to the ACCMM and ISGW models and the earlier
DELCO \cite{Delco} measurement of the $D$ semileptonic momentum
spectrum. CLEO~II finds
${\cal{B}}(b\to c\to se\nu)\, = \, (7.7 \pm 0.3 \pm 1.2)$\%
using the ACCMM model and $(8.3 \pm 0.3 \pm 1.2)$\%
for the ISGW model. Within errors, these results are consistent
with the expectations, and with the CLEO~II single lepton measurement.
% N. B. **** LEP ALSO MEASURES B->C->L ****

\subsubsection{The semileptonic branching fractions of the
$B^-$ and $\bar{B}^0$ mesons}

The semileptonic branching fractions reported so far are truly inclusive
in the sense that  no attempt is made to
distinguish different $B$ meson flavors.
Separate
semileptonic branching fractions for charged and neutral $B$ mesons,
${\cal B}(B^0\to X \ell^- \nu)$
and ${\cal B}(B^-\to X \ell^- \nu)$  have been determined
by measuring the lepton yield in events with fully or partly
reconstructed $B$ mesons.
Measurements of ${\cal B}(B^0\to X \ell^- \nu)$ were reported
by the CLEO~1.5 and ARGUS experiments\cite{Hend1l},\cite{argusdlmix}.
Simultaneous measurement of the separate branching fractions
has been accomplished by CLEO~II using its large sample of reconstructed
B mesons\cite{cleoiitptz}.
In the CLEO~II analysis
\cite{lambrecht},\cite{saulnier},
neutral B mesons are reconstructed using the modes
$\bar{B}^0\to D^{(*)+} \pi^- $,
$\bar{B}^0\to D^{(*)+} \rho^-$,
$\bar{B}^0\to D^{(*)+}  a_1^-$,
$\bar{B}^0\to
\psi K^{(*)0}$, and  partially reconstructed
$\bar{B}^0\to D^{*+} \ell^-\nu$ and $\bar{B}^0\to D^{*+}\pi^-$
yielding a total of $8456\pm 152$ $\bar{B}^0$ tags.
The modes
$B^- \to D^{(*)0} \pi^- $
$B^- \to D^{(*)0} \rho^-$
$B^- \to D^{(*)0}  a_1^-$,
and $B^-\to \psi K^{(*)-}$ are used to reconstruct
$834\pm 42$ charged B mesons.
\begin{table}[htb]
\caption{Measurements of the $B^0$ and $B^+$ Semileptonic Branching
Fractions [\%].}
\label{Tmbzbpsemi}
\begin{tabular}{lll}
Experiment& ${\cal B}(\bar{B}^0\to X \ell^- \nu)$ & ${\cal B}(B^-\to X \ell^-
\nu)$
 \\ \hline
CLEO~1.5 \cite{Hend1l}& $9.9\pm 3.0\pm 0.9$ & \\
ARGUS \cite{argusdlmix}& $9.3\pm 1.1\pm 1.15$ & \\
CLEO~II \cite{cleoiitptz}&
$10.9\pm 0.7\pm 1.1$ & $10.1\pm 1.8 \pm 1.4$\\ \hline
Average & $10.2 \pm 1.0$ & $10.1\pm 1.8\pm 1.4$
 \end{tabular}
\end{table}
The yield of
leptons above background for
 momenta above 1.4 GeV in the $\bar{B}^0$ and $B^-$
samples is extrapolated to zero momentum using the ISGW$^{**}$
model\cite{ISGW}.
A correction is then applied for $B-\bar{B}$ mixing. This
 leads to the measurements of
branching fractions given in Table~\ref{Tmbzbpsemi}.
These measurements confirm, albeit with lower statistical precision,
the other experimental indications that ${\cal B}(B\to X l\nu)< 12.5\%$, the
theoretical lower bound.

\subsubsection{Measurements of ${\cal{B}}_{sl}$ on the $Z^0$
Resonance}

The LEP experiments have determined the semileptonic $b$ branching
fraction using dilepton events. Single lepton events could be used if
Standard Model Z couplings were assumed. A $b$-enriched sample is
prepared by selecting events containing a lepton with large transverse
momentum, $p_{\perp}$. The semileptonic branching fraction
${\cal{B}}_{sl}$ is then extracted by simultaneously
fitting the lepton momentum and $p_{\perp}$ distributions. The shape of the
$B \to X\ell \nu$ spectrum is taken from the $\Upsilon(4S)$
measurements which causes the LEP results to suffer similar model
dependence.
The LEP results are summarized in Table~\ref{Tbsemilep}\cite{ALEPH1l},
\cite{OPAL1l},\cite{DELPHI1l},\cite{L31l}.
The second systematic error quoted in the individual measurements
is from model dependence. The average value
of ${\cal{B}}_{sl}\, =\, 11.3\pm0.3\pm0.4$\% is consistent with the
results from CLEO and ARGUS.
\begin{table}[htb]
\caption{
branching fractions($\%$) for inclusive semileptonic b decay from LEP.}
\label{Tbsemilep}
\begin{tabular}{ll}
Experiment & Branching Fraction \\ \hline
ALEPH & $11.40\pm 0.33\pm 0.37 \pm 0.20$ \\
OPAL & $10.5\pm 0.6\pm 0.4\pm 0.4$ \\
DELPHI & $11.41\pm 0.45 \pm 0.50 \pm 0.31$ \\
L3 & $11.73\pm 0.48 \pm 0.28\pm 0.31$ \\ \hline
LEP Average& $11.3\pm 0.3 \pm 0.4$ \\
\end{tabular}
\end{table}

Assuming the semileptonic decay width is the same
for all $b$ flavored hadrons,
the semileptonic branching ratio should be slightly different at LEP since
other $b$-particles are produced:
$$
{\cal{B}}_{sl}(\Upsilon(4S))\; = \; \frac{\Gamma_{sl}}{\Gamma_{tot}}
\; = \; \Gamma_{sl}\times\frac{(\tau_{B^+}+\tau_{B^0})}{2}
$$
while
$$
{\cal{B}}_{sl}(Z^0)\; = \; \Gamma_{sl}\times \tau_b
$$
Using the world averages for
lifetimes determined earlier this gives
\begin{eqnarray*}
{\cal{B}}_{sl}(Z^0) & = & \frac{2\tau_b}{(\tau_{B^+}+\tau_{B^0})}\times
{\cal{B}}_{sl}(\Upsilon(4S))\\
 & =&  9.77\pm 0.37\%
\end{eqnarray*}
Note that the contribution of other hadrons {\it reduces} the expected
average semileptonic branching fraction at the $Z^0$.
This prediction is
below the experimental average from LEP but the errors
are still too large to draw any significant conclusions.

\subsubsection{Measurement of $b\to X\tau \nu$}
\label{taunew}

The branching fraction for $B\to X \tau \nu^-$ has been measured
by several LEP experiments by using the $\tau \to {\rm hadron}~ \bar{\nu}$
decay mode and the large missing energy which is characteristic of
this decay mode\cite{ALEPHxtnu},\cite{ALEPHxtnu1},\cite{L3xtnu}.
After applying standard selection procedures for $Z \to b\bar{b}$ decays
events containing electrons or muons are rejected in order to remove
conventional semileptonic $B$ decays.
Comparing the remaining data with a detailed Monte Carlo simulation, which does
not include $B \to X\tau \nu$ decays, yields an excess from which the
branching fraction is determined to be
$$
{\cal{B}}(B\to X\tau\nu)\; = \; 2.75\pm 0.30 \pm 0.37\, \% \; \;, {\rm ~ALEPH}
$$
$$
{\cal{B}}(B\to X\tau\nu)\; = \; 2.4\pm 0.7 \pm 0.8\, \% \; \;, {\rm ~L3}
$$
These measurements are
consistent with the Standard Model expectation of
 $2.3\pm 0.25$\% \cite{falktaunu},\cite{seagal}.
The measurement of $B\to X~\tau~\nu$ imposes the constraint
$\tan \beta \, < \, 0.4\times m_H/$GeV at the 90\% confidence level on
(model II) charged Higgs which occur in various extensions of the
Standard Model including the
Minimal Supersymmetric Standard Model (MSSM) \cite{ALEPHxtnu1}.
The $B\to X\tau\nu$ mode is difficult to isolate at threshold experiments
because of the overlap with decay products from the second $B$ and
the low energy of the final state lepton/hadron from the $\tau$ decay.

\subsection{Exclusive Semileptonic Transitions.}
\label{exclbdstl}

The determination of Cabibbo-Kobayashi-Maskawa matrix elements is
one of the central experimental problems in heavy quark physics.
The
inclusive semileptonic branching ratio discussed in the previous section
can be used to determine the element $|V_{cb}|$ and the measurements
are now quite precise
with experimental uncertainty below the 5 \% level. However,
the conversion of the resulting
semileptonic width to $|V_{cb}|$ has a fairly large theoretical
uncertainty. Estimates of this uncertainty range from $5\%$ to $15\%$.
By contrast, it is possible that measurements of exclusive
semileptonic modes can be used to extract $|V_{cb}|$ with smaller
theoretical uncertainty. In the following sections we will summarize
the results obtained for  the decays $B\to D\ell \nu$, $B\to D^* \ell \nu$
and $B\to D^{**} \ell \nu$.

\subsubsection{Measurements of ${\cal{B}}(B\to D^*\ell \nu)$}

The mode $B\to D^* \ell \nu$ is preferred
experimentally to the mode $B\to D \ell \nu$ since the addition of the
$D^*$ constraint allows the isolation of a large and clean experimental
signal. ARGUS, CLEO~1.5, CLEO~II, and ALEPH have reported signals
 in $\bar{B^0} \to D^{*+} \ell^- \nu$ with $D^*\to D^0 \pi^+$
and $D^0 \to K^- \pi^+$.
CLEO~II can also observe the decay chain
$B^- \to D^{*0} \ell^- \nu$ with $D^* \to D^0 \pi^0$.
In threshold experiments,
the pion from the $D^*$ decay has a momentum below
 $225$ MeV in the laboratory
and is often referred to as the slow pion (denoted $\pi_s$). For the
lower portion of the momentum range accessible to the slow pion, the
large curvature of the track in the high magnetic field
and the multiple orbits of its trajectory complicate
track reconstruction.
It is possible to reconstruct slow neutral pions,
however, the combinatorial background is larger.
For high energy experiments, the charged slow pion is boosted so that
there is full acceptance for the entire momentum range.

In the decay of the $\Upsilon (4S)$ resonance,
B mesons are produced in pairs with momenta of about $330$ MeV.
 The signals at ARGUS and CLEO
are isolated using the kinematic constraints from production
at threshold. The effective mass of the neutrino in the decay
$B\to D^* \ell \nu$ is given by
$$m_{\nu}^2 = (E_B - E_{D^* l} )^2 - |p_{B}|^2 - |P_{D^* l}|^2 +
2 |p_{B}| |p_{D^* L}| \rm{\cos\Theta}$$
where $(E_{B}, p_{B})$ is the B meson 4-momentum,
($E_{D^{*} l}, p_{D^{*} l}$) is the sum of the $D^*$ and lepton 4-momenta,
and $\Theta$ is the angle between the 3-momenta $p_{D^* l}$ and $p_{B}$.
The first three terms in the expression for $m_{\nu}^2$ are the
missing mass squared, denoted $MM^2$.
The factor multiplying $\rm{\cos}\Theta$
will be denoted C.
Since the direction of the B momentum cannot be measured, a
common approximation is to set $|p_B|=0$ in the above expression
and then substitute the precisely known beam energy, $E_{beam}$ for $E_B$.
Then, the missing mass squared becomes
$$ MM^2 = (E_{beam} - E_{D^* l} )^2 -  |P_{D^* l}|^2 $$
and the signal will peak at $MM^2 = 0$ with a width determined by
the B momentum.
A variety of methods have been used to measure the signal yield. The
ARGUS experiment which first isolated a signal
in $\bar{B}^0\to D^{*+} l^- \nu$, measured the excess in the
background subtracted $MM^2$ distribution. Their fit allows
for contributions from $\bar{B}^0\to D^{*+}\ell^-\nu$
and $\bar{B}\to D^{**}\ell^-\nu$, $D^{**}\to D^{*+}(\pi)$.
The CLEO~1.5 experiment also used this method.

The CLEO~II experiment has chosen a different technique
to determine the $\bar{B}\to D^* \ell \nu$
branching fraction which uses slightly
more of the available information.
For lepton momenta above 1.4 GeV,
correctly reconstructed $B\to D^* \ell \nu$ decays
must lie in a triangular region in the plane of ${\rm MM}^2$ and C.
A cut on the $D^*-D$ mass difference is imposed, and the $D^0$ invariant
mass spectrum (shown in Figure~\ref{dstlnu1})
 is fitted to extract the number of B candidates.
The largest background is due to combinations of incorrectly
reconstructed $D^*$s and
real leptons. This background is subtracted using the sidebands of
the $D^*-D$ mass difference. There is also a small background from
uncorrelated combinations of
correctly reconstructed $D^*$s and leptons, which can be
estimated from data. A small correction for background from non-resonant
processes (continuum) and misidentified leptons is also included.
\begin{figure}[htb]
\begin{center}
\unitlength 1.0in
\begin{picture}(3.,3.0)(0,0)
\put(-0.31,0.0)
{\psfig{bbllx=0pt,bblly=0pt,width=3.3in,height=2.9in,file=bexcl_dstlnu1.ps}}
\end{picture}
\vskip 4 mm
\caption{CLEO II $D^0$ mass distributions for
(a) $\bar{B^0}\to D^{*+} \ell^- \nu$ candidates and
(b) $B^-\to D^{*0} \ell^- \nu$ candidates.}
\label{dstlnu1}
\end{center}
\end{figure}
The resulting signal
yield is due to $B\to D^* (X) \ell \nu$ events.
After removing all backgrounds, CLEO~II finds
$376\pm 27 \pm 16$ $\bar{B^0}\to D^{*+} \ell^- \nu$ events and
$302\pm 32 \pm 13$ $B^- \to D^{*0} \ell^- \nu$ events as shown in
Figure~\ref{dstlnu1}. This sample is also used to
evaluate $|V_{cb}|$ using the HQET inspired method (see section~\ref{vcbhqet}).

Larger event samples and significantly better statistical precision
can be obtained using a partial reconstruction technique as demonstrated
by the ARGUS analysis of $B\to D^{*+} \ell \nu$. In this case, only the
low momentum pion from the $D^{*+}$ decay and the lepton are detected.
The momentum of the undetected $D^0$ meson can be deduced from the
direction of the slow pion and kinematic constraints. The momentum
of the $D^*$ meson is approximately $\alpha p_{\pi}+\beta$, where
$\alpha$ and $\beta$ are constants which can be determined
from Monte Carlo (also see discussion
in section~\ref{bdbdbmix}). The signal yield
is determined using a modified
form of $MM^2$ with the estimated $D^*$ direction replacing the
measured $D^*$ direction in the expression above. The systematic error
from background subtraction, which is estimated using the wrong sign sample,
must be evaluated with care.

\begin{table}[htb]
\caption{Measurements of
branching fractions($\%$) for exclusive semileptonic B decay
with a $D$ or $D^*$ in the final state. The symbol
$\dagger$ indicates the branching ratio for this mode was measured using
a partial reconstruction technique. Due to the complexity of the analysis
procedure, those measurements marked with
a $^{*}$ cannot be renormalized to take into account the new values of
the $D$ and $D^*$ branching fractions.}
\label{Tbsemiexcl}
\begin{tabular}{lllll}
Mode & CLEO 1.5  & ARGUS & CLEO~II & ALEPH\\  \hline
$ \bar{B}^0\to D^{*+} \ell^- \nu $& $4.1\pm 0.5\pm 0.7$& $4.7\pm 0.6\pm 0.6$&
$4.49\pm 0.32\pm 0.39$& $5.36\pm 0.50\pm 0.76$\\
$ \bar{B}^0\to D^{*+} \ell^- \nu $& &
$4.5\pm 0.3\pm 0.4$ $^{\dagger}$&
                               &                       \\
${B}^- \to D^{*0}l^-\nu$& $4.1\pm 0.8^{+0.8}_{-0.9}$ $^{*}$&
$6.8\pm 1.6\pm 1.5$&
$5.13\pm 0.54\pm 0.64$& \\
$ \bar{B}^0\to D^{+} \ell^- \nu $& $1.8\pm 0.6\pm 0.3$ $^{*}$&
 $2.1\pm 0.7\pm 0.6$&                               & \\
$ \bar{B}^0\to D^{0} \ell^- \nu $& $1.6\pm 0.6\pm 0.3$ $^{*}$ &
$1.4\pm 0.6\pm 0.5$ $^{*}$ &
                              & \\
%$ \bar{B}\to D^{**} \ell^- \nu $&  & $2.7\pm 0.5\pm 0.5$& $<2.8\; $(95\% C.L.)
% & $1.5\pm **$ \\
%$ \bar{B}\to D^{**}(2420) \ell^- \nu $&  & &
% & $0.52\pm 0.15\pm 0.09$ \\
%$ \bar{B}\to D^{*+} \pi^- \ell^- \nu $&  & &
% & $****$ \\
\end{tabular}
\end{table}

By comparing the branching ratios for $\bar{B}^0\to D^{*+} \ell^- \nu$
and $B^- \to D^{*0} \ell^- \nu$ and using measurements of the ratio of
lifetimes from collider experiments, CLEO~II
obtains
the ratio of the production of $B^+ B^-$ and $B^0 \bar{B^0}$ meson
pairs at the $\Upsilon (4 S)$ resonance, $f_{+}/f_{0}=1.04\pm 0.13 (stat)
\pm 0.12 (sys) \pm 0.10$ (lifetime~ratio). This confirms
to an accuracy of about 15\% the initial
assumption that the production of charged and neutral B meson pairs
are equal. The small value of the $B^+ - B^0$ mass difference,
$0.2 \pm 0.3$ MeV, discussed in Section~\ref{bmss}
also supports this conclusion.

At LEP, the kinematic
constraints from production at threshold are not available. However,
the B direction can be measured from the vector between the production
point and the detached vertex from the B decay.
The decay products of the two B hadrons are cleanly separated into
jets. In addition, the
neutrino energy can be crudely determined from a missing energy measurement.
These features have allowed ALEPH to measure the $\bar{B^0}\to D^* l \nu$
branching fraction\cite{scott}.

The dominant systematic error in the threshold experiments
is due to the uncertainty in the slow
pion detection efficiencies while
ALEPH and the LEP experiments are limited by the sizeable uncertainty in the
number of B mesons produced in $Z^0$ decay which is used for the
normalization of the branching fraction ($\sim 8\%$).
The uncertainty from
the $D^{**}$ background is much smaller ($\sim 3\%$) \cite{scott}.

Since the publication of the CLEO 1.5 and ARGUS results, the $D$ and
$D^*$ branching ratios have changed significantly. Wherever possible,
the published values for semileptonic branching fractions have been rescaled
to  accommodate the new charm
branching fractions. For some results, which are marked with an asterix in
Table~\ref{Tbsemiexcl}, insufficient information
to perform the rescaling was provided in the
original papers. A separate systematic error for the contribution of
charm branching fraction is quoted in the world averages.

\begin{figure}[htb]
\begin{center}
\unitlength 1.0in
\begin{picture}(3.,3.0)(0,0)
\put(-0.51,0.0)
{\psfig{bbllx=0pt,bblly=0pt,width=4.5in,height=4.5in,file=aleph_dsstar.ps}}
\end{picture}
\vskip 10 mm
\caption{
Distributions from the ALEPH experiment for
 $\Delta m^*\; = \; m_{D^{*+}\pi^-} - m_{D^{*+}}$ for the right
sign (a)  and wrong sign (b) $D^*\pi$ combinations.
The right sign spectrum
is fitted to the sum of a Breit-Wigner signal and a background
shape.}
\label{aleph_dsstar}
\end{center}
\end{figure}

\subsubsection{Measurements of ${\cal{B}}(B\to D \ell \nu)$}

Measurements of the modes $B^+\to D^+ \ell^- \nu$ and
$\bar{B}^0\to D^0 \ell^- \nu$ are experimentally difficult
because of the  significant
combinatorial backgrounds in the
$D^+\to K^- \pi^+ \pi^+$ and $D^0\to K^- \pi^+$ signals
and the large backgrounds from the decay chain
$B\to D^*  \ell \nu$, $D^*\to D (\pi,\gamma)$
as well as $B\to D^{**} \ell \nu$, followed by $D^{**}\to D^* (\pi)$
or $D^*\to D (\pi,\gamma)$
where the $\pi$ (or $\gamma$)
from the $D^*$ decay is not reconstructed.
The separation of the latter two backgrounds, which yield
the same final state particles, is accomplished using their slightly
different behaviours as a function of lepton momentum
and D momentum.
In addition, the presence of an additional pion shifts
the center of the missing mass distribution.
In some analyses, additional constraints are provided
by requiring that the $D^{*(+,0)}  \ell \nu$ fraction be consistent with
the branching fraction from the dedicated exclusive measurement.

\subsubsection{Measurements of ${\cal{B}}(B\to D^{**}\ell \nu)$}

Semileptonic $B$ decay to orbitally excited
$D$ states has been searched for at LEP and by the
$\Upsilon (4S)$ experiments.
For experiments at threshold
the background from $B\to D^{**} \ell \nu \to D^* (\pi) \ell \nu$ (where the
$(\pi)$ from the $D^{**}$ decay
is not detected) is small and manageable.
The background from events with
$D^* \ell \nu$ and additional pion(s) in the final state peaks at higher
values of MM$^2$ and has a characteristically soft lepton momentum
spectrum. Thus, ARGUS and CLEO~1.5 found that this background could
be separated statistically.
ARGUS finds a sizeable signal for $B\to D^{**}\ell\nu$, $D^{**}\to D^*(\pi)$
which includes significant resonant $D_1(2420)\ell\nu$ and other
resonant modes as well as non-resonant channels.
In the recent CLEO~II analysis
of $\bar{B}\to D^*\ell\nu$,
the background  from final states with additional pions is determined
by examining $D^*$-lepton combinations
with $p$(lepton) in the range $0.8-1.4$ GeV in the portion of
C-$MM^2$ plane
which is preferentially populated by $B\to D^{**} X \ell \nu$ decays.
CLEO~II find a modest excess in this region which corresponds to
a model dependent upper
limit of ${\cal{B}}(B\to D^{**} X \ell \nu)< 2.8\% $ at the 95 \% confidence
level.

\begin{figure}[htb]
\begin{center}
\unitlength 1.0in
\begin{picture}(3.,3.5)(0,0)
\put(0.0,-0.5)
{\psfig{bbllx=0pt,bblly=0pt,width=4.0in,height=5.0in,file=opal_dstarstar_fig7.ps}}
\end{picture}
\caption{$D^{(*)}\pi \, - \, D^{(*)}$ mass difference distributions from OPAL
for a) $D^{*+}\pi^-$ combinations,
b) $D^{+}\pi^-$ combinations,
and c) $D^{0}\pi^-$ combinations.}
\label{dstarstar}
\end{center}
\end{figure}

The use of solid state vertex detectors has allowed the experiments
at LEP to isolate signals for $\bar{B}^0\to D^{*+} \pi^- X l \nu$
where the additional charged pion is vertexed with the slow pion from
the $D^*$ decay and the lepton. The ALEPH, OPAL, and DELPHI experiments
have reported signals\cite{carpinelli}; the OPAL results
are shown in Figure \ref{dstarstar}. The quantity plotted it the mass
difference between the $D^{(*)}\pi$ system and the charmed meson ($D^*$ or $D$)
which has a better experimental resolution than could be obtained
from the $D^{(*)}\pi$ invariant mass distribution.
Both ALEPH and OPAL have
reported measurements of the product
${\cal B}(b\to B)\times
{\cal B}(\bar{B}\to D(2420)X \ell^- \nu) \times
{\cal B}(D_1(2420)\to D^{*+}\pi^-)$ and quote
 ${\cal B}(\bar{B}\to D(2420)X \ell^- \nu)$ assuming ${\cal B}(b\to B)=0.37$
and ${\cal B}(D_1(2420)\to D^{*+} \pi^-)\sim 0.67$. In addition,
OPAL has reported a significant signal for
${\cal B}(\bar{B}\to D_2^*(2470) X \ell^- \nu)$ using
$D_2^* (2470)\to D^0 \pi^+$.
These signals may include $X$, one or more undetected pions.
ALEPH has also performed an inclusive
topological measurement which is sensitive to
$\bar{B}\to D^{*+}\pi^-\ell^- \nu$ in which the $D^{*+}\pi^-$ system is
non-resonant.

The results on semileptonic $B\to D^{**}$ transitions are given in
Table~\ref{Tbsemiexcit}. The LEP measurements and the CLEO~II upper limit
are marginally consistent.
The corresponding world averages can be found in
Table~\ref{Tbsemiwa}.

These measurements of the production of orbitally excited $D$
mesons in semileptonic decay show that the Shifman-Voloshin (SV) limit,
in which the $B\to D\ell\nu$ and
$B\to D^*\ell\nu$ channels saturate the semileptonic
width, is not achieved\cite{svlimit}.
Quark models also predict small rates
for the production of p-wave $D$ mesons. For example, the ISGW2
model predicts that about 7\% of the total semileptonic rate will
occur in channels with excited charmed mesons and ${\cal B}(\bar{B}\to
D_1(2420) \ell\nu)$ $\sim 0.25\%$, well below
the observed rate\cite{ISGWprime}.
Predictions from the model of Colangelo \etal ~also give rates in
the $0.1\%$ range for the largest channels\cite{colangelo}.

\begin{table}[htb]
\caption{Measurements of
branching fractions($\%$) for exclusive semileptonic B decay
with p-wave charmed mesons in the final state.
The signal may include $X$, one or more undetected pions.}
\label{Tbsemiexcit}
\begin{tabular}{lllll}
Mode & ARGUS & CLEO~II & ALEPH & OPAL \\  \hline
$ B^-\to D_1^0(2420) \ell^- X \nu $&  & &
$0.84\pm 0.24\pm 0.14$ &$0.81\pm 0.20\pm 0.19$  \\
$ B^-\to D_2^{*0}(2460) \ell^- X \nu $&  & &
& $0.35\pm 0.14\pm 0.17$  \\
$ \bar{B^0}\to D_1^+(2430) \ell^- X \nu $&  & &
&$0.78\pm 0.28\pm 0.18$  \\
$ \bar{B^0}\to D_2^{*+}(2470) \ell^- X \nu $&  & &
&$0.90\pm 0.27\pm 0.21$  \\  \hline
$ \bar{B}\to D^{**}(2420) \ell^- X \nu $&  & &
$0.84\pm 0.24\pm 0.14$ & $0.80\pm 0.24$ \\
$ \bar{B}\to D^{**}(2460) \ell^- X \nu $&  & &
 & $0.44\pm 0.14$ \\
$ \bar{B}\to D^{*+} \pi^- \ell^- X \nu $&  & &
$1.08\pm 0.3\pm 0.22$ &  \\
$ \bar{B}\to D^{**} \ell^- \nu $&  $2.9\pm 0.5\pm 0.5$& $<2.8\%$(95\% C.L.)
& &       \\
\end{tabular}
\end{table}

% Update TEB
\begin{table}[htb]
\caption{World average
branching fractions($\%$) for exclusive semileptonic B decay.
The modes marked with the symbol $\dagger$ are not included
in the sum of exclusive modes as discussed in the text.}
\label{Tbsemiwa}
\begin{tabular}{ll}
Mode & World Average \\ \hline
$ \bar{B}^0\to D^{*+} \ell^- \nu $& $4.56\pm 0.27\pm 0.25$ \\
${B}^- \to D^{*0}l^-\nu$& $5.31\pm 0.70\pm 0.41$ \\
$ \bar{B}^0\to D^{+} \ell^- \nu $& $2.1\pm 0.9\pm 0.14$ \\
$ B^-\to D^{0} \ell^- \nu $& $1.5\pm 0.5$ $^{\dagger}$ \\
$ \bar{B}\to D^{**} \ell^- \nu $&  $2.9\pm 0.5\pm 0.5$ $\dagger$\\
$ \bar{B}\to D^{**}(2420) \ell^- X \nu $&  $0.82\pm 0.18\pm 0.06$ \\
$ \bar{B}\to D^{**}(2460) \ell^- X \nu $&  $0.44\pm 0.14\pm 0.03$ \\ \hline
Sum of exclusive semileptonic ${\cal{B}}$ &  $8.05 \pm 1.7$\\
Inclusive semileptonic  ${\cal{B}}$ & $10.98\pm 0.28$ \\
\end{tabular}
\end{table}

\subsubsection{Summary of exclusive semileptonic $b\to c$ measurements}

Additional exclusive semileptonic channels
have been searched for by the ARGUS \cite{ARGUSplnu}
and L3 collaborations \cite{L3xtnu}.
These include modes which require either $s \bar{s}$ popping
at the lower part of the spectator graph\cite{ARGUSssbar} or baryon
production\cite{ARGUSplnu}.
\begin{table}[htb]
\caption{Measurements of Other Inclusive Semileptonic Branching Ratios.}
\label{Tbslother}
\begin{tabular}{lll}
Experiment & Mode  & B(\%)   \\ \hline
ARGUS   & $B\to D_s X \ell^+ \nu$ & $<0.9 (90\% ~{\rm C.L.})$ \\
ARGUS   & $B\to \bar{p} X \ell^+ \nu$ & $<0.16 (90\%~{\rm C.L.})$ \\
L3      & $B\to X \nu$ &  $2.27\pm 0.8\pm 1.5$ \\
\end{tabular}
\end{table}
No signal was
observed and the upper limits are summarized in Table~\ref{Tbslother}.

In Table~\ref{Tbsemiwa} we sum the exclusive semileptonic
modes and find a value consistent with
the inclusive semileptonic branching ratio.
The mode $B^- \to D^0 \ell^-\nu$ is omitted
since the measurements of this mode cannot
be adjusted to account for the changes
in $D$ and $D^*$ branching fractions.
The measurement of the mode $B\to D^{**} \ell\nu$ is also omitted
from the calculation of the sum of the exclusive modes
and instead the measurements of $B\to D^{**}(2420)\ell\nu$
and $B\to D^{**}(2460)\ell\nu$ are used. If the other possible approach
is taken and $B\to D^{**}\ell\nu$ is used but
$B\to D^{**}(2420)\ell\nu$ and $B\to D^{**}(2460)\ell\nu$ are omitted,
then the sum of the exclusive modes becomes $9.4\pm 1.2$. The conclusion
is the same using either approach.

To determine ${\cal B}(B\to D^{**} \ell \nu)$, ARGUS and CLEO~II
have assumed that the model of ISGW correctly predicts the relative
fractions of $B\to D_0^1(2420)\ell\nu$, $B\to D_2^*(2460)\ell\nu$ and the
other excited charm mesons. This procedure can be applied to the
world average for ${\cal B}(B\to D^{**}(2420) \ell \nu)$
which is the most precisely known of these rates for the semileptonic decay
to an excited charmed meson, this gives
${\cal B}(B\to D^{**} \ell \nu) = 1.64\pm 0.22\pm \pm 0.11\%$ which is
consistent with the ISGW$^{**}$
fit to the inclusive semileptonic spectrum which
gives ${\cal B}(B\to D^{**} \ell \nu) = 2.3\pm 0.9$.

Model dependent
extrapolations from the rate of
observed $D^{**} \ell \nu$, $D^{**}\to D^* \pi$ decays
appear to saturate the remainder of the missing portion
of the semileptonic rate. The other remaining decays may correspond to
$B\to D^{**}  \ell \nu$ where $D^{**}$ denotes a p-wave charmed
meson with a large width (e.g. the very broad but as of
now unobserved $1^{3}P_1$(2490) and $1^{3}P_0$(2440) states).
It is also possible that the other missing decays
are $B\to D \pi \ell^- \nu$ where the $D\pi$ system is non-resonant or
originates from the decay of a broad excited charm meson.
These possibilities are difficult to check experimentally.

\subsection{The Dynamics of Semileptonic $B$ Decay}
\label{bdynamics}

Since leptons are not sensitive to the strong interaction,
the amplitude for a semileptonic $B$ decay can be factorized
into two parts, a leptonic and a hadronic current. The leptonic
factor can be calculated exactly while the hadronic part
is parameterized by form factors. A simple example is the
transition $B\to D l \nu$. The differential decay rate in
this case is given by
$$ {{d \Gamma} \over {d q^2}} = {G^2 \over {24\pi^3}} |V_{cb}^2| P_D^3
f_{+}^2 (q^2)$$
where $q^2$ is the mass of the virtual W $(\ell \nu)$ and
$f_{+}(q^2)$ is the single vector form factor.
The form factor which describes semileptonic decay
 is analogous to the form factor which arises
in electron-nucleon scattering. In this case, the form factor
$|f_{+}(q^2)|^2$ gives the probability that the final state
quarks will form a $D$ meson.

The form factor is largest when the initial and final state
heavy quarks have the smallest relative velocities, that is
at maximum $q^2$. As $q^2$ decreases and the momentum transfer
increases, the modulus of the form factor decreases.

Since the $\bar{B^0}\to D^{*+} \ell^- \nu$ mode has a large
branching ratio and good signal to background ratio, it is
experimentally preferred to $\bar{B}\to D\ell \nu$ for form
factor studies. Moreover, the theoretical predictions for this
mode are thought to be reliable.
The corresponding expression
for the differential rate in
$B\to D^* \ell \nu$ is given in section~\ref{dstlnuffexp}.
In this case, there
are three form factors which correspond to the
three possible partial waves of the $B\to D^{*} \bar{W}^{'}$ system
(here $\bar{W}^{'}$ is the virtual W which becomes the lepton-antineutrino
pair).
% *** Kutschke says this is wrong **** ???
%The two vector particles may have relative
%angular momentum and combine in a s, p, or d wave.

\subsubsection{Polarization in $B\to D^* \ell \nu$ decays}
In the past, insufficient data was available to perform
a measurement of the individual form factors in $B\to D^* \ell \nu$
decay. Various integrated quantities which give information
on the form factors were determined.

For example, the polarization $\alpha$ in $B\to D^* \ell\nu$
can be determined by fitting the $D^*$ helicity angle distribution
which should be distributed as $1 + \alpha \cos^2\theta_{D^*}$.
It is also possible to measure
the forward-backward asymmetry of the
lepton in the W rest frame which gives information on the ratio of
the positive and negative helicity amplitudes.
\begin{table}[htb]
\caption{Measurements of integrated observables in
$B\to D^* \ell \nu$  Decays.}
\label{Tbinteg}
\begin{tabular}{lll}
Experiment & $\alpha$ & $A_{FB}$ \\ \hline
CLEO~1.5 & $0.65\pm 0.66\pm 0.25$ & \\
ARGUS & $1.1\pm 0.4\pm 0.2$ & $0.20\pm 0.08\pm 0.06$\\
CLEO~II & $1.48\pm 0.32\pm 0.14$ & $0.209\pm 0.034\pm 0.015$\\ \hline
World average & $1.24\pm 0.25$   & $0.208\pm  0.035$   \\
\end{tabular}
\end{table}

These results can be compared, for example, to the prediction of
Scora of $\alpha=1.32$ in the ISGW$^{'}$ model\cite{ISGWprime}
and to the HQET based predictions of Neubert $\alpha=1.37$ and
$A_{FB}=0.22$. The agreement of $A_{FB}$ in sign and magnitude
with quark model predictions has been used to deduce limits on
a hypothetical $V+A$ coupling of the b quark\cite{Sanghera}.

Other tests of models are provided by the vector to pseudoscalar
ratio which is, for example, predicted to be $2.6$ in the ISGW$^{'}$
model and $2.79$ in the HQET based model of Neubert.
Experimentally, this is found to be $2.17\pm 0.93$ where
the large error reflects the poor precision of the $\bar{B^0}\to
D^+ \ell^-\nu$ branching fraction.

The measurements of integrated observables are thus in
good agreement with models.
Form factor measurements are a  more sophisticated approach and
provide better discrimination between models. In addition, all
the available information is used and hence the statistical precision
is improved.

\begin{figure}[htb]
\begin{center}
\unitlength 1.0in
\begin{picture}(3.,2.5)(0,0)
\put(-0.01,0.0)
{\psfig{bbllx=0pt,bblly=0pt,width=3.0in,height=2.5in,file=angles_dstlnu.ps}}
\end{picture}
\vskip 10 mm
\caption{The kinematic variables used in the
CLEO~II $B\to D^* \ell \nu$ form factor analysis.}
\label{dstlnuang}
\end{center}
\end{figure}

\subsubsection{Measurement of the $B\to D^* \ell \nu$ Form Factors}
\label{dstlnuffexp}

The differential decay rate for $B\to D^* \ell \nu$ can be expressed in
terms of three $q^2$-dependent helicity amplitudes
$H_{\pm}(q^2)$ and $H_0(q^2),$
where the subscripts refer to the helicity of either
the virtual $W$ ($\ell\nu$)
or the $D^*$\cite{jdr},\cite{ks89,hagiwara,gilman,ks90}. The rate is given by
\begin{eqnarray}
{d\Gamma\over dq^2\, d\cos{\theta_{\ell\nu}}\, d\cos{\theta_V}\, d\chi}&=&
{3G_F^2|V_{cb}|^2\,P_{D^*}\,q^2\over 8(4\pi)^4 {m_B^2}}\times\nonumber\\
&&\{[(1-\cos{\theta_{\ell\nu}})^2|H_{+}(q^2)|^2+
(1+\cos{\theta_{\ell\nu}})^2|H_{-}(q^2)|^2]\sin^{2}\theta_V \nonumber\\
&&+4\sin^2\theta_{\ell\nu}\cos^2\theta_V|H_{0}(q^2)|^2\nonumber\\
&&-2\sin^2\theta_{\ell\nu}\sin^2\theta_V
\cos(2\chi)H_{+}(q^2)H_{-}(q^2)\nonumber\\
&&-4\sin\theta_{\ell\nu}(1-\cos\theta_{\ell\nu})
\sin\theta_V\cos\theta_V\cos\chi\,
H_{+}(q^2)H_{0}(q^2)\nonumber\\
&&+4\sin\theta_{\ell\nu}(1+\cos\theta_{\ell\nu})
\sin\theta_V\cos\theta_V\cos\chi\,
H_{-}(q^2)H_{0}(q^2)\},\nonumber\\
\label{eqn:decayrate}
\end{eqnarray}
where $m_B$
is the mass of $B$ meson,
$P_{D^*}$ is the momentum of the $D^*$
and is a function of $q^2$,
%and
%the factor
%$\eta$=+1 ($\eta=-1$) applies to $B$ decays ($D$ decays for $\dtokstlnu$).
and the angles $\theta_{\ell \nu}$, $\theta_V$, and $\chi$ are defined in
Fig.~\ref{dstlnuang}.
%Note that $\theta_{l\nu}$ is defined with
%respect to the direction of the virtual $W$; this accounts for the sign
%differences between our formula and
%certain others in the literature~\cite{KS}.
The helicity
amplitudes $H_{\pm}$ and $H_{0}$ can be expressed
in terms of two axial-vector form factors, $A_1 (q^2)$ and $A_2 (q^2)$,
and a vector form factor $V(q^2):$

\begin{eqnarray}
H_{\pm}(q^2)&=&(m_B+m_{D^*})A_1(q^2)\mp{2m_B\,P_{D^*}\over(m_B+m_{D^*})}
V(q^2)\nonumber\\
H_{0}(q^2)&=&{1\over 2m_{D^*}\sqrt{q^2}}\Big[(m_B^2-m_{D^*}^2-q^2)
(m_B+m_{D^*})A_1(q^2)
-{4m_B^2\,P_{D^*}^2\over(m_B+m_{D^*})}A_2(q^2)\Big],\nonumber\\
\label{eqn:helamps}
\end{eqnarray}
\noindent
where $m_{D^*}$ is the mass of the $D^*$ meson.

There are a number of important and simple qualitative features that
are present in equation~\ref{eqn:decayrate}. These are most
easily seen after integrating over the variable $\chi$,
so that the last three terms vanish.
 The helicity zero (longitudinal) component then
has a $\cos^2\theta_{D^*}$ and a $\sin^2\theta_{\ell\nu}$
dependence.  The negative and
positive helicity components have a $\sin^2\theta_{D^*}$ dependence
 as well as a $(1+\cos\theta_{\ell\nu})^2$
and $(1-\cos\theta_{\ell\nu})^2$ behaviour, respectively.

As a result of the $V-A$ coupling of the virtual
$W$, the difference between
the positive and negative helicity amplitudes,
$H_+ - H_-$, is large and negative. This feature can be
clearly observed in the scatter plot of
$\chi$ versus $\cos\theta_{D^*}$\cite{jdr}.

Other intuitive and useful features (see previous section)
can be deduced from consideration
of integrated quantities. For instance, $A_{FB}$ is controlled
primarily by the form factor ratio
$R_2 = V(q^2=q^2_{max})/A_1(q^2=q^2_{max})$ and the
$\cos\theta_{\ell\nu}$ distribution.
The  size of the second independent form factor ratio,
$R_1 = A_2(q^2=q^2_{max})/A_1(q^2=q^2_{max})$
is determined by the $\cos\theta_{D^*}$ distribution
and to a fair extent by $\alpha$, the degree of polarization.

\begin{figure}[htb]
\begin{center}
\unitlength 1.0in
\begin{picture}(4.,4.0)(0,0)
\put(-0.5,-0.1)
{\psfig{bbllx=0pt,bblly=0pt,width=4.0in,height=4.0in,file=bexcl_ff_fit.ps}}
\end{picture}
\vskip 10 mm
\caption{The experimental distributions of kinematic variables in data
compared to the
fit in the CLEO~II $B\to D^* \ell \nu$ form factor analysis for: (a)
$\cos\theta_{D^*}$ (b) $\cos\theta_{\ell\nu}$ (c) $q^2$ (d) $\chi$}
\label{dstlnu3}
\end{center}
\end{figure}

Using the measured values of
$ q^2 /q_{max}^2$, $\cos\Theta_{D^*}$, $\cos\Theta_{\ell \nu}$ and $\chi$,
a 4-dimensional unbinned maximum likelihood fit was performed
using a Monte Carlo integration technique in a manner similar
to reference~\cite{E691}.
This technique allows a multi-dimensional
likelihood fit to be performed to variables modified by experimental
acceptance and resolution, and is necessary due to the
substantial smearing of the kinematic variables from the motion of
the B meson.
The basis of the method
 is to determine the probability density function by
using the population of appropriately weighted MC
events in the four dimensional kinematic space.
This is accomplished by generating one  high statistics sample
of MC events with a known value of the form factor ratios $R_1, R_2$
and corresponding known values of the four kinematic
variables $q^2/q^2_{max}$, $\cos\Theta_{D^*}$, $\cos\Theta_W$, and
$\chi$ for each
event. The generated
events are then processed through the full detector simulation and
analysis chain.
Using the generated kinematic
variables, the accepted MC events are weighted by the ratio of the
decay distribution for the trial values of $R_1, R_2$  to that of the
generated distribution.
The accepted MC events are now, therefore,
distributed according to the
probability density corresponding
to the trial values of $R_1, R_2$.
By such weighting, a likelihood may be evaluated
for each data event for different values of the form factor ratios, and a fit
can be
performed.  The probability for each event is determined by sampling this
distribution using a search volume around each data point.
The volume size is chosen so that the systematic effect
from finite search volumes is small and the required number of MC
events is not prohibitively high\cite{rydff},\cite{tingff}.

The results of such a measurement from CLEO~II
are given in Table~\ref{Tmffsemi}.
The measurement of the form factor ratios from CLEO~II was obtained using
a larger dataset and an improved analysis technique and supersedes the
result of Ref~\cite{Sanghera}.

In the limit of pure HQET, the form factor ratios are both unity.
Including $O(\bar{\Lambda}/m_c)$ corrections gives 1.3 and 0.8 for
$R_1$ and $R_2$ respectively.
The experimental precision on the form factor ratios
is not sufficient to distinguish between these two possibilities
and the models, which are listed in Table~\ref{Tbffpred}.
However, the experimental results
do indicate that deviations from the limit of heavy quark
symmetry are not large.

For the purposes of comparison
between data and models, it should be
noted that several of the quark models quoted in Table~\ref{Tbffpred}
use a different $q^2$ dependence for their form factors
than is assumed by HQET and by the CLEO~II measurement.
This will be not a large effect given the small range in $q^2$
available in the reaction $B\to D^* l \nu$.
For instance, the value of $A_2/A_1$ in HQET
varies from $1.35$ to $1.27$ over the full kinematic range.

\begin{table}[htb]
\caption{Measurements of the form factor ratios in
$B\to D^* \ell \nu$  decays at $q^2=q^2_{max}$.}
\label{Tmffsemi}
\begin{tabular}{lccc}
Experiment& $R_1$ & $R_2$ & $\rho^2$
 \\ \hline
CLEO~II & $1.30\pm 0.36\pm 0.16$ & $0.64\pm 0.26 \pm 0.12$ &
 $1.01\pm 0.15\pm 0.09$ \\
 \end{tabular}
\end{table}

\begin{table}[htb]
\caption{Predictions for the form factor ratios from theoretical
models at $q^2=q^2_{max}$.}
\label{Tbffpred}
\begin{tabular}{lll}
Model & $R_1$ & $R_2$ \\ \hline
HQET (Neubert)\cite{neubert1} & $1.35$ & $0.79$ \\
Ball\cite{Ballff} & $1.31$ & $0.95$ \\
ISGW\cite{ISGW} & $1.01$ & $0.91$ \\
BSW\cite{WSB}  & $0.91$ & $0.85$ \\
KS\cite{KS}   & $1.09$ & $1.09$ \\
\end{tabular}
\end{table}

\subsection{Determination of $|V_{cb}|$}
\label{brsl_vcb}

\subsubsection{$|V_{cb}|$ from inclusive measurements}

The theoretical uncertainty in the determination
of $|V_{cb}|$ is currently a matter of active discussion
and no clear consensus has emerged.
The values of $|V_{cb}|$ determined from the world average
for the inclusive
semileptonic branching fraction are given in
Table~\ref{Tbvcbincl} for different theoretical models. The models
predict the decay width in the form
$\Gamma \; = \; \gamma_c\cdot |V_{cb}|^2$.
The value of $|V_{cb}|$ is then obtained from experiment using
$$
|V_{cb}|^2 \; = \; {\cal{B}}(B \to X\ell \nu)/(\tau_B \cdot \gamma_c)
$$
One way of reducing the theoretical uncertainty associated
with the $m_b^5$ dependence of the semileptonic width
was introduced by Altarelli~\etal \cite{ACCM}. In their
quark model, the spectator quark is assigned a Fermi momentum
$p_F$ and has a Gaussian momentum distribution.
Each value of $p_F$ gives a slightly different value of $m_b$,
however, the average value of $p_F$ and $m_b$
as well as the effective
spectator quark mass $m_{sp}$ can be determined by fitting
the shape of the lepton momentum spectrum. The relationship
$$m_b^2 = m_B^2 + m_{sp}^2 - 2 m_B\sqrt{p_F^2 + m_{sp}^2}$$
where $m_B$ is the B meson mass,
allows the experimental data to be used to constrain $m_b$.

Shifman, Uraltsev, and Vainshtein propose that
the dependence of the width
in the SV (Shifman-Voloshin)\cite{svlimit}
limit is proportional to $m_b-m_c$ rather than to $m_b^5$.
This will substantially reduce the uncertainty in
the extraction of $|V_{cb}|$ from inclusive decays.
Using $m_b=(4.8\pm 0.1)$~GeV from a QCD sum rule
analysis of the $\Upsilon$ system,
Shifman \etal ~find on this basis
that the theoretical uncertainty in
the determination of $|V_{cb}|$ is less than 5$\%$ and
nearly model independent.
In contrast, Neubert asserts that the model dependence
in $|V_{cb}|$ is of the order of 10\% due to the unknown higher
order corrections in the expansion
for the semileptonic width in $\alpha_s (m_Q)$, where
$m_Q$ is the mass of the heavy quark\cite{neubert2},\cite{neubert3}.
The experimental fact that $b\to c \ell\nu$ transitions
are far from the SV limit
may also affect the reliability of the claim by Shifman \etal.
Luke and Savage have also
investigated the model dependence of $|V_{cb}|$
in a HQET framework and conclude
that the determination from the inclusive measurements gives
values of $|V_{cb}|$ in the range $0.037-0.052$ which corresponds
to an uncertainty of order $14\%$.
Using a similar approach with a constraint on $m_c$ obtained from
the experimental determination of ${\cal B}(D\to X \ell \nu)$,
Ball and Nierste find a significantly
larger value $\gamma_c=54.2\pm 5$ which gives a
somewhat lower value of $|V_{cb}|$\cite{BallNierste}.
% PRD 50, 5841, 1994

\begin{table}[htb]
\caption{ Determinations of $V_{cb}$ using inclusive
semileptonic decays.}
\label{Tbvcbincl}
\begin{tabular}{lll}
Theorist &$\gamma_c$ &$|V_{cb}|$ \\ \hline
ACCMM \etal\cite{ACCM}
 &$40\pm8$&  $0.0401\pm 0.001(\rm{exp})\pm .004(\rm{theor})$ \\
Shifman \etal\cite{shifman1} &$41.3\pm 4$&
$0.03965\pm 0.001(\rm{exp})\pm .002(\rm{theor})$\\
ISGW$^{**}$\cite{ISGW}
  & $42\pm 8$& $0.0400\pm 0.001(\rm{exp}) \pm .004(\rm{theor})$\\
Ball and Nierste\cite{BallNierste} & $54.2\pm 5$ &
$0.0344\pm 0.001(\rm{exp})\pm .002(\rm{theor})$\\
\end{tabular}
\end{table}

\subsubsection{$|V_{cb}|$ from exclusive measurements}

Using measurements of the $B^+$ and $B^0$ lifetimes, and the
assumption of isospin invariance, the $\bar{B^0}\to D^{*+} \ell^- \nu$
and $B^-\to D^{*0} \ell^- \nu$ branching
fraction measurements can be combined to obtain the width
$\Gamma (B\to D^* \ell \nu) = (30.2 \pm 2.6 \pm 1.0)$ ns $^{-1}$
which is independent of the ratio of production fractions $f_{+}/f_0$.
To allow the results to be rescaled easily,
the contribution due to the uncertainty in the
average $B$ meson lifetime is separated in the error.
This determination of the width can then
be translated into a value for $|V_{cb}|$
by using models.
The values  obtained with the  models
of Isgur, Scora,
Grinstein and Wise (ISGW)\cite{ISGW}, Bauer, Stech and Wirbel (BSW)\cite{WSB},
K\"orner and Schuler (KS)\cite{KS},
and Neubert are listed in Table~\ref{Tvcbexcl1}.
\begin{table}[htb]
\caption{Values of $|V_{cb}|$ using
the world average for $\Gamma (B\to D^* \ell \nu)$
and theoretical models. The first error is the sum in
quadrature of the experimental statistical and systematic errors.
The second error is from the $B$ lifetime.}
\label{Tvcbexcl1}
\begin{tabular}{ll}
Model & $|V_{cb}|$ \\ \hline
ISGW & $ 0.0349\pm 0.0015\pm 0.0006$\\
ISGW$^{'}$ & $ 0.0347\pm 0.0015\pm 0.0006$\\
BSW & $ 0.0371\pm 0.0016\pm 0.0006$\\
KS & $ 0.0342\pm 0.0015\pm 0.0006$ \\
Neubert & $0.0323\pm 0.0014\pm 0.0006$
\end{tabular}
\end{table}
In principle,
the detection efficiencies should be determined separately
for each model. This is a small effect and the CLEO~II analysis
finds the systematic variation from this source to be less than 3\%.

In Table~\ref{Tvcbexcl2},
we also give the values of $|V_{cb}|$ determined from exclusive
models
using the ARGUS measurement of ${\cal B}(\bar{B}^0\to D^+ \ell^- \nu)$
which gives $\Gamma(\bar{B}^0\to D^+ \ell^- \nu)=
12.3\pm5.7\pm3.5$~ns$^{-1}$.
The first error is the sum in quadrature of the statistical and
experimental systematic errors. The second error is the uncertainty
due to the average $B$ meson lifetime. We note that the other measurements
of $B\to D\ell \nu$ branching fractions
should not be used as they cannot be modified to
account for the changes in the $D$ and $D^*$ branching fractions.
Improved measurements of the branching fractions
for $B^-\to D^0 \ell^- \nu$ and $ \bar{B}^0\to D^+ \ell^- \nu$ will
be useful in testing models and determining $|V_{cb}|$.
\begin{table}[htb]
\caption{Values of $|V_{cb}|$ using
the world average for $\Gamma (B\to D^+ \ell \nu)$
and theoretical models. The first error is the sum in
quadrature of the experimental statistical and systematic errors.
The second error is from the $B$ lifetime.}
\label{Tvcbexcl2}
\begin{tabular}{ll}
Model & $|V_{cb}|$ \\ \hline
ISGW$^{'}$ & $ 0.032\pm 0.008\pm 0.0005$\\
BSW & $ 0.038\pm 0.009\pm 0.0006$\\
KS & $ 0.039\pm 0.009\pm 0.0006$ \\
\end{tabular}
\end{table}

The results in Tables~\ref{Tvcbexcl1},~\ref{Tvcbexcl2}
show that the model dependence
in the determination of $|V_{cb}|$ from the total rate is below the
$10\%$ level. From the measurement of the branching fraction
for $\bar{B}\to D^* \ell\nu$, using the ISGW$^{'}$ model
to obtain the central value, gives
$$ |V_{cb}|= 0.0347\pm 0.0016({\rm exp})\pm 0.0024({\rm model})$$

This method of obtaining $|V_{cb}|$ from the total rate
has the distinct advantage that the models used
make other detailed predictions for form factors
and various other observables
which can be experimentally verified (see section~\ref{bdynamics}).
In addition, all of the data
can be used unlike the HQET inspired method (discussed below) which
is valid only for a certain kinematic regime (near zero recoil).

\subsubsection{Determination of $|V_{cb}|$ using HQET}
\label{vcbhqet}

It has recently been appreciated that there is a symmetry of QCD
that is useful in understanding systems containing one heavy quark.
This symmetry arises when the quark becomes sufficiently heavy
to make its mass irrelevant to the nonperturbative dynamics of the
light quarks. This allows the heavy quark degrees of freedom to
be treated in isolation from the the light quark degrees of freedom.
This is analogous to the canonical treatment of hydrogenic atoms,
in which the spin and other properties of the nucleus can be neglected.
The behaviour and electronic structure of the atom
are determined by the light
electronic degrees of freedom.
Heavy quark effective theory (HQET) was developed by
Isgur and Wise \cite{ISGW} who define a single universal form factor,
$\xi(v\cdot v^{'})$, known as the Isgur-Wise function. In this function $v$ and
$v^{'}$ are the four velocities of the initial and final state heavy quarks.
In the heavy quark limit all the
form factors for hadronic matrix elements such as $B\to D^*$ and
$B\to D$ can be related to this single
function. The value of this function can then be determined from a measurement
of the $B\to D^* \ell \nu$ rate as a function of $q^2$ \cite{ISGW}.
The theory also provides a framework for systematic calculations
of corrections to the heavy quark limit.

In HQET, the decay rate for
$B\to D^* \ell \nu$
as a function of y (which is $\gamma_D^* = E_{D^*}/m_{D^*}$ in the
B rest frame)
can be expressed in terms of a single unknown form
factor $\xi(y)$.
According to the celebrated result called Luke's theorem \cite{luke},
at the point of zero recoil for the $D^*$ meson (i.e. $y=1$),
this universal
form factor is absolutely normalized up to corrections of order
$1/m_Q^2$ (where $m_Q$ is the c quark or b quark mass).

The decay rate
\begin{equation}
d \Gamma/d y = {\cal G}(y) \eta_A^2 |V_{cb}|^2 \xi^2(y)
\end{equation}
where ${\cal G}(y)$ is a known function, $\eta_A=0.986\pm 0.006$
accounts for QCD corrections and $\xi(y)$ is the universal
form factor. After subtracting background and
and correcting for efficiency
the experimental distribution of
$d \Gamma/d y$ is divided
by the factor ${\cal G}(y)$ to give a distribution whose
intercept is $|V_{cb}|^2 \xi^2(1)$. In the limit of heavy quark symmetry,
the intercept is the physical quantity of interest, $|V_{cb}|^2$.
In principle, the value obtained in this manner has no model dependence.

The $d \Gamma/d y$ distribution is extracted after subtracting backgrounds
from fake $D^*$ and random $D^*$ lepton combinations. This distribution
is then corrected for efficiency. After dividing through by ${{\cal{G}} (y)}$
CLEO~II obtains the distribution shown in Figure \ref{dstlnu2}
which combines events from the
modes $\bar{B}^0\to D^{*+} \ell^- \nu$ and $B^- \to D^{*0} \ell^- \nu$.
 Experimentally, since
there are few events near the point of zero recoil, all the
available data is used
over the entire y range and then an extrapolation to $y=1$ is made.
Most of the functional forms
proposed for $\xi(y)$ are roughly linear near $y=1$.
Thus the experimental distribution is fitted
to the functional form $|V_{cb}|^2  (1- \hat{\rho}^2 (y-1) )$.
After properly
accounting for the smearing in y due to the motion of the
B meson\cite{thesis1}, the product
$\xi(1) \eta_{A} |V_{cb}|$ is determined (Table~\ref{Tbxiexp}).
The dominant experimental systematic error is the uncertainty in the
slow pion detection efficiencies.

\begin{table}[htb]
\caption{ Experimental measurements of the product $|V_{cb}|\xi(1)\eta_A$
and $\hat{\rho}^2$ in $\bar{B}\to D^* \ell \nu$.
These have been corrected for the change in
$D$ and $D^*$ branching fractions and the average $B$ meson lifetime.}
\label{Tbxiexp}
\begin{tabular}{lll}
Experiment & $\xi(1)\eta_A~|V_{cb}|$ & $\hat{\rho}^2$ \\ \hline
ARGUS  & $0.0388\pm 0.0055$  & $1.17\pm 0.23$ \\
CLEO~II & $0.0347\pm 0.0027$ & $0.84\pm 0.15$\\
ALEPH   & $0.0382\pm 0.0056$ & $0.46\pm 0.34$\\ \hline \medskip
World Average & $0.0359\pm 0.0022$ & $0.88\pm 0.12$ \\
\end{tabular}
\end{table}

\begin{table}[htb]
\caption{ Theoretical Calculations of  the Intercept of the
Isgur-Wise Function.}
\label{Tbxi}
\begin{tabular}{ll}
Theorist & $\xi(1)\eta_A$ \\ \hline
Shifman \etal & $0.89\pm 0.03$ \\
Neubert I & $0.97\pm 0.04$ \\
Neubert II  & $0.93\pm 0.03$ \\
Mannel & $0.96\pm 0.03$ \\
\end{tabular}
\end{table}

There are two significant uncertainties in the final determination
of $|V_{cb}|$ from measurements of the spectrum at zero recoil.
These arise from the model dependence in the calculation
of the $1/m_c^2$ corrections to $\xi(1)$ and the lack of knowledge of
the functional form of the function $\xi(y)$ which is used for the
extrapolation. There are now at least four calculations of
$\xi(1)$ to order $1/m_c^2$ from Neubert \cite{neubert1},\cite{neubert2},
Mannel\cite{mannel}
and from Shifman, Uraltsev and Vainshtein\cite{shifman}.
For example, using $\xi(1)\eta_A = 0.93\pm 0.03$ from Neubert\cite{neubert2},
we obtain
$$|V_{cb}|= 0.0386\pm 0.0024({\rm exp})\pm 0.0012({\rm theory})$$
where the first error is experimental and the second is
the quoted uncertainty in
$\xi(1)$. Other recent estimates of this product obtained using
QCD sum rules are given in Table~\ref{Tbxi} and references
\cite{neubert1},\cite{neubert2},\cite{mannel},\cite{shifman}.
The model dependence from the theoretical uncertainty in the
normalization is about 4\% but may be reduced in the near future.

The uncertainty from the shape
of $\xi(y)$ can be investigated using several of the
functional forms proposed
in the literature. CLEO~II finds that the systematic
error from this source is less than 5\% in $|V_{cb}|$\cite{thesis2}.

Note that the quantity $\hat{\rho^2}$ and the quantity $\rho^2$
determined in the CLEO~II form factor analysis are slightly different.
The former is calculated assuming HQS (heavy quark symmetry). An approximate
relation between the two values is $\hat{\rho}^2 \approx
\rho^2 - 0.2$\cite{neubert2}. The two values agree well.

\begin{figure}[htb]
\begin{center}
\unitlength 1.0in
\begin{picture}(3.,3.0)(0,0)
\put(-0.0,0.0)
{\psfig{bbllx=0pt,bblly=0pt,width=3.0in,height=3.0in,file=bexcl_dstlnu2.ps}}
\end{picture}
\vskip 5 mm
\caption{
Distribution of $\displaystyle
[{{d \Gamma} \over {d y}} {1\over {{\cal G}(y)}}]^{1 \over 2}$
for $\bar{B}\to D^{*} \ell^- \nu$ candidates with a fit to a linear
parameterization of $\xi(y)$, the Isgur-Wise function.}
\label{dstlnu2}
\end{center}
\end{figure}

The universal form factor $\xi(y)$ is a quantity which cannot be
derived in perturbation theory. To obtain this function, one must depend
on models or on QCD lattice calculations. For instance,
from a fit to the
$d \Gamma/d y$ spectrum with a linear functional form, CLEO~II obtains
\begin{equation}
\hat{\rho}^2 = 0.84 \pm 0.12 \pm 0.08
\end{equation}
A fit using a Taylor expansion which includes a
quadratic term for $\xi(y)$ gives slightly different
results for $\hat{\rho^2}$. The magnitude of the quadratic term is
very poorly determined.
The value of $\hat{\rho}^2$ is consistent with most
quark models, QCD sum rules, and with lattice calculations.

\subsection{$b\to u$ Transitions and  $|V_{ub}|$}

A non-zero value of $|V_{ub}|$ is necessary but not sufficient
if the Standard Model is to provide
a consistent description of the CP violation observed in the kaon
sector. A precise
measurement of $|V_{ub}|$ is required to constrain the allowed range of
CP asymmetries in the B sector.
The experimental signature for inclusive $b\to u$ transitions
is an excess of leptons beyond
the kinematic limit for the transition $b \to c \ell \nu$.
The branching ratio for the inclusive process
is large $O(10^{-3})$.  However,
there are also substantial backgrounds
from continuum, misidentified leptons and
mismeasured $b \to c$ transitions.

The first evidence for charmless semileptonic B decay
and for non-zero $|V_{ub}|$
was reported by the CLEO~1.5 experiment in 1989
from the inclusive lepton momentum spectrum.
Corroborating evidence was presented shortly afterwards
by the ARGUS experiment, who introduced
hermiticity cuts to detect the presence of a neutrino
and thus significantly reduce background levels.

To extract a value of $|V_{ub}/V_{cb}|$, the physical quantity of interest,
the yield of leptons in the signal window, which is limited to a small
portion of the Dalitz plot, must be
corrected for detection efficiency  and then
extrapolated to the full allowed kinematic range
range. The resulting width must then be converted to the ratio of
CKM matrix elements. In other words,
\begin{equation}
 {|{ {V_{ub}} \over {V_{cb}}} |}^2 = {{\Delta B_{ub}} \over
{B_{cb}~ d (p)} } \label{vubeqn}
\end{equation}
where $\displaystyle{\Delta B_{ub}} = {{N_{ub}}\over {\epsilon}}$
and $\displaystyle{d(p)= f_u(p) {{\gamma_u} \over
{\gamma_c}}}$. In principle, each factor in equation ~(\ref{vubeqn})
is model dependent.
In practice, the only factors with large model dependence
are $\gamma_u$, which relates the width and $|V_{ub}|$
via $\displaystyle\Gamma(b\to u \ell \nu)= \gamma_u |V_{ub}|^2$ and
$d(p)$ the fraction of $b\to u \ell\nu$ decays which lie in
the momentum window.

These factors have been determined
 using a variety of models which belong to
two generic classes: inclusive parton models such as the model
of Altarelli \etal\cite{ACCM} and models with exclusive final states
e.g. \cite{ISGW},  \cite{KS}, \cite{WSB},  \cite{ISGWprime}, \cite{BBD}.
Model dependence in the value of $|V_{ub}|$
is a severe systematic limitation.
In the past, as large as a factor of
two uncertainty in the value of
$|V_{ub}|$ has been assigned to this model dependence.

 In principle, the decay rate for an exclusive mode
can be translated into a less model dependent value of $|V_{ub}|$
using exclusive models.
While no exclusive decay mode accounts for more than 3.5\% to 14\%
of the inclusive rate, backgrounds in individual exclusive
modes are fairly small.
The most promising modes are $B\to \pi \ell \nu$ and the modes with
a vector meson, $B\to \rho^0 \ell \nu$, $B \to \rho^+ \ell \nu$,
 and  $B \to \omega \ell \nu$.
%In 1991, the ARGUS experiment claimed observation
%of the exclusive decay $B \to \rho^0 \ell \nu$
%with ${\cal{B}}( B\to \rho^0 \ell \nu)= (1.13
%\pm 0.36 \pm 0.27) \times 10^{-3}$.
%This claim has not been confirmed by the CLEO~II experiment.
In 1991 ARGUS reported evidence for two fully reconstructed candidates
in the $\bar{B}^0\to \pi^+\ell\nu$ and $B^-\to \omega^0\ell^-\nu$ modes
\cite{argus_exclusive}.
%One fully reconstructed event in the ARGUS data sample
%for the $\bar{B}^0\to \pi^+\ell\nu$ mode corresponds to a
%branching fraction of about $2.8\times 10^{-3}$.

\subsubsection{Inclusive Semileptonic $b\to u$ Transitions.}

In the analyses of inclusive semileptonic $b\to u$ transitions by CLEO~1.5,
ARGUS and CLEO~II,
tight track quality cuts are imposed  and special care is taken to reduce
background from mismeasured $b\to c \ell \nu$ decays
which can be smeared beyond the kinematic limit.
The largest remaining background is then
due to continuum processes i.e. non-resonant
$q \bar{q}$ production and is suppressed with event shape cuts.

In the CLEO~II analysis which has the
highest statistical precision, two
complementary analyses are carried through. One analysis
employs strict cuts which make ample use of the hermiticity
of the detector
and imposes
the requirement that an energetic neutrino be present
which is opposite in direction to the lepton. This analysis
achieves the best signal to background ratio.
There is also a second analysis with no hermiticity cuts
and less stringent requirements on event shape variables.

%Analysis with tight cuts:
%$R2<0.2$.
% Distinguishes  $B \bar{B}$ events (spherical)
%from continuum background (jetlike). The continuum
%is reduced by a factor of 25 while 44\% of the signal
%is retained.
%$|\cos\theta_{miss}|<0.9  $. Eliminate QED backgrounds
%and mismeasured tracks.
%The last two cuts take advantage of the hermiticity
%of the CLEOII detector in order to reduce continuum background.
%$4 GeV> p_{miss}> 1 GeV$. Require an energetic neutrino.
%$cos \beta < 0$. The missing neutrino and
%the lepton are nearly back to back.

\begin{figure}[htb]
\begin{center}
\unitlength 1.0in
\begin{picture}(3.,3.0)(0,0)
\put(-0.5,0.0)
{\psfig{bbllx=0pt,bblly=0pt,width=3.0in,height=3.0in,file=b2ufig.ps}}
\end{picture}
\vskip 10 mm
\caption{Lepton momentum spectra from CLEO~II
for (a) the analysis with tight
cuts and (b) the analysis with loose cuts. The filled points
with error bars represent the $\Upsilon(4S)$ data. The open circles
are the data taken below resonance, while the dashed line is the fit
to the off-resonance data. The solid histogram is a Monte
Carlo simulation of $b\to c\ell\nu$ processes.}
\label{btouinc}
\end{center}
\end{figure}

In the analysis with loose cuts, only a modest requirement on
the event shape is imposed,
$R_2<0.3$. This analysis attempts to use selection criteria
that are similar
to the cuts used in the CLEO~1.5 analysis.
The efficiency
of the hermiticity cuts used in the strict analysis will
depend somewhat on the $Q^2$ spectrum of $B\to u\ell \nu$.
On the other hand, the analysis with strict cuts
has the advantage that the allowed
phase space is restricted to the region where resonances ($\pi$, $\rho$
$\eta$, $\omega$) dominate and where exclusive models are
most reliable.

The lepton
momentum spectrum with the continuum data and a
histogram from a $b\to c$ Monte Carlo
superimposed is shown in Fig.~\ref{btouinc}.
The background is subtracted using a fit to the continuum
data, which were recorded
at an energy slightly below the $\Upsilon(4S)$ resonance.
In the loose analysis,
$128.5 \pm 26.3 \pm 15.2$ excess leptons are observed in the
momentum interval between 2.4 and 2.6 GeV, while
in the strict analysis, an excess of $ 43.0\pm 10.1 \pm 6.6$
leptons is found in the same interval.
As can be seen from Figure~\ref{btouinc}, the yield beyond the
kinematic limit for $b\to u\ell\nu $ production is consistent with zero.

For the loose analysis,  the partial branching fraction is
$\Delta B_{u} (2.4,2.6)$ $=(0.90 \pm 0.18\pm 0.12)\times 10^{-4}$ while
for the strict analysis, the corresponding branching fraction is
$\Delta B_{u} (2.4,2.6)$ $=(0.70 \pm 0.16 \pm 0.12)\times 10^{-4}$
where the ACCMM model has been used to evaluate the
detection efficiency.

The agreement found in CLEO~II between the branching fractions
from the analysis with strict cuts and the analysis with loose cuts
indicates that the choice of
$q^2$ dependent selection criteria in the analysis
with tight cuts does not introduce
significant model dependence in the final result.
The largest source of
model dependence is due to the extrapolation
from the narrow signal window to the full kinematic range (the factor
$f_u(p)$).

The central value for the branching
fraction from the CLEO~II experiment is significantly below the
previous CLEO~1.5 and ARGUS results for this momentum interval.
The CLEO~II and CLEO~1.5 measurements are consistent at the
2.5 standard deviation level.
The large values initially reported
by CLEO~1.5 and ARGUS are now
believed to be upward fluctuations. In addition, the yields
in the lower momentum bins from both the early experiments,
which must be determined after subtractions of large
$b\to c$ backgrounds,
are marginally consistent with the yield
in the high momentum ($2.4-2.6$ GeV) bin. In the CLEO~II analysis, the
branching fractions determined from the $2.3-2.4$ GeV bin and
the $2.4-2.6$ GeV bin are in good agreement.
For these reasons, we have chosen to
determine the value of $|V_{ub}/V_{cb}|$ using the branching fraction
measured by CLEO~II in the high momentum bin. Note that including
the CLEO~1.5 and ARGUS results
 gives only a small shift in the world average
since the errors for these measurements are large.

 These results can be used to deduce values for $V_{ub}$,
which are given in Table~\ref{Tvub}. Taking the central value
from the ACCMM model gives
$$ |{{V_{ub}}\over {V_{cb}}}| = 0.073\pm 0.011(exp) \pm 0.010(model)$$
and a range at the 1 standard
deviation level
 $0.095 > |{V_{ub}\over V_{cb}}| > 0.055$.
The table includes the estimate
from D. Scora for
the revised ISGW model\cite{ISGWprime} (denoted ISGW2)
but does not include the older ISGW model.
Incomplete exclusive models such as those of K\"orner and Schuler
and Wirbel-Stech-Bauer that do not calculate all the exclusive
final states which are relevant for the endpoint region are also
omitted.

\begin{table}[htb]
\caption{ Experimental measurements of the
partial branching fraction
for $b\to u \ell\nu$ transitions in the lepton
endpoint region. (*) The ARGUS value is deduced indirectly
from the value of $|V_{ub}/V_{cb}|^2$ given in their publication.}
\label{Tbuexp}
\begin{tabular}{llll}
Experiment & ${\cal B}(2.2-2.6)$ GeV&
${\cal B}(2.3-2.6)$ GeV & ${\cal B}(2.4-2.6)$ GeV \\ \hline
CLEO~1.5 & $(33\pm 8\pm 8)\times 10^{-5}$& & $(18\pm 4\pm 3)\times 10^{-5}$\\
ARGUS  & &$(32.8\pm 7)\times 10^{-5}$  $^{*}$  & $ $ \\
CLEO~II & & $(8.2\pm 1.5\pm 0.9)\times 10^{-5}$
& $(7.0\pm 1.6\pm 1.2)\times 10^{-5}$\\ \hline
World Average & &     & $(7.0\pm 2.0)\times 10^{-5}$  \\
\end{tabular}
\end{table}

\begin{table}[htb]
\caption{Determination of $|V_{ub}/V_{cb}|$
from various theoretical models for the momentum interval
$2.4-2.6$ GeV.}
\label{Tvub}
\begin{tabular}{lllll}
Model &$\gamma_u \times 10^{12}$sec
&$f(p)$ & $d(p)$& $|V_{ub}/V_{cb}|$ \\ \hline
ACCMM  &$80.4$ &$0.055$ & $0.123$ & $0.073\pm 0.011$\\
BBD  & $68.0$&$0.074$ & $0.113$ & $0.071\pm 0.011$\\
Hybrid  &$84.3$ &$0.066$ & $0.135$ & $0.065\pm 0.010$\\
ISGW2  &$58.8$ &$0.049$  & $0.0703$ & $0.083\pm 0.012$\\
\end{tabular}
\end{table}

\subsubsection{Exclusive Semileptonic $b\to u$ Transitions}

CLEO~II has searched for the exclusive decays
$B^+ \to \rho^0 \ell^+ \nu$ and
$\bar{B^0} \to \rho^+ \ell^- \nu$ and $B^+ \to \omega \ell^+ \nu$.
They require that the observed momenta
be consistent with a missing neutrino
taking advantage of production at threshold. The
 $\rho$ ($\omega$) invariant
mass spectrum is then examined after applying additional cuts.
Event shape cuts are used to suppress background
from the dominant background $e^+ e^- \to q \bar{q}$
which are jetlike in contrast to $B \bar{B}$ events
which are more spherical.
The analysis is divided into two lepton momentum ranges:
$2.3>E_{lep}>2.0$ GeV,  and
$2.6>E_{lep}>2.3$ GeV. All models
are in approximate agreement that about 60\% (52\% in WSB
to 72\% in ISGW) of the rate
is contained in the union of the two ranges.

Both the ARGUS and CLEO~II analyses of exclusive $b\to u$ transitions
make use of the
hermiticity of the detector. For example, CLEO~II requires the
missing mass of
the remainder of the event be greater than -0.2 GeV$^2$
and less than 5.0 GeV$^2$. In addition, the
missing momentum vector should  balance
the visible momentum of $Y= \rho \ell$ system.

The branching ratios for $B \to \rho $ and $B \to \omega$ decays are
related by the
quark model and isospin invariance such that
${\cal B}
(\bar{B}^0\to \rho^+  \ell  \nu) = 2 {\cal B}
(B^-\to\rho^0  \ell  \nu) = 2 {\cal B}( B^-\to \omega \ell^- \nu)$.
No signal was observed and CLEO placed an upper limit on the production
of light vector mesons ($\rho^{+(0)},\, \omega$)
$$B(B^- \to V^0 \ell \nu) <
(1.6 - 2.7) \times 10^{-4} \rm{~at~the~90\% ~C.L.}$$
where the upper end of the range corresponds to calculating the
efficiency using the ISGW model.
The results for different models and for an older ARGUS analysis
\cite{argus_exclusive} are listed in Table \ref{Tbuexcl}.

The CLEO~II results are used to obtain the constraints:
$$|V_{ub}/V_{cb}| < 0.08-0.13 \rm{~at~the~90\% ~C.L.}$$ where again
the range given corresponds to the three models considered.
These limits are
consistent with the values of $|V_{ub}/V_{cb}|$ determined
from the inclusive lepton spectrum.

ARGUS has also reported the observation of two events
in exclusive semileptonic $b\to u$ modes, $\bar{B}^0\to \pi^+ \ell^-\nu$ and
$B^-\to \omega\ell^-\nu$ opposite
fully reconstructed modes\cite{argusvub1}.
%These events correspond to branching fractions of about $(1-3)\times 10^{-3}$
%for the modes $\bar{B}^0\to \pi^+ \ell^-\nu$ and $B^-\to \omega \ell^- \nu$
%and are inconsistent with the upper limits reported below.

\begin{table}[htb]
\caption{Measurements of Exclusive Charmless
Semileptonic B Branching Ratios. Here V denotes a vector meson
as discussed in the text.}
\label{Tbuexcl}
\begin{tabular}{lllll}
Experiment & Mode & ISGW & WSB & KS  \\ \hline
ARGUS & $\bar{B}\to \pi^+ \ell  \nu$ & $<0.9 \times 10^{-3}$ \\
ARGUS & $\bar{B}\to \rho^0 \ell  \nu$ & $<1.1 \times 10^{-3}$ \\
% commented out for lkg result
%CLEO~II & $\bar{B}\to \pi^+ \ell\nu$ &
%$<3.3\times 10^{-4}$ &$4.5\times 10^{-4}$
%& $<4.7\times 10^{-4}$ \\
CLEO~II & $\bar{B}\to \rho^0 \ell  \nu$ & $<2.1\times 10^{-4}$ &
&  \\
CLEO~II & $\bar{B}\to \omega \ell  \nu$ & $<2.1\times 10^{-4}$ &
&  \\
CLEO~II & $\bar{B}\to \rho^+ \ell  \nu$ & $<4.1\times 10^{-4}$ &
&  \\
CLEO~II & $\bar{B}\to V \ell  \nu$ & $<1.6\times 10^{-4}$ &$2.7\times 10^{-4}$
& $<2.3\times 10^{-4}$
\end{tabular}
\end{table}

\begin{figure}[htb]
\begin{center}
\unitlength 1.0in
\begin{picture}(3.,3.2)(0,0)
\put(-0.5,0.0)
{\psfig{bbllx=0pt,bblly=0pt,width=3.0in,height=2.5in,file=pilnu.ps}}
\end{picture}
\vskip 20 mm
\caption{
 Beam constrained mass distributions from CLEO~II for
(a) $B\to \pi\ell\nu$
and (b) $B\to \rho\ell\nu$
candidates. The dotted line shows the background contribution
which includes $b\to c$ and other $b\to u$ decays.}
\label{pilnu}
\end{center}
\end{figure}

\subsubsection{Observation of $B \to \pi \ell \nu$ Transitions}

The first signal for an
exclusive semileptonic charmless decay mode has
been reported recently by CLEO~II
in the mode $B\to \pi\ell\nu$ \cite{lkgmoriond}.
Events in which the
neutrino momentum is well constrained from the missing energy
are used.
This allows a beam constrained mass and energy difference
to be constructed in analogy to exclusive hadronic B decays.
The effective beam constrained mass distribution
for $\bar{B}^0\to \pi^+\ell^-\nu$ and
$B^-\to \pi^0\ell^-\nu$ candidates
after a cut on the energy difference is shown in
Fig.~\ref{pilnu}. A likelihood fit to the
beam constrained mass, the energy difference,
 and $\sin^2\theta_{\ell \pi}$ distributions\cite{jdrpilnu}, shows
that the excess has a significance of $3.8$ standard deviations.
The resulting branching fraction is
$${\cal B}(B\to \pi^+ \ell\nu) = 1.19\pm 0.41\pm 0.22\pm 0.19\times
10^{-4} (ISGW)$$
$${\cal B}(B\to \pi^+ \ell\nu) = 1.70\pm 0.50\pm 0.31\pm 0.27\times
10^{-4} (BSW)$$
for the ISGW and BSW models, respectively\cite{lkgmoriond}.
No significant excess is observed in the ${\cal B}(B\to \rho\ell\nu)$
mode.

Measurements of branching fractions for exclusive
charmless modes will be an important step
towards establishing a reliable value of $|V_{ub}|$.

\subsubsection{Prospects for the Determination of $|V_{ub}|$.}

The model dependence in $|V_{ub}/V_{cb}|$, which is
larger than the experimental error,
will be reduced eventually by the detailed study of
inclusive
and then exclusive decay modes.
At present, with the inclusive sample,
rough checks of the $Q^2$ dependence in inclusive
$b\to u$ decay are possible.
In the future these may allow some
discrimination between models (see Figure~\ref{vubq2theor})
\cite{nelson},\cite{artusovub}.
For instance, the inclusive
Altarelli style model peaks at a lower $q^2$ than does the ISGW
model (the sum of exclusive final states). The early version of the
ISGW model disagrees with the observed $q^2$ spectrum at the 1.5 $\sigma$
level. With larger data samples this approach will either
rule out or severely constrain models of $b\to u$ decay.

\begin{figure}[htb]
\begin{center}
\unitlength 1.0in
\begin{picture}(3.,3.0)(0,0)
\put(-0.5,0.0)
{\psfig{bbllx=0pt,bblly=0pt,width=3.0in,height=3.0in,file=vub_model.ps}}
\end{picture}
\vskip 10 mm
\caption{$q^2$ distribution for two theoretical models of
$b\to u\ell\nu$ transitions. The solid histogram is the model
of Altarelli~\etal while the dashed histogram
is the original model of Isgur, Scora, Grinstein and Wise (ISGW).}
\label{vubq2theor}
\end{center}
\end{figure}

If the $b\to c$ backgrounds can be well constrained and the
continuum background
reduced to a sufficient degree by the use of detector hermiticity,
then the stringent lepton momentum cut may be
relaxed and a larger fraction of $b\to u \ell\nu$ events accepted.
This approach could significantly reduce the model dependence
in $|V_{ub}|$.

\begin{figure}[htb]
\begin{center}
\unitlength 1.0in
\begin{picture}(3.,3.0)(0,0)
\put(-0.5,0.0)
{\psfig{bbllx=0pt,bblly=0pt,width=3.0in,height=3.0in,file=vub_q2dist.ps}}
\end{picture}
\vskip 10 mm
\caption{Pseudo $q^2$ distribution in CLEO~II data compared
with two theoretical models of
$b\to u\ell\nu$ transitions.
The original model of ISGW
is shown as the solid histogram, while the
ACCMM model is the dashed histogram.}
\label{vubq2exp}
\end{center}
\end{figure}

Once exclusive $b\to u$ modes are observed, various integrated
quantities such as  the
vector meson polarization in $B\to V\ell\nu$
and the average lepton energy can be compared
to models. Unlike the $b\to c$ case, the differences between
models are significant.
It is also possible
to compare data on
other heavy quark to light quark transitions to models.
For example, one can compare
the form factors in $D^0 \to K^- \ell \nu$ and $D \to K^* \ell \nu$ to
various models\cite{jdr}.
The form factors for $B \to D^* \ell \nu$, where
large event samples are available, have also been checked
(see section~\ref{bdynamics}) and are consistent at the present
level of experimental precision with HQET and quark models.

As noted by Isgur and Wise, the use of HQS (heavy quark symmetry)
gives relations between various
heavy to light form factors\cite{wiserholnu}.
For instance, the $B\to \rho$ form factor can be related to
the corresponding $D\to \rho$ form factor at equal $\rho$ energies
provided one assumes that the light degrees of freedom decouple
\begin{equation}
 <\rho(k, \epsilon)| \bar{u} \gamma_u (1-\gamma_5) b | \bar{B}(v)>
= {({m_B \over m_D})}^{1/2} [{\alpha_s(m_b) \over \alpha_s(m_c)}]^{-6/25}
{}~<\rho(k, \epsilon)| \bar{u} \gamma_u (1-\gamma_5) c | \bar{D}(v)>
\label{isgweqn}
\end{equation}
when the momentum transfer to the $\rho$ meson is much less than
a heavy quark mass, i.e. $v \cdot k << m_c, m_b$.
If in addition, $SU(3)$ symmetry holds, then the $B\to \rho$ form
factor can be related to the $D\to K^*$ form factor.
Analogous relations can be derived between
the $B\to \pi$ and $D\to K$ form factors\cite{burdpilnu}.
Thus, precise
measurements of the form factors in semileptonic $D$ decays
in conjunction with measurements of exclusive charmless
semileptonic $B$ decays may be
useful in the future determination of $|V_{ub}|$\cite{dibvub}.
However,
there are several potential difficulties with this approach. The type of
relation given in equation~(\ref{isgweqn})
 may have significant $1/m_Q$ corrections.
At very small momentum transfer, other theoretical corrections
(``pole terms'') may be large.
Experimentally, the
kinematic range available in $D\to \rho$ decays is much smaller
than in the corresponding $B$ decay, so models
will still be needed
to extrapolate to the case of $B$ decays.

Akhoury, Sterman, and Yao have suggested that measurements
of the exclusive decays
$\bar{B}\to \pi \ell\nu$ and $\bar{B}\to \rho\ell\nu$
at low $Q^2$ may be suitable for the determination of
$|V_{ub}|$\cite{stermanvub}. They suggest
that the inclusive width of these decays in the range $Q^2=(0-9.2)$ GeV$^2$
can be calculated reliably using QCD factorization theorems. Then
experimental measurements of these
decays can be used to extract $|V_{ub}|$.

Another possible approach to a model independent determination
of $|V_{ub}/V_{cb}|$ has been suggested by Neubert\cite{neubertvub}.
The proposed
method makes use of the relation between the differential spectra
for $b\to u \ell\nu$ and $b\to s\gamma$ transitions to eliminate the
uncertainty from the hadronization of $b\to u$ in the endpoint region.
A simplified and intuitive description may be useful. The
width for inclusive $b\to u$ transitions is
proportional to $|V_{ub}|^2 m_b^5$
where $m_b$ is the b quark mass. On the other hand, in the rest frame
of the b quark, the average energy of the photon emitted in the
inclusive $b\to s\gamma$ decay is $m_b/2$. Thus, determination of
the average photon energy in the electromagnetic penguin provides
a way to eliminate the uncertainty from the $b$ quark mass.
It is remarkable that a slight modification of this idea
remains useful after the effect of the
cut on lepton energy in $b\to u$, QCD corrections, and the Fermi motion
of the $b$ quark have been properly
 included. Several authors have proposed similar
techniques for the
determination of $|V_{ub}|$\cite{falkvub},\cite{russkievub},\cite{korvub}.
It should be noted that if the $b\to u$
endpoint region is dominated by a single resonant
mode (e.g. $B\to \rho\ell\nu$), this technique
may not be valid\cite{falkvub}.

%%%%%\section{$B-\bar{B}$ MIXING}
%\input mixing.tex
\section{$B-\bar{B}$ MIXING}
\label{mixing}

In production
processes involving the strong or the electromagnetic interaction
neutral $B$ and $\bar{B}$ mesons can be produced.
These flavor eigenstates are not
eigenstates of the weak interaction which is responsible
for the decay of neutral mesons containing b quarks.
The strong eigenstates are
linear combinations of the weak eigenstates,
$$ |B_1> = {1\over \sqrt{2}}(|B^0>+|\bar{B}^0>)$$
$$ |B_2> = {1\over \sqrt{2}}(|B^0>-|\bar{B}^0>)$$
This feature and the small difference between the masses
and/or lifetimes of the
weak interaction eigenstates gives rise to the phenomenon
of $B-\bar{B}$ mixing. The formalism which describes
$B$ meson mixing closely follows that used to describe
$K^0-\bar{K}^0$ mixing, although the
time scale characteristic of
 $B-\bar{B}$ oscillations is much shorter.

 If a pure $|B^0>$
state is produced at time $t=0$, then at later times it will evolve
into a new state which contains an admixture of $|\bar{B^0}>$.
The weak interaction eigenstates are denoted $B_1$ and $B_2$,
and have masses $M_1, ~M_2$ and widths $\Gamma_1, ~\Gamma_2$
respectively. The difference between the masses is denoted
$\Delta M$, while the difference between the widths is $\Delta\Gamma$.
The average width will be referred to as $\Gamma$.
The probability that the state, initially produced
as $|B^0>$ will mix into $|\bar{B^0}>$ at time t is given by
$$P(B^0\to \bar{B^0}) = \exp{(-\Gamma t)} ~[1-\cos(\Delta M ~t)].$$

For convenience, the ratios $x= \Delta M/\Gamma$
and $y=\Delta \Gamma/2\Gamma$ are frequently introduced.
The contribution of $y$ to mixing is usually neglected. Its size
is determined by the fraction of final states which are common to
both $B$ and $\bar{B}$ mesons. The difference between the width for
those final states with one sign of CP and those with the opposite
sign determines the magnitude of $\Delta\Gamma$.
In contrast to the case of neutral kaons, the branching ratio for
such modes (e.g. ${\cal B}(\bar{B}^0\to \psi K^*)$) is small, so
$y<< x$ for neutral B mesons.
The magnitude of $x$ determines the frequency of the mixing oscillations.

The quark level process responsible for $B-\bar{B}$ mixing
is shown in Figures 1(g) and 1(h). The contribution to
$\Delta M$ for $B_d-\bar{B_d}$ mixing is found to be
\begin{equation}
\Delta M_d = {{G_F^2} \over {6 \pi^2}} m_B m_t^2
{}~F({{m_t^2} \over {m_W^2}}) ~\eta_{QCD} B_{B_d} f_{B_d}^2
 |V_{tb}^{*} V_{td}|^2
\end{equation}
where $G_F$ is the weak coupling constant, $m_t$ is the top quark mass,
$F$ is a slowly decreasing function
which depends on $m_{t}$ and $m_{W}$,
$\eta_{QCD}$ is a factor which accounts for QCD corrections,
$B_{B_d}$ is a constant which is used to account for the vacuum insertion
approximation, and $f_B$ is the decay constant of the $B_d$ meson.
An analogous expression for $B_s$-$\bar{B_s}$ mixing can also be obtained.
Since $|V_{ts}| >> |V_{td}|$, the rate and frequency of mixing for the
$B_s$ meson will be significantly larger than for the $B_d$ meson.

Since the mass of the top quark has been determined
and the QCD correction has been calculated to NLO
($\eta_{QCD}=0.55$)\cite{Burasmix},
the largest
uncertainties in $\Delta m_d$ arise from the product $B_{B_d}^{1/2} f_{B_d}$.
This last factor must be determined from non-perturbative methods
such as lattice calculations or QCD sum rules.

Evidence for $B_d-\bar{B_d}$ mixing was first reported in 1987 by the ARGUS
experiment from the study of like sign lepton correlations\cite{mixdiscover}.
The CLEO~1.5 experiment
later confirmed the result\cite{cleo15mix}. The observed
level of mixing was significantly larger than theoretically
expected and provided
the first suggestion that the mass of the top quark was large,
much greater than 30 GeV as was indicated by the UA1 experiment
at that time.

\begin{figure}[htb]
\begin{center}
\unitlength 1.0in
\vskip 20mm
\begin{picture}(3.,3.0)(0,0)
\put(-0.5,0.0)
{\psfig{bbllx=0pt,bblly=0pt,width=4.0in,height=4.0in,file=argusb0b0bar.eps}}
\end{picture}
\vskip 5 mm
\caption{A fully reconstructed event with mixing observed by the
ARGUS experiment.}
\label{argmix}
\end{center}
\end{figure}

The substantial mixing rate also implies that CP violation could
be large in $B$ decay to CP eigenstates, provided the amplitude for
the $B$ decay to a CP eigenstate interfers with the amplitude for
a $\bar{B}$ meson to mix and then decay to the same final state.
A relative phase between the two
amplitudes can be introduced by the CKM couplings.
That is, mixing provides a mechanism that gives rise to interfering
amplitudes and hence CP violation.

To experimentally measure mixing
 requires identification of the initial
state flavor of the neutral $B$ meson at production
as well as the final state flavor after decay of the $\bar{B}$ meson.
This ``tagging'' of the initial flavor
can be accomplished
using leptons or other partial reconstruction techniques.
In addition, the production fraction for the neutral B meson
in question must be known.

Several experimental parameters are used to measure the strength
of mixing. For example, the ratio of the time integrated number
of $B^0$ and $\bar{B^0}$ mesons is denoted
$r= {N(\bar{B}^0)\over N(B^0)}$ if the initial state is $|B^0>$.
In general, mixing is measured by studying pairs of B mesons since one of the
B hadrons is needed to tag the flavor at  production.
The ratio of the number of mixed events to the number of unmixed
events is given by
$$R= {{N(B B + \bar{B} \bar{B})} \over {
N(B \bar{B} + \bar{B} B)}}$$
At threshold, this becomes
$$R= {{N(B^0 B^0 + \bar{B}^0 \bar{B}^0)} \over {
N(B^0 \bar{B}^0 + \bar{B^0} B^0)}}$$

On the $\Upsilon(4S)$ resonance and at the threshold for
$B \bar{B}^*$ production, $B \bar{B}$ pairs are produced coherently
i.e. in a state of definite orbital angular momentum.
Quantum statistics for spin zero particles implies that the
wave function of the $B \bar{B}$ must be antisymmetric (symmetric)
for production with odd (even) orbital angular momentum.
This is the case for production at the $\Upsilon(4S)$ where
$$ R= {x^2 \over {2+ x^2}} \; ({\rm l~odd})$$
At $B \bar{B^*}$ threshold, the relative orbital angular momentum is zero,
and $$ R= {{3 x^2+ x^4} \over {2+ x^2+x^4}} \; ({\rm l~even})$$
Production of neutral B mesons
at the $Z^0$ and at hadron colliders is an incoherent
sum of these two cases.

It is also useful to define $\chi$ which is the probability
that a produced neutral B meson mixes and then decays as a
neutral $\bar{B}$ meson.
The fraction of mixed events is then $2 \chi ~(1-\chi)$.
Experiments at the $\Upsilon(4S)$ resonance extract
the mixing parameter from the observed ratio of like-sign
dileptons to opposite dileptons. This
requires knowledge of the relative production of
$B^+ B^-$ and $B^0 \bar{B}^0$ mesons pairs. For this case,
$\chi_d = (1+ \Lambda) r_{wrong}$,
where $\Lambda= f_+/f_0 (\tau^+/\tau^0)^2$, $r_{wrong}$
is the ratio of the number of like-sign dileptons to opposite sign
dileptons and
$f_+, f_0$ are the fractions of $B^0 \bar{B^0}$ and $B^+ B^-$ pairs
produced, respectively.
The current uncertainty in $\Lambda$
leads to a systematic uncertainty of about 20\% in measurements of mixing
from threshold experiments which use like sign dileptons to determine the
$b$ quark flavor.

\subsection{$B_d-\bar{B_d}$ Mixing}
\label{bdbdbmix}

The determination of the $B_d$ mixing parameter
with the best statistical precision
is obtained from measurements of the rate for
like-sign dileptons in experiments at threshold.
The yield of like-sign dileptons is found after subtracting
background contribution from non-resonant production (continuum), cascades,
$\psi^{(')}$s and misidentified leptons.
There are results from the ARGUS, CLEO1.5 and CLEOII experiments
(see Table~\ref{mixthresh}).

To reduce the systematic error from the poorly measured
fraction $\Lambda$, another technique to increase
the $B^0-\bar{B^0}$ content of the sample has been developed.
Events which contain a wrong sign lepton and a partially
reconstructed $B \to D^{*+} \ell \nu$ decay
are used. In this case, it is not necessary to detect the
decay products of the $D^0$ meson.
The method takes advantage
of the small energy release in the decay $D^{*+}\to D^0 \pi^+$
and the kinematic constraints
of production near threshold\cite{cleoiimix},\cite{argusdlmix}.

% new version
\begin{table}[htb]
\caption{Measurements of the mixing parameter $\chi_d$
from threshold experiments. The first error is statistical,
while the second is the experimental systematic error and the
third is due to the uncertainty in $\Lambda$.}
\medskip
\label{mixthresh}
\begin{tabular}{lll}
Experiment & Technique & $\chi_d$\\ \hline\medskip
ARGUS      &  $\ell^+, \ell^+$& $0.173\pm 0.038\pm 0.044^{+0.031}_{-0.023}$ \\
CLEO~1.5   &   $\ell^+, \ell^+$
&$0.142\pm0.035\pm0.034^{+0.025}_{-0.019}$ \\
CLEO~II    &    $\ell^+, \ell^+$
&$0.157\pm0.016\pm0.018^{+0.028}_{-0.021}$ \\
ARGUS      & partial $D^{*-} \ell^+ \nu,\ell^+$ &
$0.162\pm 0.043\pm 0.039$ \\
CLEO~II    &  partial $D^{*-} \ell^+ \nu,\ell^+$&
$0.149\pm0.023\pm0.019\pm 0.010$ \\ \hline \medskip
Average $\Upsilon(4S)$ &   $\ell^+, \ell^+$ &  $0.156\pm 0.020
^{+0.027}_{-0.020} (\Lambda)$
        \\
Average $\Upsilon(4S)$ & partial $D^{*-} \ell^+ \nu, \ell^+$&  $0.151\pm
0.0265\pm
0.010(\Lambda)$        \\
\end{tabular}
\end{table}

% Useful comment:
% to convert from chi_d to r, use r=chi_d /(1-chi_d)
% to convert from r to chi_d use chi_d =r /(1+r)
% while to convert from r to x, use x =sqrt(2r/(1-r))

In the CLEO~II analysis,
the energy of the $D^*$ is approximated by $(E_{\pi}/E_{\pi}^*) \times
M_{D^*}$
where $E_\pi$ is the energy of the slow $\pi$ in the laboratory
frame and
$E_{\pi}^*$ is the corresponding
energy in the center of mass frame\cite{cleoiimix},
\cite{mssmix}.
The direction of the slow pion
is a good estimator of the $D^*$ direction. A quantity analogous
to missing mass is formed:
$$ MM^2 \approx (E_{beam} - E_l - E_D^*)^2 -|p_l + p_D^*|^2$$
where $E_{D^*}$ and $p_{D^*}$  are the estimates of the $D^*$ energy
and momentum determined from the slow pion
energy and direction. For signal events, the variable $MM^2$
peaks near zero with a width of $0.9$ GeV$^2$. A similar
partial reconstruction technique has been applied by ARGUS\cite{argusdlmix}.

The measurement of $\chi_d$ using the
partial reconstruction technique substantially reduces the systematic
uncertainty from $\Lambda$ but
has a slightly larger statistical error. Note that the results from
the dilepton
and partial reconstruction analyses should
not be combined since there is
substantial overlap between the datasets used for the two analyses.

% cleo ii result dileptons chi_d
% for r_d 0.187+-0.022+-0.025+0.040-0.030
% cleo ii result partial rec chi_d=
% cleo 1.5 chi_d =
%CLEO 1.5
% argus result chi_d =

% **********new version
\begin{table}[htb]
\caption{Measurements of the mixing parameter $\Delta M_d$ in units of
ps$^{-1}$.
The notation (time) indicates that an explicitly time dependent
measurement was performed to  determine the mixing parameter.}
\medskip
\label{Tmix1}
\begin{tabular}{lll}
Experiment & Technique & $\Delta m_d$ \\ \hline \medskip
ALEPH      &  $\ell^+, \ell^+$(time)& $0.44\pm 0.05^{+0.09}_{-0.06}$     \\
OPAL       &   $\ell^+, \ell^+$(time)& $0.462^{+0.040+0.052}_{-0.053-0.035}$
             \\
DELPHI     &    $\ell^+, \ell^+$(time)& $0.53^{+0.11+0.11}_{-0.10-0.10}$  \\
ALEPH      &  $D^* \ell^+, Q_j$(time) & $0.497\pm 0.070\pm 0.036$   \\
OPAL       &   $D^* \ell^+, Q_j$(time)   &  $0.508\pm 0.075\pm 0.025$  \\
DELPHI     &    $D^* \ell^+, Q_j$ (time)& $0.456\pm 0.068\pm 0.043$ \\
OPAL       &   $D^{*+}, \ell^+$(time)    &  $0.57\pm 0.11\pm 0.02$
\\
DELPHI     &     $\ell^+, K^+, Q_j$(time)&  $0.586\pm 0.049\pm 0.062$ \\
DELPHI     &     $D^{*+}, Q_j$(time)&  $0.50\pm 0.12\pm 0.06$ \\ \hline
$\Upsilon(4S)$& partial $D^{*-} \ell^+, \ell^+$  &  $0.405\pm 0.041$       \\
LEP        &  (time)   &  $0.501\pm 0.034$       \\ \hline \smallskip
World Average &             &   $0.462\pm 0.026$      \\
\end{tabular}
\end{table}

Time integrated mixing measurements from LEP and hadron colliders
 determine
the quantity $\chi$ =$f_d \chi_d + f_s \chi_s$ where
$f_d$ and $f_s$ are the
fractions of $B_d$ and $B_s$ mesons produced.
This method gives weak constraints on $\chi_s$ and
$\chi_d$ which are not competitive with those deduced from
the time dependent measurements and from the experiments at the
$\Upsilon (4S)$ resonance. Hence these results will
not be discussed further. Additional details and a
summary of these results are available in Ref.~\cite{Venus}.

The ALEPH, DELPHI and OPAL experiments
have performed explicit measurements
of $P(B^0\to \bar{B^0})$
 as a function of time to obtain
the parameter $x_d$\cite{alephmix},\cite{delphimix},\cite{opalmix}.
The initial state $b$ quark flavor is tagged either using leptons or
jet charge, while the flavor of the final state $b$ quark
is tagged using
either $\bar{B}_{d}\to D^{*+} \ell^- X$, $\bar{B}_{d}\to D^{*+} X$,
or $\bar{B}_{d}\to \ell^- X$.
If the final state is not fully reconstructed, as is the case for
the analyses using
dileptons, then the decay time must be determined using
a topological vertexing technique where the lepton from the $B$ decay
and the other tracks in the same jet hemisphere are combined.
The boost is determined using the observed energy, missing momentum
and a correction
factor determined from a Monte Carlo simulation.

The observed fraction of
like sign leptons $N_{++}/(N_{++} + N_{+-})$ is clearly not
time independent (see Figure~\ref{alephmix})
and the beginning of one oscillation cycle is
visible. A full oscillation lasts
about 15 ps.  Due to the effects of acceptance
and resolution, additional oscillations are not seen.
The largest contributions to
the systematic errors in the measurements
of $\Delta m_d$ using time dependent $D^*-$lepton correlations
arise from the uncertainties in the decay time resolution
and the knowledge of the charged B meson fraction. For the
measurements with dileptons and those using the lepton-jet
charge tagging technique, the uncertainty in the $B_s$
fraction at the $Z^0$
also gives a significant contribution to the systematic error.

The results from the LEP experiments with silicon vertex
detectors are given in Table~\ref{Tmix1} as well
as the average of the CLEO~II and ARGUS measurements using
the partial reconstruction technique. The $\Upsilon(4S)$ value
was computed using the world average for the $B^0$ lifetime.
The results from the various
time dependent techniques used by the  LEP experiments
were averaged separately and then combined to form
the LEP average.
For the LEP dilepton measurements,
a common systematic error of $0.09$ ps$^{-1}$  was assumed,
while for the $D^{*+} l^-,\, Q_J$ results, a common systematic error
of $0.02$ ps$^{-1}$ was assumed. The systematic errors
in the different techniques
were then assumed to be uncorrelated. This
treatment of the systematic errors gives slightly
more weight to the results with jet charge tagging technique than
a simple weighted average.

\begin{figure}[htb]
\begin{center}
\unitlength 1.0in
\vskip 25mm
\begin{picture}(3.,3.0)(0,0)
\put(-1.6,-0.8)
{\psfig{bbllx=0pt,bblly=0pt,width=6.0in,height=6.0in,file=aleph_time.ps}}
\end{picture}
\caption{The fraction of wrong sign leptons as a function of time
from the ALEPH experiment.}
\label{alephmix}
\end{center}
\end{figure}

\subsection{$B_s-\bar{B_s}$ Mixing}

The measurement of the mixing parameter $x_s=\Delta M/\Gamma$ for
the $B_s$ meson is one of the goals of high energy collider experiments
and experiments planned for future facilities \cite{BELLE,BABAR,DESY}.
A measurement of $x_s$ combined with a determination of $x_d$,
the corresponding quantity for the $B_d$ meson, allows the determination
of the ratio of the CKM matrix elements $|V_{td}|^2/|V_{ts}|^2$
with significantly reduced theoretical uncertainties. The ratio of the mixing
parameters can be written as
\begin{equation}
{{x_s}\over {x_d}} =
{{(m_{B_s} \eta_{QCD} B f_{B_d}^2)}\over
{(m_{B_d} \eta_{QCD} B f_{B_s}^2)}} \,|{{V_{ts}^2} \over {V_{td}^2}}|
\, {{\tau_d}\over{\tau_s}}
\end{equation}
where the factor which multiplies the ratio of CKM matrix elements
is believed to be unity up to $SU(3)$ breaking effects.
Ali and London \cite{Alickm} ~ have estimated
$$
{{\Delta m_s}\over{\Delta m_d}} \; = \; (1.19 \pm 0.10)\,
{{|V_{ts}|^2}\over{|V_{td}|^2}}
$$
The time integrated probability that a neutral $B$ meson mixes is
$\chi= {1\over 2} {x^2 \over{1+x^2}}$.
% for $y<<x$
As x becomes large, as is expected for the $B_s$ meson,
$\chi$ asymptotically approaches 0.5.
Thus time integrated measurements
of $B_s$ mixing are insensitive to $x_s$ when mixing is maximal,
and one must make time dependent measurements in order
to extract this parameter.
These are
experimental challenging due to
the rapid oscillation rate of the $B_s$ meson.

Using dilepton events in which the tagging lepton is vertexed
with other tracks in the same hemisphere and the neutrino
energy is deduced
from the energy flow in the event, ALEPH has searched for a
high frequency component in their fit to the proper time distribution. They
find $\Delta M_s > 3.9$ ps $^{-1}$ or $x_s>5.5$\cite{alephmix}.
{}From an event sample with a lepton and a tag using
a special jet charge
where each track is weighted by its rapidity\cite{jetcharge},
ALEPH has obtained an even
 tighter constraint on the rate of $B_s$ mixing.
The jet charge tagging technique allows
a high efficiency, of order 45\%, to be achieved with mistagging
probability of about 21\%.
The upper limit on $B_s$ mixing
is determined by performing a series of Monte Carlo
experiments. Allowing for systematic error
including a 30\% uncertainty in the $B_s$ fraction, they obtain
$\Delta M_s > 6$ ps $^{-1}$ or $x_s>8.5$
at the 95\% confidence level\cite{alephmix},\cite{fortymix}.
Similarly, OPAL uses the dilepton technique and
allows for a high frequency component in their fit. They
obtain $\Delta M_s > 2.2$ ps$^{-1}$ which gives
$x_s > 3.0$ at the 95\% confidence level\cite{opalmix}.

Other tagging techniques, for example, using the fragmentation
kaon to enhance the $B_s$ content or partial reconstruction of
the $D_s$ meson are being investigated by the high energy
collider experiments.

For the $B_s$ meson, the quantity $\Delta\Gamma$ may be large enough
to be observable. Parton model calculations\cite{Hagelin}
and calculations with exclusive final states\cite{Aleksan} suggest
that the width difference may be $10-20\%$. This lifetime
difference could be determined experimentally by using decays to
final states with different CP.
For example, a measurement of a difference in
the lifetimes between $\bar{B}_s^0\to \psi\phi$ and $\bar{B}_s^0\to
D_s^- \ell^+ \nu$
would yield $\Delta\Gamma/\Gamma^2$. It has also been
suggested that such measurements could be used to constrain
$|V_{ts}/V_{td}|^2$ if parton model calculations are reliable\cite{Browpak}.

%%%%\section{INCLUSIVE B DECAY}
%\input inclusive.tex
\section{INCLUSIVE B DECAY}

\subsection{Motivation}

Due to the large mass of the $b$ quark $B$ meson decays
give rise to a large
number of secondary decay products. For instance,
CLEO finds that the charged and photon multiplicities at the
$\Upsilon (4 S)$ are:
$n_{\rm charged}=10.99 \pm 0.06 \pm 0.29$,
$n_{ \gamma}=10.00\pm 0.53 \pm 0.50$, respectively
\cite{multi}.
Similarly, ARGUS \cite{multiARG} finds $n_{\rm charged}=10.74 \pm 0.02$.
The high multiplicity of final state particles leads to a large
number of possible exclusive final states. Even with a
detector that has a large acceptance
for both charged tracks and showers, it is difficult to reconstruct
many exclusive final states because of the combinatorial backgrounds.
Furthermore the detection efficiency drops for high multiplicity final states.
Thus, to get a complete picture of $B$ meson decay, it is important to
study inclusive decay rates.

A number of theoretical calculations of inclusive $B$ decay rates have been
made using the parton model.
It is believed that measurements
of such inclusive rates can be more reliably compared to the theoretical
calculations than can measurements of exclusive decays
While this is sufficient motivation for studying the inclusive rates,
there is also a need for accurate measurements in order to model the
decays of $B$ mesons both for high energy collider experiments, and for
experiments at the $\Upsilon (4S)$. As a specific example, the inclusive
rate for $B\to\psi$ has been used to determine the $B$ meson production
cross-section at the Tevatron \cite{pppsi}.

The branching ratios for inclusive $B$ decays to particular final state
particles are determined by measuring the inclusive yields of these
particles in data taken
on the $\Upsilon (4S)$ resonance, and subtracting the non-resonant background
using data taken at energies below the
$\Upsilon (4S)$ resonance. The off-resonance data are scaled to correct
for the energy dependence of the continuum cross-section.
Results on inclusive production at the $\Upsilon (4S)$ are usually presented
as a function of the variable $x$, which is the fraction of the maximum
possible momentum carried by the particle, $p_{max}=\sqrt{E_{beam}^2 - M^2}$.
The endpoint for production in $B$ decays is at $x=0.5$.

\subsection{Inclusive $B$ Decay to Mesons}

CLEO~1.5 \cite{CLEOK} has measured
the branching fractions of inclusive $B$ decays
to light mesons, while
ARGUS has determined the average multiplicities of light mesons
in $B$ decay.

\begin{figure}[htb]
\begin{center}
\unitlength 1.0in
\begin{picture}(3.,2.)(0,0)
\put(-.4,-.2){\psfig{width=3.0in,height=2.0in,%
file=bexcl_etamom.ps}}
\end{picture}
\bigskip
\vskip 10 mm
\caption{The momentum spectra for
$B\to \eta X$ as measured in CLEO~II data.}
\label{etamom}
\end{center}
\end{figure}
\begin{figure}[htb]
\begin{center}
\unitlength 1.0in
\begin{picture}(3.,2.7)(0,0)
\put(-.7,-.8){\psfig{width=4.5in,height=4.0in,%
file=jdl.ps}}
\end{picture}
\bigskip
\vskip 10 mm
\caption{$B\to D^0 X$, $D^+ X$, and $D^{*+}X$
momentum spectra in CLEO 1.5 data. The dashed curve is
the prediction of the phenomenological model of Wirbel and Wu while
  the solid histogram is the prediction of their free quark model}
\label{Fdmomdata}
\end{center}
\end{figure}

If more than one meson of the particle type under study is
produced in a $B\bar{B}$ decay,
then the branching fraction and the multiplicity
will differ. Unless otherwise noted,
the results reported in Table \ref{Tbmulti} are averaged over $B$ and
$\bar{B}$ decay.

\begin{table}[htb]
\caption{Multiplicities or branching fractions of light mesons per $B$ meson
decay.}
\label{Tbmulti}
\begin{tabular}{lll}
Mode & CLEO 1.5 \cite{CLEOK} & ARGUS \cite{ARGUSK} \\
     & (Branching Ratio) & (Multiplicity) \\ \hline
$ B/\bar{B}\to \pi^{\pm} $ &      & $ 3.59\pm 0.03\pm0.07$   \\
(not from $K_s,\Lambda$) & & \\
$ B/\bar{B}\to \pi^{\pm} $ &      & $ 4.11\pm 0.03\pm0.08$   \\
(incl. $K_s,\Lambda$) & & \\
$ B/\bar{B}\to K^{\pm} $ & $ 0.85\pm 0.07\pm 0.09$ & $0.78\pm 0.02\pm 0.03$ \\
$ \bar{B}\to K^{-} $  & $ 0.66\pm 0.05\pm 0.07$ &  \\
$ \bar{B}\to K^{+} $  & $ 0.19\pm 0.05\pm 0.02$ &  \\
$ B/\bar{B}\to K^0/\bar{K}^0 $ & $ 0.63 \pm 0.06\pm0.06$ & $0.64\pm 0.01 \pm
0.04$   \\
$ B/\bar{B}\to K^{*0} $    &    & $0.146\pm 0.016\pm 0.020$ \\
$ B/\bar{B}\to K^{*+} $    &    & $0.182\pm 0.054\pm 0.024$  \\
$ B/\bar{B}\to \rho^0 $    &    & $0.209\pm 0.042 \pm 0.033$ \\
$ B/\bar{B}\to \omega $    &    & $< 0.41$ (90\% C.L.)       \\
$ B/\bar{B}\to f_0(975) $  &    & $<0.025$ (90\% C.L.) \\
% UPDATE TEB
$ B/\bar{B}\to \eta   $  & $0.176\pm 0.011 \pm 0.0124$ (CLEO II) &  \\
$ B/\bar{B}\to \eta ' $  &    & $<0.15$ (90\% C.L.) \\
$ B/\bar{B}\to \phi $ & $ 0.023 \pm 0.006  \pm 0.005 $
& $0.039\pm 0.003\pm 0.004 $ \\
\end{tabular}
\end{table}

In the decay $b \to c \to s$ the
charge of the kaon can be used to determine the
flavor of the $b$ quark. A first attempt to measure the tagging efficiency and
misidentification probability for this method
has been performed by ARGUS \cite{ARGUSK}.
With the large sample of
reconstructed $B^0$ and $B^+$ decays from CLEO~II it should
be possible to measure these quantities directly.
The experiments also measure the momentum spectra for the particles listed
in Table \ref{Tbmulti}. An example of such data is
the momentum spectrum for $B\to \eta$ production shown in Figure~\ref{etamom}.
These results provide important information needed
to improve Monte Carlo generators and determine tagging efficiencies for
future $B$ experiments\cite{dunietztag}.

\begin{figure}[htb]
\begin{center}
\vskip 10mm
\unitlength 1.0in
\begin{picture}(3.,2.8)(0,0)
\put(-0.7,0.3){\psfig{width=4.5in,height=3.0in,%
bbllx=0pt,bblly=0pt,bburx=567pt,bbury=567pt,file=bdds_fig9_dswithfits.ps}}
\end{picture}
\caption{$B\to D_s X$ momentum spectrum in CLEO~II data.
The solid histogram
 is the sum of the two components. The two dotted histograms
indicate the two body components from $\bar{B}\to D^{(*)} D_s^{(*)-}$
and $\bar{B}\to D^{(**)} D_s^{(*)-}$.
The dash-dotted histogram shows the
contribution of the three body process.}
\label{Fdsmomdata}
\end{center}
\end{figure}
The inclusive production of $D^0, D^+, D_{s}^+$ and $D^{*+}$ mesons
in $B$ decay has been measured
by ARGUS \cite{ARGUSD} and CLEO~1.5 \cite{CLEOD}.
Preliminary measurements of several of these inclusive
branching fractions from CLEO~II have also
become available\cite{cleodds},\cite{dpfd0}.
To improve signal to background, only the $D^0 \to K^- \pi^+$,
$D^+ \to K^- \pi^+ \pi^+$ and $D_{s}^{+} \to \phi \pi^+$ decay modes are used.
The results, rescaled for the charm branching ratios,
are given in Table~\ref{khinc}.

Other detailed properties of inclusive B decay have been determined
in addition to branching fractions.
The momentum spectrum for the inclusive decay of $B$ mesons
to $D^0$, $D^+$, and $D^{*+}$ as measured by
CLEO~1.5 are shown in Fig. \ref{Fdmomdata}.
The $D^{*+}$ spectrum is not measured for $x <0.1$ due to poor
reconstruction efficiency for slow tracks.
The polarization as a function of $x$ for $B \to D^{*+}$ has also been
measured and was found to be consistent with the predictions of Wirbel
and Wu\cite{D*pol} and of Pietschmann and Rupertsberger\cite{PR}.

Analyses of the shape of the $D_s$ momentum
spectrum (Fig. \ref{Fdsmomdata})
indicates that there is a substantial two body component.
In model dependent fits the ARGUS and CLEO~1.5
experiments find two body fractions of $(58 \pm 7 \pm 9)$\%
\cite{ARGUSD} and $(56 \pm 10)$\% \cite{CLEOD}, respectively.
CLEO~II finds a somewhat smaller
two body fraction, $45.7\pm 1.9\pm 3.7\pm 0.6$
where the last error accounts for the uncertainty due to model
dependence in the predictions for the rates for
the two body modes\cite{cleodds}. There is no uncertainty in this results
from the $D_s \to \phi \pi$ branching fraction.
Averaging the results from the three experiments we find a two body
component of $(49.4 \pm 4.4)\%$ which leads to ${\cal{B}}(B\to D_sX\;
{\rm (two~body)})\; = \; (4.8\pm 1.3)\%$.
It is important to determine
which mechanisms are responsible for the production
of the remainder, the
 lower momentum $D_s$ mesons. Two possibilities are external
$W^-$ emission with $W^-\to \bar{c} s$ or $W^-\to \bar{u} d$ with
$s \bar{s}$ quark popping.

Results on inclusive $B$ decay to final states with $\psi$ and $\psi '$ mesons
have been reported by CLEO~1.5\cite{SecondB}, ARGUS\cite{FifthB}, and
CLEO~II\cite{CLEOpsiinc}. Indirect measurements of charmonium production
have been reported by  CDF \cite{cdf_cc} and the LEP experiments \cite{lep_cc}.
Because of the large uncertainties in the composition of their
data samples, these results have not been included
in our determination of the world averages listed in
Table \ref{khinc}.
In the most recent high statistics analysis from CLEO~II, the effect of final
state radiation has been taken into account \cite{dmc}.
This effect leads
to a significant tail on the low side of the
$\psi \to e^+ e^-$ mass peak and a smaller
effect in the $\mu^+ \mu^-$ spectrum. Even with a large mass window that
extends from $2.50$ to $3.05$ GeV$/c^2$, this effect
can modify the calculated detection efficiency by more than $10\%$.
Small corrections are also made for non-resonant $\psi$ production in the
CLEO~II analysis \cite{fastpsi}.
The resulting invariant dielectron and dimuon mass distributions are
shown in Fig. \ref{Fpsi}. The theoretical predictions for charmonia
production in $B$ decay\cite{bodwin,kuehn_psi,palmstech}
will be discussed further in Section~\ref{fac-color}.

\begin{figure}[htb]
\begin{center}
\unitlength 1.0in
\begin{picture}(3.,3.)(0,0)
\put(-2.0,-.8){\psfig{height=10.0in,file=cleo_charmonium_mass.ps}}
\end{picture}
\vskip 2mm
\caption{$B\to {\rm Charmonium} ~X$ invariant
mass spectra from CLEO II: (a) $\psi \to e^+ e^-$
channel and (b) $\psi \to \mu^+ \mu^-$ channel. (c) $\psi \gamma
- \psi$ mass difference showing the $\chi_{c1}$ and $\chi_{c2}$ signals.}
\label{Fpsi}
\end{center}
\end{figure}

\begin{figure}[htb]
\begin{center}
\unitlength 1.0in
\begin{picture}(3.,3.)(0,0)
\put(-1.9,-.8){\psfig{height=10.0in,file=cleo_charmonium_mom.ps}}
\end{picture}
\vskip 2 mm
\caption{$B\to {\rm Charmonium}~X$ momentum spectra in CLEO~II data.
(a) Inclusive $B \to \psi X$ production with  contributions from  individual
decay channels overlaid.
(b) Direct $B \to \psi X$ production.
(c) $B \to  \psi$'$X$.}
\label{Fpsimomdata}
\end{center}
\end{figure}

The momentum spectrum  for $B \to \psi, \psi^{'}$ transitions
has been measured (Fig.~\ref{Fpsimomdata}).
The two body component due to $B\to \psi K$ and $B\to \psi K^*$
saturates the spectrum in the momentum range between 1.4 and 2.0 GeV.
By subtracting the contributions from $\psi$'s
originating in $\psi$' and $\chi_c$ decays
CLEO and ARGUS measure the momentum
distribution of the direct component shown in  Fig.~\ref{Fpsimomdata}(b).
The average branching ratio for  direct $\psi$ production is found to be
${\cal{B}}(B \to \psi$, where $\psi$ not from $\psi ')
\; = \; (0.82 \pm 0.08)\%$.
The two body component constitutes about 1/3 of direct $\psi$ production.

\begin{table}[htb]
\caption{$\psi$ polarization $\Gamma_L/\Gamma $
in inclusive $B$ meson decays.}
\label{Tbpsipol}
\begin{tabular}{lll}
$\psi$ momentum & CLEO II \cite{psipol} & ARGUS \cite{argpol} \\ \hline
$p_{\psi}< 0.8$ GeV/c & $0.55 \pm 0.35 $ & \\
0.8 GeV/c $<p_{\psi}< 1.4$ GeV/c & $0.49 \pm 0.32 $ & \\
1.4 GeV/c $<p_{\psi}< 2.0$ GeV/c & $0.78 \pm 0.17 $ & $1.17\pm 0.17$ \\
all $p_{\psi}< 2.0$ GeV/c & $0.59 \pm 0.15 $ &
\end{tabular}
\end{table}

The polarization $\Gamma_L/\Gamma$
as a function of momentum for $B\to \psi$ transitions
 has also been determined (see Table~\ref{Tbpsipol}).
According to ARGUS, the $\psi$ mesons in the highest momentum bin are
completely longitudinally polarized.
Since the highest momentum
bin is dominated by two body $B$ decay, the polarization measured in this bin
can be used to estimate the polarization of $B \to \psi K^*$ after
correcting for the contribution of $B \to \psi K$.
Therefore the ARGUS result indicates that the $B \to \psi K^*$
mode is dominated by a single orbital angular momentum state and hence
by a single CP eigenstate.
Integrating over the range of kinematically allowed momenta,
CLEO measures the
average polarization of $\psi$ mesons in $B$ decay
to be $\Gamma_L/\Gamma \; =\; 0.59 \pm 0.15$.
This result is consistent with the
 longitudinal polarization of 54\% predicted by Palmer and
Stech \cite{palmstech}.
Using factorization and HQET, M. Wise finds significantly more
transverse polarization, $\Gamma_L/\Gamma\; \approx 0.25$
in the inclusive process\cite{wise_pol}.

Results on
inclusive $B \to \chi_c X, \chi_c \to \gamma \psi$
decays have been reported by ARGUS \cite{arguschi}
and CLEO~II \cite{CLEOpsiinc,fastpsi}.
ARGUS assumes there is no $\chi_{c2}$ production.
CLEO~II has significantly better $\chi_c$ mass resolution
than ARGUS and allows
for both possibilities.
The branching ratio for $\chi_{c0} \to \gamma \psi$ is
$(6.6\pm 1.8) \times 10^{-3}$ so the contribution
of the $\chi_{c0}$ meson to the $\psi\gamma$
final state can be neglected.
CLEO finds evidence at the 2.5 standard deviation level
for a $B\to \chi_{c2}$ contribution which would
indicate non-factorizable contributions or higher order processes
$O(\alpha_s^2)$ in $b\to c \bar{c} s$\cite{bodwin}.

The decay of $B$ mesons to the lightest charmonium state,
the $\eta_c$, has not
yet been observed. A recent CLEO~II search  placed a
 upper limit of 0.9\% on the process $B \to \eta_c X$
at the 90\% confidence level \cite{CLEOpsiinc}.

Using the results in Table~\ref{khinc}, it is possible to isolate the
component of $B\to \psi$ production which is due to production of higher
charmonium states in B decay and the direct component.
Similiarly, the direct
$B\to \chi_{c1}$ component can be determined by removing the contribution
from $B\to \psi^{'}$, $\psi^{'}\to \chi_{c1} \gamma$.
It is assumed, that all $\psi$' mesons are directly produced.

\begin{table}[htb]
\let\tabbodyfont\scriptsize
\caption{Branching fractions [\%] of inclusive $B$ decays }
\label{khinc}
 \begin{tabular}{l|lll|l}
\multicolumn{1}{l}{Particle} &
\multicolumn{1}{l}{ARGUS} &
\multicolumn{1}{l}{CLEO 1.5} &
\multicolumn{1}{l}{CLEO II} &
\multicolumn{1}{l}{Average} \\
\hline
 $\bar{B} \rightarrow \bar{D}^0 X$ & $ 49.7 \pm 3.8  \pm 6.4  \pm 2.6  $ &
$ 59.7 \pm 3.2  \pm 3.6  \pm 3.1  $ &
$ 63.8 \pm 1.1  \pm 2.0  \pm 1.7  $ &
$ 62.1 \pm 2.0  \pm 3.2  $ \\
 $\bar{B} \rightarrow D^- X$ & $ 23.0 \pm 3.0  \pm 4.4  \pm 1.5  $ &
$ 24.9 \pm 3.3  \pm 2.0  \pm 1.6  $ &
 &
$ 24.2 \pm 3.1  \pm 1.6  $ \\
 $\bar{B} \rightarrow D^{*-} X$ & $ 26.7 \pm 2.3  \pm 4.5  \pm 1.4  $ &
$ 22.7 \pm 1.3  \pm 2.3  \pm 1.2  $ &
 &
$ 23.5 \pm 2.3  \pm 1.2  $ \\
 $\bar{B} \rightarrow D_s^- X$ & $ 7.9 \pm 1.1  \pm 0.8  \pm 1.9  $ &
$ 8.3 \pm 1.2  \pm 2.0  $ &
$ 11.5 \pm 0.4  \pm 0.8  \pm 2.9  $ &
$ 9.8 \pm 0.6  \pm 2.4  $ \\
 $\bar{B} \rightarrow \phi X$ &  &
$ 2.3 \pm 0.6  \pm 0.5  $ &
 &
$ 2.3 \pm 0.8  $ \\
 $\bar{B} \rightarrow \psi X$ & $ 1.25 \pm 0.19  \pm 0.26  $ &
$ 1.31 \pm 0.12  \pm 0.27  $ &
$ 1.13 \pm 0.04  \pm 0.06  $ &
$ 1.15 \pm 0.07  $ \\
 $\bar{B} \rightarrow \psi X$ (direct) & $ 0.95 \pm 0.27  $ &
 &
$ 0.81 \pm 0.08  $ &
$ 0.82 \pm 0.08  $ \\
 $\bar{B} \rightarrow \psi$'$ X$ & $ 0.50 \pm 0.19  \pm 0.12  $ &
$ 0.36 \pm 0.09  \pm 0.13  $ &
$ 0.34 \pm 0.04  \pm 0.03  $ &
$ 0.35 \pm 0.05  $ \\
 $\bar{B} \rightarrow \chi_{c1} X$ & $ 1.23 \pm 0.41  \pm 0.29  $ &
 &
$ 0.40 \pm 0.06  \pm 0.04  $ &
$ 0.42 \pm 0.07  $ \\
 $\bar{B} \rightarrow \chi_{c1} X$ (direct) &  &
 &
$ 0.37 \pm 0.07  $ &
$ 0.37 \pm 0.07  $ \\
 $\bar{B} \rightarrow \chi_{c2} X$ &  &
 &
$ 0.25 \pm 0.10  \pm 0.03  $ &
$ 0.25 \pm 0.10  $ \\
 $\bar{B} \rightarrow \eta_{c} X$ &  &
 &
$ <0.90 $  (90\% C.L.) &
 $ <0.90 $  (90\% C.L.) \\
  $\bar{B} \rightarrow p X$ & $ 8.2 \pm 0.5  \pm 1.2  $ &
$ 8.0 \pm 0.5  \pm 0.3  $ &
 &
$ 8.0 \pm 0.5  $ \\
 $\bar{B} \rightarrow \bar{\Lambda} X$ & $ 4.2 \pm 0.5  \pm 0.6  $ &
$ 3.8 \pm 0.4  \pm 0.6  $ &
 &
$ 4.0 \pm 0.5  $ \\
 $\bar{B} \rightarrow \Xi ^+ X$ & $ <0.51 $ (90\% C.L.)&
 $ 0.27 \pm 0.05  \pm 0.04  $ &
 &
$ 0.27 \pm 0.06  $ \\
 $\bar{B} \rightarrow \Lambda _c^- X$ & $ 7.0 \pm 2.8  \pm 1.4  \pm 2.1  $ &
$ 6.3 \pm 1.2  \pm 0.9  \pm 1.9  $ &
$ 4.2 \pm 0.5  \pm 0.6  \pm 1.3  $ &
$ 4.7 \pm 0.7  \pm 1.4  $ \\
 $\bar{B} \rightarrow \Sigma_c^0 X$ &  &
 &
$ 0.53 \pm 0.19  \pm 0.16  \pm 0.16  $ &
$ 0.53 \pm 0.25  \pm 0.16  $ \\
 $\bar{B} \rightarrow \Sigma_c^0 \bar{N}$ &  &
 &
$ <0.17 $  (90\% C.L.) &
 $ <0.17 $  (90\% C.L.) \\
  $\bar{B} \rightarrow \Sigma_c^{++} X$ &  &
 &
$ 0.50 \pm 0.18  \pm 0.15  \pm 0.15  $ &
$ 0.50 \pm 0.23  \pm 0.15  $ \\
 $\bar{B} \rightarrow \Sigma_c^{++} \bar{\Delta}^{--}$ &  &
 &
$ <0.12 $  (90\% C.L.) &
 $ <0.12 $  (90\% C.L.) \\
  $\bar{B} \rightarrow \Xi_c^+ X$ &  &
 &
$ 1.5 \pm 0.7  $ &
$ 1.5 \pm 0.7  $ \\
 $\bar{B} \rightarrow \Xi_c^0 X$ &  &
 &
$ 2.4 \pm 1.3  $ &
$ 2.4 \pm 1.3  $ \\
\end{tabular}
\let\tabbodyfont\small
\end{table}

Using the procedures outlined in Section II the results reported by the
different experiments have been rescaled to accommodate the new charm branching
ratios. The world averages for inclusive $B \to$~meson  decays are given in 
Table~\ref{khinc}.

\subsection{Inclusive $B$ Decay to Baryons }

ARGUS\cite{argusbary}
and CLEO~1.5\cite{crawbary} have observed inclusive production of $\bar{p}$,
$\Lambda$, $\Xi$ and the charmed $\Lambda_c$ baryon. Recently
CLEO~II has reported the observation 
of $B \to \Sigma_c X$\cite{sigmamz}, $B\to \Xi_c^{0} X$ and
$B\to \Xi_c^{+} X$\cite{cleocascade}.
The measured branching ratios for these decays and the world
averages can be found in Table \ref{khinc}.

The determination of branching ratios for inclusive $B$ decays to the charmed
baryons $\Lambda_c$ and $\Sigma_c$ requires knowledge of 
${\cal{B}}(\Lambda_c^+ \to pK^-\pi^+)$.
However, the uncertainty in this quantity is still large as it can only
be determined by indirect and somewhat model dependent 
methods. The results given in this review use
${\cal{B}}(\Lambda_c^+ \to pK^-\pi^+)\:=\:
(4.3 \pm 1.0 \pm 0.8)$\% \cite{crawbary}.
For modes involving $\Lambda_c$ baryons the uncertainty due to
the $\Lambda_c$ branching ratio scale is listed as a separate error.

The momentum spectrum of $B\to \Lambda_c$ transitions has been measured
by CLEO~1.5 \cite{crawbary} and ARGUS \cite{argusbary}. 
The result of a recent CLEO~II \cite{sigmamz} measurement is shown in
Fig.~\ref{lambdacmom}(a). The momentum spectrum is rather soft indicating
$\Xi_c$ production or the presence of a significant multibody component.
Similarly,
CLEO~II has found that $B\to \Sigma_c^0 X$ and $B\to \Sigma_c^{++} X$ 
decays have no two body contribution. 

In addition to the inclusive branching ratios given above, the experimental
data has been used in  attempts to disentangle which of the baryon
production mechanisms shown in  Fig.~\ref{btobaryon} dominates.
CLEO~1.5 \cite{crawbary}
and ARGUS \cite{argusbary} have investigated baryon correlations in  $B$ decay in order
to elucidate the underlying decay process.
\begin{table}[htb]
\caption{Branching fractions [\%] of 
inclusive $B$ decays to baryon pairs.}
\label{Tbbaryonp} 
\begin{tabular}{lll} 
Mode          &       CLEO 1.5 & ARGUS                     \\ \hline
$B\to p \bar{p}~X$ & $2.4\pm 0.1\pm 0.4   $ & $ 2.5\pm 0.2\pm 0.2$  \\
$B\to \Lambda \bar{\Lambda}~X$ & $<0.5$ (90\% C.L.)  & $<0.88$ (90\% C.L.) \\
$B\to \Lambda \bar{p} ~X$ & $2.9\pm 0.5\pm 0.5  $ & $2.3\pm 0.4\pm 0.3$  \\
% corrected for D0 and D* branching ratios
$B\to D^{*+} p \bar{p} ~X$ & $< 0.35 $ (90\% C.L.) & $       $       \\
$B\to D N \bar{N} ~X$ & $<5.2$ (90\% C.L.) & $       $       
\end{tabular}
\end{table}
We follow the notation of Reference\cite{crawbary} . Let N
denote baryons with $S=C=0$ (e.g. p, n, $\Delta$, $N^*$). Let Y
refer to baryons with $S=-1, C=0$ (e.g. $\Lambda$, $\Sigma^0$, $\Sigma^+$).
Let $Y_c$ refer to baryons with $S=0, C=1$ [e.g. $\Lambda_{c}^{+}$,
$\Sigma_{c}^{(+,0,++)}$] . Then the following final states can be used
to distinguish possible mechanisms for baryon production in $B$ decay 
(Fig. \ref{btobaryon}).
\begin{figure}[htb]
\begin{center}
\unitlength 1.0in
\begin{picture}(3.,2.5)(0,0)
\put(-1.1,-0.1)
{\psfig{bbllx=0pt,bblly=0pt,bburx=567pt,bbury=567pt,%
width=6.0in,height=4.0in,file=btobaryon.ps}}
\end{picture}
\vskip 10 mm
\caption{
Decay diagrams for $B$ meson decays to baryons: (a) External spectator
diagram (b) W Exchange diagram
(c) External spectator diagram which produces $D N \bar{N} X$ 
and $D Y \bar{Y} X$ final states (d) Internal spectator diagram which  
produces $DN\bar{N}X$ and $DY\bar{Y}X$ final states.}
\label{btobaryon}
\end{center}
\end{figure}
\begin{figure}[htb]
\begin{center}
\unitlength 1.0in
\begin{picture}(2.0,1.5)(0.0,0.0)
\put(-0.6,0.81){\psfig{width=3.8in,height=1.1in,file=intbaryon.ps}}
\end{picture}
%\vskip 10 mm
\caption{
Baryon production in $B$ meson decay via internal $W$ emission.
(a) $b \to c\bar{u}d$ with $q\bar{q}$ popping,
(b) $b \to c\bar{c}s$ with $q\bar{q}$ popping.}
\label{intbaryon}
\end{center}
\end{figure}
\begin{figure}[htb]
\begin{center}
\unitlength 1.0in
\begin{picture}(4.,2.7)(0,0)
\put(-1.1,-0.2)
{\psfig{bbllx=0pt,bblly=0pt,bburx=567pt,bbury=567pt,%
width=6.5in,height=5.5in,file=lambdac_momentum.ps}}
\end{picture}
\vskip 2 mm
\caption{
Momentum spectrum of $\Lambda_c$ baryons from $B$ decay (CLEO~II).
(a) The overlaid histograms show the spectra from two components
of the internal W-emission process $b\to c \bar{c} s$, 
$\bar{B}\to \Xi_c\bar{\Lambda_c}$,
and $\bar{B}\to \Xi_c^{'}\bar{\Lambda_c}$.
(b)The overlaid histograms are the
results of a Monte Carlo study assuming multibody $\Lambda_c + (n\pi)$
final states with 
different numbers of additional pions.
(c) The $\Lambda_c$ momentum spectrum for events with 
a $\Lambda_c^+$ in coincidence with a high momentum  lepton ($\ell^-$) tag.
(d) The $\Lambda_c$ momentum spectrum for events with 
a $\Lambda_c^+$ in coincidence with a high momentum  lepton ($\ell^+$) tag.
}
\label{lambdacmom}
\end{center}
\end{figure}
\begin{figure}[htb]
\begin{center}
\unitlength 1.0in
\begin{picture}(4.,4.0)(0,0)
\put(-1.1,-0.7)
{\psfig{bbllx=0pt,bblly=0pt,bburx=567pt,bbury=567pt,%
width=5.5in,height=4.5in,file=lambdac_lep_cor.ps}}
\end{picture}
\vskip 2 mm
\caption{$\Lambda_c -$ lepton correlation in $B$ decay (CLEO~II).
(a) The $p K^-\pi^+$ invariant 
mass spectrum for $\Lambda_c^+-\ell^+$ combinations.
(b) The $p K^-\pi^+$ invariant mass 
spectrum for $\Lambda_c^+-\ell^-$ combinations.}
\label{lambdalep}
\end{center}
\end{figure}
\begin{figure}[htb]
\begin{center}
\unitlength 1.0in
\begin{picture}(3.,2.5)(0,0)
\put(-1.1,-0.1)
{\psfig{bbllx=0pt,bblly=0pt,bburx=567pt,bbury=567pt,%
width=5.0in,height=5.0in,file=bksi_ksi_mass.ps}}
\end{picture}
\vskip 4 mm
\caption{Evidence for $\Xi_c$ production in $B$ decays (CLEO~II).
(a) Continuum subtracted $\Xi_c^0\to \Xi^-\pi^+$ invariant mass distribution.
(b) Continuum subtracted $\Xi_c^+\to \Xi^-\pi^+\pi^+$ 
invariant mass distribution.}
\label{bksimass}
\end{center}
\end{figure}

\begin{enumerate}
\item {$\bar{B} \to Y_c \bar{N} X$, $\bar{B} \to \Xi_c \bar{Y} X$}\\
These final states are produced by the usual $b \to c W^-$ coupling
in a spectator or exchange diagram in conjunction with the popping of
two quark pairs from the vacuum (as shown in Figs.~\ref{btobaryon}(a),(b)).
 It should be noted that the two
mechanisms can be distinguished by examination of the $Y_c$ momentum
spectrum,  since the exchange diagram will
produce two body final states (e.g. $\Lambda_c \bar{p}$ or
$\Sigma_c^{++} \bar{\Delta}^{--}$).

\item {$\bar{B}\to D N \bar{N} X$, $\bar{B} \to D Y \bar{Y} X$}\\
 The non-charmed baryon-antibaryon pair is produced from W fragmentation
after hadronization with two quark-antiquark pairs popped from
the vacuum (as shown in Figs.~\ref{btobaryon}(c),(d)). 
The $D$ meson is formed from the charm spectator quark system.
If this mechanism is significant, inclusive production of
charmless baryon-antibaryon pairs should be
observed in $B$ decay.

\item{$\bar{B} \to Y_c \bar{Y} X$,  $\bar{B} \to \Xi_c \bar{Y_c} X$}\\
 These states are produced by the internal spectator graph
with $W^- \to \bar{c} s$ in conjunction with the popping of two quark
antiquark pairs. Since ${\cal B} (W^- \to \bar{c} s)/ {\cal B} (W^- \to all)$
is about $0.15$, this mechanism may be suppressed.

\item {$\bar{B}\to D_{s}^{-} Y_c \bar{N} X$,
 $\bar{B}\to D_{s}^{-} \Xi_c \bar{Y} X$}\\
This is the same as mechanism (1) with $W^- \to \bar{c} s$. 
\end{enumerate}

The low rates for $B\to \Lambda \bar{\Lambda} X$, $\Lambda \bar{p} X$ and
$D^* p \bar{p} X$(see Table \ref{Tbbaryonp}) 
suggest that mechanism (2) is small. 
The absence of a two body component in the momentum spectra
 of $B\to \Lambda_c X$, $\Sigma_c X$ indicates that the W-exchange
mechanism is small. Thus it was thought
reasonable to assume that $\bar{B}\to Y_c \bar{N} X$ with
an external spectator  $b\to c W^-$
coupling (Fig.~\ref{btobaryon}(a)) 
 is the principal mechanism in $B$ to baryon transitions.

If $B$ decays to baryons are dominated by $\bar{B} \to \Lambda_c \bar{p} X$
and $\bar{B} \to \Lambda_c \bar{n} X$ then
measurements of the branching ratios for
 $B \to \bar{p} X$,
$B \to p \bar{p} X$ can be used to extract the absolute $\Lambda_c \to
p K^- \pi^+$ branching ratio.  The CLEO~1.5 measurements give
$B (\Lambda_c \to p K^- \pi^+) = 4.3 \pm 1.0 \pm 0.8 \%$ which can be
used to normalize all other measured $\Lambda_c$ branching ratios. 
In a similar
fashion, ARGUS finds $(4.1\pm 2.4)$\% for this branching ratio.

An alternate explanation for the absence of a two body component
in $B$ decays to baryons was recently proposed by Dunietz, Falk and
Wise\cite{dunietzbary}. These authors suggested that the primary
mechanism in such decays is 
the internal W-emission process $b\to c \bar{c} s$.
This might lead to two body final states such as 
$\bar{B}\to \bar{\Lambda_c} \Xi_c$
which would account for the softness of the $\Lambda_c$ momentum spectrum.  
CLEO has searched for the mechanism suggested by Dunietz \etal~in 
a variety of ways.
By examining $\Lambda_c$-lepton correlations, 
it is possible to constrain the size of the $b\to c \bar{c} s$ component
in $B\to {\rm baryon}$ decays. The $b\to c \bar{c} s$ component gives
rise to opposite sign $\Lambda_c^+ \ell^-$ correlations whereas the internal
process W-emission process
$b\to c u {\bar d}$ gives same sign $\Lambda_c^+ \ell^+$ correlations
(Fig.~\ref{lambdalep}(a)).
From the ratio of same sign to opposite sign $\Lambda_c$-lepton 
yields, CLEO finds $b\to c \bar{c} s/
b\to c \bar{u} d = (20\pm 13 \pm 4) \%$ for internal W-emission processes. 
This shows that $b\to c \bar{c} s$, although present, is not
the dominant mechanism operating in B decays to baryons. 

CLEO~II has measured the $\Lambda_c^+$ momentum spectrum 
separately for $\Lambda_c^+\ell^-$ and $\Lambda_c^+\ell^+$ correlations
(Figs.~\ref{lambdalep}(b), (c)). The $\Lambda_c^+$ 
momentum spectrum is somewhat softer
in events containing an additional $\ell^-$ tag. This
is consistent with the
expectation that $b\to c\bar{c}s$ transitions produce $\Lambda_c^+$
baryons accompanied by very massive $\bar{\Xi}_c$ baryon.
On the other hand, in $b\to c\bar{u}d$ transitions, the $\Lambda_c^+$ is
produced in association with a lighter nucleon or nucleon resonance, which
should result in a hard $\Lambda_c^+$ momentum spectrum.

Since the $b\to c \bar{c} s$ mechanism is present, $\Xi_c^+$
and $\Xi_c^0$ baryons should be
 produced in $B\to$baryon transitions.
However, $\Xi_c^0$ baryons can also be produced from $b\to c\bar{u}d$
transitions with $s\bar{s}$ popping.
Naively, one estimates $s\bar{s}$ popping to be approximately 15\% of
all $q\bar{q}$ popping. Thus this mechanism should contribute
$\displaystyle
\frac{B\to\Xi_c\bar{\Lambda}X}{B\to \Lambda_c N X}\approx 0.15$ to the
observed $B\to \Xi_c$ rate. A simple phase space argument gives
$\displaystyle\frac{W^- \to \bar{c}s}{W^- \to \bar{u}d}\approx 0.30$. 
Combining these two contributions, one expects a
$B\to \Xi_c$ branching ratio of  $0.45 \times {\cal{B}}(B\to \Lambda_c X)$.
Experimentally, the sum of the rates for $B\to \Xi_c^+$  and $B\to \Xi_c^0$ 
decays relative to $B\to \Lambda_c X$ is consistent with this
expectation
$$
\frac{{\cal{B}}(B\to \Xi_c X)}{{\cal{B}}(B\to \Lambda_c X)}
\; = \; 0.8 \pm 0.4
$$
However, the $\Xi_c$ absolute branching ratio scale is poorly known
and the experimental errors need to be reduced before any final conclusion
can be deduced from this ratio.

To verify whether the dominant mechanism for baryon production in
B decays is the external spectator mechanism 
with $b\to c \bar{u} d$ as was previously assumed 
by the CLEO and ARGUS analyses, 
CLEO~II has searched for evidence of $B\to \Lambda_c
\bar{N} \ell \nu$. This should give rise to several 
distinctive experimental signatures:
$\Lambda$-lepton correlations, $\Lambda_c$-lepton correlations,
and semi-exclusive $B\to \Lambda_c^+ \bar{p} \ell^- \nu$ production having a 
massing mass consistent with a B decay. No significant signals were
observed and limits (at the 90\% C.L.) of 
$(B\to \Lambda_c \bar{N} X\ell\nu)/(B\to \Lambda_c X)
$ $<5.7\%$, 
$(B\to \Lambda_c \bar{N}\ell\nu)/(B\to \Lambda_c X)
$ $<6\%$, 
$(B\to \Lambda_c \bar{p}\ell\nu)/(B\to \Lambda_c X)
$ $<10\%$, respectively, were obtained\cite{glasbary}. 
These limits indicate that
the conventional and previously
accepted picture of baryon production in $B$ decay is incorrect.

A possible explanation of all the existing data requires the
simultaneous presence
 of several production mechanisms. The internal spectator
process $b\to c \bar{u} d$ followed by $u \bar{u}$ 
or $d \bar{d}$ quark popping is
dominant. This leads to production of
a high mass excited anti-nucleon in conjunction
with a charmed baryon and accounts for the soft momentum spectrum
of charmed baryons produced in B decay as well as the absence of
$B\to \Lambda_c \bar{N} X \ell \nu$. The internal spectator process
$b\to c \bar{c} s$ with quark popping as well as the internal spectator
process $b\to c \bar{u} d$ with $s \bar{s}$ quark popping are also
operative at the 10-20\% level. The latter two mechanisms
account for the production of $\Xi_c$ baryons in B decay.

\subsection{Charm Production in $B$ Decay}
\label{charmpro}

The measurements of inclusive decay rates can be used to test the parton level
expectation that most $B$ decays proceed via a $b\to c$ transition.
If we neglect the small contributions from $b\to u$
and penguin transitions, we expect about
1.15 charm quarks to be produced per  $B$ decay.
The additional $15\%$ is due
to the fact that the virtual W forms a $s \bar{c}$ quark pair with
a probability of approximately $0.15 $. 
To verify this expectation we use the experimental results listed in 
Table~\ref{khinc} and determine the charm yield to

\begin{eqnarray*}
{\rm Charm~yield} & = & {\cal B}(B \to D^0 X) + {\cal B}(B \to D^+ X) +
{\cal B}(B \to D_s X)  \\
 & +&  {\cal B}(B \to \Lambda_c X) +  {\cal B}(B \to \Xi^+_c X) + 
 {\cal B}(B \to \Xi^0_c X) \\
 & + &2\times{\cal B}(B\to \psi X)
 + 2\times{\cal B}(B\to \psi{\rm '} X)
 + 2\times{\cal B}(B\to \chi_{c1} X)  \\
 &+& 2\times{\cal B}(B\to \chi_{c2} X)
 + 2\times{\cal B}(B\to \eta_c X~({\rm incl.~other ~c\bar{c}}))\\
 & = & 1.10 \pm 0.06 \\
\end{eqnarray*}

The factor of 2 which multiplies ${\cal{B}}(B\to c\bar{c}X)$ accounts for the 
two charm quarks produced in $b \to c\bar{c}s$ transitions. Wherever possible
the branching fractions for direct production are used.
The contribution of $B\to \eta_c X$ and
other charmonia is generously taken to be at the CLEO 90\% confidence
level upper limit for the process $B\to \eta_c X$.

Another interesting quantity is the fraction of $B$ decays in which two charm
quarks are produced.
In a parton level calculation, Palmer and Stech \cite{palmstech}
find that ${\cal{B}}(B \to X_{c \bar{c}}) = 19 \pm 1 \%$
where the theoretical error is the uncertainty due to the choice
of quark masses. This can be 
compared to the sum of the experimental measurements
\begin{eqnarray*}
{\cal{B}}(B \to X_{c \bar{c}}) & = & 
{\cal{B}}(B \to D_s X) + {\cal{B}}( B \to \psi X) + {\cal{B}}(B\to \psi' X) \\
 & +& {\cal {B}}(B\to \chi_{c1} X) + {\cal {B}}(B\to \chi_{c2} X) +
{\cal {B}}(B\to \Xi_c X)  \\
 & +& {\cal {B}}(B\to \eta_c X~({\rm incl.~other ~\bar{c}})) \\
 &= & (15.6 \pm 2.7)\% \\
\end{eqnarray*}
where the direct $B \to \psi$ and  $B \to \chi_{c1}$ branching fraction have
been used. 
The contribution from $B\to \Xi_c^0 X$ is reduced by 1/3 to take
into account the fraction that is not produced by the 
$b\to c \bar{c} s$ subprocess but by
$b\to c\bar{u}d \, + \, s \bar{s}$ quark popping. 

With the addition of these recent
experimental results the understanding of baryon production in B decay is 
improving. In contrast
to meson production in $B$ decay, $B \to {\rm baryon}$
transitions proceed predominantly through the internal W-emission process
$b\to c\bar{u}d$ followed by light quark pair popping.
In a parton level calculation with diquark correlation taken into
account, Palmer and Stech \cite{palmstech}
have performed a calculation of the total rate
for inclusive $B$ decay to charmed baryons.
They find ${\cal{B}}(B \to$ charmed baryons) $\approx 6\%$.
In order to compare this prediction with experimental data,
we will
assume most $B$ to charmed baryon decays proceed through a $\Lambda_c$
baryon but correct for the small fraction of $\Xi_c$ baryons produced
by $b \to c\bar{u}d$ transitions combined with $s\bar{s}$-popping.
This gives
\begin{center}
\begin{tabular}{rcl}
${\cal B}(B \to {\rm charmed~baryons})$ & = & 
${\cal{B}}(B\to \Lambda_c X) + 1/3 \times {\cal{B}}(B\to \Xi_c^0)$\\
 & =  & $(5.5 \pm 1.6)\% $
\end{tabular}
\end{center}
\begin{table}[htb]
\caption{CLEO II results on exclusive branching ratios for $B \to $baryon
transitions \protect\cite{exclbaryon}.}
\label{exclbaryons}
\begin{tabular}{lll}
$B$ mode & Events observed & ${\cal{B}}$ [\%]\\ \hline
$\bar{B}^0 \to \Lambda_c^+\bar{p}$ & $<2.3$ & $< 4.4 \times 10^-2$\\
$\bar{B}^0 \to \Lambda_c^+\bar{p}\pi^0$ & $<2.3$ & $< 0.076$ \\
$\bar{B}^0 \to \Lambda_c^+\bar{p}\pi^+\pi^- $& $15.0\pm4.7$ & 
$0.187 \pm 0.059 \pm 0.056 \pm 0.045$\\
$\bar{B}^0 \to \Lambda_c^+\bar{p}\pi^+\pi^-\pi^0$ & $<11.6$ & $< 0.76$\\
$\bar{B}^0 \to \Lambda_c^+\bar{p}\pi^+\pi^-\pi^+\pi^-$ & $<6.4$ & $< 0.34$\\
\hline
$B^- \to \Lambda_c^+\bar{p}\pi^-$ & $<6.4$ & $< 0.084 $\\
$B^- \to \Lambda_c^+\bar{p}\pi^-\pi^0$ & $<8.7$ & $< 0.36$ \\
$B^- \to \Lambda_c^+\bar{p}\pi^-\pi^+\pi^-$ & $<14.7$ & $< 0.55$ \\
$B^- \to \Lambda_c^+\bar{p}\pi^-\pi^+\pi^-\pi^0$ & $<15.6$ & $< 2.17 $
\end{tabular}
\end{table}

The experimental result for the charm yield per $B$ decay
is consistent with the naive expectation
that $1.15$ charm quarks are produced per $b$ decay. However,
it does not support a number of  proposals
which suggest that at least $1.3$ quarks should
be produced per $b$ decay. 
In these recent theoretical efforts, large charm quark yields 
are a consequence of modifying
the heavy quark masses in order to explain
the discrepancy between 
theoretical calculations and experimental measurements of
the inclusive semileptonic rate, ${\cal B}(B \to X\ell \nu)$ 
(see Section~\ref{baffle})\cite{falk_baffle}.

The data are not yet sufficiently precise 
to convincingly rule out the possibility of a larger
charm yield. In addition, there
are several possible systematic flaws in the computation of the yield of
charm quarks.
The charm meson absolute branching
fractions can contribute a systematic uncertainty, although the 
errors from this source
 have been significantly reduced by the recent precise determinations
of ${\cal B}(D^0\to K^-\pi^+)$\cite{DKpi}
and ${\cal B}(D^+\to K^-\pi^+\pi^+)$. However, the absolute branching
fraction scales for the $D_s$ meson and $\Lambda_c$ baryons are
still quite uncertain. 
Since the inclusive branching ratios to
these particles are small, a substantial change to the branching ratio
scale would be required to significantly modify the charm yield.

There could also
be a large contribution to the inclusive rate that has not been measured.
 It has been suggested by Palmer and Stech\cite{palmstech},
that $b \to c \bar{c} s$ followed by $c \bar{c} \to \rm{gluons}$,
 which in turn hadronize into a final state with no charm, has a large
branching ratio. The charm content for this mechanism would not be properly
taken into account.
Another related suggestion is that the rate for the hadronic penguin
diagram $b\to sg$ is larger than expected\cite{kaganbsg}.

\section{EXCLUSIVE B DECAY TO BARYONS}
The first exclusive $B\to$baryon decay has been observed by
CLEO~II\cite{exclbaryon}. 
A small signal was reconstructed in the
mode $\bar{B}^0\to \Lambda_c \bar{p} \pi^+ \pi^-$ corresponding to
a branching ratio of $0.187\pm 0.059\pm 0.056\pm 0.045\% $.
In addition, CLEO ~II has 
set limits on other exclusive modes which are given
in Table~\ref{exclbaryons}.

%%%%%\section{EXCLUSIVE B DECAY TO D MESONS}
%\input bd.tex 
\begin{figure}[p]
\begin{center}
\unitlength 1.0in
\begin{picture}(3.,3.)(0,0)
\put(-.35,0.0){\psfig{width=2.5in,height=2.5in,%
file=bexcl_dstr1.ps}}
\end{picture}
\bigskip
\bigskip
\vskip 15 mm
\caption[]{Beam constrained mass
 distributions from CLEO~II for: 
(a) $B^- \to D^{*0} \pi^-$ decays, 
 (b) $B^- \to D^{*0} \rho^-$ decays,
 (c) $\bar{B}^0 \to
D^{*+} \pi^-$ decays,  and
  (d) $\bar{B}^0
\to D^{*+} \rho^-$ decays.}
\label{dspi}
\end{center}
%\end{figure}

%\begin{figure}[htb]
\vskip 2 mm
\begin{center}
\unitlength 1.0in
\begin{picture}(2.2,2.2)(0.0,0.0)
\put(-1.,-0.9){\psfig{width=4.in,height=4.in,file=bexcl_fig11.ps}}
\end{picture}
\vskip 10 mm
\caption[]{Resonant substructure for $B\to D^* \rho^-$ from CLEO~II for:
 (a) the
$\pi^0\pi^-$ invariant mass spectrum for the 
$ \bar{B}^0 \to D^{*+} \pi^0\pi^-$ decay mode in data.
 (b) the
$\pi^0\pi^-$ invariant mass spectrum for the 
$ \bar{B}^0 \to D^{*+} \pi^0\pi^-$ decay mode in data.}
\label{subs}
\end{center}
\end{figure}

%\begin{figure}[htb]
%\unitlength 1.0in
%\vskip 10 mm
%\begin{tabular}{cc}
%\begin{picture}(2.0,2.3)(0.0,0.0)
%\put(.6,0.0){\psfig{width=1.8in,height=1.8in,file=bexcl_fig15.ps}}
%\end{picture}
%&
%\begin{picture}(2.0,2.3)(0.0,0.0)
%\put(1.7,0.0){\psfig{width=1.8in,height=1.8in,file=bexcl_fig12.ps}}
%\end{picture}
%\end{tabular}
%\vskip 15 mm
%\caption{a) $\bar{B^0}\to  D^{*+}\pi^-\pi^-\pi^+$, where the
%$\pi^-\pi^-\pi^+$ invariant mass is required to be consistent with the
%$a_1^-$ mass.
%b) $\bar{B^0}\to D^{*+}\pi^-\pi^-\pi^+$, where the
%$\pi^-\pi^-\pi^+$ invariant mass is required to be in the
%$a_1^-$ mass sidebands.}
%\label{FBmaone}
%\end{figure}

\begin{figure}[p]
\unitlength 1.0in
\vskip 10 mm
\begin{picture}(3.,2.0)(0,0)
\put(1.4,-1.1){\psfig{width=3.0in,height=2.6in,%
file=bexcl_dstr2.ps}}
\end{picture}
\bigskip
\caption[]{Beam constrained mass
 distributions from CLEO~II for: 
(a) $B^- \to D^{*0} a_{1}^{-}$ and 
 (b) $\bar{B}^0 \to D^{*+} a_{1}^{-}$.}
\label{FBmaone}
%\end{figure}

%\begin{figure}[htb]
\unitlength 1.0in
\vskip 18 mm
\begin{center}
\begin{picture}(2.2,2.4)(0.0,0.0)
\put(-0.7,-0.8){\psfig{bbllx=0pt,bblly=0pt,bburx=567pt,bbury=567pt%
,width=3.7in,height=3.3in,file=bexcl_ma1dst.ps}}
\end{picture}
\vskip 15 mm
\caption[]{Resonant substructure of $\bar{B^0}\to D^{*+} a_1$ from
CLEO~II:
(a) The $\pi ^- \pi ^- \pi ^+$ invariant mass spectrum from
a Monte Carlo simulation of $\bar {B}^0 \to D^{*+} a_1^-$
(b) The $\pi ^- \pi ^- \pi ^+$ invariant mass spectrum from
Monte Carlo simulation for $\bar {B}^0 \to D^{*+} (\pi ^- \rho ^0)_{NR}$
(c) The $\pi ^- \pi ^- \pi ^+$ mass spectrum from data
after $B$ mass sideband subtraction. The fit
to the sum of (a) and (b) is superimposed.}
\label{Fmaonea}
\end{center}
\end{figure}

\begin{figure}[htb]
\unitlength 1.0in
\vskip 10 mm
\begin{center}
\begin{picture}(3.0,3.4)(0.0,0.0)
\put(-.35,.3){\psfig{width=2.5in,height=2.5in,file=bexcl_d24202460.ps}}
\end{picture}
\vskip 15 mm
\caption[]{
Beam constrained mass distributions from CLEO~II
 for:  (a) $B^- \to D^{**0}(2420)
\pi^-$ where $D^{**0}(2420) \to D^{*+} \pi^-$, (b) $B^- \to D^{**0}(2460)
\pi^-$ where $D^{**0}(2460) \to D^{*+} \pi^-$,  (c) $B^- \to
D^{**0}(2420) \pi^-\pi^0$ where 
$D^{**0}(2420) \to D^{*+}\pi^-$,  (d) $B^- \to D^{**0}(2460) \pi^-\pi^0$
where $D^{**0}(2460) \to D^{*+} \pi^-$}
\label{dsspi}
\end{center}
\end{figure}

\begin{figure}[htb]
\unitlength 1.0in
\vskip 10 mm
\begin{center}
\begin{picture}(3.0,3.5)(0.0,0.0)
\put(-0.35,0.3){\psfig{width=2.5in,height=2.5in,file=bexcl_fig22.ps}}
\end{picture}
\vskip 15 mm
\caption[]{Angular distributions (efficiency corrected)
from CLEO~II for
(a) the helicity angle from $D^{*+} \to D^0 \pi ^+$ 
in  $\bar {B^0} \to D^{*+} \rho ^-$ and
(b)the helicity angle from $\rho ^- \to \pi ^- \pi ^0$
in  $\bar {B^0} \to D^{*+} \rho ^-$  (c) the helicity angle from
$D^{*+} \to D^0 \pi ^+$ in  $\bar {B^0} \to D^{*+} \pi ^-$}
\label{helrho}
\end{center}
\end{figure}

%\begin{figure}[htb]
%\unitlength 1.0in
%\begin{center}
%\begin{picture}(2.2,1.6)(0.0,0.0)
%\put(-.55,-.3){\psfig{width=2.8in,height=1.8in,file=bexcl_csbar.ps}}
%\end{picture}
%\bigskip
%\caption[]{$B$ meson decay diagrams
%with emission of $c$ $\bar{s}$ quarks: (a)
%external spectator and (b) color suppressed.}
%\label{Fcsdiag}
%\end{center}
%\end{figure}

%\begin{figure}[htb]
%\unitlength 1.0in
%\vskip 10 mm
%\begin{center}
%\begin{picture}(4.0,3.0)(0.0,0.0)
%\put(-.6,.3){\psfig{width=\textwidth,height=8.0in,file=bdds_fig.ps}}
%\end{picture}
%\caption[]{
%Beam constrained mass distributions for
%$\bar{B}^0 \to D^{(*)+} D_s^{(*)-}$ and
%$B^- \to D^{(*)0} D_s^{(*)-}$ decays from CLEO~II.}
%\label{bddszero}
%\end{center}
%\end{figure}

\begin{figure}[htb]
\unitlength 1.0in
\vskip 10 mm
\begin{center}
\begin{picture}(3.0,3.5)(0.0,0.0)
\put(-0.3,.3){\psfig{width=4.in,height=4.2in,file=bdds_bzero.ps}}
\end{picture}
\caption[]{
Beam constrained mass distributions for
$\bar{B}^0 \to D^{(*)+} D_s^{(*)-}$ from CLEO~II.}
\label{bdds_zero}
\end{center}
\end{figure}

\begin{figure}[htb]
\unitlength 1.0in
\vskip 10 mm
\begin{center}
\begin{picture}(3.0,3.5)(0.0,0.0)
\put(-.3,.3){\psfig{width=4.in,height=4.2in,file=bdds_bplus.ps}}
\end{picture}
\caption[]{
Beam constrained mass distributions for
$B^- \to D^{(*)0} D_s^{(*)-}$ from CLEO~II.}
\label{bdds_plus}
\end{center}
\end{figure}

\section{EXCLUSIVE B DECAY TO D MESONS}
\label{BDpiDrho}

\subsection{Measurements of $D (n \pi)^-$ Final States}

The decay modes
$ \bar{B^0}\to D^+ \pi^-$, $ \bar{B^0} \to D^+ \rho^-$,
$ B^-\to D^0 \pi^-$, and $ B^- \to D^0 \rho^-$ are reconstructed
following the procedures outlined in Section \ref{B-recon}.
The beam constrained mass distributions from CLEO II
are shown in Fig.~\ref{dpi}, while
the experimental branching ratios are given in Tables ~\ref{kh1}  and
\ref{kh2} .

To select $ \bar{B} \to D \rho^-$ candidates additional requirements are
imposed on the $\pi^-\pi^0$ invariant mass and the $\rho$ helicity angle.
The CLEO~II analysis requires $ |m(\pi^- \pi^0) - 770|< 150$~MeV and
$|\cos\Theta_{\rho}|>0.4$.
For the $B \to D \rho^-$ modes there are
events which are consistent with both $B \to D \rho^-$ and with
$ B \to D^{*} \pi^-$, followed by $ D^{*} \to D \pi^0$. These
events are removed from the $B \to D \rho^-$ sample using a cut on
the $D^{*} - D$ mass difference.
By fitting the $\pi^- \pi^0$ mass spectrum and the helicity angle distribution,
CLEO~II finds that
at least 97.5\% of the $B \to D \pi^-\pi^0$ rate is described by the
decay $B \to D \rho^-$\cite{mcdd}.
ARGUS\cite{ThirdB} also finds that the $\pi^- \pi^0$ mass spectrum 
is consistent
with the dominance of $\rho$ production.

\subsection{Measurements of $D^*(n\pi)^-$ Final States}

We now consider final states containing a $D^*$ meson and one, two or
three pions. These include the $B \to D^* \pi^-$ ,
$B \to D^* \rho^-$, and $B \to D^* a_1^-$ decay channels.
The results for the decays $\bar{B^0} \to D^{*+} \pi^-$,
$\bar{B^0} \to D^{*+} \rho^-$ and $\bar{B^0} \to D^{*+} \pi^-\pi^-\pi^+$
are listed in Table~\ref{kh2}, and the results for
$B^- \to D^{*0} \pi^-$, $B^- \to D^{*0} \rho^-$ and
$B^- \to D^{*0} \pi^-\pi^-\pi^+$ are given in Table ~\ref{kh1}.

The CLEO II $B^-$ and $\bar{B}^0$ signals in the $D^* \pi$ and $D^* \rho$
decay channels are shown in Fig. \ref{dspi}.
They find that $B \to D^* \pi^-\pi^0$ is saturated by the
decay $B \to D^* \rho^-$ (Fig. \ref{subs}) and
set a tight upper limit of $<9$\% at
90\% C.L. on a possible non-resonant contribution \cite{mcdrho}.
This disagrees with an ARGUS analysis that finds about 50\%
of $\bar{B}^0 \to D^{*+} \pi^- \pi^0$ decays
do not contain a $\rho^-$ meson \cite{FifthB}.

The CLEO~II data suggest that the signal in
$B\to D^{*}\pi^-\pi^-\pi^+$ arises dominantly
from $B\to D^{*} a_1^-$. 
Taking into account the $a_1 \to \pi^-\pi^-\pi^+$ branching fractions it follows
that ${\cal{B}}(B\to D^{*} a_1^-) = 2 \times
{\cal{B}}(B\to D^{*}\pi^-\pi^-\pi^+)$. 
In Fig.~\ref{FBmaone} we show the $M_B$ distributions
when the $\pi^-\pi^-\pi^+$ invariant mass is required to be in the interval
$1.0 <\pi^-\pi^-\pi^+ < 1.6$ GeV.
Fig.~\ref{Fmaonea} shows a fit to the
$\pi^-\pi^-\pi^+$ mass distributions with contributions from
$B \to D^{*+} a_1^-$ and a $B \to D^{*+}\pi^- \rho^0$ non-resonant background.
The $a_1$ meson has been parameterized
as a Breit-Wigner resonance shape
with $m_{a_1} = 1182 $ MeV and $\Gamma_{a1} = 466$ MeV.
This fit gives an upper limit of 13\% on
the non-resonant component in this decay.
This conclusion differs from
CLEO~1.5 which attributed $(35\pm 15\pm 8)$\% of their
$\bar{B^0} \to D^{*+} \pi^-\pi^-\pi^+$ signal to non-resonant
$\bar{B^0} \to D^{*+} \pi^-\rho^0$ decays \cite{anotherB}.
ARGUS also finds a significant non-$a_1$ component in this decay
but does not quote a quantitative result \cite{FifthB}.

The Cabibbo suppressed decay modes such as 
$B\to D K$ should also be observed and studied in the future. 
These modes, in particular, 
$B^+\to D^0 K^+$ and $B^+ \to \bar{D}^0 K^+$ with
$D^0\to |f_{CP}>$ (where $|f_{CP}>$ denotes a CP eigenstate) 
will be used at B factories to constrain one of the three angles
of the unitary triangle.

\subsection{Polarization in $B \to D^{*+}\rho^-$ Decays}
\label{pol-D*-rho}

The sample of fully
reconstructed $ \bar{B^0} \to D^{*+}\rho^-$ decays from CLEO~II
has been used to  measure the $D^{*+}$ and $\rho^-$ polarizations.
By comparing the measured polarizations in $\bar{B^0} \to D^{*+}\rho^-$
with the expectation from the corresponding semileptonic
B decay a test of the factorization hypothesis
can be performed (see Sec.~\ref{fac-ang-cor}).
The polarization is obtained from the distributions of the helicity angles
$\Theta_{\rho}$ and $\Theta_{D^*}$. The $D^{*+}$ helicity angle,
$\Theta_{D^*}$, is the angle between the $D^0$ direction
in the $D^{*+}$ rest frame and the $D^{*+}$ direction
in the rest frame of the $B$ meson.
After integration over $\chi$, the angle between the
normals to the $D^{*+}$ and the
$\rho^-$ decay planes, the helicity angle distribution can be expressed
as
\begin{equation}
{d^2\Gamma\over{d\cos\Theta_{D^*}d\cos\Theta_{\rho}}}
\propto
{1\over{4}}\sin^2\Theta_{D^*}\sin^2\Theta_{\rho}(|H_{+1}|^2+|H_{-1}|^2)
+\cos^2\Theta_{D^*}\cos^2\Theta_{\rho}|H_{0}|^2 \label{polar3d}
\end{equation}
where $H_{i}$ are the amplitudes for the various possible
$D^*$ helicity states.
The fraction of  longitudinal polarization is defined by
\begin{equation}
 {{\Gamma_L}\over{\Gamma}}
 ~ = ~ {{|H_0|^2}\over{|H_{+1}|^2 + |H_{-1}|^2 + |H_{0}|^2}} \label{ratiohel}
\end{equation}
If $\Gamma_L$ is large both
the $D^{*+}$ and the $\rho^{-}$ helicity angles will
follow a $\cos^{2}\Theta$ distribution, whereas a large transverse
polarization, $\Gamma_T$, gives a $\sin^2\Theta$ distribution for both
helicity angles.

To measure the polarization the helicity angle distributions in the $B$
signal region are corrected by
subtracting the distributions from a properly scaled mass sideband.
The resulting helicity angle distributions, corrected for efficiency,
are fitted to the functional form:
\begin{equation}
{d\Gamma\over{d\cos\Theta}} =
 N ~  \left[ \cos^2 \Theta ~ + ~{{1}\over{2}} {{\Gamma_{T}}\over{\Gamma}}
 (1 -3 \cos^2 \Theta) \right]. \label{fithel}
\end{equation}
This form is derived from the angular distribution given above. It
is well behaved for large longitudinal polarization.
From the fit to the $D^{*+}$ helicity angle distribution, they find
$\Gamma_{L}/\Gamma =(85 \pm 8) \% $, while a fit to
the $\rho$ helicity angle distribution gives
$\Gamma_{L}/\Gamma = (97\pm 8)\%$. The results of the
fit are shown in Fig.~\ref{helrho}(a) and (b).
As a consistency check they have verified that the $D^{*+}$ mesons in
$\bar{B}^0 \to D^{*+}\pi^-$ are completely longitudinally polarized, as
expected from angular momentum conservation (Fig. \ref{helrho}(c)).

The statistical errors can be reduced by taking
advantage of the correlation between the two helicity angles.
An unbinned two dimensional likelihood fit
to the joint $(\cos\Theta_{D^*}, \cos\Theta_{\rho})$ distribution gives
\begin{equation}
(\Gamma_{L}/\Gamma)_{\bar{B^0} \to D^{*+} \rho^-}\; =\; 93 \pm 5 \pm 5 \%
\end{equation}

\subsection{Measurements of $D^{**}$ Final States}
\label{B->D**}

In addition to the production of $D$ and $D^*$ mesons,
the charm quark and spectator antiquark can hadronize as a $D^{**}$ meson.
The $D^{**0}(2460)$ has been observed experimentally and identified
as the J$^P=2^+$ state, while the
$D^{**0}(2420)$ has been identified as the $1^+$ state. These states have
full widths of approximately 20 MeV. Two other states, a $0^+$ and another
$1^+$ are predicted but have not yet been observed, presumably because of their
large intrinsic widths.
There is evidence for $D^{**}$ production in semileptonic $B$
decays\cite{Dssin}, and $D^{**}$ mesons have also been seen in hadronic
decays. However, early experiments did not have sufficient data to
separate the two narrow $D^{**}$ states and hence reported branching
ratios only for the combination of the two (see results listed under
$B \to D_J^{(*)0}$ in Tables~\ref{kh1} -- \ref{kh4}).

In order to search for $D^{**}$ mesons from $B$ decays the
final states $B^- \to D^{*+} \pi^- \pi^-$ and
$B^- \to D^{*+} \pi^- \pi^- \pi^0$ are studied.
These decay modes are not expected to occur via
a spectator diagram in which the $c$ quark and the spectator
antiquark form a $D^*$ rather than a $D^{**}$ meson.
The $D^{*+}$ is combined with a $\pi ^-$ to form a $D^{**}$ candidate.
If the $D^{**}$ candidate is within one full width of the nominal mass of
either a $D^{**0}(2420)$ or a $D^{**0}(2460)$, it is
combined with a  $\pi^-$ or $\rho^-$ to form a
$B^-$ candidate.
CLEO~II has also looked for $D^{**}$ production in the channels
$B^-\to D^+ \pi^- \pi^-$ and $\bar{B^0}\to D^0 \pi^- \pi^+$.
Since $D^{**0}(2420)\to D \pi$ is forbidden, only the
$D^{**0}(2460)$ is searched for in the $D \pi \pi$ final state.

Fig.~\ref{dsspi} shows candidate $B$ mass distributions
obtained by  CLEO~II for the four combinations of
$D^{**0}(2460)$ or $D^{**0}(2420)$, and $\pi^-$ or $\rho^-$. In the
$D^{**0}(2420) \pi^-$ mode, there is
a significant signal of 8.5 events on a background of 1.5 events.
In this channel CLEO~II quotes the
branching ratio given in Table~\ref{kh1}, while
for the other three channels, they give upper limits.
ARGUS has also found evidence for $B \to D^{**}(2420) \pi^-$ using
a partial reconstruction technique in which they observe a fast and slow pion
from the $D^{**}$ decay but 
do not reconstruct the $D^0$ meson\cite{Krieger}.

Other final states with higher pion multiplicities should be systematically
studied in the future. 
For example, due to the large combinatorial background,
there is little information available on $B\to D^{(*)}\pi\pi\pi\pi$.

\begin{table}[htb]
\caption{$D_s$ decay channels used to reconstruct $B\to DD_s$ decays.}
\label{TDinf}
\hfill{
\begin{tabular}{lll}
ARGUS \cite{ARGUSDDs} & CLEO 1.5 \cite{DDcleo}& CLEO~II\\ \hline
$D_s^+ \to \phi \pi^+ $& $D_s^+ \to \phi \pi^+ $ & $D_s^+ \to \phi \pi^+ $\\
$D_s^+ \to \phi \pi^+ \pi^0 $& & $D_s^+ \to \phi \pi^+ \pi^0 $ \\
$D_s^+ \to \phi \pi^+ \pi^+ \pi^- $& & \\
$D_s^+ \to K_s K^+ $& $D_s^+ \to K_s K^+ $ & $D_s^+ \to K_s K^+ $\\
$D_s^+ \to K_s K^{*+}$ & & \\
$D_s^+\to\bar{K}^{*0} K^+$&$D_s^+\to\bar{K}^{*0} K^+$ & $D_s^+\to\bar{K}^{*0}K^+$\\
$D_s^+ \to K^{*0} \bar{K}^{*+} $& $D_s^+ \to \bar{K}^{*0} K^{*+}$ &\\
 & & $D_s^+ \to \eta  \pi^+$ \\
 & & $D_s^+ \to \eta  \rho^+$ \\
$D_s^+ \to \eta ' \pi^+$ & & 
\end{tabular}}
\hfill
\end{table}

\subsection{Exclusive Decays to $D$ and $D_s$ Mesons}
\label{doubledees}

Another important class of modes are decays to two charmed mesons.
As shown in Fig. ~\ref{Fdiag} (a)
the production of an isolated pair of charmed mesons
($D_s^{(*)}$ and $D^{(*)}$) proceeds through a Cabibbo favored
spectator diagram in which
the $s\overline{c}$ pair from the virtual $W^-$ hadronizes into a
$D_s^-$ or a $D_s^{*-}$ meson and the remaining spectator quark and the
$c$ quark form a $D^{(*)}$ meson. 
These modes have been observed by the CLEO~1.5\cite{DDcleo},
ARGUS\cite{ARGUSDDs} and CLEO~II\cite{cleodds} experiments.
The decay channels listed in Table~\ref{TDinf} are used to form
$D_s$ meson candidates.
B mesons are then reconstructed in eight decay modes:
$D_s^-D^+$, $D_s^-D^0$,
$D_s^{*-}D^+$, $D_s^{*-}D^0$,
$D_s^-D^{*+}$, $D_s^-D^{*0}$,
$D_s^{*-}D^{*+}$, and $D_s^{*-}D^{*0}$
(See
figs.~\ref{bdds_zero},\ref{bdds_plus}).
The sum of the exclusive modes, averaged over $B^-$ and $\bar{B}^0$
decays, is 
%%% OLD $4.73 \pm 0.71  \%$. This can be compared
$4.93 \pm 0.72  \%$. This can be compared
to the branching fraction of the 
two body component found in the fit to the inclusive $D_s$ momentum spectrum
of $4.5\pm 1.2$. The error is dominated by the uncertainty in
${\cal{B}}(D_s \to \phi \pi)$. The remaining contribution to 
the inclusive production of $D_s$ mesons must be due to the decay modes
$B\to D_s^{**} D^{(*)}$, $B\to D_s^{(*)} D^{(*)} (n\pi)$ or
$D_s^{(*)} D \pi$.

Partial reconstruction techniques 
are also being 
investigated to improve the size of the signals in
$B\to D^{(*)} D_s^{(*)+}$. Larger samples not only  reduce the
statistical error in the branching ratio measurements but will
also allow the polarization in $B\to D^* D_s^{*+}$ decays to be determined.
Comparsion of the yield in partially reconstructed 
and fully reconstructed $B\to D^* D_s^{(*)+}$
events will also give a model independent 
measurement  of ${\cal B}(D_s\to \phi\pi^+)$ which sets the
scale for the $D_s$ branching fractions.
Branching fractions and background levels for CP eigenstates
such as $\bar{B}^0\to D^{(*)+} D^{(*)-}$ will also be studied.

%%%%\section{COLOR SUPPRESSED B DECAY}
%\input colorsuppressed.tex 
\begin{figure}[htb]
\unitlength 1.0in
\begin{center}
\begin{picture}(3.0,2.5)(0.0,0.0)
\put(-.35,.8){\psfig{width=3.5in,height=3.0in,file=cleo_b2psi_excl.ps}}
\end{picture}
\vskip 15 mm
\caption[]{
Beam-constrained mass from CLEO~II for: (a) $B^-\to\psi K^-$,  (b)
$\bar{B^0}\to\psi\bar{K^0}$,  (c) $B^- \to\psi\bar{K}^{*-}$, (d)
$\bar{B^0}\to\psi K^{*0}$, 
(e) $B^-\to\psi' K^-$,  (f)
$\bar{B^0}\to\psi'\bar{K^0}$,  (g) $B^-\to\psi'\bar{K}^{*-}$, and 
(h) $\bar{B}^0\to\psi' K^{*0}$.}\label{bpsipk}
\label{bpsik}
\end{center}
\end{figure}

\begin{figure}[htb]
\unitlength 1.0in
\begin{center}
\begin{picture}(3.0,3.0)(0.0,0.0)
\put(-.1,.3){\psfig{width=2.5in,height=3.0in,file=psipol_final_fig.ps}}
\end{picture}
\vskip 15 mm
\caption[]{
Distributions of the efficiency corrected
$\psi$ and $K^*$ helicity angles in
$B \to \psi K^*$ decays from CLEO~II.
The overlaid smooth curves are projections of the
unbinned maximum likelihood fit described in the text.}
\label{expol}
\end{center}
\end{figure}

\begin{table}[htb]
\caption{Upper limits (90\% C.L) on color suppressed $B$ decays.}\label{Tbrcol}
\begin{tabular}{lcc}
Decay Mode &  Events &   U. L. (\%)  \\ \hline
$\bar{B^0} \to D^{0} \pi^0$    & $<20.7  $   & $<0.048$   \\
$\bar{B^0} \to D^{0} \rho^0$   & $<19.0$     & $<0.055$   \\
$\bar{B^0} \to D^{0} \eta$     & $<9.5$      & $<0.068$    \\
$\bar{B^0} \to D^{0} \eta^{'}$ & $<3.5 $     & $<0.086$   \\
$\bar{B^0} \to D^{0} \omega $  & $<12.7 $     & $<0.063$    \\
$\bar{B^0} \to D^{*0} \pi^0$   & $<11.0 $     & $<0.097$   \\
$\bar{B^0} \to D^{*0} \rho^0$  & $<8.1$      & $<0.117$    \\
$\bar{B^0} \to D^{*0} \eta$    & $<2.3 $     & $<0.069$   \\
$\bar{B^0} \to D^{*0} \eta^{'}$  & $<2.3 $   & $<0.27$    \\
$\bar{B^0} \to D^{*0} \omega$    & $<9.0 $     & $<0.21$
\end{tabular}
\end{table}

\begin{table}[htb]
\caption{Upper limits on ratios of branching fractions
for color suppressed to normalization modes.}\label{Tratcol}
\begin{tabular}{cc}
Ratio of Branching Ratios & CLEO~II (90\% C.L.)  \\ \hline
${\cal B}(\bar{B^0} \to D^0 \pi^0)/{\cal B}(B^- \to D^0 \pi^-)$
                                                    & $< 0.09$   \\
${\cal B}(\bar{B^0} \to D^0 \rho^0)/{\cal B}(B^- \to D^0 \rho^-)$
                                                    & $< 0.05 $   \\
${\cal B}(\bar{B^0} \to D^0 \eta)/{\cal B}(B^- \to D^0 \pi^-)$
                                                    & $< 0.12 $   \\
${\cal B}(\bar{B^0} \to D^0 \eta^{'})/{\cal B}(B^- \to D^0 \pi^-)$
                                                    & $< 0.16 $   \\
${\cal B}(\bar{B^0} \to D^0 \omega)/{\cal B}(B^- \to D^0 \rho^-)$
                                                    & $< 0.05 $   \\
${\cal B}(\bar{B^0} \to D^{*0}\pi^0)/{\cal B}(B^- \to D^{*0} \pi^-)$
                                                    & $< 0.20 $   \\
${\cal B}(\bar{B^0} \to D^{*0} \rho^0)/{\cal B}(B^- \to D^{*0} \rho^-)$
                                                    & $< 0.07 $   \\
${\cal B}(\bar{B^0} \to D^{*0} \eta)/{\cal B}(B^- \to D^{*0} \pi^-)$
                                                    & $< 0.14 $   \\
${\cal B}(\bar{B^0} \to D^{*0} \eta^{'})/{\cal B}(B^-\to D^{*0}\pi^-)$
                                                    & $< 0.54 $   \\
${\cal B}(\bar{B^0} \to D^{*0} \omega)/{\cal B}(B^- \to D^{*0} \rho^-)$
                                                    & $< 0.09 $
\end{tabular}
\end{table}

\section{COLOR SUPPRESSED B DECAY}
\label{B->psi-K(*)}

\subsection{Exclusive $B$ Decays to Charmonium}
\label{intro-B->psi-K(*)}

In $B$ decays to charmonium the $c$ quark from the
$b$ combines with a $\bar{c}$ quark from the virtual $W^-$ to
form a charmonium state. This process is described by the color suppressed
diagram shown in Fig.~\ref{Fdiag}(b).
By comparing $B$ meson decays to different final states with
charmonium mesons the dynamics of this decay mechanism can be investigated.

The decay modes $\bar{B^0} \to \psi K^0$ and  $\bar{B^0} \to \psi' K^0$ are
of special interest since the final states are
CP eigenstates. These decays are of great importance
for the investigation of
one of the three CP violating angles accessible to study in $B$ decays.
It is also possible to use the decay
$\bar{B^0} \to \psi K^{*0}$, $K^{*0} \to K^0 \pi^0$ which has a
somewhat higher branching ratio, but this final state 
consists of a mixture of CP eigenstates.
It has even CP if the
orbital angular momentum L is 0 or 2 and odd CP for L=1.
If both CP states are present the CP asymmetry will be diluted.
A measurement of CP violation in this channel is only possible if one of the
CP states dominates, or if a detailed moments analysis of the various decay
components is performed \cite{Idunit}.
Recent measurements of the polarization in the decay $\bar{B^0}
\to\psi \bar{K^{*0}}$  allow us to determine the
fractions of the two CP states.

B meson candidates are formed by combining a charmonium
and a strange meson candidate. CLEO~1.5 and ARGUS have observed
signals for some of these modes.
Using the procedures outlined in Sec.~\ref{B-recon} the
beam constrained mass distributions shown
in Fig.~\ref{bpsik} are obtained by CLEO~II.
CLEO~II has also reported a signal in the Cabibbo suppressed decay
$B^- \to \psi \pi^-$,
The branching ratios are listed in Tables~\ref{kh1}
and \ref{kh2} . Recently, CDF has reported signals
for $B\to \psi K^{*0}$ and $B\to \psi K^-$ (see Fig.~\ref{cdfbd})
and measurements of polarization in $B\to \psi K^*$ decays\cite{cdfpolar}.
Averaging over $B^-$ and $\bar{B}^0$ decays
%\footnote{For this calculation we
%have assumed ${\cal{B}}(\bar{B}^0 \to \psi \pi^0) \; = \; 1/2 \;
%{\cal{B}}(B^- \to \psi \pi^-)$.}
we determine the sum of the  exclusive two-body decays to ${\cal{B}}(B\to \psi
\:K(K^*,\:\pi)) \; = \; 0.258 \pm 0.030\%$ and 
${\cal{B}}(B\to \psi$'$\:K(K^*,\:\pi)) \; = \; 0.22 \pm 0.09\%$.
The first results represents about 1/4 of the inclusive rate for direct
$B\to \psi$ production. The experimental investigation of the remaining
fraction is important, since any additional 
quasi-two body channel open to $B\to \psi$ transitions
could be useful for future studies of CP violation, 
Lower momentum $\psi$  mesons could originate from 
multibody final states or from two body decays involving heavier
$K^{(*)}$ resonances.

Evidence for the decay mode $B\to \chi_{c1} K$ has been reported by CLEO~II
\cite{fastpsi,SixthB}
and  ARGUS \cite{FifthB}. The average branching fraction is 
${\cal B}(B^-\to \chi_c K^-) = (0.104\pm 0.040) \%$.
The CLEO~II collaboration has also placed upper limits on $\chi_{c1}K^0$
and $\chi_{c1}K^*$ production in $B$ decay.

\subsection{Polarization in $B\to\psi K^*$ }

The polarization in $B\to\psi K^*$ is studied using the methods described
for the 
$\bar{B^0}\to D^{*+}\rho^-$ polarization measurement
in Section \ref{pol-D*-rho}.
After integration over the azimuthal angle between the $\psi$ and the
$K^*$ decay planes, the angular distribution in $B \to \psi K^*$ decays
can be written as
\begin{equation}
 {d^2\Gamma\over{d\cos\Theta_{\psi}d\cos\Theta_{K^*}}}
\propto {1\over{4}}\sin^2\Theta_{K^*}
(1+\cos^2\Theta_{\psi})(|H_{+1}|^2+|H_{-1}|^2)
+\cos^2\Theta_{K^*}\sin^2\Theta_{\psi}|H_{0}|^2 , \label{psipolar}
\end{equation}
where the $K^*$ helicity angle $\Theta_{K^*}$ is the angle between
the kaon direction in the $K^*$ rest frame and the $K^*$ direction in the
$B$ rest frame and $\Theta_{\psi}$ is the corresponding $\psi$ helicity angle,
and $H_{\pm1,0}$ are the helicity amplitudes.
The fraction of longitudinal polarization in $B \to \psi K^*$
is determined by an unbinned fit to the $\psi$ and $K^*$ helicity angle
distributions. The results obtained by the CLEO~II, ARGUS and
CDF collaborations are listed in Table \ref{Tpsipolex}.
\begin{table}[htb]
\caption{Longitudinal polarization of $\psi$ mesons from $B \to \psi K^*$
decays.}
\label{Tpsipolex}
\begin{tabular}{cc}
Experiment & ${\left({\Gamma_L\over{\Gamma}}\right)}$\\ \hline
CLEO II & $ 0.80\pm 0.08 \pm 0.05$ \\
ARGUS \cite{argpol} & $0.97 \pm 0.16\pm 0.15$ \\
CDF \cite{cdfpolar} & $ 0.66 \pm 0.10^{+0.08}_{-0.10}$\\
\hline
Average & $ 0.78 \pm 0.07$\\
\end{tabular}
\end{table}
The efficiency corrected distributions in
each of the helicity angles $\cos\Theta_{\psi}$ and $\cos\Theta_{K^*}$
are shown in Fig.~\ref{expol} (CLEO II).
Assuming that the systematic errors from the various
experiments are uncorrelated, these three results can be averaged to obtain
\begin{equation}
{\Gamma_L\over \Gamma} = 0.78 \pm 0.07  \label{psikstavg}
\end{equation}

Although the decay mode $B \to \psi K^*$  may not be completely
polarized,  it is still dominated by a single CP eigenstate.
This mode will therefore be useful for measurements of CP violation.

\subsection{Exclusive Decays to a $D^{0 (*)}$ and a Neutral Meson.}
\label{color-supress}

We now discuss searches for $B$ decays which can occur
via an internal W-emission graph but which
do not yield charmonium mesons in the final
state. Naively, one expects that
these decays will be suppressed relative to decays which
occur via the external W-emission graph.
For the internal graph, in the absence
of gluons, the colors of the quarks from the virtual $W$ must
match the colors of the $c$ quark
and the accompanying spectator antiquark.
In this simple picture, one expects that the suppression
factor should be  $1/18$ in rate for decays involving $\pi^0$, $\rho^0$
and $\omega$ mesons\cite{Dpi}.
In heavy quark decays the effects
of gluons cannot be neglected, and QCD based calculations
\cite{Neubie} predict suppression factors of order $1/50$.
If color suppressed $B$ decay modes are not greatly suppressed
then these modes could also be useful for CP violation studies\cite{Dunietz}.

CLEO~II has searched for color suppressed decay modes of $B$ mesons which
contain a single $D^0$ or $D^{*0}$ meson in the final state\cite{wex}.
The relevant color suppressed modes are listed in Table~\ref{Tbrcol}.
The decay channels used are $\eta \to \gamma \gamma$,
$\omega \to \pi^+ \pi^- \pi^0$ and $\eta^{'} \to
\eta \pi^+ \pi^-$, followed by $\eta \to \gamma \gamma$\cite{BReta}.
For decays of a pseudoscalar meson into a final state containing a
pseudoscalar and a vector meson (V), a helicity angle cut of
$|\cos \Theta_{V}| \; > \; 0.4$ is used\cite{omeg}.
No signals were observed.
Upper limits \cite{PDGul} on the
branching ratios for color suppressed modes are given in Table~\ref{Tbrcol}.
Upper limits on
the ratios of color suppressed modes to normalization modes are given in
Table~\ref{Tratcol}.
These limits show that there is color suppression of these $B$ decay modes.

\begin{table}[htb] 
\caption{$B^-$ Branching fractions [\%]} 
\label{kh1} 
 \begin{tabular}{llll} 
Mode & ARGUS & CLEO 1.5 & CLEO II \\ 
\hline 
 $B^- \rightarrow D^0 \pi ^-$ & $ 0.22 \pm 0.09  \pm 0.06  \pm 0.01 $ & 
$ 0.56 \pm 0.08  \pm 0.05  \pm 0.02 $ & 
$ 0.53 \pm 0.04  \pm 0.05  \pm 0.02  $\\ 
 $B^- \rightarrow D^0 \rho ^-$ & $ 1.41 \pm 0.43  \pm 0.39  \pm 0.06 $ & 
 & 
$ 1.31 \pm 0.12  \pm 0.14  \pm 0.04  $\\ 
 $B^- \rightarrow D^{0} \pi ^+ \pi ^- \pi ^-$ &  & 
$ 1.24 \pm 0.31  \pm 0.14  \pm 0.05 $ & 
 \\ 
 $B^- \rightarrow D^{*0} \pi ^-$ & $ 0.38 \pm 0.13  \pm 0.10  \pm 0.02 $ & 
$ 1.00 \pm 0.25  \pm 0.18  \pm 0.04 $ & 
$ 0.49 \pm 0.07  \pm 0.06  $\\ 
 $B^- \rightarrow D^{*0} \rho ^-$ & $ 0.94 \pm 0.56  \pm 0.35  \pm 0.04 $ & 
 & 
$ 1.59 \pm 0.20  \pm 0.26  \pm 0.05  $\\ 
 $B^- \rightarrow D_J^{(*)0} \pi ^-$ & $ 0.13 \pm 0.06  \pm 0.03  \pm 0.01 $ & 
$ 0.13 \pm 0.07  \pm 0.01  \pm 0.01 $ & 
 \\ 
 $B^- \rightarrow D^{*+} \pi ^- \pi ^- \pi ^0$ & $ 1.64 \pm 0.64  \pm 0.37  \pm 0.07 $ & 
 & 
 \\ 
 $B^- \rightarrow D_J^{(*)0} \rho ^-$ & $ 0.32 \pm 0.19  \pm 0.07  \pm 0.01 $ & 
 & 
 \\ 
 $B^- \rightarrow D^{*0} \pi ^- \pi ^- \pi ^+$ &  & 
 & 
$ 0.92 \pm 0.20  \pm 0.17  \pm 0.01  $\\ 
 $B^- \rightarrow D^{*0} a_1 ^-$ &  & 
 & 
$ 1.83 \pm 0.39  \pm 0.33  \pm 0.02  $\\ 
 $B^- \rightarrow D^+ \pi^- \pi ^- $ &  & 
 & 
$ <0.14 $  (90\% C.L.)\\ 
  $B^- \rightarrow D^{*+} \pi ^- \pi ^-$ & $ 0.24 \pm 0.13  \pm 0.05  \pm 0.01 $ & 
$ <0.37$  (90\% C.L.)& 
 $ 0.18 \pm 0.07  \pm 0.03  \pm 0.01  $\\ 
 $B^- \rightarrow D^{**0}(2420) \pi^- $ & $ 0.30 \pm 0.08  \pm 0.06  \pm 0.01 $ & 
 & 
$ 0.11 \pm 0.05  \pm 0.02  \pm 0.01  $\\ 
 $B^- \rightarrow D^{**0}(2420) \rho^- $ &  & 
 & 
$ <0.13 $  (90\% C.L.)\\ 
  $B^- \rightarrow D^{**0}(2460) \pi^- $ &  & 
 & 
$ <0.13 $  (90\% C.L.)\\ 
  $B^- \rightarrow D^{**0}(2460) \rho^- $ &  & 
 & 
$ <0.45 $  (90\% C.L.)\\ 
  $B^- \rightarrow D^0 D_s^-$ & $ 1.69 \pm 0.85  \pm 0.27  \pm 0.41 $ & 
$ 1.66 \pm 0.70  \pm 0.13  \pm 0.40 $ & 
$ 1.11 \pm 0.20  \pm 0.23  \pm 0.28  $\\ 
 $B^- \rightarrow D^0 D_s^{*-}$ & $ 1.13 \pm 0.85  \pm 0.20  \pm 0.27 $ & 
 & 
$ 0.79 \pm 0.25  \pm 0.15  \pm 0.19  $\\ 
 $B^- \rightarrow D^{*0} D_s^-$ & $ 0.79 \pm 0.55  \pm 0.11  \pm 0.19 $ & 
 & 
$ 1.27 \pm 0.39  \pm 0.32  \pm 0.31  $\\ 
 $B^- \rightarrow D^{*0} D_s^{*-}$ & $ 1.89 \pm 0.98  \pm 0.28  \pm 0.46 $ & 
 & 
$ 2.82 \pm 0.80  \pm 0.59  \pm 0.68  $\\ 
 $B^- \rightarrow \psi K^-$ & $ 0.08 \pm 0.04  \pm 0.01 $ & 
$ 0.09 \pm 0.02  \pm 0.02 $ & 
$ 0.110 \pm 0.015  \pm 0.009  $\\ 
 $B^- \rightarrow \psi ' K^-$ & $ 0.20 \pm 0.09  \pm 0.04 $ & 
$ <0.05$  (90\% C.L.)& 
 $ 0.061 \pm 0.023  \pm 0.009  $\\ 
 $B^- \rightarrow \psi K^{*-}$ & $ 0.19 \pm 0.13  \pm 0.03 $ & 
$ 0.15 \pm 0.11  \pm 0.03 $ & 
$ 0.178 \pm 0.051  \pm 0.023  $\\ 
 $B^- \rightarrow \psi ' K^{*-}$ & $ <0.53 $ (90\% C.L.)& 
 $ <0.38$  (90\% C.L.)& 
 $ <0.30 $  (90\% C.L.)\\ 
  $B^- \rightarrow \psi K^- \pi ^+ \pi ^-$ & $ <0.19 $ (90\% C.L.)& 
 $ 0.14 \pm 0.07  \pm 0.03 $ & 
 \\ 
 $B^- \rightarrow \psi ' K^- \pi ^+ \pi ^-$ & $ 0.21 \pm 0.12  \pm 0.04 $ & 
 & 
 \\ 
 $B^- \rightarrow \chi_{c1} K^-$ & $ 0.22 \pm 0.15  \pm 0.07 $ & 
 & 
$ 0.097 \pm 0.040  \pm 0.009  $\\ 
 $B^- \rightarrow \chi_{c1} K^{*-}$ &  & 
 & 
$ <0.21 $  (90\% C.L.)\\ 
  $B^- \rightarrow \psi \pi ^-$ &  & 
 & 
$ 0.0047 \pm 0.0024  \pm 0.0004  $\\ 
\end{tabular} 
\end{table} 
\begin{table}[htb] 
\caption{$\bar{B}^0$ Branching fractions in [\%]} 
\label{kh2}
 \begin{tabular}{llll} 
Mode & ARGUS & CLEO 1.5 & CLEO II \\ 
\hline 
 $\bar{B}^0 \rightarrow D^+ \pi ^-$ & $ 0.48 \pm 0.11  \pm 0.08  \pm 0.03 $ & 
$ 0.27 \pm 0.06  \pm 0.03  \pm 0.02 $ & 
$ 0.29 \pm 0.04  \pm 0.03  \pm 0.02  $\\ 
 $\bar{B}^0 \rightarrow D^+ \rho ^-$ & $ 0.90 \pm 0.50  \pm 0.27  \pm 0.06 $ & 
 & 
$ 0.81 \pm 0.11  \pm 0.12  \pm 0.05  $\\ 
 $\bar{B}^0 \rightarrow D^+ \pi ^- \pi ^- \pi ^+$ &  & 
$ 0.81 \pm 0.21  \pm 0.09  \pm 0.05 $ & 
 \\ 
 $\bar{B}^0 \rightarrow D^{*+} \pi ^-$ & $ 0.25 \pm 0.08  \pm 0.03  \pm 0.01 $ & 
$ 0.45 \pm 0.11  \pm 0.05  \pm 0.02 $ & 
$ 0.25 \pm 0.03  \pm 0.04  \pm 0.01  $\\ 
 $\bar{B}^0 \rightarrow D^{*+} \rho ^-$ & $ 0.64 \pm 0.27  \pm 0.25  \pm 0.03 $ & 
$ 2.13 \pm 0.90  \pm 1.24  \pm 0.09 $ & 
$ 0.70 \pm 0.09  \pm 0.13  \pm 0.02  $\\ 
 $\bar{B}^0 \rightarrow D^{*+} \pi ^- \pi ^- \pi ^+$ & $ 1.09 \pm 0.27  \pm 0.32  \pm 0.04 $ & 
$ 1.77 \pm 0.31  \pm 0.30  \pm 0.07 $ & 
$ 0.61 \pm 0.10  \pm 0.11  \pm 0.02  $\\ 
 $\bar{B}^0 \rightarrow D^{*+} a_1^-$ &  & 
 & 
$ 1.22 \pm 0.19  \pm 0.22  \pm 0.04  $\\ 
 $\bar{B}^0 \rightarrow D^{0} \pi ^+ \pi^- $ &  & 
 & 
$ <0.16 $  (90\% C.L.)\\ 
  $\bar{B}^0 \rightarrow D^{**+}(2460) \pi^- $ &  & 
 & 
$ <0.21 $  (90\% C.L.)\\ 
  $\bar{B}^0 \rightarrow D^{**+}(2460) \rho^- $ &  & 
 & 
$ <0.47 $  (90\% C.L.)\\ 
  $\bar{B}^0 \rightarrow D^+ D_s^-$ & $ 1.05 \pm 0.80  \pm 0.35  \pm 0.26 $ & 
$ 0.54 \pm 0.31  \pm 0.03  \pm 0.13 $ & 
$ 0.82 \pm 0.23  \pm 0.19  \pm 0.20  $\\ 
 $\bar{B}^0 \rightarrow D^+ D_s^{*-}$ & $ 1.67 \pm 1.05  \pm 0.52  \pm 0.41 $ & 
 & 
$ 0.95 \pm 0.33  \pm 0.21  \pm 0.23  $\\ 
 $\bar{B}^0 \rightarrow D^{*+} D_s^-$ & $ 0.83 \pm 0.59  \pm 0.11  \pm 0.20 $ & 
$ 1.17 \pm 0.66  \pm 0.09  \pm 0.28 $ & 
$ 0.85 \pm 0.21  \pm 0.15  \pm 0.21  $\\ 
 $\bar{B}^0 \rightarrow D^{*+} D_s^{*-}$ & $ 1.54 \pm 0.83  \pm 0.24  \pm 0.37 $ & 
 & 
$ 1.85 \pm 0.46  \pm 0.33  \pm 0.45  $\\ 
 $\bar{B}^0 \rightarrow \psi K^0$ & $ 0.09 \pm 0.07  \pm 0.02 $ & 
$ 0.07 \pm 0.04  \pm 0.02 $ & 
$ 0.075 \pm 0.024  \pm 0.008  $\\ 
 $\bar{B}^0 \rightarrow \psi ' K^0$ & $ <0.30 $ (90\% C.L.)& 
 $ <0.16$  (90\% C.L.)& 
 $ <0.08 $  (90\% C.L.)\\ 
  $\bar{B}^0 \rightarrow \psi \bar{K}^{*0}$ & $ 0.13 \pm 0.06  \pm 0.02 $ & 
$ 0.13 \pm 0.06  \pm 0.03 $ & 
$ 0.169 \pm 0.031  \pm 0.018  $\\ 
 $\bar{B}^0 \rightarrow \psi ' \bar{K}^{*0}$ & $ <0.25 $ (90\% C.L.)& 
 $ 0.15 \pm 0.09  \pm 0.03 $ & 
$ <0.19 $  (90\% C.L.)\\ 
  $\bar{B}^0 \rightarrow \psi K^{-} \pi ^+$ &  & 
$ 0.12 \pm 0.05  \pm 0.03 $ & 
 \\ 
 $\bar{B}^0 \rightarrow \psi ' K^- \pi ^+$ & $ <0.11 $ (90\% C.L.)& 
  & 
 \\ 
 $\bar{B}^0 \rightarrow \chi_{c1} K^0$ &  & 
 & 
$ <0.27 $  (90\% C.L.)\\ 
  $\bar{B}^0 \rightarrow \chi_{c1} \bar{K}^{*0}$ &  & 
 & 
$ <0.21 $  (90\% C.L.)\\ 
  $\bar{B}^0 \rightarrow \psi \pi ^0$ &  & 
 & 
$ <0.0069 $  (90\% C.L.)\\ 
 \end{tabular} 
\end{table} 
\begin{table}[htb] 
\caption{World average $B^-$ branching fractions [\%]} 
\label{kh3} 
\begin{tabular}{ll} 
Mode & Branching Fraction \\ 
\hline 
$B^- \rightarrow D^0 \pi ^-$ & $0.48 \pm 0.05 \pm 0.02 $ \\ 
$B^- \rightarrow D^0 \rho ^-$ & $1.32 \pm 0.17 \pm 0.05 $ \\ 
$B^- \rightarrow D^{0} \pi ^+ \pi ^- \pi ^-$ & $1.24 \pm 0.34 \pm 0.05 $ \\ 
$B^- \rightarrow D^{*0} \pi ^-$ & $0.50 \pm 0.08 \pm 0.02 $ \\ 
$B^- \rightarrow D^{*0} \rho ^-$ & $1.47 \pm 0.29 \pm 0.06 $ \\ 
$B^- \rightarrow D_J^{(*)0} \pi ^-$ & $0.13 \pm 0.05 \pm 0.01 $ \\ 
$B^- \rightarrow D^{*+} \pi ^- \pi ^- \pi ^0$ & $1.64 \pm 0.73 \pm 0.07 $ \\ 
$B^- \rightarrow D_J^{(*)0} \rho ^-$ & $0.32 \pm 0.20 \pm 0.01 $ \\ 
$B^- \rightarrow D^{*0} \pi ^- \pi ^- \pi ^+$ & $0.92 \pm 0.26 \pm 0.04 $ \\ 
$B^- \rightarrow D^{*0} a_1 ^-$ & $1.83 \pm 0.51 \pm 0.07 $ \\ 
$B^- \rightarrow D^+ \pi^- \pi ^- $ & $<0.14 $  (90\% C.L.)\\ 
$B^- \rightarrow D^{*+} \pi ^- \pi ^-$ & $0.19 \pm 0.06 \pm 0.01 $ \\ 
$B^- \rightarrow D^{**0}(2420) \pi^- $ & $0.15 \pm 0.05 \pm 0.01 $ \\ 
$B^- \rightarrow D^{**0}(2420) \rho^- $ & $<0.13 $  (90\% C.L.)\\ 
$B^- \rightarrow D^{**0}(2460) \pi^- $ & $<0.13 $  (90\% C.L.)\\ 
$B^- \rightarrow D^{**0}(2460) \rho^- $ & $<0.45 $  (90\% C.L.)\\ 
$B^- \rightarrow D^0 D_s^-$ & $1.24 \pm 0.27 \pm 0.30 $ \\ 
$B^- \rightarrow D^0 D_s^{*-}$ & $0.83 \pm 0.28 \pm 0.20 $ \\ 
$B^- \rightarrow D^{*0} D_s^-$ & $1.06 \pm 0.38 \pm 0.26 $ \\ 
$B^- \rightarrow D^{*0} D_s^{*-}$ & $2.37 \pm 0.71 \pm 0.58 $ \\ 
$B^- \rightarrow \psi K^-$ &$ 0.102 \pm 0.014 $ \\ 
$B^- \rightarrow \psi ' K^-$ &$ 0.070 \pm 0.024 $ \\ 
$B^- \rightarrow \psi K^{*-}$ &$ 0.174 \pm 0.047 $ \\ 
$B^- \rightarrow \psi ' K^{*-}$ & $<0.30 $  (90\% C.L.)\\ 
$B^- \rightarrow \psi K^- \pi ^+ \pi ^-$ &$ 0.140 \pm 0.078 $ \\ 
$B^- \rightarrow \psi ' K^- \pi ^+ \pi ^-$ &$ 0.207 \pm 0.127 $ \\ 
$B^- \rightarrow \chi_{c1} K^-$ &$ 0.104 \pm 0.040 $ \\ 
$B^- \rightarrow \chi_{c1} K^{*-}$ & $<0.21 $  (90\% C.L.)\\ 
$B^- \rightarrow \psi \pi ^-$ &$ 0.0047 \pm 0.0024 $ \\ 
\end{tabular} 
\end{table} 
\begin{table}[htb] 
\caption{World average $\bar{B}^0$ branching fractions [\%]} 
\label{kh4} 
 \begin{tabular}{ll} 
Mode & Branching Fraction \\ 
\hline 
$\bar{B}^0 \rightarrow D^+ \pi ^-$ & $0.30 \pm 0.04 \pm 0.02 $ \\ 
$\bar{B}^0 \rightarrow D^+ \rho ^-$ & $0.82 \pm 0.16 \pm 0.05 $ \\ 
$\bar{B}^0 \rightarrow D^+ \pi ^- \pi ^- \pi ^+$ & $0.81 \pm 0.23 \pm 0.05 $ \\ 
$\bar{B}^0 \rightarrow D^{*+} \pi ^-$ & $0.27 \pm 0.04 \pm 0.01 $ \\ 
$\bar{B}^0 \rightarrow D^{*+} \rho ^-$ & $0.70 \pm 0.15 \pm 0.03 $ \\ 
$\bar{B}^0 \rightarrow D^{*+} \pi ^- \pi ^- \pi ^+$ & $0.77 \pm 0.13 \pm 0.03 $ \\ 
$\bar{B}^0 \rightarrow D^{*+} a_1^-$ & $1.22 \pm 0.29 \pm 0.05 $ \\ 
$\bar{B}^0 \rightarrow D^{0} \pi ^+ \pi^- $ & $<0.16 $  (90\% C.L.)\\ 
$\bar{B}^0 \rightarrow D^{**+}(2460) \pi^- $ & $<0.21 $  (90\% C.L.)\\ 
$\bar{B}^0 \rightarrow D^{**+}(2460) \rho^- $ & $<0.47 $  (90\% C.L.)\\ 
$\bar{B}^0 \rightarrow D^+ D_s^-$ & $0.71 \pm 0.21 \pm 0.17 $ \\ 
$\bar{B}^0 \rightarrow D^+ D_s^{*-}$ & $1.02 \pm 0.37 \pm 0.25 $ \\ 
$\bar{B}^0 \rightarrow D^{*+} D_s^-$ & $0.88 \pm 0.22 \pm 0.21 $ \\ 
$\bar{B}^0 \rightarrow D^{*+} D_s^{*-}$ & $1.75 \pm 0.47 \pm 0.43 $ \\ 
$\bar{B}^0 \rightarrow \psi K^0$ &$ 0.075 \pm 0.021 $ \\ 
$\bar{B}^0 \rightarrow \psi ' K^0$ & $<0.08 $  (90\% C.L.)\\ 
$\bar{B}^0 \rightarrow \psi \bar{K}^{*0}$ &$ 0.153 \pm 0.028 $ \\ 
$\bar{B}^0 \rightarrow \psi ' \bar{K}^{*0}$ &$ 0.151 \pm 0.091 $ \\ 
$\bar{B}^0 \rightarrow \psi K^{-} \pi ^+$ &$ 0.117 \pm 0.058 $ \\ 
$\bar{B}^0 \rightarrow \psi ' K^- \pi ^+$ & $<0.11 $  (90\% C.L.)\\ 
$\bar{B}^0 \rightarrow \chi_{c1} K^0$ & $<0.27 $  (90\% C.L.)\\ 
$\bar{B}^0 \rightarrow \chi_{c1} \bar{K}^{*0}$ & $<0.21 $  (90\% C.L.)\\ 
$\bar{B}^0 \rightarrow \psi \pi ^0$ & $<0.007 $  (90\% C.L.)\\ 
\end{tabular} 
\end{table} 

\clearpage

%%%%%\section{THEORETICAL INTERPRETATION OF HADRONIC B DECAY}
%\input theory.tex
\section{THEORETICAL INTERPRETATION OF HADRONIC B DECAY}

\subsection{Introduction}

The simple spectator diagram for two-body hadronic $B$ meson decays
that occur through the CKM favored $b\to c$ transition
is described by the Hamiltonian\cite{vud}:
\begin{equation}
H ={G_F\over \sqrt 2}V_{cb}
\left\{\left[(\bar d u)+(\bar s c)\right]
(\bar c b)\right\} \label{Eraw}
\end{equation}
where $(\bar q_i q_j)=\bar q_i\gamma_{\mu}(1-\gamma_5)q_j$,
$G_F$ is the Fermi coupling constant, and $V_{cb}$ is the CKM matrix element.

The spectator diagram is modified by hard gluon exchange
between the initial and final quark lines. The effect of these exchanges can
be taken into account by use of the renormalization group, with the
result that the original Hamiltonian of equation~(\ref{Eraw})
is replaced by one which now
contains two pieces, the original term multiplied by
a coefficient $c_1(\mu)$, and an additional term multiplied by $c_2(\mu)$:
\begin{equation}
H_{eff}={G_F\over \sqrt 2}V_{cb} \left\{c_1(\mu)\left[(\bar d
u)+(\bar s c)\right] (\bar c b)+ c_2(\mu)\left[(\bar c u)(\bar d b)+(\bar c
c)(\bar s b)\right] \right\} \label{Eheff}
\end{equation}
The $c_i$ are Wilson coefficients that can be calculated from QCD.
However, the calculation is inherently uncertain  because it is unclear
at what mass scale, $\mu$,  these coefficients should be evaluated.
The usual scale is taken to be $\mu \sim m_b^{2}$.
Defining
\begin{equation}
c_{\pm}(\mu)=c_1(\mu)\pm c_2(\mu) \label{Ecees}
\end{equation}
the leading-log approximation gives\cite{Neubie}
\begin{equation}
c_{\pm}(\mu)=\left({\alpha_s(M_{W}^{2})\over\alpha_s(\mu)}\right)
^{\displaystyle {-6\gamma_{\pm}\over (33-2n_f)}} \label{Ecpmcal}
\end{equation}
where $\gamma_-=-2\gamma_+=2$, and $n_f$ is the
number of active flavors, which is taken to be five in this case.

The additional term in the Hamiltonian in Eq.\ (\ref{Eheff}) corresponds to the
``color suppressed'' diagram.
The quark pairings in this diagram are different from those in the spectator
diagram, and lead to the decay modes discussed in section \ref{B->psi-K(*)}.
From the observation of the $B \to \psi X_s$ decays, where $X_s$ is a strange
meson, the magnitude of the color-suppressed term can be deduced.
In $B^-$ decays, both spectator and color-suppressed
diagrams are present and can interfere.
By comparing the rates for $B^-$ and $\bar{B^0}$ decays, both the
size and the relative sign of the color-suppressed term can be determined
(see Sec.~\ref{a1-a2}).

For comparisons between theoretical models and data we will use the
following values for couplings and lifetimes:
\begin{eqnarray}
|V_{cb}|& = & 0.0386 \pm 0.0027 \\ \nonumber
\frac{|V_{ub}|}{|V_{cb}|} & =  &0.073 \pm 0.011 \pm 0.010 \\ \nonumber
\tau_B^{0} & = & (1.621 \pm 0.067) {\; \rm ps}\\ \nonumber
\tau_B^{+} & = & (1.646 \pm 0.063) {\; \rm ps}\\ \nonumber
<\tau_B> & = & (1.634 \pm 0.046) {\; \rm ps}
\label{theo_parms}
\end{eqnarray}
$<\tau_B>$ is the average lifetime for a sample consisting  of
equal numbers of $B^-$ and $\bar{B}^0$ mesons.

\subsection{Factorization}

Factorization is the assumption that two body hadronic decays of $B$ mesons
can be expressed as the product of two independent hadronic currents, one
describing the formation of a meson from the converted $b$ quark and the
light spectator quark, and the other describing the production of a meson by
the hadronization of the virtual $W^-$. This description is expected to be
valid for the external spectator decays where the large energy carried by the $W^-$
causes the products of the $W^-$ to be well separated from the spectator
quark system \cite{Bjorken},\cite{DG}. It has also been used to calculate color-suppressed and
penguin diagrams, although it is not known whether factorization is a correct
assumption for these diagrams.

There are number of tests of the factorization hypothesis that can be
made by comparing rates and polarizations for semileptonic and hadronic
$B$ decays. These will be discussed in section \ref{test-factor}.
If factorization holds, then measurements of hadronic $B$ decays
can be compared to the theoretical models, and
used to extract fundamental parameters of the Standard Model.
For instance the CKM matrix element $|V_{ub}|$ could be obtained from
$\bar{B^0} \to \pi^+\pi^-$ or $ \bar{B^0} \to D_s^- \pi^+$,  and the decay
constant $f_{D_s}$ could be determined from $\bar{B^0} \to D_s^- D^{*+}$.

\subsection{Phenomenological Models of Hadronic $B$ Decay}

Several groups have developed models of hadronic $B$ decays
based on the factorization approach.
To compute rates 
for all hadronic $B$ decays the magnitude and sign of the color
amplitude must also be known. It is difficult to calculate this
amplitude from first principles in QCD.
Instead a
phenomenological approach was adopted by Bauer, Stech and Wirbel
\cite{Stech}, in which two undetermined coefficients were assigned to
the effective charged current,  $a_1$,
and the effective neutral current, $a_2$, parts of the $B$ decay
Hamiltonian. In reference \cite{Stech} these coefficients were determined from
a fit to a subset of the experimental data on charm decays.
The values of $a_1$ and $a_2$ can be related to the QCD coefficients
$c_1$ and $c_2$ by
\begin{equation}
a_{1} = c_{1} + \xi c_{2},~~~ a_{2} = c_{2} + \xi c_{1} \label{a1defs}
\end{equation}
where $\xi= 1 /N_{\rm color}$.
The values 
\begin{equation}
a_1 = 1.26,~~~a_2 = -0.51 \label{a1charm}
\end{equation}
that give the best fit to the experimental data on charm decay
correspond to $1/N_{\rm color} \sim 0$ \cite{Neubie}.
However, there is no convincing
theoretical justification for this choice of $N_{\rm{color}}$.
In section \ref{a1-a2} we will discuss the 
experimental determination of the values of $a_1$
and $a_2$ from a fit to the $B$ meson decay data.

\subsection{Heavy Quark Effective Theory}

The evaluation of amplitudes for hadronic decays requires not only the
assumption of factorization, but also the input of hadronic form factors and
meson decay constants. As a result of the development of HQET it is now
believed that many of the hadronic form factors for $b \to c $ transitions
can be calculated quite
well in an essentially model-independent way. This has been done by
several groups \cite{Neubie},\cite{Bari}. The comparison of these theoretical
predictions with the experimental results can be used to test the range of
validity of HQET, and the extent to which $1/M_Q$ corrections to the heavy
quark symmetry are needed.

%%%%\section{TESTS OF THE FACTORIZATION HYPOTHESIS }
%\input factorization.tex
\section{TESTS OF THE FACTORIZATION HYPOTHESIS }
\label{test-factor}

\subsection{Branching Ratio Tests}

The large samples of reconstructed hadronic $B$ decays
have made possible the precise measurements of branching ratios
discussed in section \ref{BDpiDrho}. As an example 
of the use of these results
to test the factorization hypothesis we will
consider the specific case of $\bar{B^0}\to D^{*+}\pi^-$.
The amplitude for this reaction is
\begin{equation}
A ={G_F\over \sqrt 2}V_{cb}V_{ud}^*
\langle \pi^- | (\bar{d} u) | 0 \rangle
\langle D^{*+} | (\bar c b) | \bar{B^0} \rangle.\label{EHeffDP}
\end{equation}
The CKM factor $|V_{ud}|$ arises from the $W^-\to\bar u d$ vertex. The first
hadron current that creates the $\pi^-$ from the vacuum is related to
the pion decay constant, $f_{\pi}$, by:
\begin{equation}
\langle \pi^-(p) | (\bar d u) | 0 \rangle = -if_{\pi}p_{\mu}.\label{Efpi}
\end{equation}
The other hadron current can be found from the
semileptonic decay $\bar{B^0}\to D^{*+}\ell^- \bar{\nu_{\ell}}$.
Here the amplitude is the product of a lepton current and the hadron
current that we seek to insert in Eq.~(\ref{EHeffDP}).
Factorization can be tested experimentally by
verifying whether the relation
\begin{equation} {\Gamma\left(\bar{B^0}\to
D^{*+}\pi^-\right)\over\displaystyle{d\Gamma\over
\displaystyle dq^2}
\left(\bar{B^0}\to D^{*+} \ell ^- \bar{\nu_{l}}
\right)\biggr|_{q^2=m^2_{\pi}}} =
6\pi^2{ c_1^2}
f_{\pi}^2|V_{ud}|^2 ,\label{Efact}
\end{equation}
is satisfied. Here
$q^2$ is the four momentum transfer from the $B$ meson to the $D^*$
meson. Since $q^2$
is also the mass of the lepton-neutrino system, by
setting $q^2 = m_{\pi}^2=0.019 ~ GeV^2$
we are requiring that the lepton-neutrino system has
the same kinematic properties as does the
pion in the hadronic decay. $V_{ud}$ and $f_{\pi}$ have well
measured values of 0.975 and 131.7~MeV respectively.
For the coefficient $c_1$ we will use the value $1.12\pm 0.1$ deduced from
perturbative QCD \cite{qcd}. The error in $c_1$ reflects the uncertainty
in the mass scale at which the coefficient $c_1$ should be evaluated.
In the original test of equation~(\ref{Efact}),  Bortoletto and Stone
\cite{Bort} found that the equation was satisfied for $c_1$=1.
In the following discussion we will denote the left hand side of
Eq.~(\ref{Efact}) by
$R_{Exp}$ and the right hand side by $R_{Theo}$.

This type of factorization  test
can be extended to larger $q^2$ values by using
other $\bar{B^0}\to D^{*+} X^-$ decays, e.g. $ X^- =\rho^-$ or
$a_1^-$. For the $\rho^-$ case Eq.~(\ref{Efact}) becomes:
\begin{equation}
 R = { {\Gamma(\bar{B}^0 \to D^{*+} \rho^-)}\over{
{{d\Gamma}\over{dq^2}} {(B\to D^{*} ~l ~\nu)|}_{q^2=m_{\rho}^2}} }
 = {6 \pi^2  c_1^2 f_{\rho}^2  |V_{ud}|^2}  \label{Efctr}
\end{equation}
where the semileptonic decay
is evaluated at $ q^2 = m_{\rho}^2=0.60$  GeV$^2$.
The decay constant on the right hand side of this equation can be
determined from $ e^+ e^- \to \rho^0$ which gives  $ f_\rho=215 \pm 4$ MeV.
A second method uses the relation
$\Gamma(\tau^- \to \nu \rho^-)=~
0.804 {G_F^2 \over{16 \pi}} |V_{ud}^2| M_{\tau}^3 f_\rho^2$, where the
$\rho$ width has been taken into account \cite{Pham}. 
This gives f$_{\rho} = 212.0 \pm 5.3$ MeV \cite{narrow}.
%With a similar
%approach but using more precise $\tau$ branching ratio measurements
%and including interference effects caused by the $\rho$'
%meson Weinstein \cite{alan} found $f_{\rho}\; = \; (223 \pm 1)$~MeV.
For the factorization test with $\bar{B^0} \to D^{*+} a_1^-$ 
we use $f_{a_1} = 205 \pm 16$ MeV \cite{ir} determined from $\tau$ decay.
%The the significant change in the 
%$a_1$ decay constant\footnote{We previously used $f_{a_1} = 205 \pm 16$ MeV
%determined by Isgur and Reader \cite{bhp}\cite{ir}.}
%is caused by a large increase in the $\tau \to a_1 \nu$ branching fraction.

To derive numerical predictions
for the left hand side of equation (\ref{Efact}),
we must interpolate the observed differential
$q^2$ distribution \cite{width} for
$B \to D^* \ell ~\nu$ to $q^2=m_\pi^2$, $m_\rho^2$, and $m_{a_1}^2$,
respectively. 
Until this distribution is measured more precisely we have to use
theoretical models to perform this interpolation.
The $d\Gamma/dq^2$ distribution obtained 
in a recent CLEO~II analysis \cite{dstlnu} is shown in Fig.~\ref{dlnu}.
The solid and dashed lines represent fits to different models.
The differences between the extrapolations using
models for $B \to D^* \ell ~ \nu$ are 
small, on the order of 10-20\%.
The measurement of this differential distribution from CLEO~II
can be combined with the earlier results from the ARGUS and
CLEO 1.5 experiments\cite{Bort,bhp}.
The values of $d\Gamma/dq^2(B\to D^*\ell \nu)$  used for the
factorization test are given in Table~\ref{TFactst}.
The statistical and systematic errors have been combined in quadrature; the
uncertainty due to the $D^0$ branching ratios cancels in the ratio.
\begin{figure}[htb]
\unitlength 1.0in
\vskip 10 mm
\begin{center}
\begin{picture}(3.0,3.5)(0.0,0.0)
\put(-1.,-0.8){\psfig{width=4.0in,height=1.5in,file=dstlnu_q2_new.ps}}
\end{picture}
\vskip 2 mm
\caption{The $d\Gamma/dq^2$ distribution for 
$\bar{B} \rightarrow D^*~\ell~\bar{\nu}$ decays from CLEO~II
data (from Ref.~\protect\cite{dstlnu}).
The solid lines represent fits to the CLEO~II data based on HQET.  
The upper solid line corresponds to a
linear extrapolation near the endpoint, 
and the lower solid line corresponds to a quadratic extrapolation.
The dashed line shows $d\Gamma/dq^2$ for the ISGW model,
the dotted line is the BSW model.}
\label{dlnu}
\end{center}
\end{figure}
\begin{table}[htb]
\caption{Ingredients for Factorization Tests.}\label{TFactst}
\begin{tabular}{cc}
$ \vert c_1 \vert $ & $1.12 \pm 0.1$  \\
$ f_{\pi}$ & $131.74 \pm 0.15 $ MeV   \\
$ f_{\rho}$ & $215\pm 4$ MeV \\
$ f_{a_1}$ & $205\pm 16$ MeV   \\
$ V_{ud}$ \protect{\cite{PDG}} & $0.9744 \pm 0.0010 $   \\
$ {{d {\cal B}}\over{dq^2}}(B \to D^* \ell ~\nu)\vert_{q^2=m_{\pi}^2} $ &
$(0.237 \pm 0.026)\:\times 10^{-2} $ GeV$^{-2}$ \\
%0.0023 GeV$^{-2}$   \\
$ {{d {\cal B}}\over{dq^2}}(B \to D^* \ell ~\nu)\vert_{q^2=m_{\rho}^2} $ & 
$(0.250 \pm 0.030)\:\times 10^{-2} $ GeV$^{-2}$ \\
%0.0025 GeV$^{-2}$   \\
$ {{d {\cal B}}\over{dq^2}}(B \to D^* \ell ~\nu)\vert_{q^2=m_{a_1}^2})$ &
$(0.335 \pm 0.033)\:\times 10^{-2} $ GeV$^{-2} $ \\
%0.0032 GeV$^{-2}$   \\
$ {{d {\cal B}}\over{dq^2}}(B \to D^* \ell ~\nu)\vert_{q^2=m_{D_s}^2} $ & 
$(0.483 \pm 0.033)\:\times 10^{-2} $ GeV$^{-2} $\\
$ {{d {\cal B}}\over{dq^2}}(B \to D^* \ell ~\nu)\vert_{q^2=m_{D_s^*}^2} $ & 
$(0.507 \pm 0.035)\:\times 10^{-2} $ GeV$^{-2} $
\end{tabular}
\end{table}
Using the information listed in Table~\ref{TFactst}
we obtain from Eqs.~(\ref{Efact}) and
(\ref{Efctr}) the results given in Table ~\ref{Tfactc}.
\begin{table}[htb]
\caption{Comparison of $R_{Exp}$ and $R_{Theo}$}\label{Tfactc}
\begin{tabular}{lcc}
 & $R_{Exp}$ (GeV$^2$) & $R_{Theo}$ (GeV$^2$)  \\ \hline
$\bar{B}^0 \to D^{*+}\pi^- $
& $1.14\pm 0.21$
  & $1.22 \pm 0.15$   \\
$\bar{B}^0 \to D^{*+}\rho^- $
& $2.80\pm 0.69$ 
& $3.26 \pm 0.42$   \\
$ \bar{B}^0 \to D^{*+} a_1^- $
& $3.6\pm 0.9$  
& $3.0 \pm 0.50$
\end{tabular}
\end{table}
Some of the systematic uncertainties in $R_{Exp}$
cancel if we form ratios of branching fractions, as does the
QCD coefficient $c_1$ in $R_{Theo}$. Thus in the case of
$D^{*+}\rho^-$/$D^{*+}\pi^-$,
the expectation from factorization is given by
$R_{Theo}(\rho)$/$R_{Theo}(\pi)$ times the ratio of the
semileptonic branching ratios evaluated at the appropriate $q^2$ values.
In Table~\ref{Tfacthh} we show the comparison between the measured ratios
and two theoretical predictions by Reader and Isgur \cite{ir}, and
the revised BSW model \cite{Neubie}.
\begin{table}[htb]
\caption{Ratios of $B$ decay widths.}\label{Tfacthh}
\begin{tabular}{lcccc}
 & Exp. & Factorization & RI Model & BSW Model \\ \hline
${\cal B}(\bar{B}^0 \to D^{*+}\rho^-) /
{\cal B}(\bar{B}^0 \to D^{*+}\pi^-)$
& $2.59 \pm 0.67$ & $2.81 \pm 0.46$ & 2.2 -- 2.3 & 2.8  \\
${\cal B}(\bar{B}^0 \to D^{*+} a_1^-) /
{\cal B}(\bar{B}^0 \to D^{*+}\pi^-)$
& $4.5 \pm 1.2$ & $3.4 \pm 0.6$ & 2.0 -- 2.1 &3.4
\end{tabular}
\end{table}
At the present level of precision, there is good
agreement between the experimental results and the expectation from
factorization for the $ q^2$ range $ 0 < q^2 < m_{a_1}^2$.
Note that it is  possible that factorization will be a poorer
approximation for decays will smaller energy release or larger $q^2$.
Factorization tests can be extended to higher $q^2$  using
$B\to D^{*} D_s^{(*)}$ 
decays as  will be discussed in section \ref{facapply} .

\subsection{Factorization and Angular Correlations}
\label{fac-ang-cor}

More subtle tests of the factorization hypothesis can be performed by examining
the polarization in $B$ meson decays into two vector mesons, as
suggested by K\"orner and Goldstein\cite{Kg}.
Again, the underlying principle is to compare the hadronic decays to the
appropriate semileptonic decays evaluated at a fixed value in $q^2$.
For instance, the ratio of longitudinal to transverse polarization
($\Gamma_{L}/\Gamma_{T}$)
in $\bar{B^0} \to D^{*+} \rho^{-}$ should be equal to the corresponding ratio
for $B\to D^{*}\ell \nu$
 evaluated at $ q^2={m_\rho}^2=0.6~ \rm{GeV}^2$.
\begin{equation}
 {{\Gamma_{L}}\over{\Gamma_{T}}} ({\bar{B^0} \to D^{*+} \rho^{-}})
= {{\Gamma_{L}}\over{\Gamma_{T}}}
{(B\to D^*\ell\nu)|}_{q^2=m_{\rho}^2}
\end{equation}
The advantage of this method is that it is not affected by QCD
corrections \cite{lepage}.

For $B \to D^*\ell\nu$ decay, longitudinal polarization
dominates at low $q^2$, whereas near
$ q^2= q^2_{\rm max}$ transverse polarization dominates. There is a simple
physical argument for the behaviour of the form factors
near these two kinematic limits. Near $ q^2=q^2_{\rm max}$,
the $D^*$ is almost at rest and its small velocity is uncorrelated with the
$D^*$ spin, so all three $D^*$ helicities
are equally likely and we expect $\Gamma_T / \Gamma_L$ = 2.
At $q^2=0$, the $D^*$ has the maximum possible momentum, while the lepton and
neutrino are collinear and travel in the direction opposite to the $D^*$.
The lepton and neutrino helicities are aligned to give
$S_z= 0$, so near $q^2=0$ longitudinal polarization is dominant.

For $\bar{B^0} \to D^{*+} \rho^-$,
we expect $88\%$ longitudinal polarization from the argument
described above \cite{Rosfac}. Similar results have been obtained by
Neubert\cite{neub}, Rieckert\cite{ricky}, and Kramer \etal \cite{Kramfac}.

\begin{figure}[htb]
\unitlength 1.0in
\vskip 15 mm
\begin{center}
\begin{picture}(2.5,2.5)(0.0,0.0)
\put(-.35,-0.2){\psfig{width=3.2in,height=3.2in,file=bexcl_fig33.ps}}
\end{picture}
\vskip 15 mm
\caption[]{
The differential branching ratio for 
$\bar{B^0} \to D^{*+} \ell \bar{\nu}_{\ell}$. The curves show the theoretical
prediction for producing transversely (dashed) and longitudinally (dash-dotted)
polarized $D^*$ mesons, as well as the total decay rate (solid) (from
Ref.~\protect\cite{neub}).}\label{neuba}
\end{center}
\end{figure}

Fig.~\ref{neuba} shows the prediction of Neubert for transverse and
longitudinal polarization in $B \to D^*\ell\nu$ decays.
Using this figure we find $\Gamma_L /\Gamma$ to be
85\% at $q^2={m_\rho}^2=0.6$.
The agreement  between these predictions and the
experimental result (Sec.~\ref{pol-D*-rho})
\begin{equation}
\Gamma_L /\Gamma \; = \;  90 \pm 7 \pm 5 \%
\end{equation}
supports the factorization hypothesis in hadronic $B$ meson decay
for $q^2$ values up to $m_{\rho}^2$.

Factorization breaks down in the charm sector due to the presence of final
state interactions, ``FSI". The strength of these long distance effects in the
B system can be determined by performing an isospin analysis of related decay
channels such as
$B^- \to D^0\pi^-$,
$\bar{B}^0 \to D^0\pi^0$, and
$\bar{B}^0 \to D^+\pi^-$ as was done in the past
for the $D\to K \pi$ and $D \to K^* \pi$
systems. At the present level of experimental precision, there
is no evidence for non-zero isospin phase shifts in B decay
From a maximum likelihood fit to the observed branching
fractions, Yamamoto finds
that $\cos\delta^* > 0.82$ at the 90\% confidence level, 
where $\delta^*$ is the phase shift
 for the $B\to D \pi$ system
and comparable constraints,
$\cos\delta^* > 0.57 (0.92)$, for the $B\to D^* \pi$ ($B\to D\rho$) isospin
multiplets\cite{hitoshi}. 
In $B$ decays to two vector mesons, such as $B \to D^*\rho$, the  presence
of final state interaction could also be probed by studying the
angle, $\chi$, between the $D^*$ and $\rho$ decay planes. FSI would cause
a phase shift between the helicity amplitudes and break 
the symmetry of the $\chi$ distribution. The
presence of FSI would lead to a angular distribution
proportional to $\sin\chi$ or $\sin 2\chi$\cite{hitoshichi}.

Until the $D_s$ decay constant, $f_{D_s}$, 
is measured more precisely, e.g. in $D_s \to
\mu\nu$, tests of the factorization hypothesis based on  
branching fractions 
cannot be applied to $B\to D^* D_s$ decays. However, CLEO~II has
accumulated about 20 events in the $B^- \to D^{*0}D_s^{*-}$ and 
$\bar{B}^0 \to D^{*+}D_s^{*-}$ modes. 
As the data sample increases, it will become
possible to measure the polarization in these decay modes and investigate
whether factorization is still a valid assumption at $q^2=m^2_{D_s}$.

\subsection{Tests of Spin Symmetry in HQET}
\label{spin-sym}

In HQET the effect of the heavy quark magnetic moment does not enter to
lowest order \cite{Mannel}, and the assumption of factorization leads to the
following predictions based on the spin symmetry of HQET:
\begin{equation}
 \Gamma (\bar{B^0} \to D^+ \pi^-) = \Gamma (\bar{B^0} \to D^{*+}\pi^-)
\end{equation}
and
\begin{equation}
 \Gamma (\bar{B^0} \to D^+ \rho^-) =
 \Gamma (\bar{B^0} \to D^{*+}\rho^-).
\end{equation}
After correcting for phase space and deviations from
heavy quark symmetry it is predicted that
${\cal B}(\bar{B^0} \to D^+ \pi^-) = 1.03~ {\cal B}(\bar{B^0} \to D^{*+}\pi^-)$
and ${\cal B}(\bar{B^0} \to D^+ \rho^-) = 0.89 ~
{\cal B}(\bar{B^0} \to D^{*+} \rho^-)$.
A separate calculation by Blok and Shifman using a QCD sum rule approach
predicts that
${\cal B}(\bar{B^0} \to D^+ \pi^-) = 
1.2 {\cal B}(\bar{B^0} \to D^{*+} \pi^-)$.
This differs from the HQET prediction due to the presence of non-factorizable
contributions \cite{BS}.

From the experimental data we find
\begin{equation}
{{{\cal B}(\bar{B^0} \to D^+ \pi^-)}\over{{\cal B}(\bar{B^0} \to D^{*+}
\pi^-)}}
\; = \; 1.11 \pm  0.22 \pm 0.08
\end{equation}
and
\begin{equation}
{{{\cal B}(\bar{B^0} \to D^+ \rho^-)}\over{ {\cal B}(\bar{B^0} \to D^{*+}
\rho^-)}}\; = \; 1.06 \pm 0.27 \pm 0.08
\end{equation}
The second error is due to the uncertainty in the D branching
fractions.
The two ratios of branching fractions are consistent with the
expectations from HQET spin symmetry as well as with the prediction
from Blok and Shifman that includes non-factorizable contributions.

Mannel \etal \cite{Mannel} observe that by using a combination of HQET,
factorization, and data on $B\to D^*~\ell~\nu$,
they can obtain model dependent predictions for
${\cal B} (\bar{B^0}\to D^+ \rho^-)/ {\cal B}(\bar{B^0} \to D^+ \pi^-)$.
Using three parameterizations of the universal Isgur-Wise form factor
\cite{param}, they predict this ratio to be 3.05, 2.52, or 2.61.
From the measurements of the branching ratios we obtain
\begin{equation}
{{{\cal B}(\bar{B^0}\to D^+ \rho^-)}\over{ {\cal B}(\bar{B^0} \to D^{+}
\pi^-)}}\; = \; 2.7 \pm 0.6 
\end{equation}
The systematic errors from the $D$ branching fractions
cancel in this ratio.
Again we find good agreement with the prediction from HQET
combined with factorization. 

Similar comparisons can be performed 
for $B\to D^{(*)} D_s^{(*)}$ decay modes.
Using isospin invariance to combine the $\bar{B}^0\to D^+ D_s^{(*-)}$
and $B^-\to D^0 ~D_s^{(*-)}$ decay modes, we obtain
%%%\begin{equation}
%%%{{{\cal B}(\bar{B}\to D D_s^{-})}\over{ {\cal B}(\bar{B} \to D^*
%%%D_s^{-})}}\; = \; 1.00 \pm 0.37 
%%%\end{equation}
\begin{equation}
{{{\cal B}(\bar{B}\to D D_s^{-})}\over{ {\cal B}(\bar{B} \to D^*
D_s^{-})}}\; = \; 0.94 \pm 0.35 
\end{equation}
The predicted range for this ratio is $1.35-1.56$\cite{cleodds}.

Similarly, 
%%%\begin{equation}
%%%{{{\cal B}(\bar{B} \to D D_s^{*-})}\over{ {\cal B}(\bar{B} \to D^{*}
%%%D_s^{*-})}}\; = \; 0.52 \pm 0.17
%%%\end{equation}
\begin{equation}
{{{\cal B}(\bar{B} \to D D_s^{*-})}\over{ {\cal B}(\bar{B} \to D^{*}
D_s^{*-})}}\; = \; 0.46 \pm 0.19
\end{equation}
In this case, the additional helicity states available leads to the
expectation that this ratio will lie in the range $0.33-0.39$\cite{cleodds}.

\subsection{Applications of Factorization}
\label{facapply}

If factorization holds, hadronic $B$ decays can be used to extract information
about semileptonic
decays. For example, we can determine the poorly measured  
rate $B\to D^{**}(2420)~\ell~\nu$ from
the branching ratio of $B\to D^{**}(2420)\pi$.
By  assuming that
the rate for $B\to D^{**}(2420)\pi$ is related to
$d\Gamma/dq^2 (B \to D^{**}(2420) \ell \nu)$ evaluated at $q^2 = m_{\pi}^2$.
Using the model of Colangelo \etal \cite{Bari}
to determine the shape of 
the form factors we obtain the ratio
$$\frac{\Gamma(B \to D^{**}(2420) ~\ell ~\nu)}
{\Gamma(B \to D^{**}(2420)\pi)}= 3.2$$
Combining this 
with the experimental result,
${\cal {B}}(B^- \to D^{**0}(2420)\pi^- )\, = \, 0.15 \pm 0.05\, \%$,
(Table~\ref{kh3}) we predict
${\cal B} (D^{**}(2420) \ell \nu ) = 0.48 \pm 0.16 \%$. This
is consistent with the average of 
recent direct measurements by OPAL and ALEPH (Table~\ref{Tbsemiexcit}),
${\cal B} (D^{**}(2420) \ell \nu ) = 0.82 \pm 0.18 \pm 0.06\%$.

A second application of factorization is the determination of $f_{D_s}$
using the decays $B \to D^*D_s$.
The rate for $\bar{B^0}\to D^{*+}D_s$ is related
to the differential rate for
$\bar{B^0}\to D^{*+}\ell^-\nu$ at $q^2 = m_{D_s}^2$ if factorization continues
to be valid at larger values of $q^2$:
\begin{equation} {\Gamma\left(\bar{B^0}\to
D^{*+} D_{s}^{-}\right)\over\displaystyle{d\Gamma\over
\displaystyle dq^2}
\left(\bar{B^0}\to D^{*+}\ell^-\nu\right)\biggr|_{q^2=m^2_{D_s}}} =
6\pi^2 \delta ~{ c_1^2}
f_{D_s}^2|V_{cs}|^2 ,\label{Efacts}
\end{equation}
The factor $\delta = 0.37$ accounts for the different form factors
which enter in  $B \to D^* D_s$ and $B\to D^*\ell\nu$ \cite{Neubie}.

Using the value listed in Table \ref{TFactst}
for $d\Gamma/dq^2(B\to D^*\ell \nu)$
at $q^2\: = \: m_{D_s}^2$ and the average branching ratio
for ${\cal B}(B\to D^{*} D_{s}^-)=0.93 \pm 0.25 \%$, we obtain
$$
f_{D_s} = (271 \pm 77) \sqrt{3.7\%/B(D_s \to \phi \pi^+)} ~\rm{MeV}
$$
and with ${\cal B}(B\to D^{*} D_{s}^{*-})=1.95 \pm 0.52 \%$, we find 
($\delta = 1$)
$$ 
f_{D_s^*} = (248 \pm 69) \sqrt{3.7\%/B(D_s \to \phi \pi^+)} ~\rm{MeV}
$$
This result can be compared to the value
$$ f_{D_s} = (344 \pm 37 \pm 52) \sqrt{B(D_s \to \phi \pi^+)/3.7\%} ~\rm{MeV}$$
that was obtained from a
direct measurement of $D_s\to \mu \nu$ decays in continuum charm events
\cite{CLNS9314}. Both values of $f_{D_s}$ are consistent with 
theoretical predictions which are in the range $f_{D_s}=200-290$~MeV
\cite{Lattices}, \cite{Potentials}, \cite{QCDsum}.
If both the $D_s^+ \to \phi \pi^+$ branching fraction and $f_{D_s}$
are measured more precisely, then measurements of the
branching ratios of $B\to D^* D_s$ decays can be used
to test factorization in $B$ decay at $q^2 = m_{D_s}^{2}$. 
In the near future,
it will also be possible to test factorization in this $q^2$ range
by measuring $\Gamma_{L}/\Gamma$ in $B \to D^* D_{s}^*$ decays.

\subsection{Factorization in Color Suppressed Decay}
\label{fac-color}

It is not obvious whether the factorization hypothesis 
will be satisfied in decays which proceed via  internal W-emission
e.g $B\to \psi K^{(*)}$. Two observables have been compared
to phenomenological
models based on the factorization hypothesis: the ratio of vector
to pseudoscalar modes and the polarization in $B\to \psi K^*$ decays.

The ratio of vector to pseudoscalar meson production
\begin{equation}
{{\cal B}(B \to \psi K^*)\over{{\cal B}
(B \to \psi K)}} = 1.68 \pm 0.33
\end{equation}
can be calculated using factorization and the ratio of the
$B\to K^*$ and $B\to K$ form factors. 
The revised BSW model of Neubert \etal\cite{Neubie}
predicts a value of 1.61 for this ratio, which is close to the
experimental value. Another test is the corresponding ratio
for $\psi^{'}$ decays:
\begin{equation}
{{\cal B}(B \to \psi' K^*)\over{{\cal B}
(B \to \psi' K)}} = 2.1 \pm 1.5
\end{equation}
This can be compared to the revised BSW model which predicts 1.85
for this ratio. 
Gourdin \etal \cite{gkpeta}
argue, that the ratio 
${{\cal B}(B \to \eta_c  K^*)/{{\cal B}
(B \to \eta_c K)}} $ would provide a good test of the
factorization hypothesis in internal spectator decays. However,
this will require a significantly larger data sample than is
available at present before this ratio can be measured with sufficient
precision. Other ratios of decay rates in modes with charmonium mesons
may also be used to test for the violation of factorization\cite{gkpother}.

The experimental results on $\psi K^*$ polarization
 can be compared to the theoretical predictions of
Kramer and Palmer\cite{Krampalm}
which depend on the assumption of factorization and on the
unmeasured $B\to K^*$ form factor. Using the BSW model to estimate the form
factors, they find $\Gamma_{L}/\Gamma= 0.57$. 
Using HQET to extrapolate
from the E691 measurements of the $D\to K^*$ form factor, they obtain
$\Gamma_{L}/\Gamma=0.73$. 
The group of Gourdin, Kamal and Pham as well as the collaboration of 
Aleksan, Le Yauoanc, Oliver, P\`ene, and Raynal 
have noted that there is no set of experimental
or theoretical form factors that can  simultaneously reproduce the
measured values of $\Gamma_{L}/\Gamma$ and ${\cal B}(B\to \psi K^*)/
{\cal B}(B\to \psi K)$ \cite{gkp},\cite{ayopr}. They conclude that
there is either a
fundamental problem in heavy to light form factors or a
breakdown of factorization for this class of decay modes.
Kamal and Santra have suggested that all the measured
observables in exclusive $B\to \psi$ can be accommodated with a
single non-factorizable amplitude\cite{kamalpsi}.

CLEO also finds evidence at the 2.5 standard deviation level
for $B\to \chi_{c2}$ transitions at a branching ratio
of $ 0.25\pm 0.10\pm 0.03\%$. If confirmed, this
would indicate the presence of either
non-factorizable color octet
contributions which are neglected in the usual treatment of
hadronic $B$ decays
or higher order processes
$O(\alpha_s^2)$ in $b\to c \bar{c} s$ decays\cite{bodwin}.

%%%%%\section{DETERMINATION OF THE COLOR SUPPRESSED AMPLITUDE}
%\input a1a2.tex
\section{DETERMINATION OF THE COLOR SUPPRESSED AMPLITUDE}
\label{eff-color-supp}

\subsection{Color Suppression in $B$ Decay}

In the decays of charmed mesons the effect of color suppression
is obscured by the effects of final state interactions (FSI), and
soft gluon effects which enhance $W$ exchange diagrams.
Table~\ref{Tcolsuprat} gives
ratios of several charmed meson decay modes with approximately
equal phase space factors where the mode in the numerator is color suppressed
while the mode in the denominator is  an external spectator decay.
With the exception of the
 decay $D^0\to \bar{K}^0\rho^0$ it is clear that the color
suppressed decays do not have significantly smaller branching ratios.
\begin{table}[htb]
\caption{Measured Ratios of color suppressed to external spectator branching
fractions.}\label{Tcolsuprat}
\begin{tabular}{cc}
Mode & Branching fraction \cite{PDG} \\ \hline
${\cal B}(D^0 \to \bar{K^0}\rho^0) / {\cal B}(D^0 \to K^- \rho^+)$
         & $0.08 \pm 0.04$   \\
${\cal B}(D^0 \to K^0 \pi^0) / {\cal B}(D^0 \to K^- \pi^+)$
         & $0.57 \pm 0.13$   \\
${\cal B}(D^0\to \bar{K^{*0}} \pi^0) / {\cal B}(D^0 \to K^{*-} \pi^+)$
         & $0.47\pm 0.23$  \\
${\cal B}(D^0 \to \pi^0 \pi^0) / {\cal B}(D^0 \to \pi^- \pi^+)$
         & $0.77 \pm 0.25$   \\
${\cal B}(D_s^{+} \to \bar{K^{*0}} K^+) / {\cal B}(D_s \to \phi \pi^+)$
         & $0.95\pm 0.10$   \\
${\cal B}(D_s^{+} \to \bar{K^0} K^+) / {\cal B}(D_s \to \phi \pi^+)$
         & $1.01 \pm 0.16$ \\
\end{tabular}
\end{table}

When the BSW model is used to fit the data on charm decays
it gives values of $a_1=1.26$ and $a_2 = -0.51$.
The BSW model assumes that the values of the coefficients can be
extrapolated from $\mu = m_{c}^2$ to $\mu = m_{b}^2$
taking into account the evolution of the strong coupling constant
$\alpha_s$. This extrapolation gives the predictions $a_1=1.1$ and
$a_2=-0.24$ for $B$ decays.
The smaller magnitude of $a_2$ means that in contrast to the charm sector
one expects to find a more consistent pattern of color suppression in $B$ meson
decays.
Another approach uses the factorization hypothesis, HQET
and model dependent form factors (RI model)\cite{ir}. In this approach,
$a_1$ and $a_2$ are determined from QCD (with
$ 1 /N_{\rm color} =1 /3$), and color suppressed $B$ decays are
expected to occur at about $1/1000$ the rate of unsuppressed decays.
%The observation of color suppressed $B$ decays at a much greater
%level would indicate the breakdown of the factorization hypothesis.
In Section ~\ref{color-supress} we obtained upper limits for color suppressed
$B$ decays with a $D^0$ or $D^{*0}$ meson in the final state.
In Table~\ref{Tbrcolcomp} these results are compared to the predictions
of the BSW and the RI models.

\begin{table}[htb]
\caption{Branching fractions of color suppressed $B$ decays
and comparisons with models.}\label{Tbrcolcomp}
\begin{tabular}{lcccc}
Decay Mode & U. L. (\%) & BSW (\%) &
 $\cal{B}$ (BSW) & RI~model(\%)  \\ \hline
$\bar{B^0} \to D^{0} \pi^0$     &$<0.048$ & $0.012$
 & $0.20 a_2^{2} (f_{D}/220 \rm{MeV})^2$ & $0.0013 - 0.0018$   \\
$\bar{B^0} \to D^{0} \rho^0$    &$<0.055$ & $0.008$
 & $0.14 a_2^{2} (f_{D}/220 \rm{MeV})^2$ & $0.00044$   \\
$\bar{B^0} \to D^{0} \eta$      &$<0.068$ & $0.006$
& $0.11 a_2^{2} (f_{D}/220 \rm{MeV})^2 $             &              \\
$\bar{B^0} \to D^{0} \eta^{'}$  &$<0.086$ & $0.002$
& $ 0.03 a_2^{2}(f_{D}/220 \rm{MeV})^2$  &              \\
$\bar{B^0} \to D^{0} \omega $   &$<0.063$ & $0.008$
& $0.14 a_2^{2}(f_{D}/220 \rm{MeV})^2$   &              \\
$\bar{B^0} \to D^{*0} \pi^0$    &$<0.097$ & $0.012$
& $ 0.21 a_2^{2}(f_{D*}/220 \rm{MeV})^2$ & $0.0013-0.0018$   \\
$\bar{B^0} \to D^{*0} \rho^0$   &$<0.117$  & $0.013$
& $ 0.22 a_2^{2}(f_{D*}/220 \rm{MeV})^2$ & $0.0013 -0.0018$   \\
$\bar{B^0} \to D^{*0} \eta$     &$<0.069$ & $0.007$
& $0.12 a_2^{2}(f_{D*}/220 \rm{MeV})^2$   &   \\
$\bar{B^0} \to D^{*0} \eta^{'}$ &$<0.27$  & $ 0.002$
& $0.03 a_2^{2}(f_{D*}/220 \rm{MeV})^2$   &   \\
$\bar{B^0} \to D^{*0} \omega$   &$<0.21$  & $0.013$
& $ 0.22 a_2^{2}(f_{D*}/220 \rm{MeV})^2$  &
\end{tabular}
\end{table}

In contrast to charm decay, color suppression seems to be operative
in hadronic decays of $B$ mesons. The limits on the color suppressed
modes with $D^{0(*)}$ and neutral mesons are still above the level
expected by the two models, 
but we can already exclude a prediction
by Terasaki \cite{tera} that
${\cal{B}}(\bar{B^0} \to D^0 \pi^0) 
\approx 1.8 {\cal{B}}(\bar{B^0} \to D^+\pi^-)$.
To date, the only color suppressed $B$ meson decay modes
that have been observed are final states
which contain charmonium mesons e.g. $B\to \psi K$ and 
$B\to \psi K^*$
\cite{psicomment}.

\subsection{Determination of $|a_1|$, $|a_2|$ and
the Relative Sign of ($a_2/a_1$)}
\label{a1-a2}

In the BSW model \cite{Neubie}, the branching
fractions of the $\bar{B}^0$ normalization 
modes are proportional to $a_1^2$ while
the branching fractions of the
$B\to\psi$ decay modes depend only on $a_2^2$.
 A fit to the
branching ratios for the modes
$\bar{B^0}\to D^+\pi^-$, $D^+\rho^-$, $D^{*+}\pi^-$ and $D^{*+}\rho^-$
using the model of Neubert \etal\ yields
\begin{equation}
|a_1| = 1.03 \pm 0.04  \pm  0.06
\label{normal_a1}
\end{equation}
and a fit to
the modes with $\psi$ mesons in the final state gives
\begin{equation}
|a_2| = 0.23 \pm 0.01 \pm 0.01
\label{psi_a2}
\end{equation}
The first error
on $|a_1|$ and $|a_2|$ includes the
uncertainties from the charm or charmonium branching ratios,
 the experimental systematics associated with detection
efficiencies and background subtractions as well as the statistical
errors from the branching ratios.
The second  error quoted is the uncertainty due to
the $B$ meson production fractions and lifetimes.
 We have assumed that the ratio of $B^+ B^-$ and $B^0 \bar{B^0}$ production
at the $\Upsilon(4S)$ is one \cite{dstlnu}, 
and assigned an uncertainty of 10\% to it. 

The magnitude of the amplitude for external spectator processes,
$|a_1|$ can also be determined from $B\to D^{(*)}D_s^{(*)}$ decays. 
Since these transitions are not subject to interference with the internal
spectator amplitude we can combine $B^-$ and $\bar{B}^0$ decays to reduce
the statistical error. Using the average branching fractions given in
Tables~\ref{kh3},~\ref{kh4} we obtain
\begin{equation}
|a_1|_{DD_s} = 0.93 \pm 0.06 \pm 0.04
\label{dds_a1}
\end{equation}
It is interesting to note that this value of $|a_1|$ agrees with the
result of the fit to the $B\to D^{(*)} \pi$ and $B\to D^{(*)}\rho$ modes
(see~\ref{normal_a1}).
In general, $|a_1|$ could be different
for exclusive $b\to c \bar{u} d$ and $b\to c \bar{c} s$ processes.

\begin{table}[htb]
%
%   Rescaling:
%
%	BSW are normalized to  (tau_b/1.18)^0.5 |Vcb| = 0.045
%		so far we have been lucky to match this with the
%		new numbers:
%			(1.634/1.18)^0.5 0.0386 = 0.045
%
%	Deandrea et al list their prediction as Gamma = xxxx|Vcb|^2
%		so we multiply with the current Vcb value squared and the
%		lifetime.
%
\label{Tbswcol}
\caption{Predicted branching fractions in terms of BSW parameters $a_1$, $a_2$.
The coefficients have been rescaled to accommodate the new $B$ lifetime and
$|V_{cb}|$ values given in equation \protect{\ref{theo_parms}} and 
$f_D\, = \, f_{D^*}\, = 220$~MeV.}
\begin{tabular}{lcc}
Mode & Neubert \etal \cite{Neubie} & Deandrea \etal \cite{DBGN} \\ \hline

$\bar{B}^0 \to D^+ \pi^- $        &$ 0.264 a_1^2 $ & $0.278 a_1^2$ \\
$\bar{B}^0 \to D^+ \rho^-$        &$ 0.621 a_1^2 $ & $0.717 a_1^2$ \\
$\bar{B}^0 \to D^{*+} \pi^-$      &$ 0.254 a_1^2 $ & $0.278 a_1^2$ \\
$\bar{B}^0 \to D^{*+} \rho^-$     &$ 0.702 a_1^2 $ & $0.949 a_1^2$ \\
$\bar{B}^0 \to D^+ D_s^- $        &$ 1.213 a_1^2 $ & $ 1.094 a_1^2$ \\
$\bar{B}^0 \to D^+ D_s^{(*-)}$    &$ 0.859 a_1^2 $ & $ 0.745 a_1^2$ \\
$\bar{B}^0 \to D^{*+} D_s^-$      &$ 0.824 a_1^2 $ & $ 0.768 a_1^2$ \\
$\bar{B}^0 \to D^{*+} D_s^{(*-)}$ &$ 2.203 a_1^2 $ & $ 2.862 a_1^2$ \\
$ B^- \to D^0 \pi^- $             &
$ 0.265 [a_1 +1.230 a_2 ( f_D/220)]^2 $ &
$ 0.278 [a_1 +1.12655 a_2 ( f_D/220)]^2 $  \\
$ B^- \to D^0 \rho^-$             &
$  0.622 [a_1 + 0.662 a_2 ~( f_D/220)]^2 $ &
$ 0.717 [a_1 +0.458 a_2 ( f_D/220)]^2 $  \\
$ B^- \to D^{*0} \pi^-$           &
$  0.255 [a_1 +1.292 a_2 ~( f_{D^*}/220)]^2$ &
$ 0.278 [a_1 + 1.524 a_2 ( f_{D^*}/220)]^2 $  \\
$ B^-  \to D^{*0} \rho^-$         &
 $  0.703 [a_1^2 + 1.487 a_1 a_2 ~( f_{D^*}/220) $   &
$ 0.949 [a_1^2 + 1.31 a_1 a_2 ~( f_{D^*}/220) $  \\
 & $+0.635 a_2^2 (f_{D^*}/220)^2]$ &
  $+ 0.53 a_2^2 ( f_{D^*}/220)^2]$ \\
$B^- \to D^0 D_s^- $        &$ 1.215 a_1^2 $ & $ 1.094 a_1^2$ \\
$B^- \to D^0 D_s^{(*-)}$    &$ 0.862 a_1^2 $ & $ 0.745 a_1^2$ \\
$B^- \to D^{*0} D_s^-$      &$ 0.828 a_1^2 $ & $ 0.768 a_1^2$ \\
$B^- \to D^{*0} D_s^{(*-)}$ &$ 2.206 a_1^2 $ & $ 2.862 a_1^2$ \\
$ \bar{B^0} \to \psi \bar{K}^0$    &$ 1.817 a_2^2 $ & $ 1.652 a_2^2$ \\
$ \bar{B^0} \to \psi \bar{K}^{*0} $&$ 2.927 a_2^2 $ & $ 2.420 a_2^2$ \\
$ \bar{B^0} \to \psi$'$ \bar{K}^0$    &$ 1.065 a_2^2 $ & $ 0.559 a_2^2$ \\
$ \bar{B^0} \to \psi$'$ \bar{K}^{*0} $&$ 1.965 a_2^2 $ & $ 1.117 a_2^2$ \\
$ B^- \to \psi K^- $               &$ 1.819 a_2^2 $ & $ 1.652 a_2^2$ \\
$ B^- \to  \psi K^{*-}$            &$ 2.932 a_2^2 $ & $ 2.420 a_2^2$ \\
$ B^- \to \psi$'$ K^- $            &$ 1.068 a_2^2 $ & $ 0.559 a_2^2$ \\
$ B^- \to  \psi$'$ K^{*-}$         &$ 1.971 a_2^2 $ & $ 1.117 a_2^2$ \\
\end{tabular}
\end{table}

\begin{table}[htb]
\caption{Ratios of normalization modes to determine the sign of
$a_2/a_1$. The magnitude of $a_2/a_1$ is the value in the 
BSW model which agrees with our result for $B\to \psi$ modes.}\label{Tbswexpc}
\begin{tabular}{ccccc}
Ratio &$a_2/a_1 =-0.23 $ & $a_2/a_1 =0.23 $ & Experiment & RI~ model \\ \hline
$R_1 $& 0.51  & 1.64 & $1.60 \pm 0.30$ &$1.20-1.28$  \\
$R_2 $& 0.72  & 1.33 & $1.61 \pm 0.39$ &$1.09-1.12$  \\
$R_3 $& 0.49  & 1.68 & $1.85 \pm 0.40$ &$1.19-1.27$  \\
$R_4 $& 0.68  & 1.37 & $2.10 \pm 0.61$ &$1.10-1.36$
\end{tabular}
\end{table}

By comparing branching ratios of $B^-$ and $\bar{B^0}$ decay modes it is
possible to determine the the sign of $a_2$ relative to $a_1$.
The BSW model,~\cite{Neubie} predicts the following ratios:
\begin{equation}
R_1 = {{\cal B}(B^- \to D^0 \pi^-) \over {\cal B}(\bar{B^0}\to D^+ \pi^-)}
                = (1 + 1.23 a_2/a_1)^2  \label{colrate1}
\end{equation}
\begin{equation}
R_2 = {{\cal B}(B^- \to D^0 \rho^-)
\over {\cal B}(\bar{B^0} \to D^+ \rho^-)}
                = (1 + 0.66 a_2 /a_1)^2  \label{colrate2}
\end{equation}
\begin{equation}
R_3 = {{\cal B}(B^- \to D^{*0} \pi^-)
         \over {\cal B}(\bar{B^0} \to D^{*+} \pi^-)}
                     =(1 + 1.29 a_2/a_1)^2  \label{colrate3}
\end{equation}
\begin{equation}
R_4 = {{\cal B}(B^- \to D^{*0} \rho^-)
          \over{\cal B}(\bar{B^0} \to D^{*+} \rho^-)}
                     \approx (1 + 0.75 a_2/a_1)^2   \label{colrate4}
\end{equation}

Table~\ref{Tbswexpc} shows a comparison between the
experimental results and
the two allowed solutions in the BSW model.
In the experimental ratios the systematic errors due to
detection efficiencies partly cancel.
In the ratios $R_3$ and $R_4$ the $D$ meson branching ratio uncertainties
do not contribute to the systematic error.

A least squares fit to the ratios $R_1$ - $R_3$ gives
\begin{equation}
a_2/a_1 = 0.25 \pm 0.07 \pm 0.06
\label{a2a1_ratio}
\end{equation}
where we have ignored uncertainties in the
theoretical predictions.
$R_4$ is not included in the fit since
the model prediction in this case is not thought to be reliable \cite{volkie}.
The second error is due to the uncertainty in
the $B$ meson production fractions and lifetimes
which enter into the determination of $a_1/a_2$ in the combination
$(f_+  \tau_{+}/ f_{0} \tau_{0})$. 
 As this ratio increases,
the value of $a_2/a_1$ decreases. 
The allowed range of $(f_+  \tau_{+}/ f_{0} \tau_{0})$ 
excludes a negative value of $a_2/a_1$.

Other uncertainties in the magnitude\cite{fdvari}
 of $f_D$, $f_{D^*}$ and in the hadronic form
factors can change the magnitude of $a_2/a_1$ but not its sign.
The numerical factors which
multiply $a_2/a_1$ include the ratios of $B \to \pi$($B\to\rho$)
to $B\to D$ ($B\to D^*$) form
factors, as well as the ratios of the meson decay constants. We
assume values of 220~MeV for $f_D$ and $f_{D^*}$ \cite{rosfd}.
To investigate the model dependence of the result we have recalculated
$|a_1|$, $|a_2|$, and $a_2/a_1$ in the model of Deandrea \etal\ We find
$|a_1| = 0.97 \pm 0.04 \pm 0.06$, 
$|a_2| = 0.24 \pm 0.01 \pm 0.01$, and
$a_2/a_1 = 0.25 \pm 0.07 \pm 0.05$, consistent with the results discussed
above. A different set of $B \to \pi$ form factors can be calculated using
QCD sum rules. Using the form factors determined by Belyaev, Khodjamirian 
and R\"uckl \cite{brueckl} and 
by Ball \cite{Ballff}, $a_2/a_1$ changes by 0.04. Kamal and Pham
have also considered the effect of uncertainties in form factors,
the effects of final state interactions, and annihilation terms. They
conclude that these may change the magnitude of $a_2/a_1$ but 
not its sign \cite{KPham}. Systematic uncertainties in the ratio 
of $D$ branching fractions could also modify its magnitude.

\begin{table}[htb]
\caption{Predicted (BSW) and measured ratios of widths of
$D^+$ and $D^0$ modes in charm decay.}\label{Tbcharm}
\begin{tabular}{cccc}
Mode &$a_2/a_1 =-0.40 $ & $a_2/a_1 =0.40 $ & Ratio of widths (exp)\cite{PDG} \\ \hline
$D^+ \to \bar{K}^0\pi^+ /D^0\to K^- \pi^+$ &
   0.26           &  2.2   &  $0.28  \pm 0.05 $   \\
$D^+\to \bar{K}^0\rho^+ /D^0\to K^-\rho^+$ &
   0.58        &  1.5  & $ 0.36 \pm 0.10   $   \\
$D^+\to \bar{K}^{*0}\pi^+ /D^0\to K^{*-}\pi^+$ &
   0.05       &   3.2   & $0.17\pm  0.07 $   \\
$D^+\to \bar{K}^{*0}\rho^+ /D^0\to K^{*-}\rho^+ $ &
   0.34       &   2.0   & $0.25 \pm 0.12 $
\end{tabular}
\end{table}

The magnitude of $a_2$ determined from this fit 
to the ratio of $B^-$ and $B^0$ modes is consistent with the value
of $a_2$ determined from the fit to the $B\to\psi$ decay modes.
The sign of $a_2$ disagrees with the theoretical
extrapolation from the fit to charmed meson decays using
the BSW model\cite{oldfit}. It is also disagrees with the 
expectation from the $1/N_{c}$ rule\cite{BS},\cite{halperin}.
Table~\ref{Tbcharm} compares the corresponding charm decay ratios 
to the theoretical expectations for positive and negative values of $a_2/a_1$.
The result may be consistent with the expectation of perturbative QCD
\cite{Burasa1a2}.

%%%%\section{THE BAFFLING SEMILEPTONIC BRANCHING RATIO}
%\input bsl_problem.tex 
\subsection{The Sign of a$_2$/a$_1$ and 
the Anomalous Semileptonic Branching Ratio}
\label{baffle}

A relative plus sign between the coefficients $a_1$ and $a_2$
indicating constructive interference in $B^-$ decays came somewhat
as a surprise since destructive interference is observed in charm
decay. 
Although constructive interference has been observed in all the
$B^-$ modes studied so far these only comprise a small fraction of the
total rate. It is therefore important to broaden the experimental base
and to measure $a_1$ and $a_2$ for a large variety of decay modes.
One approach would be to compare inclusive $B^- \to D^0_{\rm direct}$
with $\bar{B}^0 \to D^+_{\rm direct}$
production.
It is intriguing that $a_1$ determined from $B\to D^{(*)}\pi , \,
D^{(*)} \rho$
modes agrees well with the value of $a_1$ extracted from 
$B \to DD_s$ decays. 
The observation of  color suppressed decays such as $\bar{B}^0 \to D^0 \pi^0$
would certainly help to clarify this picture since they give another
measure of $|a_2|$ complementary to $B \to $ Charmonium decays .

Keum \cite{keumi} has suggested that the relative sign of $a_1$ and $a_2$
could be determined from a measurement of the polarization in 
$B^- \to D^{*0} \rho^-$ decays. For $a_2/a_1 > 0 $ the amount of
longitudinal polarization should be less than 88\% and vice versa.

The experimentally measured semileptonic branching ratio is determined to be
$(10.35\pm 0.17 \pm 0.35)$\% in the model independent dilepton analysis
\cite{Cleo2l}.
Comparable but more precise rates are also obtained from the analysis of the
single lepton spectrum.
These measurements are significantly below the theoretical lower
bound  ${\cal{B}}_{sl}>12.5 $\% from  QCD calculations within the
parton model\cite{bbsl}.

It is possible to 
understand simply the origin of the theoretical limit.
In the absence of QCD corrections, the virtual $W$ emitted by the
b quark can decay into a lepton-antineutrino pair, a $\bar{u}-d$ quark
pair, or $\bar{c}-s$ quark pair. For the decay into a quark pair,
there are three possible color states which are equally probable.
In addition, corrections must be made for the reduction in phase space
in the $W\to \tau \nu$ and $W\to \bar{c} s$ decays.
Then the semileptonic fraction, ${\cal B}_{SL}$ is given by
\begin{equation}
 {\cal B}_{SL} = {{f_c} \over {5 f_c + 3 f_{\bar{c} s} +  f_{c \tau}} }
\end{equation}
Using the phase space factors, $f_c=0.45$, $f_{\bar{c} s} \approx f_{c \tau}
=0.12$ gives ${\cal B}_{SL} = 16.5\%$. Including QCD corrections, modifies
the hadronic contributions to the width and gives ${\cal B}_{SL} = 14.3\%$.
The theoretical lower limit of $12.5\%$
is obtained by varying the quark masses and
QCD scale to their lower limits.

Several explanations of this discrepancy have been proposed and await
experimental confirmation:
\begin{itemize}
\item An increased $b\to c\bar{c}s$ component of the 
$B$ meson hadronic width 
\cite{bbsl}, \cite{palmstech},\cite{dunietz}. However, recent
experimental data rule out 
the mechanism suggested by reference \cite{dunietz}
 as a major contributor to $B \to$ baryon decays.
\item Higher order contributions might reduce the theoretical expectation
or the assumption of duality may not hold for b quark decay\cite{falk}.
The former
has been advocated by Bagan, Ball, Braun, and Gosdzinsky
who find results consistent 
with the experimental result\cite{bagan1},\cite{bagan2} but
also predict $N_{c}=1.28\pm 0.08$ for the number of charm quarks
produced per $b$ decay due to higher order enhancements
of the $b\to c \bar{c} s$ channel\cite{bagan2}.
\item Constructive interference in $B^-$ decays would reduce the theoretical
expectation for the semileptonic branching ratio.
A small contribution from W exchange to $\bar{B}^0$ decays would
keep the lifetime ratio close to unity and satisfy the experimental
constraints on this quantity\cite{hsw}.
\end{itemize}
Increasing the $b\to c\bar{c}s$ component would increase the average number of
c quarks produced per b quark 
decay and lead to another interesting problem:
 the predicted number of charm quarks per b decay would increase 
to 1.3 while the current experimental world average  
for this number is $1.10\pm 0.06$ (see section~\ref{charmpro}).

There could also
be a large contribution to the hadronic width that has not been measured.
 It has been suggested by Palmer and Stech\cite{palmstech},
that $b \to c \bar{c} s$ followed by $c \bar{c} \to \rm{gluons}$,
 which in turn hadronize into a final state with no charm, has a large
branching ratio. 
Another related suggestion is that the rate for the hadronic penguin
diagram $b\to sg$ is much larger than expected\cite{kaganbsg}.
These possibilities will lead to significant production of high
multiplicity charmless final states and 
are difficult to distinguish experimentally.

A systematic study of inclusive hadronic $B$ decays
to mesons and baryons will be required to resolve this problem.

%%%%%\section{RARE HADRONIC DECAYS}
%\input rare.tex
\section{RARE HADRONIC DECAYS}

There are hadronic $B$ meson decays that cannot be produced
by the usual $b\to c$ transition. The results of the
experimental search for these rare
decay modes provides important information on the mechanisms of $B$ meson
decay and significant progress is being made
with the collection of large samples of $B$ mesons by the CLEO II experiment.
As an indication of this we will discuss the 
first observation of radiative penguin decay as well as new
experimental results on the decays $\bar{B^0}\to\pi^+\pi^-$ and
$\bar{B^0}\to K^-\pi^+$ where a statistically significant signal has been
observed in the sum of the two modes.

Decays of the kind $B\to D_s X_u$, where the $X_u$ system hadronizes as
pions, can occur via a $b\to u$ spectator diagram where the $W$ forms a
$c\bar{s}$ pair. Since other contributing diagrams are expected to be
negligible these decays may provide a clean environment in which to measure
$V_{ub}$ in hadronic decays. Decays of the kind $\bar{B^0}\to D_s^+ X_s^-$,
where $X_s$ is a strange meson, are also interesting since they are
associated with a $W$ exchange diagram.

\begin{figure}[htb]
\begin{center}
\vskip 15 mm
\unitlength 1.0in
\begin{picture}(2.,1.2)(0,0)
\put(-1.9,-3.4){\psfig{width=6.5in,height=5.8in,%
file=rarefeyn.ps}}
\end{picture}
%\vskip 5 mm
\caption{Rare $B$ meson decay diagrams: (a)
$b \to u$ spectator and (b) gluonic penguin.}
\label{rarefeyn}
\end{center}
\end{figure}

Charmless hadronic decays
 such as $\bar{B^0}\to\pi^+ \pi^-$, $B^-\to\pi^-\pi^0$,
$\bar{B^0}\to\pi^{\pm}\rho^{\mp}$ and $B^-\to\pi^0 \rho^-$, are expected to be
produced by the $b\to u$ spectator diagram (Fig. \ref{rarefeyn}(a)), although there is a possible small
contribution from a $b\to d$ penguin diagram (Fig. \ref{rarefeyn}(b)). The decay
$\bar{B^0}\to \pi^+\pi^-$ has been discussed as a possible place to observe
CP violation in the $B$ meson system \cite{CPpipi}. The final state is
a CP eigenstate, and CP violation can arise from interference between the
amplitude for the direct decay via the $b\to u$ spectator diagram, and the
amplitude for the decay following $B^0\bar{B^0}$ mixing. In this decay the
CP violating angle is different from the one accessible in
$\bar{B^0}\to \psi K_s$, so the measurement is complementary.
There is a possible complication if the $b\to d$ penguin contribution
to the amplitude is significant. This could be resolved if measurements are
made on other rare hadronic decay modes to determine the role of the penguin
amplitude in any observed CP violating effect \cite{CPpipi}.

Decays to charmless hadronic final states containing an $s$ quark are expected
to have a significant contribution from a $b\to s$ penguin diagram, although
they can also occur through a CKM suppressed $b\to u$ spectator diagram.
The inclusive rates for the hadronic penguin diagrams $b\to sg$
and $b\to sq\bar{q}$ are estimated
to be of order 0.1\% from the parton model, but predictions for the
hadronization into exclusive final states are uncertain because the simple
assumptions about factorization of the amplitude used for the spectator
diagram may not be valid for loop diagrams.

Gronau, Rosner and London have  observed that precise
measurements of the branching fractions for hadronic charmless decay modes
will provide sufficient information to determine the 
CKM complex phase\cite{gronros}.
SU(3) symmetry gives the relationship 
\begin{equation}
\sqrt{2} A (B^+\to \pi^0 K^+) + A(B^+\to \pi^+ K^0)
= \tilde{r_u} \sqrt{2} A(B^+\to \pi^+ \pi^0) \label{roseqn}
\end{equation}
where $\tilde{r_u}=f_K/f_\pi |V_{us}/V_{ud}|$ accounts for
$SU(3)$ breaking.
The weak phase $\gamma$ enters only in the charmless decay
modes proportional to $|V_{ub}|$ but not in those which are proportional
to $|V_{ts}|$. By taking appropriate linear combinations of the rates
for the above decay modes and their charge conjugates, it is then possible
to solve for $\gamma$. Of order at least 100 reconstructed decays in
each of the modes would be required to complete the determination.
However, it has been recently noted that
a possible contribution of electroweak penguins to the amplitudes
for these decays may invalidate equation~(\ref{roseqn})~ \cite{deshiso}.
Other systematic uncertainties due to contributions
from loops with $c$ and $u$ quarks may also be problematic.

\subsection{Decays to $D_s$ Mesons}

These decays have recently been searched for by ARGUS \cite{ARGUSDspi}
and CLEO~II \cite{CLEODspi}. The upper limits 
are given in Table~\ref{TABDspi}
along with theoretical predictions by Choudury~\etal \cite{CISS},
and Deandrea~\etal \cite{DBGN}.

\begin{table} [hbt]
\caption{Theoretical predictions and experimental upper limits (90\% C.L.)
for $B$ decays to $D_s$.
All numbers quoted are branching fractions $\times 10^{5}$}
\label{TABDspi}

\begin{center}
\begin{tabular}{lcccc}
$B$ Decay          & Choudury & Deandrea & ARGUS    & CLEO II  \\ \hline
$D_s^+\pi^-$       &   1.9    &    8.1   & $<$170.0 & $<$27.0  \\
$D_s^{*+}\pi^-$    &   2.7    &    6.1   & $<$120.0 & $<$44.0  \\
$D_s^+\rho^-$      &   1.0    &    1.2   & $<$220.0 & $<$66.0  \\
$D_s^{*+}\rho^-$   &   5.4    &    4.5   & $<$250.0 & $<$74.0  \\
$D_s^+\pi^0$       &   1.8    &    3.9   & $<$90.0  & $<$20.0  \\
$D_s^{*+}\pi^0$    &   1.3    &    3.0   & $<$90.0  & $<$32.0  \\
$D_s^+\eta$        &          &    1.1   &          & $<$46.0  \\
$D_s^{*+}\eta$     &          &    0.8   &          & $<$75.0  \\
$D_s^+\rho^0$      &    0.5   &    0.6   & $<$340.0 & $<$37.0  \\
$D_s^{*+}\rho^0$   &    2.8   &    2.4   & $<$200.0 & $<$48.0  \\
$D_s^+\omega$      &          &    0.6   & $<$340.0 & $<$48.0  \\
$D_s^{*+}\omega$   &          &    2.4   & $<$190.0 & $<$68.0  \\
$D_s^+ K^-$        &          &          & $<$170.0 & $<$23.0  \\
$D_s^{*+} K^-$     &          &          & $<$120.0 & $<$17.0  \\
$D_s^+ K^{*-}$     &          &          & $<$460.0 & $<$97.0  \\
$D_s^{*+} K^{*-}$  &          &          & $<$580.0 & $<$110.0 \\
\end{tabular}
\end{center}
\end{table}
The experimental limits are still at least a factor of three above the
theoretical predictions. If these limits are compared to the predictions of
Deandrea~\etal then the best constraint on $|V_{ub}/V_{cb}|$ will
come from the CLEO~II limit on $\bar{B^0}\to D_s^+\pi^-$, but that this model
dependent limit is still above the range
$0.06<|V_{ub}/V_{cb}|<0.10$ allowed by the recent semileptonic
 data \cite{btoulnu}.
Combining several $D_s X_u$ modes the sensitivity to $V_{ub}$ can be slightly 
improved. For example, using the BSW model CLEO obtains an upper limit
of $|V_{ub}/V_{cb}|< 0.15$ (90\% C.L.) \cite{CLEODspi}.

\subsection{Charmless Hadronic $B$ Decay}

Predictions of branching ratios for charmless hadronic decays were made by
Bauer, Stech and Wirbel \cite{Stech} using the $b\to u$ spectator diagram and
the assumption of factorization. The possible contributions from penguin
diagrams were neglected. These predictions have recently been updated
by Deandrea \etal \cite{DBGN} using new estimates of the hadronic form factors.
We compare their results to 
the experimental upper limits in Table \ref{TABbsw}.
Recently, the LEP experiments with silicon vertex detectors
have also contributed to the 
search for these charmless decay modes. These experiments are
also able to set limits on rare decays of the $B_s$ meson
(see Table~\ref{bsrare}), which
are not produced at threshold experiments.
%
% TEB note:
% since the Lambda_b has not yet been reconstructed
% decided to omit the ALEPH limits on rare Lambda_b decays
% Lambda_b\to p \pi^- < 16 \times 10^{-5}
% Lambda_b\to p  K^-  < 16 \times 10^{-5}
% Lambda_b\to \Lambda \gamma  < 56 \times 10^{-5}

\begin{table} [hbt]
\caption{Theoretical predictions and experimental upper limits (90\% C.L.)
for charmless hadronic $B$ decays. All numbers quoted are branching fractions
$\times 10^{5}$.}
\label{TABbsw}
\begin{center}
\begin{tabular}{lccccccc}
$B$ Decay     &Deandrea &  ARGUS  &  CLEO 1.5 & CLEO II & DELPHI 
& ALEPH & OPAL \\ \hline
$\pi^+\pi^-$  &  1.8  & $<$13.0 & $<$7.7 &$<2.2$  & $<5.5$ & $<7.5$ & $<4.7$ \\
$\pi^{\pm}\rho^{\mp}$ & 5.2 & $<$52.0 &  &$<9.5$  & & &\\
$\rho^+\rho^-$&  1.3  &         &        &        & & & \\
$\pi^{\pm}a_1^{\mp}$ &   & $<$90.0 & $<$49.0 &    & & & \\
$\pi^0\pi^0$  &  0.06 &         &        & $<1.0$ & & & \\
$\pi^0\rho^0$ &  0.14 & $<$40.0 &        & $<2.9$ & & & \\
$\rho^0\rho^0$&  0.05 & $<$28.0 & $<$29.0 &       & & & \\
$\pi^-\pi^0$  &  1.4  & $<$24.0 &        & $<2.3$ & & & \\
$\pi^-\rho^0$ &  0.7  & $<$15.0 & $<$17.0& $<4.1$ &$<26$& & \\
$\pi^0\rho^-$ &  2.7  & $<$55.0 &        &        & & & \\
$\rho^-\rho^0$&  0.7  & $<$100.0 &       &        & & & \\
$\pi^-\pi^+\pi^-$&    &         &       &        & $<22$& & \\
$\pi^+\pi^+\pi^-\pi^-$&    &         &       &        & $<28$& & 
%$p\bar{p}\pi^-$&      & ($52\pm 14\pm 19$)& $<$14.0 & & & &
\end{tabular}
\end{center}
\end{table}
In addition to the results given in Table~\ref{TABbsw}, L3
has set limits on two rare modes with all neutral final states
$B^0\to \eta\pi^0$ and $B\to \eta\eta$ modes
of $84\times 10^{-5}$ and $210\times 10^{-5}$, respectively.

%We have included in Table \ref{TABbsw} the reported observation of the decay
%$B^-\to p\bar{p}\pi^-$ by ARGUS \cite{ppbarpi}, even though this has not
%been confirmed by later data from either ARGUS or CLEO \cite{ppbarpi2}.

There are two recent sets of theoretical predictions by
Deshpande \etal \cite{Desh} and Chau \etal \cite{Chau} that
take into account both penguin and spectator contributions and
make predictions for a large number of charmless hadronic $B$ decays.
A selection of these predictions are shown in table~\ref{TABbsg}. Large
contributions from the penguin amplitude are expected in decays such as
$B\to K^{(*)}\phi$ and $B\to K^{(*)}\pi$.
However, the decays $B\to K\rho$ are predicted to
have very small penguin amplitudes due to cancellations in the contributions
to the amplitude \cite{Desh}.

\begin{table} [hbt]
\caption{Theoretical predictions and experimental upper limits (90\% C.L.)
for $b\to s$ decays. All numbers quoted are branching fractions
$\times 10^{5}$}
\label{TABbsg}
\begin{center}
\begin{tabular}{lcccccccc}
$B$ Decay & Deshpande & Chau & ARGUS & CLEO 1.5 & CLEO II & DELPHI & 
ALEPH & OPAL \\ \hline
$K^-\pi^+$   & 1.1 &  1.7  & $<$18.0 & $<$7.7& $<1.9$ & $<9$ &$<7.5$ & $<8.1$\\
$K^-\rho^+$  &  0  &  0.2  &         &         & $<4.3$ & & & \\
$K^- {a_1}^+$  &   &       &         &         &        & $<39$& &\\
$K^- \pi^+ \pi^-$  &    &  &         &         &        & $<40$& &\\
$K^- \pi^- \pi^+ \pi^+$  &    &  &         &            &    & $<21$& &\\
$K^0\pi^0$   & 0.5 &  0.6  &         &         & $<6.3$ & & &\\
$K^0\rho^0$  & 0.01&  0.04 & $<$16.0 & $<$50.0 &        & & &\\
$K^{*-}\pi^+$& 0.6 &  1.9  & $<$62.0 & $<$38.0 & $<23.8$& & &\\
$K^{*0}\pi^0$& 0.3 &  0.5  &         &         & $<3.5$ & & &\\
$K^-\pi^0$   & 0.6 &  0.8  &         &         & $<3.2$ & & &\\
$K^-\rho^0$  & 0.01&  0.06 & $<$18.0 & $<$8.0  & $<2.6$ & $<19$& &\\
$K^0\pi^-$   & 1.1 &  1.2  & $<$9.6  & $<$10.0 & $<6.8$ & & &\\
$K^0\rho^-$  &  0  &  0.03 &         &         &        & & &\\
$K^{*0}\pi^-$& 0.6 &  0.9  & $<$17.0 & $<$15.0 & $<6.0$ & & &\\
$K^{*-}\pi^0$& 0.3 &  0.9  &         &         &        & & &\\
$K^0\phi$    & 1.1 &  0.9  & $<$36.0 & $<$42.0 & $<10.7$& & &\\
$K^{*0}\phi$ & 3.1 &  0.9  & $<$32.0 & $<$38.0 & $<3.9$ & & &\\
$K^-\phi$    & 1.1 &  1.4  & $<$18.0 & $<$9.0  & $<1.4$ & $<44$& &\\
$K^{*-}\phi$ & 3.1 &  0.8  & $<$130.0&         & $<9.0$ & & & \\
$\phi \phi  $&     &       &         &         & $<4.8$ & & &\\
$K^- K^+ K^-$&     &       &         &         &    & $<31$& &\\
\end{tabular}
\end{center}
\end{table}

New upper limits have been presented for
$\bar{B}^0\to\pi^+ \pi^-$ \cite{PRLkpi} 
and $\bar{B^0}\to\pi^{\pm}\rho^{\mp}$
\cite{CLEOglasrare}.
 The CLEO~II search for $\bar{B}^0\to\pi^+ \pi^-$ is
discussed in detail in the next section.
CLEO~II also has a new limit on $\bar{B^0}\to K^-\pi^+$ \cite{PRLkpi}, and
preliminary results on $\bar{B^0}\to K^-\rho^+$ \cite{CLEOglasrare} 
as well as the
$B\to K^{(*)}\phi$ modes \cite{CLEOglasrare}. The CLEO~II limits on
$\bar{B^0}\to K^-\pi^+$ and $B^-\to K^-\phi$,
which are expected to have a large penguin amplitude, are close
to the theoretical predictions.

\begin{table}[htb]
\begin{center}
\label{bsrare}
\caption{Upper limits on branching fractions for
rare $B_s$ decay modes in unit of $10^{-5}$.}
\vspace{0.5cm}
\begin {tabular}{l c c c}
$\bar{B}_s$ Mode	        &  ALEPH & OPAL & DELPHI\\ \hline
$\bar{B}_s\to \pi^+\pi^-$   &        $<25$         &                 &  \\
$\bar{B}_s\to K^+\pi^-$     &        $<25$         &   $<26$         & $<9$ \\
$\bar{B}_s\to K^+K^-$       &        $<11$         &   $<14$         & $<12$ \\
$\bar{B}_s\to p \bar{p}$    &        $<11$         &                 &  \\
\end{tabular}
\end{center}
\end{table}

The experimental sensitivities to branching ratios
have now reached the $10^{-5}$ range.
Since the theoretical predictions for several $B$ decay modes
are in this range, it is possible that some signals will be observed soon.
By measuring a sufficient number of charmless $B$ decay modes
(e.g. $\bar{B}^0 \to \pi^- \pi^+$, $B^- \to \pi^- \pi^0$, 
$\bar{B}^0 \to \pi^0 \pi^0$) it may be possible to isolate the spectator
and penguin contributions.

\subsection{New Experimental Results on $\bar{B^0}\to \pi^+\pi^-$ and
$\bar{B^0}\to K^- \pi^+$}
\label{newpipi}

The decay modes
$\bar{B^0}\to\pi^+\pi^-$, $\bar{B^0}\to K^-\pi^+$, and $\bar{B^0}\to K^+ K^-$
\cite{bkk}, have been searched for by CLEO~II using a 
data sample of 2.0~fb$^{-1}$ 
taken on the $\Upsilon$(4S)\cite{PRLkpi}. A sample of
0.9~fb$^{-1}$ taken just below the resonance is used to study the continuum
background. Since $B$ mesons are produced nearly at rest on the $\Upsilon$(4S),
the final state has two nearly back-to-back tracks with momenta 
about 2.6~GeV/c. 
Candidates for 
$B$ meson decays are distinguished
from continuum background using the difference, $\Delta E$, 
between the total energy of the two tracks and the beam
energy, and the beam-constrained mass,
$M_B$. The r.m.s. resolutions on 
$\Delta E$ and $M_B$ are 25~MeV and 2.5~MeV respectively.

Separation between $\pi^-\pi^+$, $K^-\pi^+$ 
and $K^-K^+$ events is provided by the
$\Delta E$ variable, and by $dE/dx$ information from the 51-layer main drift
chamber. The $\Delta E$ shift between $K\pi$ and $\pi\pi$ events is 42~MeV
if $E_1$ and $E_2$ are determined using the pion mass. This is
1.7$\sigma_{\Delta E}$. The $dE/dx$ separation between kaons and pions at
2.6~GeV/c is found to be $(1.8\pm 0.1)\sigma$ from a study of a sample of
$D^{*+}$-tagged $D^0\to K^-\pi^+$ decays. Thus, in the CLEO II experiment
the total separation between $K\pi$ and $\pi\pi$ events is $2.5\sigma$.

The background arises entirely from the continuum where the two-jet
structure of the events can produce high momentum, back-to-back tracks.
These events can be discriminated against by calculating the angle, $\theta_T$,
between the thrust axis of the candidate tracks, and the thrust axis of the
rest of the event. The distribution of $\cos\theta_T$ is peaked at $\pm$1 for
continuum events, and is nearly flat for $B\bar{B}$ events. A cut is made at
$|\cos\theta_T|<0.7$. Additional discrimination is provided by a Fisher
discriminant
 \cite{CLNSKpi},\cite{Fisher}, $\cal{F}$ = $\sum_{i=1}^{n}\alpha_i y_i$.
The inputs
$y_i$  are the direction of the candidate thrust axis, the $B$ meson flight
direction, and nine variables measuring the energy flow of the rest of the
event. The coefficients $\alpha_i$ are chosen to maximize the separation
between $B\bar{B}$ signal events and continuum background events.
The optimal cut on the Fischer discriminant is
84\% efficient for signal and 40\% efficient for background.

Two approaches are used to evaluate the amount of signal in the data sample.
In the first approach a cut is made on ${\cal {F}}$ and 
events are classified as
$\pi\pi$, $K\pi$ or $KK$ according to the
most probable hypothesis from the $dE/dx$ information.
The signal and background numbers are given in
Table~\ref{rareKpi}. The efficiency for the correct identification of a signal
event in this analysis is 19\%.
The background is estimated using sidebands in the continuum and on-resonance
data and scaling factors 
from Monte Carlo studies. There is no $B\bar{B}$ background
in the signal region.

%
% KPI/PIPI UPDATE TABLE FROM GLASGOW
%
\begin{table}[htb]
\begin{center}
\label{rareKpi}
\caption{Updated results for 
the branching fractions of $B^0\rightarrow K^+\pi^-$, 
$B^0\rightarrow \pi^+\pi^-$, and $B^0\rightarrow K^+K^-$.  
Upper limits are at the 90\% confidence level.} 
\vspace{0.5cm}
\begin {tabular}{l c c c}
Mode	        & Event Yield           & ${\cal{B}}~(10^{-5})$ &Theoretical 
Predictions $(10^{-5})$ \\ \hline
$\pi^+\pi^-$	& $8.5^{+4.9}_{-4.0}$	& $<2.2$	 & 1.0-2.6 \\
$K^+\pi^-$	& $7.1^{+4.2}_{-3.4}$	& $<1.9$	 & 1.0-2.0 \\
$K^+ K^-$	& $0.0^{+1.6}_{-0.0}$	& $<0.7$	 & $-$	  \\
 & & & \\
$\pi^+\pi^-$ + $K^+\pi^-$ & $15.7^{+5.3}_{-4.5}$ 
                             & $1.8^{+0.6}_{-0.5}\pm0.2$ &  \\
\end{tabular}
\end{center}
\end{table}

\begin{figure}[hbt]

\vspace{-1.0cm}
\centerline{\psfig{figure=contour.ps,height=5in,bbllx=0bp,bblly=0bp,bburx=600bp,bbury=700bp,clip=}}
\vspace{-3.5cm}
\vskip 4mm
\caption{Likelihood contours in the CLEO~II analysis
for the fit to $N_{\pi\pi}$ and $N_{K\pi}$.
The best fit is indicated by the cross, the 1, 2, 3, and 4$\sigma$ contours by solid lines,
and the $1.28\sigma$ contour by the dotted line.}
\label{contour}
\end{figure}

To increase the efficiency of the search and to exploit the information
contained in the distributions of the $\Delta E$, $M_B$, $\cal{F}$ and $dE/dx$
variables a second analysis is performed.
The cuts described above are removed, and
an unbinned maximum-likelihood fit is made. In this fit the signal and
background distributions are defined by probability density functions
derived from Monte Carlo studies. The fit determines the relative contributions
of $\pi^-\pi^+$, $K^-\pi^+$ and $K^-K^+$ to the signal
 and background. The best fit values
for the signal yields $N_{\pi\pi}$, $N_{K\pi}$ and $N_{KK}$, are given
in Table~\ref{rareKpi}. Fig.~\ref{contour} shows the $n\sigma$ contours in the
plane $N_{\pi\pi}\; vs.\; N_{K\pi}$, 
and Fig.~\ref{Kpiproj} shows the projections
of the likelihood fit onto the $M_B$ and $\Delta E$ axes compared to the
events observed. The efficiency for
a signal event to be included in the likelihood analysis is 38\%.

\begin{figure}[hbt]

\vspace{-1.5cm}
\centerline{\psfig{figure=mde.ps,height=5in,bbllx=0bp,bblly=0bp,bburx=600bp,bbury=700bp,clip=}}
\vspace{-2.6cm}
\vskip 4mm
\caption{CLEO~II results on $B\to \pi^+ \pi^-$ and $K^+ \pi^-$.
Comparison of on-resonance data (histogram) with projections of
the likelihood fit (solid curve). (a) Projection onto $M_B$ after cuts on
$\Delta E$ and $\cal{F}$ (b) Projection onto $\Delta E$ after cuts on $M_B$
and $\cal{F}$. The shaded portions of the histogram are $\pi\pi$ events, the
unshaded are $K\pi$ events. The dotted and dot-dashed lines in (b) indicate the
fit projections for $K\pi$ and $\pi\pi$ separately.}
\label{Kpiproj}
\end{figure}

The best fit value shown in Fig. \ref{contour} is more than 4$\sigma$ away
from the point $N_{\pi\pi} = N_{K\pi} = 0$. After including the effect
of systematic errors on the sum of $N_{\pi\pi}$ and $N_{K\pi}$ \cite{CLNSKpi},
it has been concluded that the significance of the sum is sufficient to claim
the observation of a signal for charmless hadronic $B$ decays.
It should be emphasized that
the present data do not have sufficient statistical precision
to allow any conclusion to be reached about the relative importance of the two
decays. While the
 CLEO~II experiment does not measure signals for the
individual decays $\bar{B^0}\to\pi^+\pi^-$ and $\bar{B^0}\to K^-\pi^+$,
 it does set stringent upper limits (Table \ref{rareKpi}).

Studies have been made of the amount of additional data that might be required
to measure signals in the individual modes, and it is estimated that a sample
of about 4 fb$^{-1}$ may be sufficient, 
assuming that the best fit continues to give
the same yields for $N_{\pi\pi}$ and $N_{K\pi}$. Note that the
separation between $\pi\pi$ and $K\pi$ provided by $\Delta E$ and $dE/dx$
is adequate for this analysis, as can be seen from the nearly circular
form of the contours in Fig.~\ref{contour}.

\subsection{Inclusive/Semi-Inclusive $b\to s g$ transitions}

It is major experimental challenge to measure the rate for
the inclusive process $b \to s$ gluon, where
the virtual gluon hadronizes into a $q \bar{q}$ pair. At least
two methods have
been proposed to determine the rate for such inclusive
transitions. Since the coupling of gluons to quark-antiquark
pairs is flavor independent, it is expected that except for modifications
due to phase space $b\to s \bar{s} s$ will be comparable to
$b\to s \bar{u} u$, $b\to s \bar{d} d$. Experimentally, one searches
for inclusive $B\to \phi$ transitions with the $\phi$ momentum in
the range beyond the kinematic limit for $b\to c$ transitions,
or $B\to \phi X_s$ where the 
$X_s$ system contains a kaon and additional pions. 
For example, Deshpande and He find
${\cal B}(B\to X_s\phi) =
(0.6 -2)\times 10^{-4}$\cite{Deshbsg},\cite{ciuchini}.
Alternately,
one may attempt to reconstruct $B\to K^-$ with additional pions.

Several authors have proposed that $b\to s g$ transitions are
enhanced in order to explain the anomalously low value of the
B semileptonic branching fraction. In addition, it is 
conceivable that new physics 
could modify $b\to s g$ without 
significantly modifying the rates for $b\to s \gamma$ or
$b\to s l^+ l^-$.

%%%%%\section{ELECTROMAGNETIC PENGUIN DECAYS}
%\input penguin.tex
\section{ELECTROMAGNETIC PENGUIN DECAYS}

\subsection{Observation of $B\to K^* (892)\gamma$}

The first observation of the electromagnetic decay $B\to K^*\gamma$
has been reported by CLEO~II\cite{PRLbsg}.
A data sample of 1.38 fb$^{-1}$ taken on the $\Upsilon$(4S) resonance
was searched for both
 $\bar{B}^0\to \bar{K}^{*0}\gamma$ and $B^-\to K^{*-}\gamma$,
where the $\bar{K}^{*0}$ was detected 
in its $K^-\pi^+$ decay mode, and the $K^{*-}$
in both the $K^-\pi^0$ and $K_s\pi^-$ decay modes.
If a $K^*$ candidate is within 75~MeV of the known $K^*$ mass then
it is combined with an isolated photon with an energy between 2.1 and 2.9~GeV.
The photon candidate must not be matched to a charged track, and
must have a shower shape consistent with an isolated
photon.
If the photon candidate forms a $\pi^0$($\eta$) meson when combined with
any another photon with energy greater than 30(200)~MeV it is rejected.

Candidates for $B$ meson decays are identified using the variables
$\Delta E = E_{K^*} + E_{\gamma} - E_{beam}$ and $M_B$.
%A slight improvement in the resolution on $M_B$ can be obtained by
%assuming that all the error in $\Delta E$ comes from the measurement of
%$E_{\gamma}$ and scaling the value of $E_{\gamma}$ to the value corresponding
%to $\Delta E = 0$ before calculating
%$M_B = \sqrt{E_{beam}^2 - ({p}_{K^*} + {p}_{\gamma})^2}$. 
The r.m.s.
resolutions on $\Delta E$ and $M_B$ are 40~MeV and 2.8~MeV respectively.
\begin{table} [hbt]

\caption{Summary of results for $B\to K^*\gamma$}
\label{TABksg}

\begin{center}
\begin{tabular}{lccc}
&$\bar{B}^0\to K^{*0}\gamma$&  \multicolumn{2}{c}{$B^-\to K^{*-}\gamma$}\\
&$\bar{K}^{*0}\to K^-\pi^+$&$K^{*-}\to K_s\pi^-$&$K^{*-}\to K^-\pi^0$\\ \hline
Signal Events        &  8  &  2  &  3  \\
Continuum Background & 1.1$\pm$0.2 & 0.05$\pm$0.03 & 0.8$\pm$0.3 \\
$B\bar{B}$ Background& 0.30$\pm$0.15 & 0.01$\pm$0.01 & 0.10$\pm$0.05 \\
Detection Efficiency & (11.9$\pm$1.8)\% & (2.0$\pm$0.3)\% & (3.1$\pm$0.5)\%\\
Branching Ratio & (4.0$\pm$1.7$\pm$0.8)$\times 10^{-5}$&
\multicolumn{2}{c}{(5.7$\pm$3.1$\pm$1.1)$\times 10^{-5}$}\\
\end{tabular}
\end{center}
\end{table}

There are two main sources of background from the continuum, $q\bar{q}$ jets
and initial state radiation (ISR). These backgrounds are suppressed by
applying cuts on the shape variables $R_2<0.5$, $|\cos\theta_T|<0.7$,
and $0.25<s_{\perp}<0.60$. The upper restriction on $s_{\perp}$ is useful
for rejecting ISR background.  
By transforming the event into the frame where the photon is at
rest, and defining new shape variables $R_2^{'}$ and $\cos\theta_T^{'}$ in this frame,
the ISR background can be further suppressed.
There is a small amount of background
to $B\to K^*\gamma$ from other $B\bar{B}$ events. 
The size of this background was determined from a high statistics
Monte Carlo study. This study includes a feeddown
from other $b\to s\gamma$ decays, which was 
estimated using the theoretical models for $b\to s\gamma$ discussed in the
next section. 
The remaining background is mainly due to continuum $e^+e^-$ annihilation. 
This contribution has been determined using $\Delta E, \; M_B$ sidebands in 
both the $\Upsilon(4S)$ and continuum data and
 scaling factors determined from Monte
Carlo studies.

Supporting evidence that the events in the signal region are due to
the decay $B\to K^*\gamma$ comes from a likelihood analysis similar to
the one described in section \ref{newpipi}.
In this analysis the distributions of the events in the variables
$M_B$, $\Delta E$, $M_{K^*}$, $\cos\Theta_{K^*}$ (the $K^*$
helicity angle), $\cos\theta_B$,
$R_2$, $R_2'$, $s_{\perp}$, and $\cos\theta_T$
are compared to the distributions expected from Monte Carlo
samples of signal and continuum background events \cite{CLNSksg}.
This analysis gives results completely consistent with the signal
and background yields given in Table~\ref{TABksg}.

\begin{figure}[hbt]
\vspace{-1.5cm}
\centerline{\psfig{figure=ksg.ps,height=4.5in,bbllx=0bp,bblly=0bp,bburx=600bp,bbury=700bp,clip=}}
\vspace{-2.5cm}
\vskip 4mm
\caption{The beam-constrained mass distribution
from CLEO~II for $B\to K^*\gamma$
candidates: $K^-\pi^+\gamma$ solid, $K^-\pi^0\gamma$ shaded, $K_s\pi^-\gamma$
unshaded}
\label{FIGksg}
\end{figure}

The eight $\bar{K}^{*0}\gamma$ and five $K^{*-}\gamma$
events in the signal region, $|\Delta E|<90$~MeV and $5.274<M_B<5.286$~GeV, are
a clear signal for the decay $B\to K^*\gamma$ (Fig.~\ref{FIGksg}).
The yields in the observed modes are consistent.
Assuming that $\bar{B}^0\to \bar{K}^{*0}\gamma$ and $B^-\to K^{*-}\gamma$ are
equal, the average branching ratio is $(4.5\pm 1.5\pm 0.9)\times 10^{-5}$.
This is in agreement with theoretical predictions from the
electromagnetic penguin diagram \cite{Ali}.

\subsection{Search for exclusive $b\to d \gamma$ transitions}
% TEB update
%
CLEO has also searched for exclusive $b\to d \gamma$ decay modes
including $\bar{B}^0\to \rho^0 \gamma$, $B^- \to \rho^- \gamma$ and
$B\to \omega \gamma$\cite{cleobrhog}. 
In these modes, the largest background arises from continuum production with
significant contributions 
from the $b\to s \gamma$ process $B\to K^* \gamma$, $K^*\to K^- \pi^+$
with the charged kaon misidentified as a pion. 
Some discrimination between $B\to K^{*0} \gamma$ and $B\to \rho^0 \gamma$ 
is provided by $\Delta E$, the vector decay angle, and by
the constraint that $\pi \pi $
mass lie in the $\rho $ mass region. This information is used 
by a neural network to 
reduce the background from $B\to K^* \gamma$ by a factor of 20 while
retaining 50\% of the $B\to \rho \gamma$ signal.
No signals are observed and upper limits at the 90\% C.L. of
${\cal B}(B^-\to \rho^- \gamma) < 2.0\times 10^{-5}$, 
${\cal B}(\bar{B}^0\to \rho^0 \gamma) < 2.4\times 10^{-5}$, and
${\cal B}(\bar{B}^0\to \omega \gamma) < 1.1\times 10^{-5}$ are 
obtained.

\subsection{Experimental Constraints on the $ b \to s \gamma$ Inclusive Rate}

At present, 
due to the uncertainties in the hadronization, 
only the inclusive $b \to s \gamma$
rate can be reliably compared with theoretical calculations.
This rate can be measured from the endpoint of the
inclusive photon spectrum in $B$ decay.
The signal for $b\to s\gamma$ is expected to peak in the region
$2.2<E_{\gamma}<2.7$~GeV, with only about 15\% of the rate expected to lie
outside this range \cite{Ali}.

\begin{figure}[hbt]
\vspace{+0.5cm}
\centerline{\psfig{figure=brec_bsg.ps,height=4.5in,bbllx=0bp,bblly=0bp,bburx=600bp,bbury=700bp,clip=}}
\vspace{-4.5cm}
\vskip 4mm
\caption{(a) The on-resonance data is shown as the solid
histogram, the scaled off resonance data is the dashed histogram,
and the sum of off-resonance data and background from $b\to c$ and $b\to u$
decays are the squares with error bars. 
(b) The photon energy distribution for $b\to s \gamma$
from CLEO~II for the B reconstruction analysis after subtraction
of all backgrounds.}
\label{FIGbsg1}
\end{figure}

\begin{figure}[hbt]
\vspace{+0.5cm}
\centerline{\psfig{figure=net_bsg.ps,height=4.5in,bbllx=0bp,bblly=0bp,bburx=600bp,bbury=700bp,clip=}}
\vspace{-4.5cm}
\vskip 4mm
\caption{
(a) The on-resonance data is shown as the solid
histogram, the scaled off resonance data is the dashed histogram,
and the sum of off-resonance data and background from $b\to c$ and $b\to u$
decays are the squares with error bars. 
(b) The  photon energy distribution for $b\to s \gamma$
from CLEO~II for the event shape analysis 
after subtraction of all backgrounds.
The points with error bars are the background subtracted data while
the solid curve is the Monte Carlo prediction for the shape
of the $b\to s \gamma$ signal.}
\label{FIGbsg2}
\end{figure}

Two experimental methods are employed to suppress the large background
from non-resonant $e^+ e^-\to q \bar{q}$ and initial state radiation
i.e. $e^+ e^- \to q \bar{q} \gamma$. 

One method uses partial reconstruction
of the decay products of the kaonic resonance which recoils against
the energetic photon. This is referred to as the B reconstruction
analysis. 
A kaon candidate (either charged or neutral),
up to four charged tracks, and two or fewer $\pi^0$s are
combined with the photon candidate and are required to be consistent
with the B mass. The resulting photon spectrum after event shape cuts
is examined.
This technique does not require that the final state
kaonic resonance be correctly reconstructed and is primarily 
designed to suppress the continuum background.

A complementary method using a neural network is also used to 
distinguish $b\to s\gamma$ signal and background. 
This is referred to as the event shape
analysis. The network uses
the event shape variables $R_2$, $S_{\perp}$, $R_{2}^{'}$, 
$\cos\theta_{T}^{'}$ as well as the energy deposited in cones
within 20$^{\circ}$ and 30$^{\circ}$ of the photon direction
and in similiar cones in the direction opposite to the photon.
The output of the neural network is a value between $-1$ and $1$
which measures the degree to which an event resembles signal.
The network is trained using a large sample of continuum Monte Carlo
events. There is good agreement between the network output 
from Monte Carlo simulations  and various data samples 
(e.g. continuum data, $B\to X\mu \nu$
candidates).

The experimental photon energy spectra from the two methods 
are shown in Figs.~\ref{FIGbsg1},~\ref{FIGbsg2}.
In the CLEO~II data there is an excess of 
$2.2<E_{\gamma}<2.7$~GeV in the B reconstruction analysis and an excess
of $263\pm 104$ events from $B$ decays in this region
for the event shape analysis\cite{jaethesis}.
The detection efficiencies are 9\% and 32\% respectively. However,
the signal to noise ratio in the B reconstruction analysis is a factor
of 4 higher so the sensitivities are comparable.
These correspond to branching ratios of 
$(1.88\pm 0.74) \times 10^{-4}$ for the B reconstruction analysis and
$(2.75\pm 0.67) \times 10^{-4}$ for the event shape analysis. 
The errors quoted are statistical only. The two results
are consistent at the 1.1 standard deviation level. 
The model dependence introduced when extrapolating 
from the partial branching fraction in the signal
window to the branching fraction
for the entire range of photon energies is evaluated using a
parton model calculation. The largest uncertainty arises
from the error in $m_{b}$, the b quark mass. This parameter
is allowed to vary in the range $m_{b}= 4.87 \pm 0.10$. 
The resulting
($10$\%) change in the branching ratio is 
incorporated in the systematic error.
The results of the two analyses can be combined allowing for 
statistical correlations and separating the correlated
and independent components of the systematic error to give,
$${\cal B}(b\to s \gamma) = (2.32 \pm 0.57 \pm 0.35) \times 10^{-4}$$
where the first error is statistical and the second is systematic.

For the purposes of constraining extensions of the Standard Model, it is
useful to derive upper and lower limits at the 95\% confidence level from
this measurement. This gives:
$$ 1.0 \times 10^{-4} < {\cal B} (b\to s \gamma) < 4.2 \times 10^{-4}
~({\rm at~the~95\% ~c.l.})$$

An alternative approach to measuring the inclusive rate is to use the
observed exclusive rate for $B\to K^*\gamma$.
However, the fraction of the inclusive rate that hadronizes to a particular
exclusive final state is not very well understood. Ali \etal \cite{Ali}
predict the mass distribution of the $X_s$ system using an estimate of the
Fermi momentum of the 
spectator quark ($p_{F}=300$ MeV).
By integrating this spectrum up to
1~GeV and assuming this region is dominated by the $K^*$ resonance,
the fraction of $K^*(892)\gamma$ is estimated to be (13$\pm$3)\%.
Other authors have made predictions between 5\% and 40\%
for the fraction of $K^*(892)\gamma$ \cite{ksg}. A reasonable estimate that
covers most of the theoretical predictions is (13$\pm$6)\%.
%Combining this number with the measured branching ratio
%for $B\to K^*\gamma$ a lower limit can be obtained for the inclusive rate.
%This approach would become more useful if additional exclusive channels with
%higher mass $X_s$ systems were to be observed. 
Examination of the observed mass of the ($X_s$ system) particles which
accompany the high energy photon in $b\to s \gamma$ indicates that 
there are states other than $B\to K^* \gamma$ which contribute.
Note that the apparent $X_s$ mass spectrum shown in Fig.~\ref{FIGbsg3}
is not corrected for efficiency, which decreases rapidly as a function of
mass nor for misreconstruction of high multiplicity channels.

\begin{figure}[hbt]
\vspace{+0.5cm}
\centerline{\psfig{figure=mxs_bsg.ps,height=4.5in,bbllx=0bp,bblly=0bp,bburx=600bp,bbury=700bp,clip=}}
\vspace{-4.5cm}
\vskip 4mm
\caption{
Apparent $X_s$ invariant mass spectrum for $b\to s \gamma$ candidates
after background subtraction.}
\label{FIGbsg3}
\end{figure}

%Using the inclusive measurement for the upper limit, and the observation
%of $B\to K^*\gamma$ for the lower limit, the present 95\% C.L. limits on the
%inclusive rate are \cite{PCK}:
%\begin{equation}
%0.8\times 10^{-4} < {\cal{B}}(b\to s\gamma) < 5.4\times 10^{-4}
%\end{equation}

Searches have also been made for $b\to s\gamma$ processes
at LEP. The L3 experiment has set an upper limit of $1.2\times 10^{-3}$
(90\% C.L.) on the inclusive $b\to s\gamma$ rate \cite{L3shit}.
The exclusive decays $\bar{B}^0\to \bar{K}^{*0}\gamma$ and $B_s\to \phi\gamma$
have been searched for by the DELPHI experiment using
the particle identification capabilities of the RICH detector. Upper limits
of $3.6\times 10^{-4}$ and $19.0\times 10^{-4}$ are obtained for
these two decays \cite{Battaglia}. ALEPH has searched
for $B_s\to \phi\gamma$ and $\Lambda_b\to \Lambda\gamma$ and
obtains limits of $29\times 10^{-5}$ and $56\times 10^{-5}$
respectively.

\subsection{Theoretical Implications of $b \to s \gamma$}

There has been recent interest in $b\to s\gamma$ as a probe of physics beyond
the standard model\cite{Joanne}, 
\cite{Hewett,SUSY}. There are possible additional
contributions to the loop from a charged Higgs boson and from supersymmetric
particles. Hewett \cite{Hewett} has considered two Higgs doublet models and
shown that contributions comparable to the standard model are expected for a
charged Higgs mass of order 100~GeV. In supersymmetric models there
are also contributions from loops containing charginos, neutralinos and
squarks that tend to cancel the
charged Higgs and standard model contributions (in unbroken supersymmetry all
contributions to the loop diagram cancel exactly)\cite{cancel}. 
Several recent papers
\cite{SUSY} investigate the parameter space allowed by $b\to s\gamma$
for particular models of the breaking of the supersymmetry. For most of the
parameter space the charged Higgs contribution is the dominant one, and
the present CLEO~II upper limit on $b\to s\gamma$ constrains the charged Higgs
mass to be greater than 244~GeV. 
This is more restrictive than constraints from direct searches at existing
high energy colliders.
The limit on the charged Higgs mass can be avoided in some 
supersymmetric models
if the stop mass is small
since this leads to a large
negative contribution from the chargino-stop loop. For this case the rate for
$b\to s\gamma$ could even become smaller than the standard model prediction.

Other constraints on new physics have been  derived from the bounds
on $b \to s \gamma$.
If there are anomalous $W- W- \gamma$ couplings, these
can significantly modify the rate for $b\to s\gamma$. 
The CLEO measurements exclude
 certain regions of the parameter space of
anomalous dipole and quadrupole couplings of the W boson that cannot
be explored by direct studies of $W^+ -\gamma$ production at hadron colliders
\cite{Chiawwg}. It has also been suggested that these results
constrain most supersymmetric dark matter candidates to such
an extent that they will not produce significant counting rates 
in dedicated
dark matter WIMP searches planned in the near future\cite{wimps}.

% begin additional rarer sections

\subsection{$b\to s \ell ^+\ell ^-$ Decays}

The $b\to s\gamma$ diagram can be modified by replacing the real photon by a
virtual photon or by a virtual $Z^0$ or  other neutral boson
that produces a lepton pair (see Fig.~\ref{kstgll}). 
 This penguin diagram leads to both
$B\to K \ell ^+ \ell ^-$ and $B\to K^* \ell ^+\ell ^-$ decays, since the $B\to K$ transition is
no longer forbidden by angular momentum conservation as it was for
$b\to s\gamma$. Although 
the penguin amplitude for $b\to s \ell ^+\ell ^-$ is smaller
than $b\to s\gamma$ the final states can be identified easily,
and are particularly favorable for study at hadron colliders.
As in the radiative penguin decay discussed previously, the 
process $b\to s \ell ^+ \ell ^-$
is sensitive to high mass physics including charged Higgs bosons
and non-standard neutral particles. 
These modes also do not have significant 
QCD corrections which may be a useful feature when constraining 
new physics.
Ali, Mannel, and 
Guidice have noted that the constraints imposed by the combinations of
measurements of $b\to s\gamma$ and $b\to s\ell^+\ell^-$ can
severely constrain new physics including SUSY models\cite{Alickm}.

\begin{figure}[htb]
\begin{center}
\unitlength 1.0in
%\vspace{-0.5cm}
\begin{picture}(3.,0.8)(0,0)
\put(-1.1,-3.2){\psfig{bbllx=0pt,bblly=0pt,bburx=567pt,bbury=567pt,width=4.0in,height=3.5in,file=kstgll.ps}}
\end{picture}
\vskip 15 mm
\caption{Diagrams for the decays $B\to K^{(*)} \ell ^+ \ell ^-$.}
\label{kstgll}
\end{center}
\end{figure}

The penguin amplitude has been calculated by a number of authors
\cite{PengKll}, with results for the inclusive $b\to s e^+ e^-$ rate
of $ (1-2)\times 10^{-5}$ and for the $b\to s \mu^+ \mu^-$ rate of
$ (4-8) \times 10^{-6}$.
 The exclusive channels $K^* \ell ^+\ell ^-$
and $K \ell ^+\ell ^-$ are expected to comprise $5-30\%$
of the inclusive rate.  However, the theoretical
description of $b\to s\ell ^+\ell ^-$ is more complicated
than $b\to s\gamma$, since the 
final states $K^{(*)}\ell ^+\ell ^-$ can also be produced
via ``long distance'' contributions from the hadronic
decay $B\to K^{(*)}\psi$ followed by $\psi\to \ell ^+\ell ^-$ where
$\psi$ stands for a real or virtual charmonium state \cite{LongD}. 
Ali, Mannel and Morozumi \cite{AliKll} have performed
 an analysis of $b\to s \ell^+\ell^-$
including both the penguin and the long distance contributions.
Their predictions for the inclusive $b\to s \ell^+ \ell^-$ rate are in the
range $(2-6)\times 10^{-6}$ excluding the regions
close to the $\psi$ and $\psi'$ mass where the long distance contributions
dominate. There is interference between the penguin and long distance
amplitudes over a wide range of dilepton masses. Ali \etal\  point out that the
sign of the interference is controversial, and that information about the
interference can be obtained both from the dilepton mass distribution, and
from the forward-backward asymmetry of the lepton pair.

The $B\to K^* e^+ e^-$ mode has significant contributions from
both virtual $Z^0$s and virtual photons. At low $m^2$, the
virtual photon contribution is dominant and
has a pole. This must be properly
taken into account when computing experimental efficiency,
since there is usually a cut on $m_{e^+ e^-}^2$ to eliminate
conversions\cite{CakirKll}. 

Experimental searches have
been made by CLEO~1.5, CLEO~II
 and ARGUS at the $\Upsilon (4S)$, and by UA1 and CDF in $p\bar{p}$
collisions. The CLEO and ARGUS analyses 
\cite{CLEOKll,CLEOIIKll,ARGUSKll} make a simple
veto on dilepton masses consistent with a real $\psi$ or $\psi'$, and see
almost no background in their beam-constrained mass plots. 
CDF has searched for $B^+\to K^{+} \mu^+\mu^-$ and $B^0\to 
K^{*0} \mu^+ \mu^-$ using the mode
$B\to \psi K^+$ for normalization\cite{CDFKll}. 
The CDF analysis requires that the dilepton mass lie
in the range $3.3-3.6$ GeV or in the range $3.8-4.5$ GeV. This avoids
contamination from modes with 
$\psi$ and $\psi^{'}$ mesons and reduces the combinatorial background.
The UA1 analysis\cite{UA1Kll} selects 
a range $3.9 < M(\ell ^+\ell ^-) < 4.4$~GeV which is believed
to have small long distance contributions and no radiative tail from the
$\psi$. 
 UA1 performs both an exclusive
search for $\bar{B}^0\to \bar{K}^{*0}\mu^+\mu^-$ and an inclusive search for
$B\to X_s\mu^+\mu^-$. The upper limits derived from the hadron collider
searches using a small fraction of the allowed dilepton mass range e.g.
for example the CDF search considers about 25\%. The limits on the
partial branching fraction
are extrapolated to the full dilepton mass range 
using a theoretical model.
The upper limits from all the experimental measurements
are summarized in Table~\ref{Kll}. These upper limits are all well above
the theoretical predictions
The CLEO~II limit on $\bar{B^0}\to K^{*0} e^+ e^-$ 
is within a factor of 3 of the branching ratio predicted by the
Standard Model.
These limits suggest that $b\to s \ell ^+\ell ^-$ decays
will eventually be observed at either hadron colliders, or by
$\Upsilon (4S)$ experiments.

\begin{table} [hbt]
% TEB Update
\caption{Experimental upper limits (90\% C.L.)
for $b\to s\ell ^+\ell ^-$ decays. All numbers quoted are branching fractions
$\times 10^{-5}$}
\label{Kll}

\begin{center}
\begin{tabular}{lcccccc}
$B$ Decay       &   ARGUS  &  CLEO I  & CLEO 1.5& CLEO~II & UA1 & CDF\\ \hline
$K^0e^+e^-$     &  $<$15.0 & $<$56.0  &         & &       & \\
$K^- e^+e^-$     &  $<$9.0  & $<$24.0  & $<$5.7 & $<1.2$ & &       \\
$K^0\mu^+\mu^-$ &  $<$26.0 & $<$39.0  &         & &        &\\
$K^- \mu^+\mu^-$ &  $<$22.0 & $<$36.0  & $<$17.0& $<0.9$ &  & $<3.5$ \\
$\bar{K}^{*0}e^+e^-$  &  $<$29.0 &          & $<$6.9& $<1.6$  & &       \\
$K^{*-}e^+e^-$  &  $<$63.0 &          &          &   &   &   \\
$\bar{K}^{*0}\mu^+\mu^-$&$<$34.0 &    & $<$16.0  & $<3.1$&$<$2.3&$<5.1$ \\
$K^{*-}\mu^+\mu^-$&$<$110.0&          &          &       & \\
$X_s\mu^+\mu^-$ &          &          &          &       &$<$5.0 \\
\end{tabular}
\end{center}
\end{table}

%%%%\section{PURELY LEPTONIC B DECAY}
%\input leptonic.tex
\section{PURELY LEPTONIC B DECAY}

\subsection{$B$ Decays to Two Leptons}

The Standard Model allows $B^0$ and $B_s$ mesons to decay to $e^+e^-$
$\mu^+\mu^-$ or $\tau^+\tau^-$ via box diagrams or loop diagrams involving
both $W$ and $Z$ propagators (see Fig.~\ref{dilepfig})
 \cite{RareAli}. The largest branching fraction
is predicted to be $4\times 10^{-7}$ for $B_s\to\tau^+\tau^-$, and the
smallest $2\times 10^{-15}$ for $B^0\to e^+e^-$. The decays to the lighter
leptons are suppressed by a helicity factor which is proportional to
$m_{\ell} ^2$, and the $B^0$ decays are suppressed
relative to the $B_s$ decays by the factor $|V_{td}/V_{ts}|^2$.
Decays to the final states $e^{\pm}\mu^{\mp}$, $e^{\pm}\tau^{\mp}$ and
$\mu^{\pm}\tau^{\mp}$ are all forbidden in the Standard Model by lepton
family number conservation.

A search for $B^0$ decays to two leptons has been made by CLEO~II
\cite{CLEOll}, and there are also searches for
$B^0\to\mu^+\mu^-$ by the UA1 and CDF collaborations at hadron colliders
\cite{UA1Kll,CDFll}. The 90\% C.L. upper limits on the allowed processes are
$5.9\times 10^{-6}$ for $B^0\to e^+e^-$ (CLEO~II), and
$3.2\times 10^{-6}$ (CDF), $5.9\times 10^{-6}$ (CLEO~II) and
$8.3\times 10^{-6}$ (UA1) for $B^0\to\mu^+\mu^-$. The hadron collider
experiments will set similar limits on $B_s\to\mu^+\mu^-$, and
presumably have not done so because the $B_s$ mass
was unknown until recently (see section \ref{Bs-mass} ) .

\begin{figure}[htb]
\begin{center}
\unitlength 1.0in
%\vspace{-1.5cm}
\begin{picture}(3.,0.5)(0,0)
\put(-1.1,-3.0){\psfig{bbllx=0pt,bblly=0pt,bburx=567pt,bbury=567pt,width=5.0in,height=4.0in,file=dilepfig.ps}}
\end{picture}
\bigskip
\vskip 5 mm
\caption{Diagrams for the dilepton decays
 of $B$ mesons.}
\label{dilepfig}
\end{center}
\end{figure}

CLEO~II also sets limits on the lepton-flavor changing decays of
$5.9\times 10^{-6}$ for $B^0\to e^{\pm}\mu^{\mp}$,
$7.9\times 10^{-4}$ for $B^0\to e^{\pm}\tau^{\mp}$  and
$1.2\times 10^{-3}$ for $B^0\to\mu^{\pm}\tau^{\mp}$.
Upper limits on the
lepton flavor violating decays $B^- \to K^- e^{\pm} \mu^{\mp}$
and $\bar{B^0}\to \bar{K}^{*0} e^{\pm} \mu^{\mp}$ of
$< 1.2 \times 10^{-5}$ and $2.7\times 10^{-5}$ have also been set.

Several recent papers consider the relative sensitivity of
various lepton-flavor changing decays to non-Standard Model couplings 
\cite{SherYuan},\cite{Campbell} .
Sher and Yuan 
argue that larger Yukawa couplings 
are expected for third generation quarks,
and that these larger couplings not only enhance the sensitivity of the decays,
but also make them less dependent on the detailed parameterization of the new
couplings\cite{SherYuan}. 
They make a comparison of $B$ and $K$ decays which suggests that
$B_s\to\tau\mu$ has the best sensitivity, although it is unclear how to
search for this channel experimentally. The more accessible channel
$B_s\to\mu e$ could also have better sensitivity than the
equivalent decay $K_{L}\to\mu e$, even 
though the upper limit on the latter is now
in the $10^{-11}$ range.
$B^0$ decays are less sensitive than $B_s$ decays but are still of
interest because they can be 
searched for in experiments at the $\Upsilon (4S)$.

\subsection{The Decays $B\to\tau\nu$, $B\to\mu\nu$ and $B\to e\nu$.}

The decay $B^+\to\tau^+\nu$ proceeds through the annihilation of the
constituent quarks in analogy to the $\pi^+\to\mu^+\nu$ decay.
The branching fraction is given by:
$$ {\cal{B}}(B^+\to\tau^+\nu) = \frac{G_F^2m_Bm_\tau^2}{8\pi}
\left(1-\frac{m_\tau^2}{m_B^2}\right)
f_B^2|V_{ub}|^2\tau_B $$
All the parameters in this equation are well known except the
decay constant $f_B$ and the CKM matrix element $V_{ub}$.
Given a more accurate knowledge of $V_{ub}$ from other measurements
and the experimental observation of the decay $B^+\to\tau^+\nu$, it
would be possible to determine a value for $f_B$. The measurement of
this decay constant is of fundamental importance for $B$ physics
since it enters into many other $B$ decay measurements, including
most notably $B\bar{B}$ mixing \cite{mixing}.

The present theoretical estimates of $f_B$ from lattice QCD and QCD sum rules
are in the range $f_B = (180\pm 50)$~MeV \cite{fB}.
Using this value of $f_B$ and our standard values of $V_{ub}$ and $\tau_B$,
we obtain a prediction of ${\cal{B}}(B^+\to\tau^+\nu) = 4.0\times 10^{-5}$.
The decays $B^+\to\mu^+\nu$ and $B^+\to e^+\nu$ have smaller branching ratios
of $1.4\times 10^{-7}$ and $3.3\times 10^{-12}$ respectively. The decays to the
muon and electron are suppressed  relative to the tau decay by a helicity
factor proportional to the square of the lepton mass. The radiative
decays $B^+\to \mu \nu \gamma$ and $B^+ \to e \nu\gamma$
are less suppressed and occur at rates comparable to their
purely leptonic counterparts\cite{burdgw}.

% TEB update
%
CLEO~II has searched for
$B^+\to\tau^+\nu$ followed by  
$\tau \to l \nu \bar{\nu}$. In this case, 
the observed showers and tracks, apart from the lepton,
must originate from the other B meson. No additional leptons
are allowed. Constraints on the 
missing energy and momentum are used to isolate the signal.
No significant excess is found.
This leads to a 90\% C.L. upper limit of
${\cal{B}}(B^+\to\tau^+\nu) < 2.2\times 10^{-3}$ \cite{cleobtaunu}. 
Using the same
 technique as in their analysis of the mode
 $B\to \tau\nu X$ (see section \ref{taunew}),
but requiring additional missing energy,
ALEPH finds 
${\cal{B}}(B^+\to\tau^+\nu) < 1.8 \times 10^{-3}$ \cite{ALEPHxtnu1}. 

CLEO~II has also searched for $B^+\to\mu^+\nu$ as well as $B^+\to e^+\nu$.
 The $B$ meson decays almost at rest into a $\mu^+$ (or $e^+$)
and a neutrino which are back-to-back and have energies of about 2.65~GeV.
The muon is well identified and has little background. The neutrino is
``detected'' by calculating the missing momentum $p_{miss}$ of the whole event.
If all the decay products of the other $B^-$ have been measured by the CLEO~II
detector $p_{miss}$ will be a good estimator of the neutrino momentum.
Then the analysis proceeds as if this were a fully reconstructed $B$ decay,
with the calculation of the energy difference, $\Delta E$, and the
beam-constrained mass, $M_B$. 
The analysis is almost
background free, and gives a 90\% C.L. upper limit of
${\cal{B}}(B^+\to\mu^+\nu) < 2.1\times 10^{-5}$. The sensitivity
in the electron mode is comparable,
 ${\cal B}(B^+\to\ e^+\nu) < 1.5 \times 10^{-5}$\cite{cleobtaunu}.

The limits on $B^+\to\tau^+\nu$ and $B^+\to\mu^+\nu$ are both
two orders of magnitude above the theoretical predictions, 
corresponding to the  rather
uninteresting limit on $f_B$ 
of about 2.6 GeV for $|V_{ub}/V_{cb}|=0.073$.

%%%%%\section{CONSTRAINTS ON THE CKM MATRIX} 
%\input ckm.tex
\section{CONSTRAINTS ON THE CKM MATRIX} 

\subsection{Introduction}
One of the primary goals of the B physics program is the
determination of the values of V, the CKM 
(Cabibbo-Kobayashi-Maskawa) couplings. The experimental
results were discussed in previous sections. We now summarize their
implications for the CKM matrix.

The usual form of
the CKM  matrix is given below.
 \begin{equation}
  V =\pmatrix{V_{ud}&V_{us}&V_{ub}\cr
                V_{cd}&V_{cs}&V_{cb}\cr
                V_{td}&V_{ts}&V_{tb}\cr} \label{ckmmat}
  \end{equation}
The matrix $V$ can be expressed approximately as
   \begin{equation}
   V \simeq \pmatrix{1-\lambda^2/2&\lambda&A\lambda^3(\rho-i\eta)\cr
                      -\lambda&1-\lambda^2/2&A\lambda^2\cr
                      A\lambda^3(1-\rho-i\eta)&-A\lambda^2&1\cr}+
                      O(\lambda^4) \label{wolfpar}
   \end{equation}
This empirical parameterization, suggested by Wolfenstein, 
is correct to terms of order $\lambda^4$ with
$\lambda = \sin{\theta_{Cabibbo}}\approx 0.22$\cite{Wolfie}.
As noted by Buras, Lautenbacher, and Ostermaier, in the
future as the precision
of measurements improves, the above approximation will
have to be extended to be correct up to order $\lambda^6$\cite{BLO}.
This can be accomplished by adding the correction
   \begin{equation}
   \Delta V \simeq \pmatrix{-\lambda^4/8& 0     & 0\cr
                      A^2\lambda^5({1\over 2}-\rho-i\eta)&
                      -({A^2\over 2}+1/8)\lambda^4&0\cr
                      {A^2\over 2}\lambda^5(\rho+i\eta)&
                       A\lambda^4({1\over 2}-\rho-i\eta)&
                      -A^2/\lambda^6\cr}+
                      O(\lambda^6) \label{wolfcorr}
   \end{equation}

\begin{figure}[htb]
\begin{center}
\unitlength 1.0in
\begin{picture}(3.,1.5)(0,0)
\put(-0.01,0.0)
{\psfig{width=3.3in,height=1.5in,file=bjorken.eps}}
\end{picture}
%\vskip 4 mm
\caption{Representation of the Bjorken or Unitarity triangle in the
complex plane.}
\label{triangle}
\end{center}
\end{figure}

\subsection{The CKM element $|V_{cb}|$}
The value of $V_{cb}$, the fundamental weak interaction coupling constant,
may be determined from the semileptonic width:
$$ \Gamma(b\to c l \nu) = \gamma_c^2 |V_{cb}|^2 $$
where $\gamma_c$ is a constant determined from theory e.g. quark model
calculations. 
The semileptonic width is obtained from measurements
of the semileptonic branching fraction and the 
appropriate average of the charged
and neutral B meson lifetimes, ($< \tau_{B} >$)
$$ \Gamma(b\to c l \nu) ={{{\cal B}(b\to c l \nu)}\over {<\tau_{B}>}}.$$
The width has a $m_b^5$ dependence on the 
b quark mass although Shifman \etal~argue 
that in a certain limit, the dependence is much weaker
and is proportional 
to $m_b-m_c$\cite{shifman2},~\cite{shifman}. Their calculation and
the world average for the semileptonic branching fraction give,
$$ |V_{cb}|_{{\rm Inclusive}} 
=0.03965\pm 0.001({\rm exp})\pm 0.002({\rm theo})$$
where the first error is experimental and the second is the
quoted theoretical uncertainty. 
The theoretical uncertainty in the determination
of $|V_{cb}|$ from inclusive semileptonic decays
is currently a matter of active discussion
and no clear consensus has emerged\cite{neubert2}.

Two other methods are used to determine the value of $|V_{cb}|$.
These are measurements of the total widths of exclusive final
states from branching fractions and measurement of the $B\to D^*\ell\nu$
rate at zero recoil. 
The former method of obtaining $|V_{cb}|$ from the total rate 
has the distinct advantage that the models used
make other detailed predictions for form factors 
and various other observables
which can be experimentally verified. In addition, all of the data
can be used unlike the HQET inspired method which
is valid only near zero recoil. Using the world average of the
branching fraction for $\bar{B} \to D^* \ell \nu$ and the ISGW'
model to obtain the central values, we find
$$ |V_{cb}|_{{\rm Exclusive}} 
=0.0347\pm 0.0016({\rm exp})\pm 0.0024({\rm theo})$$

For the HQET method, which requires experimental measurements
of the differential spectrum of $\bar{B}\to D^* \ell\nu$ decays,
there are two significant uncertainties in the final determination
of $|V_{cb}|$ from measurements of the spectrum at zero recoil. 
These arise from the model dependence in the calculation
of the $1/m_c^2$ corrections to $\xi(1)$ and the lack of knowledge of
the functional form of the function $\xi(y)$ which is used for the
extrapolation. Using the value of $\xi(1)$  
recently calculated by Neubert\cite{neubert3} 
and the world average for the experimental intercept gives 
\begin{equation}
|V_{cb}|_{{\rm HQET}}= 0.0386\pm 0.0024({\rm exp})\pm 0.0012({\rm theory}) 
\end{equation}
where the first error is experimental and the second is 
the theoretical uncertainty in $\xi(1)$.
The model dependence from the theoretical uncertainty in the
normalization is about 4\% but may be reduced in the near future.

A precise determination of $|V_{cb}|$ constrains the product $A\lambda^2$,
following the notation of Wolfenstein. Since $\lambda$ is well
determined from measurements of kaon decays, the parameter
 $A$ is determined from $|V_{cb}|$.
Using $\lambda=0.2205\pm 0.0018$, the world average computed by
the PDG group, and the value of $|V_{cb}|$ obtained using HQET, gives
\begin{equation}
   A= 0.794 \pm 0.049 \pm 0.025 \label{Apar}
\end{equation}
A very precise value of $|V_{cb}|$ 
is desirable in order to  check the unitarity of
the CKM matrix as well as for a number of phenomenological applications.
For example, one of the largest uncertainties in the determination of
the location of the vertex in the $\rho,\eta$ plane using
$\epsilon$, the CP violation parameter in kaon decay is the parameter A.
In other words, in order to interpret CP violation in the kaon sector and
predict the magnitude of CP asymmetries for B mesons,
a precise measurement of $|V_{cb}|$ is required.
In addition, as emphasized
by Buras\cite{Burasckm}, rates for certain rare kaon decays such as 
$K_{L}\to \pi^0 \ell^+ \ell^-$, $K_{L}\to \pi^0 
\nu \bar{\nu}$ depend on $|V_{cb}|^4$. It will be worthwhile to
test the Standard Model  and verify that the value of $|V_{cb}|$ from such
loop induced decays in the kaon sector
agree with the value of $|V_{cb}|$ from tree
level semileptonic $B$ decays.

\subsection{The CKM element $|V_{ub}|$}

In the Wolfenstein parameterization,
the value of $|V_{ub}/V_{cb}|$ is 
approximately $ \lambda ~|\rho - i \eta|$. 
Thus the measured value of this ratio
constrains the vertex of the Bjorken triangle (see Fig.~\ref{triangle})
 to lie on 
a circle of radius $\lambda ~\sqrt{\rho^2 + \eta^2}$
in the $\rho-\eta$ plane. 

The value of this ratio is determined
from measurements of inclusive $b\to u\ell\nu$ decays.
Using the central value from the ACCMM model gives
\begin{equation}
 |{{V_{ub}}\over {V_{cb}}}|= 0.073\pm 0.011({\rm exp})\pm 0.01({\rm model})
\end{equation}
where the first error is the sum in quadrature of the experimental
statistical and systematic errors and the second error is due to
model dependence.
Quantifying model dependence is difficult. An alternate way is 
to give the allowed
 range, $$ 0.055< |V_{ub}/V_{cb}| < 0.095,$$
which corresponds to a one standard deviation 
variation on each of the models considered.
The measurement of $|V_{ub}/V_{cb}|$ gives the constraint
$ ~\sqrt{\rho^2 + \eta^2} = 0.331\pm 0.067.$
At present, a large
uncertainty in the radius of this circular region 
in the $\rho, \eta$
plane is due to the model dependence in the extraction of $|V_{ub}|$. 
This may be improved with additional theoretical work 
as well as the observation
of exclusive charmless semileptonic decays.

\subsection{The CKM element $|V_{td}|$}

In the Standard Model, 
$\Delta M_d$ for $B_d-\bar{B_d}$ mixing is 
$$ \Delta M_d = {{G_F^2} \over {6 \pi^2}} m_B m_t^2
~F({{m_t^2} \over {m_W^2}}) ~\eta_{QCD} B_{B_d} f_{B_d}^2
 |V_{tb}^{*} V_{td}|^2$$
where $G_F$ is the weak coupling constant, $m_t$ is the top quark mass,
$F$ is a slowly decreasing function
which depends on $m_{t}$ and $m_{W}$,
$\eta_{QCD}$ is a factor which accounts for QCD corrections, 
$B_{B_d}$ is a constant which is used to account for the vacuum insertion
approximation, and $f_B$ is the decay constant of the $B_d$ meson.

Since the mass of the top quark has been determined ($m_t=179\pm 10$), 
and the QCD correction has recently
been calculated to NLO by Buras, Jamin, and Weisz\cite{Burasmix}
 ($\eta_{QCD}=0.55$), the largest remaining
uncertainties in $\Delta M_d$ arise from the product $B_{B_d}^{1/2} f_{B_d}$.
This last factor must be determined from non-perturbative methods
such as lattice QCD, QCD sum rules, or potential models. One
estimate is $B_{B_d}^{1/2} f_{B_d}=180\pm 50$ MeV. This covers the
range found in the lattice calculations by the 
UKQCD, ELC, and Bernard \etal~groups\cite{Lattices}. 
However, this estimate
should be regarded with considerable caution and
the assigned error may be an underestimate.

Using these parameters and the world average
for $\Delta M_{B_d} =0.468\pm 0.026$(ps)$^{-1}$ gives 
$$|V_{td}|=(0.92\pm 0.03\pm 0.09\pm 0.24)\times 10^{-2}$$
where the first error is statistical, the second is due to the top
quark mass, and the third is the uncertainty in the product
$B_{B_d}^{1/2} f_{B_d}$. At present, the experimental limits 
on $B\to \tau \nu$ from ALEPH and CLEO~II
give the constraint $f_{B_d} < 2.6 $ GeV. 
A factor of ten improvement in sensitivity is required to 
reach the range of interest for $f_B$. A complementary approach
is to measure the decay constants of charmed mesons (i.e. $f_{D_s}$,
$f_{D^+}$) which can be used to verify the lattice calculations
and then be scaled to the B mass.

Since $$V_{td}= A \lambda^3 (1 - \rho -i \eta)$$ in the usual
parameterization,
the modulus is
 $$|V_{td}|^2 = A^2 \lambda^6 \{(1 - \rho)^2 +\eta^2 \}.$$
Thus, a precise determination of
$|V_{td}|$ constrains the vertex of the Bjorken triangle
to lie on a  circle centered at $\rho=1$,
$\eta=0$ with radius $A \lambda^3$ and further reduces the allowed
range of CP asymmetries in the Standard Model. At present,
the experimental measurements give the constraint,
$$ \sqrt{(1-\rho^2)+\eta^2} = 1 \pm 0.3.$$

\subsection{The CKM element $|V_{ts}|$}

The CKM parameter $|V_{ts}|$ can be extracted from a measurement
of the branching fraction for the electromagnetic penguin.
There are many calculations of the inclusive rate for $b\to s\gamma$
\cite{Ali}, \cite{bsg}.
The rate has a logarithmic dependence on the top quark mass, $m_t$, and is
proportional to the product of CKM matrix elements $|V_{ts}V_{tb}|^2$.
Large leading order 
QCD corrections increase the rate by a factor of about 3.5.
Using the measurement $ m_t=179\pm 10$~GeV, and allowing the
range of mass scales, $\mu$, at which the QCD corrections are evaluated,
to vary between $m_b/2$ and $2m_b$, 
Buras \etal~calculate the inclusive
rate to be $(2.8\pm 0.8)\times 10^{-4}$\cite{Burasbsg}.
This prediction is completely consistent with the experimental results
discussed in the previous section. The theoretical uncertainty 
from the scale dependence ($\mu$) should
be significantly reduced when a calculation including next to leading
order QCD corrections is completed.
Ali and London have used the new
experimental bounds to determine the range of possible values
for the ratio CKM matrix element $|V_{ts}/V_{cb}|$\cite{Alickm}:
$$ 0.62 < |V_{ts}/V_{cb}| < 1.1$$
which is expected from unitarity to be close to 1.

Using heavy quark symmetry, the $D\to K^*$ and $B\to K^*$ form factors
can be related at certain kinematic points.
Models are then used to extrapolate and
obtain form factors for the entire kinematic range in $B\to K^*$ decays.
Using experimentally measured form factors
for $D\to K^* \ell \nu$, the measured
branching ratio for $B\to K^* \gamma$ 
can then be used to determine a value of
$|V_{ts}|$. Griffin, Masip and McGuigan have carried out this program
and find $|V_{ts}|=0.026\pm 0.006({\rm exp}) \pm 0.011 ({\rm theo})$ 
where the first error
is from experimental data and the second arises from theoretical
uncertainties\cite{gmm}.

\subsection{The ratio $|V_{ts}/V_{td}|$}

In the future, measurements of $B_s-\bar{B_s}$ mixing may allow the
determination of $|V_{ts}/V_{td}|$ in a manner which is
fairly independent of hadronic uncertainties. This would
constrain the quantity  $|1-\rho-i\eta|$ and circumvent
the problems associated with hadronic uncertainties in 
$B_d-\bar{B_d}$ mixing.
The existing limit from ALEPH on $B_s-\bar{B_s}$ mixing implies
$${{\Delta m_s} \over {\Delta m_d}} = (1.2\pm 0.1) 
|{V_{ts}\over V_{td}}|^2 > 7.9 \Longrightarrow |{V_{ts}\over V_{td}}| > 3.0.$$
This gives the constraint 
$$\sqrt{(1-\rho)^2+\eta^2}<1.5$$ 
with minimal 
uncertainties from non-perturbative physics. This is slightly
better than the bound from the unitarity of the CKM matrix.

Using the constraint
$ (\sqrt{(1-\rho^2)+\eta^2} = 1 \pm 0.3 )$ obtained from 
$B_d$ mixing and the relation $x_d/x_s = (1/f) \lambda^2 |(1-\rho-i\eta)|^2$
gives $x_s/x_d = 13.2\pm 7.9$ where $f$
takes into account $SU(3)$ breaking and is assumed to be 
$1.25\pm 0.1$\cite{rosfd}, \cite{breaking}.
This implies
$ x_s = 19\pm 11$.
The Standard Model parameters preferred in a recent fit
by Ali and London\cite{Alickm}, 
which uses all available experimental constraints, 
also indicate that $x_s$ is large,
$$x_s =19.4\pm 6.9$$ for $f_{B_s}\sqrt{B_{B_s}}= 230$ MeV.
Such rapid time dependent oscillations of the $B_s$ meson
will be extremely difficult to measure in future experiments
at high energy colliders or asymmetric B factories.

The ratio 
$|V_{td}/V_{ts}|$ may be determined
 from a comparison of the decay rates 
for $B\to\rho\gamma$ (or $B\to \omega \gamma$)
and $B\to K^*\gamma$. In this ratio many of the theoretical uncertainties are
expected to cancel. 
$${{{\cal B}(B^-\to \rho^- \gamma)}\over {{\cal B}(B^- \to K^{*-} \gamma)}}
={{{\cal B}(B^0\to \rho^0 \gamma)+ {\cal B}(B^0\to \omega \gamma)}
\over {{\cal B}(B^- \to K^{*-} \gamma)}} = \xi |{V_{td} \over V_{ts}}|^2 $$
where the factor $\xi$ accounts for SU(3) breaking effects.
This gives limits on $|V_{td}/V_{ts}|$ between 0.64 and 0.75
for different models of $SU(3)$ breaking in the form factors.
The possible contribution of long distance effects in this ratio
is still in dispute\cite{soni},\cite{cheng},\cite{pakvasa},\cite{longdesh}.
For example,  Cheng\cite{cheng} finds that the decay $B\to \rho \gamma$
is dominated by the short distance penguin 
and gives a possible 10-20\% contribution to the amplitude from 
long distance effects. If these long distance effects can be shown
to be manageable, then 
this ratio will also provide useful constraints in the future.

\subsection{CP Violation}

The three internal angles of the 
Bjorken triangle can be expressed in terms of CKM elements
\[ 
\alpha\equiv arg
\left( \begin{array}{c}
\frac{V_{ud}V^*_{ub}}{V_{td}V^*_{tb}}
\end{array} \right)
,~~\beta\equiv arg
\left( \begin{array}{c}
\frac{V_{cd}V^*_{cb}}{V_{td}V^*_{tb}}
\end{array} \right) 
,~~ \gamma\equiv arg
\left( \begin{array}{c}
\frac{V_{cd}V^*_{cb}}{V_{ud}V^*_{ub}}
\end{array} \right).
\]

These angles can be measured indirectly once each of the CKM 
elements is precisely determined. The allowed values for the
upper vertex of the unitarity triangle is shown in Figure~\ref{ckmfig}
\cite{Alickm}.
It is also possible to determine the angles
directly from observations of 
time dependent CP asymmetries in $\bar{B}^0$ decay.
The ultimate goal is to measure both the angles and the CKM couplings
to high precision and overconstrain the Standard Model. If a inconsistency
is found, this would provide an indication for New Physics.

\begin{figure}[htb]
\begin{center}
\unitlength 1.0in
\begin{picture}(3.,3.0)(0,0)
\put(-0.31,0.0)
{\psfig{bbllx=0pt,bblly=0pt,width=3.3in,height=3.3in,file=region_ckm.ps}}
\end{picture}
%\vskip 4 mm
\caption{The allowed region in the $\rho, \eta$ plane as determined
from measurements of $B$ decays and $\epsilon$ from kaon decay.
The dashed circles correspond to the constraints from
the limits for $B_s$ mixing
with different assumptions on $SU(3)$ breaking 
\protect\cite{Alickm}.}
\label{ckmfig}
\end{center}
\end{figure}

Large CP violating asymmetries in the $B_d$ system are generated by
$B-\bar{B}$ mixing. The simplest case to consider is a process where
the final state is a CP eigenstate,$|f_{CP}>$. The amplitude for
the direct decay and the amplitude for the process
where the $B$ mixes to a $\bar{B}$ which
then decays to the same final state cannot be distinguished. If the
two amplitudes have some relative phase, then a measurable interference
effect will be generated. As noted by A. Sanda, this is analogous
to the double slit interference experiment of classical physics.
The CP violating asymmetry is due 
to the analogue of the path difference, which in this case 
is introduced by $B-\bar{B}$ mixing.

The time dependent rates for an initially pure state to decay to
a CP eigenstate $|f_{CP}>$ is given by
$$ \Gamma(B^0(t) \to |f>) \propto |A|^2 \exp^{-\Gamma t}
(1\pm Im (\lambda) \sin{(\Delta M_d t)})$$
where the plus sign obtains for $B^0$ and the minus sign for
$\bar{B^0}$. This gives a time dependent CP asymmetry
$$A(t)= 2 Im \lambda \sin{(\Delta M_d t)}$$
For the case $|f_{CP}>=\psi K_s$, Im$\lambda = -\sin{(2 \beta)}$
the expected asymmetry is of order 0.6. Similarly, modes such as
$\bar{B}^0\to \pi^+ \pi^-$ may give asymmetries proportional
to $\sin{(2 \alpha)}$ which will also probably be order 0.2-1.0 .
These asymmetries are  considerably
larger than the characteristic scale of asymmetries in the kaon sector,
which are typically of order $10^{-3}$. However, the branching fraction
of the $B_d$ meson
to CP eigenstates is small and it is difficult to produce large numbers
of $B$ mesons. 

An additional complication arises when considering production
 at the $\Upsilon(4S)$ resonance where the $B\bar{B}$ meson pairs
are produced in a coherent state. The restrictions of quantum statistics
lead to a CP asymmetry which depends on the difference
of the decay times of the $B^0$ and $\bar{B}^0$ mesons. This also has
the unfortunate side effect that time integrated CP asymmetries for
$B_d$ mesons vanish\cite{UMass}.

Several solutions to this difficulty have been proposed. One alternative
is to operate the experiment at a center of mass
energy just above the threshold where $\bar{B} B^*$ pairs
are produced. In this case, time integrated asymmetries no longer vanish
but the cross section is lower by at least a factor of five.
Or one can operate at the $\Upsilon(4S)$ resonance with asymmetric
energy beams. The center of mass frame will be boosted and the B decay
lengths will be dilated to measurable distances.
The latter solution has been chosen by the SLAC and KEK laboratories.
Another possibility is to take advantage of the large cross sections
for hadronic production of B mesons at either hadron collider of
fixed target experiments. In this case, B mesons should be produced
incoherently but it is quite challenging to trigger and operate the
experiment in a very high rate environment. The high
luminosity $B$ factory projects
are discussed in detail elsewhere\cite{BABAR},\cite{BELLE},
\cite{DESY},\cite{CDFB},\cite{LHC}.
 
%%%%\section{~CONCLUSIONS}
%\input conclusion.tex
\section{~CONCLUSIONS}

 Significant progress in
the physics of $B$ mesons has been made in the last several years.
Improved measurements of branching fractions for
semileptonic decays in conjunction with more precise
measurements of exclusive B lifetimes from LEP and CDF have
improved the knowledge of the CKM matrix elements
$|V_{cb}|$ and $|V_{ub}|$. A number of 
experimental and theoretical approaches to a model 
independent determination of $|V_{ub}|$ have been proposed.
These will be one of the foci of experimental efforts 
in the future.

Improved determinations of the 
$B_{d}-\bar{B_d}$ mixing parameters from CLEO~II and the LEP experiments,
as well as the determination of the top quark 
mass at CDF and D0 have reduced the allowed
range for the CKM element $|V_{td}|$. The LEP experiments
have provided the first evidence for time dependent 
oscillations of neutral $B$ mesons.
These experiments have also provided useful constraints
on the $B_s-\bar{B_s}$ mixing parameter, 
which can be to constrain $|V_{ts}|/V_{td}|$. 
Measurement of the 
$B_s-\bar{B}_s$ oscillation frequency
is a major experimental challenge 
for the high energy collider experiments.

Results from CLEO~II have significantly modified
our understanding of hadronic B decay.  
The data and measurements of branching fractions
are now of sufficient quality to perform non-trivial
tests of the factorization hypothesis including comparisons
of rates for $\bar{B}^0\to D^{*+} X^-$
(where $X^-=\pi^- ,\rho^-$, or $a_1^-$)
with rates for  $D^{*+} \ell ^- \bar{\nu}$
at $q^2=M_X^2$, as well as comparisons of  the polarizations in 
$\bar{B}^0\to D^{*+}\rho^-$ with 
$\bar{B}^0 \to D^{*+} \ell^-\bar{\nu}_\ell$. In all cases, the
factorization hypothesis is consistent with the data at the present
level of experimental precision and for $q^2 < m_{a_1}^2$.

Improved measurements of branching ratios of two-body
decays with a final state $\psi$ meson
have been reported from ARGUS and CLEO~II. 
The decay $B \to \psi K^*$ is strongly polarized with 
$\Gamma_L / \Gamma = (78 \pm 7)$ \%. Therefore, this mode
will be useful for measuring CP violation.

There is no evidence for 
color suppressed decays to a charmed meson and light neutral
hadron in the final state.
The most stringent limit, 
${\cal B}(\bar{B^0}\to D^0\pi^0) / {\cal B}(\bar{B^0}\to D^+\pi^-) < 0.07$
from CLEO~II, is
still above the level where these 
color suppressed $B$ decays are expected in most models.
The observation of $B \to \psi$ modes
shows that color suppressed decays are present. Using results on exclusive
$B \to \psi$ decays from CLEO~1.5, CLEO~II and ARGUS, 
 we find a value of the BSW parameter
$|a_2|\; = \; 0.23 \pm 0.01 \pm  0.01$. We also report a new value for
the BSW parameter
$|a_1|\; = \; 1.03 \pm 0.04 \pm 0.06$.
By comparing rates for $B^-$ and $\bar{B}^0$ modes, it has
 been shown that the sign of
$a_2/a_1$ is positive, in contrast to what is found in charm decays.

There has  been dramatic progress in the study of rare decays.
CLEO~II has reported evidence for charmless
hadronic $B$ decay in the sum of $B\to K^+ \pi^-$ and $B\to \pi^+ \pi^-$
and has observed the first direct evidence 
for the radiative penguin decay $B \to K^{*} \gamma$
with a branching fraction of $(4.5\pm 1.5 \pm 0.9) \times 10^{-5}$
consistent with Standard Model expectations for a heavy top quark.
CLEO~II has also succeeded in observing
the inclusive process, $b\to s \gamma$ and finds
${\cal{B}}(b\to s\gamma) = (2.32\pm 0.57 \pm 0.35) \times 10^{-4}$ .
These results restrict the allowed range for $|V_{ts}|$ and constrain
physics beyond the Standard Model.

Large samples of reconstructed hadronic decays will be obtained
in the next few years by the CLEO~II collaboration as a result
of further improvements in the luminosity of CESR, and in the
performance of the CLEO~II detector. There will also
be significant increases in the size of data samples available to the
CDF experiment.
These will permit accurate tests of the factorization hypothesis
over the full $q^2$ range. The large tagged sample at CLEO can be used
to study inclusive properties of $B^+$ and $B^0$ decays and 
constrain $f_{B}$ via $B^+\to \tau^+ \nu$.  
Measurements of additional decays to final states with charmonium 
mesons will be performed and 
other color suppressed decays will be observed.

Larger data samples
should allow further results to be obtained on rare $B$ decays
including the observation of $B^0\to \pi^+ \pi^-$, $B^0\to K^+ \pi^-$
and a measurement of the inclusive process $b\to s ~{\rm gluon}$. 
The measurement of several rare hadronic decays would 
provide information on the relative importance of the
penguin and spectator amplitudes. 
Additional electromagnetic
penguin decays such as $B\to \rho (\omega) \gamma $, $B\to K^{**} \gamma$,
and $B\to K^* \ell ^+ \ell ^-$ 
may be observed. These provide further constraints on 
the Standard Model parameters $|V_{ts}|$ and $|V_{td}|$, 
as well as on extensions of the Standard Model.

The ultimate goal of the study of B mesons is to 
 measure the large CP asymmetries 
predicted by the Standard Model in decay modes such as
$\bar{B}^0\to \psi K^0$, $\bar{B}\to \pi^+ \pi^-$ and $B^-\to D^0 K^-$. 
In order to throughly test
the consistency of the Standard Model's description of
CP violation in these decays, the mechanisms
of $B$ decay must be well understood. This review shows that
rapid progress is being made in this program.

\acknowledgements

We acknowledge the essential contributions of Dr. S. Playfer to an earlier
version of this review. We have also benefitted from the reviews of
S. Stone and the review of J.D. Richman and P. Burchat on semileptonic
charm and beauty decays.
We thank  H. Albrecht, V. Barger, C. Bebek, D. Cassel, P. Colangelo, 
J.E. Duboscq, F. DeJongh, I. Dunietz, E. Golowich,
J. Hewett, A. Jawahery, Y.Y. Keum, K. Kinoshita, S. Menary,
M. Neubert, C. Nixdorf, X.Y. Pham, A. Ryd, J. D. Richman,
V. Rieckert, J. Rodriguez, S. Pakvasa, W. Palmer, K. Schubert,
B. Stech, R. Wanke, and M. Zoeller
for useful discussions and help in preparation of this review.
We thank our colleagues from the CLEO,
ARGUS, CDF, ALEPH, OPAL, DELPHI and L3 experiments for their contributions
to the work discussed in this review.
We thank  the Department of Energy, the University of Hawaii and
Ohio State University for their unwavering support.

%\input references.tex

%\clearpage
%\newpage
%\input appendix.tex
%\section{APPENDIX}
\centerline{\Large{\bf APPENDIX}}
\addcontentsline{toc}{section}{APPENDIX}
Tables \ref{kh1} and \ref{kh2} in the body of the paper
contain the $B$ meson branching fraction
as measured by the ARGUS, CLEO 1.5 and CLEO II experiments. In this appendix
we list more technical information
 found in the ARGUS \cite{ThirdB} -\cite{FifthB},
\cite{ARGUSDDs}, \cite{arguschi} and CLEO \cite{SecondB}, \cite{DDcleo},
\cite{fastpsi}, \cite{SixthB}
publications. This includes the number of signal events and the reconstruction
efficiencies. 
Note that different experiments used different procedures to obtain branching
ratios in modes where several $D$ or $\psi$ decay channels were used (see Sec.
\ref{thatsit}.
The  information provided here will be useful 
to estimate the signal yields for future $B$ experiments and also
to rescale the $B$ meson branching ratios when more precise measurements
of the charmed meson branching fraction become available.

\begin{table}[htb] 
\caption{Detailed $B^-$ branching ratios. Experiment: ARGUS} 
\begin{tabular}{lll} 
\label{argus_bm} 
$ B^- $ decay & Signal events & Branching ratio [\%]\\ 
\hline 
$B^- \rightarrow D^0 \pi ^-$ & $  12 \pm  5 $& $ 0.22 \pm 0.09 \pm 0.06 \pm 0.01 $ \\ 
$B^- \rightarrow D^0 \rho ^-$ & $  19 \pm  6 $& $ 1.41 \pm 0.43 \pm 0.39 \pm 0.06 $ \\ 
$B^- \rightarrow D^{*0} \pi ^-$ & $   9 \pm  3 $& $ 0.38 \pm 0.13 \pm 0.10 \pm 0.02 $ \\ 
$B^- \rightarrow D^{*0} \rho ^-$ & $   7 \pm  4 $& $ 0.94 \pm 0.56 \pm 0.35 \pm 0.04 $ \\ 
$B^- \rightarrow D_J^{(*)0} \pi ^-$ & $   6 \pm  3 $& $ 0.13 \pm 0.06 \pm 0.03 \pm 0.01 $ \\ 
$B^- \rightarrow D^{*+} \pi ^- \pi ^- \pi ^0$ & $  26 \pm 10 $& $ 1.64 \pm 0.64 \pm 0.37 \pm 0.07 $ \\ 
$B^- \rightarrow D_J^{(*)0} \rho ^-$ & $   5 \pm  3 $& $ 0.32 \pm 0.19 \pm 0.07 \pm 0.01 $ \\ 
$B^- \rightarrow D^{*+} \pi ^- \pi ^-$ & $  11 \pm  6 $& $ 0.24 \pm 0.13 \pm 0.05 \pm 0.01 $ \\ 
$B^- \rightarrow D^0 D_s^-$ & $ 4.4 \pm 2.2 $& $ 1.69 \pm 0.85 \pm 0.27 \pm 0.41 $ \\ 
$B^- \rightarrow D^0 D_s^{*-}$ & $ 2.3 \pm 1.8 $& $ 1.13 \pm 0.85 \pm 0.20 \pm 0.27 $ \\ 
$B^- \rightarrow D^{*0} D_s^-$ & $ 2.0 \pm 1.4 $& $ 0.79 \pm 0.55 \pm 0.11 \pm 0.19 $ \\ 
$B^- \rightarrow D^{*0} D_s^{*-}$ & $ 4.8 \pm 2.5 $& $ 1.89 \pm 0.98 \pm 0.28 \pm 0.46 $ \\ 
$B^- \rightarrow \psi K^-$ & $   6 $& $ 0.08 \pm 0.04 \pm 0.01 $\\ 
$B^- \rightarrow \psi ' K^-$ & $   5 $& $ 0.20 \pm 0.09 \pm 0.04 $\\ 
$B^- \rightarrow \psi K^{*-}$ & $   2 $& $ 0.19 \pm 0.13 \pm 0.03 $\\ 
$B^- \rightarrow \psi ' K^{*-}$ & $ < 3.9  $& $ < 0.53 $ at $90 $\% C.L. \\ 
$B^- \rightarrow \psi K^- \pi ^+ \pi ^-$ & $ <   8  $& $ < 0.19 $ at $90 $\% C.L. \\ 
$B^- \rightarrow \psi ' K^- \pi ^+ \pi ^-$ & $   3 $& $ 0.21 \pm 0.12 \pm 0.04 $\\ 
$B^- \rightarrow \chi_{c1} K^-$ & $   4 \pm  2.0 $& $ 0.22 \pm 0.15 \pm 0.07 $\\ 
\end{tabular} 
\end{table} 
\begin{table}[htb] 
\caption{Detailed $\bar{B}^0$ branching ratios. Experiment: ARGUS} 
\begin{tabular}{lll} 
\label{argus_b0} 
$ \bar{B}^0 $ decay & Signal events & Branching ratio [\%]\\ 
\hline 
$\bar{B}^0 \rightarrow D^+ \pi ^-$ & $  22 \pm  5 $& $ 0.48 \pm 0.11 \pm 0.08 \pm 0.03 $ \\ 
$\bar{B}^0 \rightarrow D^+ \rho ^-$ & $   9 \pm  5 $& $ 0.90 \pm 0.50 \pm 0.27 \pm 0.06 $ \\ 
$\bar{B}^0 \rightarrow D^{*+} \pi ^-$ & $  12 \pm  4 $& $ 0.25 \pm 0.08 \pm 0.03 \pm 0.01 $ \\ 
$\bar{B}^0 \rightarrow D^{*+} \rho ^-$ & $  19 \pm  9 $& $ 0.64 \pm 0.27 \pm 0.25 \pm 0.03 $ \\ 
$\bar{B}^0 \rightarrow D^{*+} \pi ^- \pi ^- \pi ^+$ & $  26 \pm  7 $& $ 1.09 \pm 0.27 \pm 0.32 \pm 0.04 $ \\ 
$\bar{B}^0 \rightarrow D^+ D_s^-$ & $ 2.4 \pm 1.8 $& $ 1.05 \pm 0.80 \pm 0.35 \pm 0.26 $ \\ 
$\bar{B}^0 \rightarrow D^+ D_s^{*-}$ & $ 3.2 \pm 2.0 $& $ 1.67 \pm 1.05 \pm 0.52 \pm 0.41 $ \\ 
$\bar{B}^0 \rightarrow D^{*+} D_s^-$ & $ 2.6 \pm 1.8 $& $ 0.83 \pm 0.59 \pm 0.11 \pm 0.20 $ \\ 
$\bar{B}^0 \rightarrow D^{*+} D_s^{*-}$ & $ 3.9 \pm 2.0 $& $ 1.54 \pm 0.83 \pm 0.24 \pm 0.37 $ \\ 
$\bar{B}^0 \rightarrow \psi K^0$ & $   2 $& $ 0.09 \pm 0.07 \pm 0.02 $\\ 
$\bar{B}^0 \rightarrow \psi ' K^0$ & $ < 2.3  $& $ < 0.30 $ at $90 $\% C.L. \\ 
$\bar{B}^0 \rightarrow \psi \bar{K}^{*0}$ & $   6 $& $ 0.13 \pm 0.06 \pm 0.02 $\\ 
$\bar{B}^0 \rightarrow \psi ' \bar{K}^{*0}$ & $ < 3.9  $& $ < 0.25 $ at $90 $\% C.L. \\ 
$\bar{B}^0 \rightarrow \psi ' K^- \pi ^+$ & $ < 2.3  $& $ < 0.11 $ at $90 $\% C.L. \\ 
\end{tabular} 
\end{table} 
\eject 
\newpage 
\begin{table}[htb] 
\caption{Detailed ${B}^-$ branching ratios. Experiment: CLEO 1.5} 
\begin{tabular}{llllll} 
\label{cleo15_bm} 
$B^- $ decay & Signature & Signal & Eff. & BR [\%] & Branching ratio [\%]\\ 
\hline 
$B^- \rightarrow D^0 \pi ^-$ & & & & & $ 0.56 \pm 0.08 \pm 0.05 \pm 0.02 $ \\ 
 & $D^0 \rightarrow K^- \pi ^+$ &$  19 \pm  5 $ & 0.42 &$ 0.50 \pm 0.12 $ & \\ 
 & $D^0 \rightarrow K^- \pi ^+ \pi ^+\pi ^-$ &$  25 \pm  6 $ & 0.27 &$ 0.50 \pm 0.10 $ & \\ 
 & $D^0 \rightarrow \bar{K}^0 \pi ^+ \pi ^-$ &$  10 \pm  4 $ & 0.05 &$ 1.51 \pm 0.63 $ & \\ 
$B^- \rightarrow D^{0} \pi ^+ \pi ^- \pi ^-$ & & & & & $ 1.24 \pm 0.31 \pm 0.14 \pm 0.05 $ \\ 
 & $D^0 \rightarrow K^- \pi ^+$ &$  34 \pm  8 $ & 0.32 &$ 1.24 \pm 0.31 $ & \\ 
$B^- \rightarrow D^{*0} \pi ^-$ & & & & & $ 1.00 \pm 0.25 \pm 0.18 \pm 0.04 $ \\ 
 & $D^0 \rightarrow K^- \pi ^+$ &$   9 \pm  3 $ & 0.13 &$ 0.97 \pm 0.35 $ & \\ 
 & $D^0 \rightarrow K^- \pi ^+ \pi ^+\pi ^-$ &$  12 \pm  4 $ & 0.08 &$ 0.95 \pm 0.36 $ & \\ 
$B^- \rightarrow D_J^{(*)0} \pi ^-$ & & & & & $ 0.13 \pm 0.07 \pm 0.01 \pm 0.01 $ \\ 
 & $D^0 \rightarrow K^- \pi ^+$ &$ 2.2 \pm 1.5 $ & 0.22 &$ 0.15 \pm 0.10 $ & \\ 
 & $D^0 \rightarrow K^- \pi ^+ \pi ^+\pi ^-$ &$ 1.8 \pm 1.5 $ & 0.13 &$ 0.12 \pm 0.09 $ & \\ 
$B^- \rightarrow D^{*+} \pi ^- \pi ^-$ & & & & & $<0.37 $ \\ 
 & $D^0 \rightarrow K^- \pi ^+$ &$ <  8 $ & 0.22 &$ < 0.54 $ & \\ 
 & $D^0 \rightarrow K^- \pi ^+ \pi ^+\pi ^-$ &$ <3.5 $ & 0.11 &$ < 0.24 $ & \\ 
$B^- \rightarrow D^0 D_s^-$ & & & & & $ 1.66 \pm 0.70 \pm 0.13 \pm 0.40 $ \\ 
 &   &$ 5.0 \pm 2.2 $ & 0.07 &$ 1.66 \pm 0.70 $ & \\ 
$B^- \rightarrow \psi K^-$ & & & & & $ 0.09 \pm 0.02 \pm 0.02 $ \\ 
 & $\psi  \rightarrow \mu ^+ \mu ^- , e^+ e^-$ &$  11 \pm  3 $ & 0.41 &$ 0.09 \pm 0.02 $ & \\ 
$B^- \rightarrow \psi ' K^-$ & & & & & $<0.05 $ \\ 
 & $\psi ' \rightarrow \mu ^+ \mu ^- , e^+ e^-$ &$ <2.3 $ & 0.53 &$ < 0.10 $ & \\ 
 & $\psi ' \rightarrow \psi \pi ^+ \pi ^-$ &$ <2.3 $ & 0.23 &$ < 0.11 $ & \\ 
$B^- \rightarrow \psi K^{*-}$ & & & & & $ 0.15 \pm 0.11 \pm 0.03 $ \\ 
 & $\psi  \rightarrow \mu ^+ \mu ^- , e^+ e^-$ &$   2 \pm  1 $ & 0.05 &$ 0.15 \pm 0.11 $ & \\ 
$B^- \rightarrow \psi ' K^{*-}$ & & & & & $<0.38 $ \\ 
 & $\psi ' \rightarrow \mu ^+ \mu ^- , e^+ e^-$ &$ <2.3 $ & 0.08 &$ < 0.70 $ & \\ 
 & $\psi ' \rightarrow \psi \pi ^+ \pi ^-$ &$ <2.3 $ & 0.03 &$ < 0.82 $ & \\ 
$B^- \rightarrow \psi K^- \pi ^+ \pi ^-$ & & & & & $ 0.14 \pm 0.07 \pm 0.03 $ \\ 
 & $\psi  \rightarrow \mu ^+ \mu ^- , e^+ e^-$ &$   6 \pm  3 $ & 0.14 &$ 0.14 \pm 0.07 $ & \\ 
\end{tabular} 
\end{table} 
\eject 
\newpage 
\begin{table}[htb] 
\caption{Detailed $\bar{B}^0$ branching ratios. Experiment: CLEO 1.5} 
\begin{tabular}{llllll} 
\label{cleo15_b0} 
$\bar{B}^0 $ decay & Signature & Signal & Eff. & BR [\%] & Branching ratio [\%]\\ 
\hline 
$\bar{B}^0 \rightarrow D^+ \pi ^-$ & & & & & $ 0.27 \pm 0.06 \pm 0.03 \pm 0.02 $ \\ 
 & $D^+ \rightarrow K^- \pi ^+ \pi ^+$ &$  17 \pm  4 $ & 0.33 &$ 0.23 \pm 0.06 $ & \\ 
 & $D^+ \rightarrow \bar{K}^0 \pi ^+$ &$   4 \pm  2 $ & 0.09 &$ 0.66 \pm 0.36 $ & \\ 
$\bar{B}^0 \rightarrow D^+ \pi ^- \pi ^- \pi ^+$ & & & & & $ 0.81 \pm 0.21 \pm 0.09 \pm 0.05 $ \\ 
 & $D^+ \rightarrow K^- \pi ^+ \pi ^+$ &$  27 \pm  9 $ & 0.22 &$ 0.40 \pm 0.19 $ & \\ 
 & $D^+ \rightarrow \bar{K}^0 \pi ^+$ &$  11 \pm  4 $ & 0.06 &$ 3.20 \pm 1.19 $ & \\ 
$\bar{B}^0 \rightarrow D^{*+} \pi ^-$ & & & & & $ 0.45 \pm 0.11 \pm 0.05 \pm 0.02 $ \\ 
 & $D^0 \rightarrow K^- \pi ^+$ &$   8 \pm  3 $ & 0.34 &$ 0.36 \pm 0.14 $ & \\ 
 & $D^0 \rightarrow K^- \pi ^+ \pi ^+\pi ^-$ &$   9 \pm  3 $ & 0.19 &$ 0.39 \pm 0.13 $ & \\ 
$\bar{B}^0 \rightarrow D^{*+} \rho ^-$ & & & & & $ 2.13 \pm 0.90 \pm 1.24 \pm 0.09 $ \\ 
 & $D^0 \rightarrow K^- \pi ^+$ &$   2 \pm  1 $ & 0.02 &$ 1.35 \pm 0.90 $ & \\ 
 & $D^0 \rightarrow K^- \pi ^+ \pi ^+\pi ^-$ &$   4 \pm  2 $ & 0.02 &$ 1.93 \pm 0.96 $ & \\ 
$\bar{B}^0 \rightarrow D^{*+} \pi ^- \pi ^- \pi ^+$ & & & & & $ 1.77 \pm 0.31 \pm 0.30 \pm 0.07 $ \\ 
 & $D^0 \rightarrow K^- \pi ^+$ &$  18 \pm  4 $ & 0.15 &$ 0.17 \pm 0.05 $ & \\ 
 & $D^0 \rightarrow K^- \pi ^+ \pi ^+\pi ^-$ &$  18 \pm  5 $ & 0.08 &$ 1.83 \pm 0.58 $ & \\ 
$\bar{B}^0 \rightarrow D^+ D_s^-$ & & & & & $ 0.54 \pm 0.31 \pm 0.03 \pm 0.13 $ \\ 
 &   &$ 3.0 \pm 1.7 $ & 0.10 &$ 0.65 \pm 0.36 $ & \\ 
$\bar{B}^0 \rightarrow D^{*+} D_s^-$ & & & & & $ 1.17 \pm 0.66 \pm 0.09 \pm 0.28 $ \\ 
 &   &$ 3.0 \pm 1.7 $ & 0.05 &$ 1.17 \pm 0.56 $ & \\ 
$\bar{B}^0 \rightarrow \psi K^0$ & & & & & $ 0.07 \pm 0.04 \pm 0.02 $ \\ 
 & $\psi  \rightarrow \mu ^+ \mu ^- , e^+ e^-$ &$   3 \pm  2 $ & 0.15 &$ 0.07 \pm 0.04 $ & \\ 
$\bar{B}^0 \rightarrow \psi ' K^0$ & & & & & $<0.16 $ \\ 
 & $\psi ' \rightarrow \mu ^+ \mu ^- , e^+ e^-$ &$ <2.3 $ & 0.18 &$ < 0.30 $ & \\ 
 & $\psi ' \rightarrow \psi \pi ^+ \pi ^-$ &$ <2.3 $ & 0.07 &$ < 0.35 $ & \\ 
$\bar{B}^0 \rightarrow \psi \bar{K}^{*0}$ & & & & & $ 0.13 \pm 0.06 \pm 0.03 $ \\ 
 & $\psi  \rightarrow \mu ^+ \mu ^- , e^+ e^-$ &$   7 \pm  3 $ & 0.21 &$ 0.13 \pm 0.06 $ & \\ 
$\bar{B}^0 \rightarrow \psi ' \bar{K}^{*0}$ & & & & & $ 0.15 \pm 0.09 \pm 0.03 $ \\ 
 & $\psi ' \rightarrow \mu ^+ \mu ^- , e^+ e^-$ &$   2 \pm  1 $ & 0.25 &$ 0.19 \pm 0.13 $ & \\ 
 & $\psi ' \rightarrow \psi \pi ^+ \pi ^-$ &$   1 \pm  1 $ & 0.10 &$ 0.11 \pm 0.11 $ & \\ 
$\bar{B}^0 \rightarrow \psi K^{-} \pi ^+$ & & & & & $ 0.12 \pm 0.05 \pm 0.03 $ \\ 
 & $\psi  \rightarrow \mu ^+ \mu ^- , e^+ e^-$ &$   7 \pm  3 $ & 0.19 &$ 0.12 \pm 0.05 $ & \\ 
\end{tabular} 
\end{table} 
\eject 
\newpage 
\begin{table}[htb] 
\caption{Detailed ${B}^-$ branching ratios. Experiment: CLEO II} 
\begin{tabular}{llllll} 
\label{cleoii_bm} 
$B^- $ decay & Signature & Signal & Eff. & BR [\%] & Branching ratio [\%]\\ 
\hline 
$B^- \rightarrow D^0 \pi ^-$ & & & & & $ 0.53 \pm 0.04 \pm 0.05 \pm 0.02 $ \\ 
 & $D^0 \rightarrow K^- \pi ^+$ &  $ 76.3 \pm 9.1 $ &  0.43 & $ 0.48 \pm 0.06 $& \\ 
  & $D^0 \rightarrow K^- \pi ^+ \pi ^0$ &  $ 134 \pm 15 $ &  0.19 & $ 0.63 \pm 0.07 $& \\ 
  & $D^0 \rightarrow K^- \pi ^+ \pi ^+\pi ^-$ &  $  94 \pm 11 $ &  0.22 & $ 0.51 \pm 0.06 $& \\ 
 $B^- \rightarrow D^0 \rho ^-$ & & & & & $ 1.31 \pm 0.12 \pm 0.14 \pm 0.04 $ \\ 
 & $D^0 \rightarrow K^- \pi ^+$ &  $  80 \pm  9 $ &  0.16 & $ 1.40 \pm 0.18 $& \\ 
  & $D^0 \rightarrow K^- \pi ^+ \pi ^0$ &  $  42 \pm  9 $ &  0.04 & $ 1.05 \pm 0.23 $& \\ 
  & $D^0 \rightarrow K^- \pi ^+ \pi ^+\pi ^-$ &  $ 90.4 \pm 12.1 $ &  0.08 & $ 1.38 \pm 0.18 $& \\ 
 $B^- \rightarrow D^{*0} \pi ^-$ & & & & & $ 0.49 \pm 0.07 \pm 0.06 \pm 0.00 $ \\ 
 & $D^0 \rightarrow K^- \pi ^+$ &  $ 13.3 \pm 3.8 $ &  0.16 & $ 0.36 \pm 0.13 $& \\ 
  & $D^0 \rightarrow K^- \pi ^+ \pi ^0$ &  $ 37.7 \pm 6.9 $ &  0.08 & $ 0.64 \pm 0.12 $& \\ 
  & $D^0 \rightarrow K^- \pi ^+ \pi ^+\pi ^-$ &  $ 20.0 \pm 4.9 $ &  0.08 & $ 0.47 \pm 0.12 $& \\ 
 $B^- \rightarrow D^{*0} \rho ^-$ & & & & & $ 1.59 \pm 0.20 \pm 0.26 \pm 0.05 $ \\ 
 & $D^0 \rightarrow K^- \pi ^+$ &  $ 25.7 \pm 5.4 $ &  0.06 & $ 1.74 \pm 0.37 $& \\ 
  & $D^0 \rightarrow K^- \pi ^+ \pi ^0$ &  $ 43.8 \pm 7.8 $ &  0.03 & $ 2.27 \pm 0.41 $& \\ 
  & $D^0 \rightarrow K^- \pi ^+ \pi ^+\pi ^-$ &  $ 16.9 \pm 4.6 $ &  0.03 & $ 1.07 \pm 0.32 $& \\ 
 $B^- \rightarrow D^{*0} \pi ^- \pi ^- \pi ^+$ & & & & & $ 0.92 \pm 0.20 \pm 0.17 \pm 0.01 $ \\ 
 & $D^0 \rightarrow K^- \pi ^+$ &  $ 5.5 \pm 2.9 $ &  0.05 & $ 0.51 \pm 0.26 $& \\ 
  & $D^0 \rightarrow K^- \pi ^+ \pi ^0$ &  $ 27.7 \pm 7.2 $ &  0.02 & $ 1.76 \pm 0.46 $& \\ 
  & $D^0 \rightarrow K^- \pi ^+ \pi ^+\pi ^-$ &  $  15 \pm  5 $ &  0.03 & $ 1.14 \pm 0.33 $& \\ 
 $B^- \rightarrow D^{*0} a_1 ^-$ & & & & & $ 1.83 \pm 0.39 \pm 0.33 \pm 0.02 $ \\ 
 & $D^0 \rightarrow K^- \pi ^+$ &  $ 5.5 \pm 2.9 $ &  0.05 & $ 1.02 \pm 0.52 $& \\ 
  & $D^0 \rightarrow K^- \pi ^+ \pi ^0$ &  $ 27.7 \pm 7.2 $ &  0.02 & $ 3.46 \pm 0.91 $& \\ 
  & $D^0 \rightarrow K^- \pi ^+ \pi ^+\pi ^-$ &  $  15 \pm  5 $ &  0.03 & $ 2.28 \pm 0.67 $& \\ 
 $B^- \rightarrow D^+ \pi^- \pi ^- $ & & & & & $<0.14 $ \\ 
 & $D^+ \rightarrow K^- \pi ^+ \pi ^+$ & $ < 10.3$ &  0.11 & $ <0.14 $& \\ 
 $B^- \rightarrow D^{*+} \pi ^- \pi ^-$ & &$14.1 \pm 5.4$ & & & $ 0.18 \pm 0.07 \pm 0.03 \pm 0.01 $ \\ 
$B^- \rightarrow D^{**0}(2420) \pi^- $ & & & & & $ 0.11 \pm 0.05 \pm 0.02 \pm 0.01 $ \\ 
 & $D^{**0} \rightarrow D^{*+} \pi^- $ &  $ 8.5 \pm 3.8 $ &  & $ 0.11 \pm 0.05 $& \\ 
 $B^- \rightarrow D^{**0}(2420) \rho^- $ & & & & & $<0.13 $ \\ 
 & $D^{**0} \rightarrow D^{*+} \rho^- $ &  $ 3.4 \pm 2.1 $ &  & $ <0.13 $& \\ 
 $B^- \rightarrow D^{**0}(2460) \pi^- $ & & & & & $<0.13 $ \\ 
 & $D^{**0} \rightarrow D^{*+} \pi^- $ &  $ 3.5 \pm 2.3 $ &  & $ <0.27 $& \\ 
  & $D^{**0} \rightarrow D^+ \pi^- $ & $ < 5.6$ &  0.21 & $ <0.13 $& \\ 
 $B^- \rightarrow D^{**0}(2460) \rho^- $ & & & & & $<0.45 $ \\ 
 & $D^{**0} \rightarrow D^{*+} \pi^- $ &  $ 3.2 \pm 2.4 $ &  & $ <0.48 $& \\ 
  & $D^{**0} \rightarrow D^+ \pi^- $ & $ < 6.1$ &  0.08 & $ <0.45 $& \\ 
 $B^- \rightarrow D^0 D_s^-$ & &$58.4 \pm 10.0$ & \multicolumn{2}{c}{$ 2.00 \times 10^{-3}$ ($\epsilon $BR)} & $ 1.11 \pm 0.20 \pm 0.23 \pm 0.28 $ \\ 
$B^- \rightarrow D^0 D_s^{*-}$ & &$16.1 \pm 5.0$ & \multicolumn{2}{c}{$ 0.83 \times 10^{-3}$ ($\epsilon $BR)} & $ 0.79 \pm 0.25 \pm 0.15 \pm 0.19 $ \\ 
$B^- \rightarrow D^{*0} D_s^-$ & &$13.5 \pm 4.1$ & \multicolumn{2}{c}{$ 0.43 \times 10^{-3}$ ($\epsilon $BR)} & $ 1.27 \pm 0.39 \pm 0.32 \pm 0.31 $ \\ 
$B^- \rightarrow D^{*0} D_s^{*-}$ & &$14.9 \pm 4.2$ & \multicolumn{2}{c}{$ 0.21 \times 10^{-3}$ ($\epsilon $BR)} & $ 2.82 \pm 0.80 \pm 0.59 \pm 0.68 $ \\ 
\end{tabular} 
\end{table} 
\eject 
\newpage 
\begin{table}[htb] 
\caption{Detailed ${B}^-$ branching ratios. Experiment: CLEO II} 
\begin{tabular}{llllll} 
\label{cleoii_bm2} 
$B^- $ decay & Signature & Signal & Eff. & BR [\%] & Branching ratio [\%]\\ 
\hline 
$B^- \rightarrow \psi K^-$ & & & & & $ 0.110 \pm 0.015 \pm 0.009 $ \\ 
 & $\psi  \rightarrow \mu ^+ \mu ^- , e^+ e^-$ &  $ 58.7 \pm 7.9 $ &  0.47 & $ 0.11 \pm 0.01 $& \\ 
 $B^- \rightarrow \psi ' K^-$ & &$7.0 \pm 2.6$ & & & $ 0.061 \pm 0.023 \pm 0.009 $ \\ 
$B^- \rightarrow \psi K^{*-}$ & & & & & $ 0.178 \pm 0.051 \pm 0.023 $ \\ 
 & $K^{*-} \rightarrow K^- \pi ^0$ &  $ 6.0 \pm 2.4 $ &  0.07 & $ 0.22 \pm 0.09 $& \\ 
  & $K^{*-} \rightarrow K_s \pi ^-$ &  $ 6.6 \pm 2.7 $ &  0.17 & $ 0.13 \pm 0.06 $& \\ 
 $B^- \rightarrow \psi ' K^{*-}$ & & & & & $<0.30 $ \\ 
 & $K^{*-} \rightarrow K^- \pi ^0$ &  $   1 \pm  1 $ &  & $ <0.56 $& \\ 
  & $K^{*-} \rightarrow K_s \pi ^-$ &  $   1 \pm  1 $ &  & $ <0.36 $& \\ 
 $B^- \rightarrow \chi_{c1} K^-$ & & & & & $ 0.097 \pm 0.040 \pm 0.009 $ \\ 
 & $\chi_{c1} \rightarrow \gamma \psi $ &  $ 6.0 \pm 2.4 $ &  0.20 & $ 0.10 \pm 0.04 $& \\ 
 $B^- \rightarrow \chi_{c1} K^{*-}$ & & & & & $<0.21 $ \\ 
 & $K^{*-} \rightarrow K^- \pi ^0$ &  0 &  0.03 & $ <0.67 $& \\ 
  & $K^{*-} \rightarrow K_s \pi ^-$ &  0 &  0.11 & $ <0.30 $& \\ 
 $B^- \rightarrow \psi \pi ^-$ & & & & & $ 0.005 \pm 0.002 \pm 0.000 $ \\ 
 & $\psi  \rightarrow \mu ^+ \mu ^- , e^+ e^-$ &  $   5 \pm  1 $ &  0.33 & $ 0.005 \pm 0.002 $& \\ 
 \end{tabular} 
\end{table} 
\eject 
\newpage 
\begin{table}[htb] 
\caption{Detailed $\bar{B}^0$ branching ratios. Experiment: CLEO II} 
\begin{tabular}{llllll} 
\label{cleoii_b0} 
$\bar{B}^0 $ decay & Signature & Signal & Eff. & BR [\%] & Branching ratio [\%]\\ 
\hline 
$\bar{B}^0 \rightarrow D^+ \pi ^-$ & & & & & $ 0.29 \pm 0.04 \pm 0.03 \pm 0.02 $ \\ 
 & $D^+ \rightarrow K^- \pi ^+ \pi ^+$ &  $ 80.6 \pm 9.8 $ &  0.32 & $ 0.29 \pm 0.04 $& \\ 
 $\bar{B}^0 \rightarrow D^+ \rho ^-$ & & & & & $ 0.81 \pm 0.11 \pm 0.12 \pm 0.05 $ \\ 
 & $D^+ \rightarrow K^- \pi ^+ \pi ^+$ &  $ 78.9 \pm 10.7 $ &  0.12 & $ 0.81 \pm 0.11 $& \\ 
 $\bar{B}^0 \rightarrow D^{*+} \pi ^-$ & & & & & $ 0.25 \pm 0.03 \pm 0.04 \pm 0.01 $ \\ 
 & $D^0 \rightarrow K^- \pi ^+$ &  $ 19.4 \pm 4.5 $ &  0.35 & $ 0.22 \pm 0.05 $& \\ 
  & $D^0 \rightarrow K^- \pi ^+ \pi ^0$ &  $ 31.9 \pm 6.4 $ &  0.14 & $ 0.30 \pm 0.06 $& \\ 
  & $D^0 \rightarrow K^- \pi ^+ \pi ^+\pi ^-$ &  $ 20.5 \pm 5.2 $ &  0.15 & $ 0.24 \pm 0.06 $& \\ 
 $\bar{B}^0 \rightarrow D^{*+} \rho ^-$ & & & & & $ 0.70 \pm 0.09 \pm 0.13 \pm 0.02 $ \\ 
 & $D^0 \rightarrow K^- \pi ^+$ &  $ 21.9 \pm 5.2 $ &  0.12 & $ 0.71 \pm 0.17 $& \\ 
  & $D^0 \rightarrow K^- \pi ^+ \pi ^0$ &  $ 39.8 \pm 7.2 $ &  0.05 & $ 1.09 \pm 0.20 $& \\ 
  & $D^0 \rightarrow K^- \pi ^+ \pi ^+\pi ^-$ &  $ 14.6 \pm 4.6 $ &  0.05 & $ 0.47 \pm 0.15 $& \\ 
 $\bar{B}^0 \rightarrow D^{*+} \pi ^- \pi ^- \pi ^+$ & & & & & $ 0.61 \pm 0.10 \pm 0.11 \pm 0.02 $ \\ 
 & $D^0 \rightarrow K^- \pi ^+$ &  $ 13.5 \pm 3.9 $ &  0.10 & $ 0.58 \pm 0.17 $& \\ 
  & $D^0 \rightarrow K^- \pi ^+ \pi ^0$ &  $ 21.7 \pm 5.9 $ &  0.04 & $ 0.68 \pm 0.18 $& \\ 
  & $D^0 \rightarrow K^- \pi ^+ \pi ^+\pi ^-$ &  $ 13.9 \pm 4.4 $ &  0.04 & $ 0.59 \pm 0.17 $& \\ 
 $\bar{B}^0 \rightarrow D^{*+} a_1^-$ & & & & & $ 1.22 \pm 0.19 \pm 0.22 \pm 0.04 $ \\ 
 & $D^0 \rightarrow K^- \pi ^+$ &  $ 13.5 \pm 3.9 $ &  0.10 & $ 1.16 \pm 0.34 $& \\ 
  & $D^0 \rightarrow K^- \pi ^+ \pi ^0$ &  $ 21.7 \pm 5.9 $ &  0.04 & $ 1.36 \pm 0.36 $& \\ 
  & $D^0 \rightarrow K^- \pi ^+ \pi ^+\pi ^-$ &  $ 13.9 \pm 2.4 $ &  0.04 & $ 1.17 \pm 0.34 $& \\ 
 $\bar{B}^0 \rightarrow D^{0} \pi ^+ \pi^- $ & & & & & $<0.16 $ \\ 
 & $D^0 \rightarrow K^- \pi ^+$ & $ < 10.1$ &  0.19 & $ <0.16 $& \\ 
 $\bar{B}^0 \rightarrow D^{**+}(2460) \pi^- $ & & & & & $<0.21 $ \\ 
 & $D^{**+} \rightarrow D^0 \pi^+ $ & $ < 5.6$ &  0.26 & $ <0.21 $& \\ 
 $\bar{B}^0 \rightarrow D^{**+}(2460) \rho^- $ & & & & & $<0.47 $ \\ 
 & $D^{**+} \rightarrow D^0 \pi^+ $ & $ < 5.1$ &  0.11 & $ <0.47 $& \\ 
 $\bar{B}^0 \rightarrow D^+ D_s^-$ & &$19.7 \pm 5.5$ & \multicolumn{2}{c}{$ 1.03 \times 10^{-3}$ ($\epsilon $BR)} & $ 0.82 \pm 0.23 \pm 0.19 \pm 0.20 $ \\ 
$\bar{B}^0 \rightarrow D^+ D_s^{*-}$ & &$10.3 \pm 3.6$ & \multicolumn{2}{c}{$ 0.47 \times 10^{-3}$ ($\epsilon $BR)} & $ 0.95 \pm 0.33 \pm 0.21 \pm 0.23 $ \\ 
$\bar{B}^0 \rightarrow D^{*+} D_s^-$ & &$18.4 \pm 4.5$ & \multicolumn{2}{c}{$ 0.87 \times 10^{-3}$ ($\epsilon $BR)} & $ 0.85 \pm 0.21 \pm 0.15 \pm 0.21 $ \\ 
$\bar{B}^0 \rightarrow D^{*+} D_s^{*-}$ & &$17.7 \pm 4.4$ & \multicolumn{2}{c}{$ 0.38 \times 10^{-3}$ ($\epsilon $BR)} & $ 1.85 \pm 0.46 \pm 0.33 \pm 0.45 $ \\ 
\end{tabular} 
\end{table} 
\eject 
\newpage 
\begin{table}[htb] 
\caption{Detailed $\bar{B}^0$ branching ratios. Experiment: CLEO II} 
\begin{tabular}{llllll} 
\label{cleoii_b02} 
$\bar{B}^0 $ decay & Signature & Signal & Eff. & BR [\%] & Branching ratio [\%]\\ 
\hline 
$\bar{B}^0 \rightarrow \psi K^0$ & & & & & $ 0.075 \pm 0.024 \pm 0.008 $ \\ 
 & $\psi  \rightarrow \mu ^+ \mu ^- , e^+ e^-$ &  $ 10.0 \pm 3.2 $ &  0.34 & $ 0.08 \pm 0.02 $& \\ 
 $\bar{B}^0 \rightarrow \psi ' K^0$ & & 0 & & & $<0.08 $ \\ 
$\bar{B}^0 \rightarrow \psi \bar{K}^{*0}$ & & & & & $ 0.169 \pm 0.031 \pm 0.018 $ \\ 
 & $\psi  \rightarrow \mu ^+ \mu ^- , e^+ e^-$ &  $ 29.0 \pm 5.4 $ &  0.23 & $ 0.17 \pm 0.03 $& \\ 
 $\bar{B}^0 \rightarrow \psi ' \bar{K}^{*0}$ & &$ 4.2 \pm 2.3$ & & & $<0.19 $ \\ 
$\bar{B}^0 \rightarrow \chi_{c1} K^0$ & & & & & $<0.27 $ \\ 
 & $\chi_{c1} \rightarrow \gamma \psi $ &  $   1 \pm  1 $ &  0.14 & $ <0.27 $& \\ 
 $\bar{B}^0 \rightarrow \chi_{c1} \bar{K}^{*0}$ & & & & & $<0.21 $ \\ 
 & $\chi_{c1} \rightarrow \gamma \psi $ &  $ 1.2 \pm 1.5 $ &  0.13 & $ <0.21 $& \\ 
 $\bar{B}^0 \rightarrow \psi \pi ^0$ & & & & & $<0.0069 $ \\ 
 & $\psi  \rightarrow \mu ^+ \mu ^- , e^+ e^-$ &  $   1 \pm  1 $ &  0.22 & $ <0.01 $& \\ 
 \end{tabular} 
\end{table} 
\eject 
\newpage 

\begin{references}
\bibitem{only4s}{The integrated 
luminosities quoted here are for the $\Upsilon (4S)$ data
samples only.}
\bibitem{mbrange}{The resolution in beam constrained mass
has a weak dependence on the decay channel.
It varies between 2.6 MeV to 3.3 MeV.}
\bibitem{cfrac}{These are the fractions determined in the 
CLEO~II analysis.
The lowest fraction is for the
mode $\bar{B^0}\to D^{*+}\rho^-$ while the highest
fraction is for the mode $B^- \to D^0 \pi^-$.}
\bibitem{ARGUSback}{In the ARGUS experiment the
beam constrained mass distributions were fitted
to a slightly different background function with a kinematic threshold.
The functional form of the background shape used was
$$f(x)= c_1x\left(\sqrt{(1-({x\over E_{beam}})^2}
\right)e^{c_2(1-({x\over E_{beam}})^2)}$$
CLEO~1.5 fitted the beam constrained mass distributions to the sum
of a gaussian signal of width 2.5 MeV (r.m.s) and a background function
of the form $w_1 ~ p + w_2~ p^2$ where $p= \sqrt{E_{beam}^2 - m^2}$
and the parameters $w_1$, $w_2$ are 
determined from the $\Delta E$ sidebands.}
\bibitem{aleph_bs_plb}{D. Bukuslic \etal (ALEPH Collaboration),
\PRref\PLB&322&275(1994). Note the value given in this paper for
the $B_s$ lifetime
is slightly different from the value quoted at the ICHEP94 conference.}
\bibitem{gy}
{The known quantity ${\cal G}(y)$ is given by
$$
{G_F^2 \over {48 \pi^3}} m^3_{D^{*}} (m_B-m_{D^*})^2 
\times
\sqrt{y^2-1} [ 4 y (y +1) {{1- 2yr + r^2} \over {(1-r)^2}} + (y+1)^2 ]
$$
where $r= m_{D^*}/m_B$.}
\bibitem{jdrpilnu}{The angle between the $\pi$ and the lepton
in the W center of mass frame is expected to follow a 
$\sin^2\theta_{\ell \pi}$ distribution for a pseudoscalar to
pseudoscalar semileptonic decay.}
\bibitem{jetcharge}{The usual jet charge is defined by
$Q_{jet} = {{\Sigma q_i p_{||}^\alpha }\over {p_{||}^\alpha}}$
where $\alpha$ is a tunable parameter near 1 and $p_{||}$ is the
momentum component parallel to the jet direction.
To avoid bias from
the high momentum lepton, the ALEPH analysis replaces the weighting
by the momentum parallel to the jet axis with the rapidity of the
i-th particle. Another variation is the OPAL analysis which uses
momentum dependent weighting but uses $\alpha=1$ for the opposite side
(tagging) jet and the $\alpha=0$ for the jet which contains the 
$D^{*+} ~l^-\nu$ candidate.}
\bibitem{mcdd}{Two models are considered: non-resonant
 $B\to D \pi^- \pi^0$ and $ B\to D^{**}(2460) \pi^-$. Both  give very similar
$\pi^- \pi^0$ mass spectra and comparable limits on the non-rho contamination
in the signal region.}
\bibitem{mcdrho}{CLEO~II considers two models: (1) The $\pi^- \pi^0$ system
is produced non-resonantly following a phase space distribution or
(2) the $D^{*+} \pi^- \pi^0$ is produced from the decay
$D^{**} \pi^-$. Both give similar limits on the non-$\rho$ fraction.}
\bibitem{Dpi}{The requirement that the colors match gives a factor of
$1/3$ in the amplitude. Moreover, the $W$ couples to the $d \bar{d}$ part of
the $\pi^0$ wave function so there is an additional factor of $1/\sqrt{2}$ in
the amplitude. This is also  true for the $\rho^0$ and the $\omega$ mesons. The
numerical factor is different for the case of the
$\eta/ \eta^{'}$ system.}
\bibitem{wex}{These modes are also accessible via the W-exchange graph which is
expected to be small in B decay.}
\bibitem{BReta}{We use the PDG values for the $\eta$, $\eta^{'}$
and $\omega$ branching ratios.}
\bibitem{omeg}{In the mode $\bar{B^0}\to D^0 \omega$, $\omega \to
\pi^- \pi^+ \pi^0$, $\Theta_{N}$, the angle between the normal
to the $\omega$ decay plane and the $D^0$ direction calculated in
the $D^0$ rest frame, must satisfy $|\sin \Theta_{N}|>0.6$.}
\bibitem{vud}{We make the approximation that $V_{ud},\; V_{cs} \approx 0.975$}
\bibitem{qcd}{The error is due to
the uncertainty in the scale at which to evaluate the Wilson
coefficient.}
\bibitem{narrow}{A value f$_\rho=196$ MeV, often quoted in the literature, is derived from tau decays
using the narrow width approximation i.e. by assuming the $\rho$
resonance is a delta function.}
\bibitem{width}{Since the form factor for $B\to D^* \ell \nu$ is slowly
varying, the width of the $\rho^-$ meson does not significantly modify the
result.}
\bibitem{Dstbr}{The $q^2$ spectrum has been 
corrected for the new values of the
CLEO~II $D^*$ and $D$  branching ratios.\cite{DKpi} }
\bibitem{param}{Three parameterizations of the Isgur-Wise function
$ \xi(v v^{'})$ are considered:  (a) $ \xi ( v v^{'}) = 1 + 1/4 a (v -
v^{'})^2 (v + v^{'})^2$ with
$ a = 0.54 \pm 0.01$, (b) $ \xi ( v v^{'}) = 1 /[1- (v - v^{'})^2/w_{0}^2]$
with 
$ w_0 = 1.12 \pm 0.17$, due to Rosner\cite{Rosfac} ,
 and   (c) $ \xi (v v^{'}) = exp [ b( v-v^{'})^2]$ with 
$b= 0.91 \pm 0.03$.  All of these parameterizations are constrained by fitting
to the average of the CLEO~1.5 and ARGUS data on 
$ d \Gamma/ dq^2 (B\to D^* \ell\nu)$.}
\bibitem{psicomment}{The branching ratio for the modes $B\to \psi K$ and 
$B\to \psi K^{*}$ can be accommodated by the value $\xi \sim 0$ while 
$\xi \sim 1/3$ gives a branching ratio that is about a factor of 4 too low (see
Ref.~\cite{Shifman}.}
\bibitem{volkie}{Private communication, Volker Rieckert; The result of a fit
including $R_4$ is not significantly different. We find in this case $a_2/a_1
\; = \; 0.282 \pm 0.07 \pm 0.06$.}
\bibitem{bkk}{The decay $\bar{B}^0 \to K^+K^-$ cannot be produced 
by a $b \to u$ spectator diagram 
or a $b \to s$ penguin transition. Thus it is expected to be small
compared to $\bar{B}^0 \to \pi^+\pi^-$ and $\bar{B}^0 \to K^-\pi^+$.}
\end{references}
\end{document}